\documentclass[smallcondensed]{svjour3}     
\smartqed  
\usepackage{graphicx,xcolor}
\graphicspath{{./img/}}
\usepackage{amsmath, amssymb, amsfonts}

\usepackage[pdftex, unicode, colorlinks, bookmarks=true, citecolor=blue, link color=blue, urlcolor=blue]{hyperref}

\usepackage{longtable}
\usepackage{booktabs}
\usepackage{multirow}

\usepackage{pdflscape}
\usepackage{afterpage} 

\newcommand{\bra}[1]{\langle#1|}
\newcommand{\ket}[1]{|#1\rangle}
\newcommand{\vb}[1]{\pmb{#1}}

\newcommand{\mean}[1]{\left\langle#1\right\rangle}

\newcommand{\abs}[1]{\left|\kern0.1em#1\kern0.1em\right|}
\newcommand{\atan}[1]{{\rm arctan}\left(#1\right)}

\newcommand{\til}[1]{\widetilde{#1}}

\newcommand{\arccot}{\mathrm{arccot}}

\def\apj{\ref@jnl{ApJ}}                 
\def\apjl{\ref@jnl{ApJ}}                
\def\apjs{\ref@jnl{ApJS}}               
\def\aap{\ref@jnl{A\&A}}                
\def\aapr{\ref@jnl{A\&A~Rev.}}          
\def\aaps{\ref@jnl{A\&AS}}              
\def\pra{\ref@jnl{Phys.~Rev.~A}}        
\def\prb{\ref@jnl{Phys.~Rev.~B}}        
\def\prc{\ref@jnl{Phys.~Rev.~C}}        
\def\prd{\ref@jnl{Phys.~Rev.~D}}        
\def\pre{\ref@jnl{Phys.~Rev.~E}}        
\def\prl{\ref@jnl{Phys.~Rev.~Lett.}}    
\def\pla{\ref@jnl{Phys.~Lett.~A}}    

\newcommand{\eg}{{\it e.g.\,}}  
\newcommand{\etc}{{\it etc}}  
\newcommand{\hamilt}{\hat{\mathcal{H}}} 
\newcommand{\fulld}[2]{\dfrac{d#1}{d#2}} 
\newcommand{\fulldd}[2]{\dfrac{d^2#1}{d#2^2}} 
\newcommand{\partd}[2]{\dfrac{\partial#1}{\partial#2}}  

\usepackage{xifthen}
\newcommand{\vq}[2][]{                
  \ifthenelse{\isempty{#1}}           %
    { \hat{\pmb{#2}} }                
    { \hat{\pmb{#2}}_\mathrm{#1} }    
}
\newcommand{\vqtil}[2][]{             
  \ifthenelse{\isempty{#1}}           %
    { \hat{\til{\pmb{#2}}} }                
    { \hat{\til{\pmb{#2}}}_\mathrm{#1} }    
}
\newcommand{\tq}[2][]{                
  \ifthenelse{\isempty{#1}}           %
    { \mathbb{#2} }                   
    { \mathbb{#2}_\mathrm{#1} }       
}
\newcommand{\vs}[2][]{                
  \ifthenelse{\isempty{#1}}           %
    { \mathbf{#2} }                   
    { \mathbf{#2}_\mathrm{#1} }       
}

\newcommand{\col}[2]{
  \begin{bmatrix}
    #1 \\
    #2
  \end{bmatrix}
}

\newcommand{\matr}[4]{ 
  \begin{bmatrix}
    #1 & #2 \\
    #3 & #4
  \end{bmatrix}
}

\newcommand{\SD}[1]{{\color{black}#1}}

\begin{document}

\title{Advanced quantum techniques for future gravitational-wave detectors.%
}


\author{Stefan L. Danilishin         \and
        Farid Ya. Khalili		\and
        Haixing Miao 
}


\institute{Stefan L. Danilishin \at
              Institut f\"ur Theoretische Physik, Leibniz Universit\"at Hannover and Max-Planck-Institut f\"ur Gravitationsphysik (Albert-Einstein-Institut), Callinstra\ss e 38, D-30167 Hannover, Germany \\
              Tel.: +49 511 762 14674\\
              Fax: +49 (0) 511 762-2784\\
              \email{stefan.danilishin@itp.uni-hannover.de}           
           \and
           Farid Ya. Khalili \at
              Faculty of Physics, M.V. Lomonosov Moscow State University, 119991 Moscow, and Russian Quantum Center, 143025  Skolkovo, Russia\\
              \and
           Haixing Miao \at
           	School of Physics and Astronomy and Institute of Gravitational Wave Astronomy, University of Birmingham, Birmingham B15 2TT, United Kingdom   
}

\date{Received: date / Accepted: date}

\maketitle

\begin{abstract}
Quantum fluctuation of light limits the sensitivity of advanced laser interferometric gravitational-wave detectors. It is one of the principal obstacles on the way towards the next-generation gravitational-wave observatories. The envisioned significant improvement of the detector sensitivity requires using quantum non-demolition measurement and back-action evasion techniques, which allow us to circumvent the sensitivity limit imposed by the Heisenberg uncertainty principle. In our previous review article: ``Quantum measurement theory in gravitational-wave detectors'' [Living Rev. Relativity 15, 5 (2012)], we laid down the basic principles of quantum measurement theory and provided the framework for analysing the quantum noise of interferometers. The scope of this paper is to review novel techniques for quantum noise suppression proposed in the recent years and put them in the same framework. Our delineation of interferometry schemes and topologies is intended as an aid in the process of selecting the design for the next-generation gravitational-wave observatories.
\end{abstract}

\tableofcontents

\section{Introduction}
\label{intro}
The second generation of ground-based gravitational-wave (GW) interferometers, Advanced LIGO \cite{TheLIGOScientific:2014jea} and Advanced Virgo \cite{TheVirgo:2014hva}, with significantly improved sensitivities, superseded the initial generation in 2015, which led to a Nobel Prize-winning first direct observation of GWs from the binary black hole (BBH) coalescence on September 14, 2015 \cite{GW_Discovery_Paper_PhysRevLett.116.061102}. This has marked the start of the new era of GW astronomy.

Contrary to the predictions based on the previous X-ray observations \cite{2016_PhysRevX.6.041015_LVC}, the first detected GW signal has come from an unexpectedly massive BBH with the mass of components $\sim 30M_\odot$ and the final BH with mass $\sim 60M_\odot$. The following detections \cite{2016_PhysRevLett.116.241103_LSC_Detection,2017_PhysRevLett.118.221101_LSC_Detection,2017_PRL.119.141101_LVC,2017_ApJ.851.L35_LVC_detection} have not only confirmed the existence of this new population of massive black holes but also highlighted the importance of sensitivity improvement at low frequencies ($<30$ Hz) for better parameter estimation and more quantitative analysis of the nature of these exotic objects. 

However, massive BBHs are not the only reason for low-frequency improvement. With all three detectors of the LIGO-Virgo network being online, the sky localisation is dramatically improved (see Sec. 4.2. in \cite{Aasi:2013wya}) enabling multi-messenger astronomy of compact binaries \cite{2017_ApJ.848L.12A_LVC}. The longer lead times before the merger necessary for directing electromagnetic (EM) telescopes to the right sky location depend directly on the low-frequency sensitivity where the spectral components of the inspiral stage of the binary evolution are most prominent \cite{Harry:2018hke}. We observed this situation when LIGO and Virgo had detected a GW signal from the final stages of evolution of the binary neutron-star (BNS) system \cite{2017_PRL.119.1101A_LVC_BNS_Discovery} before the coalescence and merger that has produced a chain of follow-on electromagnetic (EM) counterparts detected by the EM partners of LIGO \cite{2017_ApJ.848L.12A_LVC}. 

This fascinating discovery has also revealed the significance of enhancing the GW detector sensitivity in the relatively high-frequency band, from 1 to 5 kHz, which hosts the spectrum of the merger and the ringdown phases of the BNS system. It is the precise measurement of the GW signal shape emitted in these two phases that promise to unveil many details about the physics of nuclear matter and also to shed light on the physical mechanisms of short gamma-ray bursts \cite{2017ApJ...848L..13A}.

\begin{figure}
\centering  \includegraphics[width=.85\textwidth]{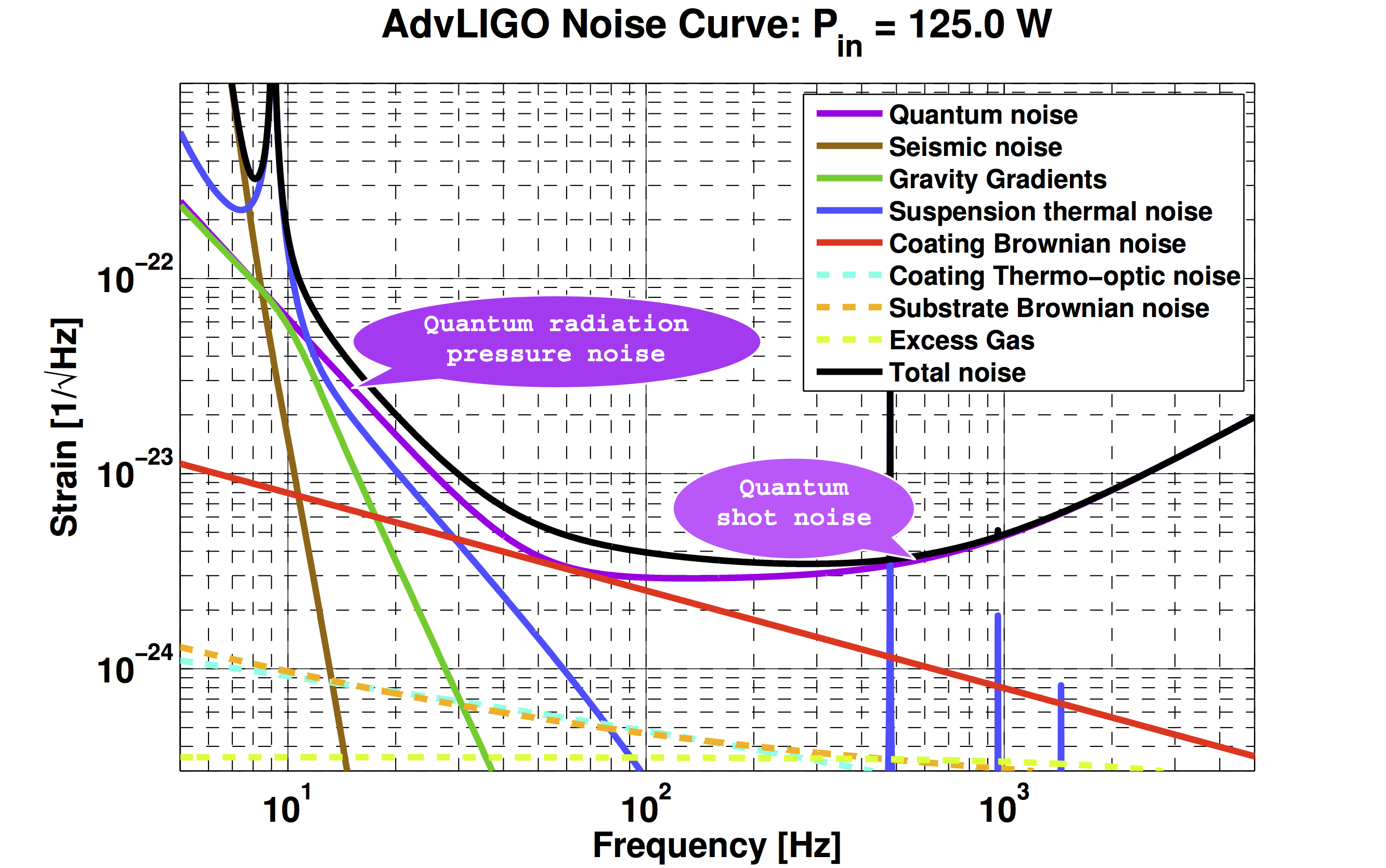}
\caption{Design sensitivity of Advanced LIGO interferometer with major noise sources.}
\label{fig1}       
\end{figure}

And this brings us to the point of this review. As we can see from the Advanced LIGO design sensitivity shown in Fig.~\ref{fig1}, the fundamental quantum fluctuations of light are limiting the sensitivity of the current generation of GW detectors in the most of its detection band, above $\sim 10\ \mathrm{Hz}$. The dominant noises below $10\ \mathrm{Hz}$ comprise seismic and gravity gradient fluctuations \cite{Harms2015} together with suspension thermal noise \cite{ISWP2017}, while at medium frequencies around $\sim50\ \mathrm{Hz}$ the mirror coating thermal fluctuations come close to the level of projected quantum noise. There is an active research going on to suppress the \SD{low-frequency noise sources} further in the next generation facilities \cite{2013_nat.Photon.7.8.644_Crystallyne_coatings}. With these classical noises suppressed, we need to reduce the quantum noise to further improve the detector 
sensitivity. Similarly for the next-generation GW interferometers \cite{CQG.27.19.194002_2010_Punturo,2011_CQG.28.9.094013_3rd_gen_design_study,2017_CQG.34.004001_CEref}, to go beyond their design sensitivity goal of at least an order of magnitude better sensitivity than in Fig.~\ref{fig1}, we will need to incorporate the advanced techniques of quantum noise suppression that this review is about. 

Quantum noise (QN) comes from quantum fluctuations of the phase and amplitude of the light, which are two conjugate canonical observables. As such they do not commute with each other and, due to Heisenberg uncertainty principle, cannot have vanishing uncertainties simultaneously. For the ground-based GW detectors, the GW signal is inferred from the relative phase difference between the two light beams that propagate in the arms of the Michelson interferometer. One might expect that only the quantum fluctuations of the phase, known as Quantum Shot Noise (QSN), shall limit their sensitivity. However, this is not the case. The beating between the strong carrier \SD{field} circulating in the arm cavities with the vacuum quantum fields from the detection port creates a random differential radiation pressure force, which shakes the freely suspended mirrors and manifests as the low-frequency component of the QN. It is called the Quantum Radiation Pressure Noise (QRPN) or quantum back-action noise, in the context of quantum measurement theory. Its domination at low frequencies comes from the strong frequency dependence of the response of the test mass' centre of mass motion to the external force.

Hence, to reach the aforesaid objective and suppress the QN in the entire detection band, one has to suppress the uncertainties of both non-commuting observables in parallel, which seemly violates the Heisenberg uncertainty relation. It sounds impossible, at a first glance. Yet, there are actually many approaches \SD{that seek to perhaps not violate (it's impossible indeed), but} circumvent the limitations imposed by the uncertainty principle. In this review, we will focus on those \SD{of these techniques applicable to interferometric GW detection}.

The quantum noise-mitigation techniques we consider in this review include (1) techniques well tested and already applied in the large scale GW detectors, 
such as squeezed light injection \cite{1981_PRD.23.1693_Caves,2011_Nat.Phys.7.962_LSC,Aasi2013NatPhot,Schnabel2017}, (2) techniques that are at the stage of prototyping, e.g., speed meters \cite{1990_PLA.147.251_Braginsky_SM,Chen2003,Purdue2001,Purdue2002,Chen2003,04a1Da,PhysRevD.86.062001,2014_CQG.31.215009_Graef,2015arXiv150301062V} and frequency-dependent squeezing \cite{Oelker2016,2013_OE.21.30114_Loss_in_FC_Isogai}, and (3) recently proposed ones, which would require quite some research and development, before one could implement them in a real detector, like conditional frequency-dependent squeezing \cite{Ma_NPhys_13_776_2017,2017_PhysRevD.96.062003_GEO_EPR_squeezing} or white-light-cavity based schemes \cite{Wicht1997,Zhou2015,Ma2015,Peano2015,Korobko2017,Miao2015a,Page2017,Miao2017c}.

Experience shows that it takes more than tens of years from concepts to the implementation of some advanced techniques in the large-scale GW detector facility. Most of the methods in this review are not targeted at short, or medium-term upgrades of Advanced LIGO and Advanced Virgo, rather at the next-generation instruments and beyond. It is quite difficult to predict what parameters these future detectors will have and what the level of classical noise sources will be. In this review, we decide to present only the QN in all the sensitivity curves for considered configurations, and adopt the set of nominal parameters listed in Table~\ref{tab:benchmark_3G} as the common ground. 

\begin{table*}
		\begin{tabular}{ l c c }
		\hline
		  Parameter                     & Notation                  & Value      \\
		  \hline
		  Mirror mass, kg               & $M$                       &         200 \\
		  Arm length, km                & $L$                       &          20  \\
		  laser wavelength, nm    & $\lambda_p$ & 1550\\
		  Optical power in each arm, MW               & $P_c/2$                     &         4.0\\
		  Effective detector bandwidth, Hz & $\gamma$             &           100 \\
\hline
		\end{tabular}
  \caption{Parameters for all configurations considered in the paper, unless explicitly specified otherwise.}\label{tab:benchmark_3G}
\end{table*}

The structure of the review is the following. In the next section, we give a brief introduction into the physics of quantum noise and how it manifests in GW interferometers. In Sec.~\ref{sec:2}, we consider the general limitations that arise in precision interferometry due to constraints that quantum mechanics imposes on the magnitude of quantum fluctuations of light. In Sec.~\ref{sec:3}, we review the concept of quantum noise mitigation using squeezed light injection, including frequency-dependent squeezing. Sec. ~\ref{sec:4} is devoted to the suppression of quantum noise through quantum non-demolition measurement of speed and to a myriad of different ways of realising this principle in GW detectors. In Sec.~\ref{sec:5}, the enhancement of the interferometer response to GW signal by modifying test masses' dynamics is investigated and different variations based on optical rigidity also sometimes referred to as dynamical back-action are analysed. Sec.~\ref{sec:6} deals with proposals which consider active elements, such as atomic spin ensembles and unstable optomechanical filters, for the mitigation of quantum noise both at low and at high frequencies. In Sec.~\ref{sec:7}, we give some concluding remarks and outlook.

It is worth emphasising that this review is by no means a replacement of the previous one under the title``Quantum measurement theory in GW detectors'' \cite{Liv.Rv.Rel.15.2012}, but rather a natural continuation thereof. The previous review defined the framework of and provided the tools for the analysis of quantum noise in this special regime of continuous quantum-limited interferometric measurements. This one builds up heavily on these materials by applying the tools and methods to the multitude of novel schemes and configurations developed recently. The main objective we had in mind is to give common ground to all of these various configurations and to facilitate the upcoming selection of the optimal design of the next generation instruments.

\ifpdf
  \renewcommand{\arraystretch}{1.3}
  \begin{longtable}{l p{9cm}}
\else
  \begin{table}[htbp]
\fi
\caption[Notations and conventions]{Notations and conventions, used in
  this review.}
\label{table:notations}
\ifpdf\\\else
  \begin{tabular}{l p{9cm}}
\fi
\toprule
\textbf{Notation and value} & \textbf{Comments} \\
 \midrule
\ifpdf
  \endfirsthead
  \multicolumn{2}{c}{\small\textbf{\tablename~\thetable{}} -- \emph{Continued}}
  \\[4mm]
  \toprule
  \textbf{Notation and value} & \textbf{Comments} \\ 
  \midrule
  \endhead
\fi
         $L$ & length of the arms of the interferometer\\
	 $\tau=L/c$     & light travel time at distance $L$ \\
	 $\omega$  &  optical frequencies \\
	 $\omega_0$ &  interferometer resonance frequency \\
	 $\omega_p$ &  optical pumping frequency (laser frequency)\\
	 $\Omega = \omega-\omega_p$ & modulation sideband frequency w.r.t. laser frequency $\omega_p$\\
	 $\Delta=\omega_p-\omega_0$ & optical pump detuning from the cavity resonance frequency $\omega_0$ \\
	 $\mathcal{E}_{0} = \sqrt{\dfrac{4\pi\hbar\omega_p}{\mathcal{A}c}}$ & normalisation constant of the second quantisation of a monochromatic light beam\\
	 $A^{in} = \sqrt{\dfrac{2P^{in}}{\hbar\omega_p}}$ & classical quadrature amplitude of the incident light beam with power $P^{in}$\\
	 $T\,(R)$ & power transmissivity (reflectivity) of the mirror\\
	$\gamma_{\rm arm} = cT/4L$  & arm cavity half-bandwidth for input mirror transsmissivity $T$ and perfect end mirror\\ 
	$\delta_{\rm arm}$  & arm cavity detuning/differential detuning of the arms of Fabry-Perot--Michelson interferometer\\ 
	$\gamma$  &  interferometer effective half-bandwidth\\
	 
	 $\beta(\Omega)$ &  phase shift acquired by sidebands in the interferometer \\
	 $\mathcal{K}(\Omega)$ & optomechanical coupling factor (Kimble factor) of the interferometer\\
	$P^{in}$ & incident light beam power\\
	$P_c = 2P_{\rm arm}$ & total power, circulating in both arms of the interferometer (at the test masses)\\
	$M$ & Mass of the mirror\\
	$m$ & reduced mass of the signal mechanical mode of the interferometer (\textit{e.g.} dARM mode)\footnote{Here we follow the same definition of the dARM mechanical mode as we adopted in \cite{Liv.Rv.Rel.15.2012}, \textit{i.e.} $x_{\rm dARM} = (x_{N}-x_{E})/2$, where $x_{N,E}$ are the corresponding elongations of the arms of the interferometer. When so defined, the \textit{dARM}-mode has the same reduced mass as a single test mass, $m=M$. Another popular definition of the \textit{dARM} as $\tilde x_{\rm dARM} = (x_{N}-x_{E})$ leads to the new reduced mass equal to $m=M/4$ and to the correspondent redefinition of the SQL.}
	\\
	
	$\Theta = \dfrac{4\omega_p P_c}{mcL}$  & Normalised intracavity power \\
	$h_{\rm SQL} = \sqrt{\dfrac{8\hbar}{mL^2\Omega^2}}$ & Standard Quantum Limit of a free mass for GW strain\\
	$x_{\rm SQL} = \sqrt{\dfrac{2\hbar}{m\Omega^2}}$ & Standard Quantum Limit of a free mass for displacement\\

  \bottomrule
\ifpdf
\end{longtable}
\else\end{tabular}
\end{table}
\fi

\section{Quantum noise} 
\label{sec:1}
\begin{figure}
\centering  \includegraphics[width=.85\textwidth]{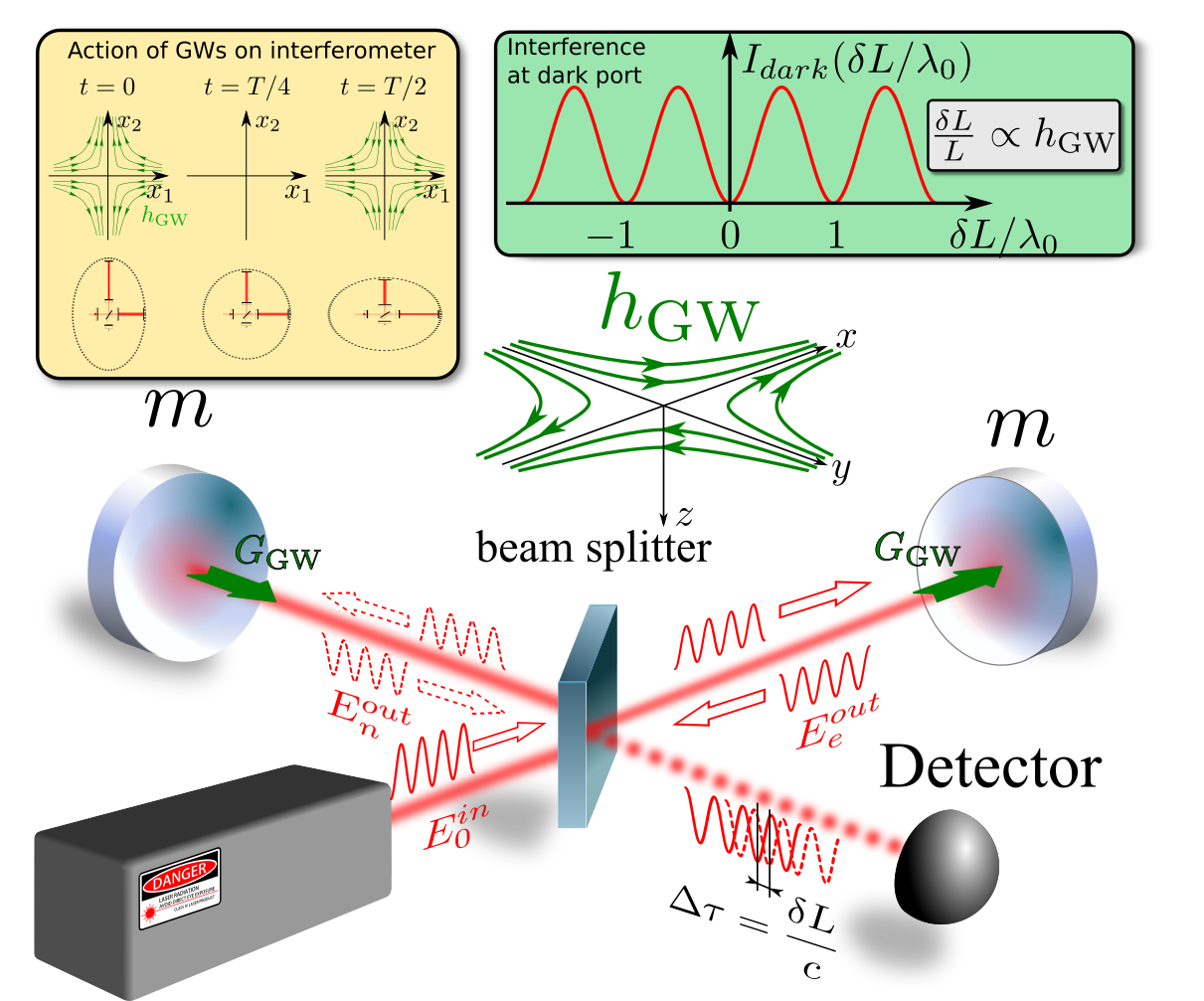}
\caption{Schematic of the working principle of a GW interferometer.}
\label{fig2}       
\end{figure}

Laser interferometric GW detectors (see Fig.~\ref{fig2}) use interference of two (almost) monochromatic light waves travelling in their arms to measure a tiny relative phase shift induced by the GW. Laser light in two orthogonal arms experiences opposite variations of the effective optical length of the arms (see yellow inset box in Fig.~\ref{fig2}), which makes the light beams reflected off the arms to recombine at the beam splitter with a slight mismatch in phase. This violates the destructive interference condition at the beam splitter and a small fraction of carrier field makes it to the photodetector at the detection (readout) port. The green inset box in Fig.~\ref{fig2} shows how the intensity of light at the photodetector would depend on the effective difference of the optical path lengths of the arms $\delta L$. 

In a nutshell, every interferometer is a device that uses interference to measure the relative phase of one beam to the other. It detects variations of intensity of the interference pattern caused by this phase shift. The precision of this procedure is dependent on many factors, which can be decomposed by source in a noise budget [cf. Fig.~\ref{fig1}]. The one, which we are focusing on, in this review is rooted in the very nature of light as a quantum field, \textit{i.e.} the quantum fluctuation of optical phase and amplitude.

\subsection{Two-photon formalism and input-output relations}
\label{ssec:2-1}

As shown by Caves and Schumaker in \cite{85a1CaSch,85a2CaSch}, the quantum noise of light in any linear optical device can be conveniently described within the framework of the two-photon formalism. Namely, noise can be considered as tiny stochastic variations in the quadratures of the optical field travelling through the device. Any variations of interferometer parameters induced by the signal, {\it e.g.} differential arm length change, also lead to variations of the quadratures of the outgoing field, which can be described using the same formalism.

In the two-photon formalism, one starts with writing down the ingoing and outgoing optical fields of the interferometer at some fixed location in terms of {\it sine} and {\it cosine} quadratures:
	\begin{eqnarray}\label{eq:Elin}
	  \hat{E}^{in}(t) &=& \mathcal{E}_0\left[(A^{in}+\hat{a}^{in}_c)\cos\omega_p t+\hat{a}^{in}_s\sin\omega_pt\right]\,,\\
	  \hat{E}^{out}(t) &=& \mathcal{E}_0\left[(B_{c}^{out}+\hat{b}^{out}_c)\cos\omega_p t+(B_s^{out}+\hat{b}^{out}_s)\sin\omega_pt\right]\,.
	\end{eqnarray}	
Here $\mathcal{E}_{0} = \sqrt{4\pi\hbar\omega_{p}/(\mathcal{A}c)}$ is a normalisation constant defined in the second quantisation of a monochromatic light beam with the carrier frequency $\omega_{p}$, optical power $P^{in}$ and cross-sectional area $\mathcal{A}$; $A^{in} = \sqrt{2 P^{in}/(\hbar\omega_{p})}$ ($B^{out}$) is classical mean amplitude of the input (output) light at frequency $\omega_p$;  $\hat{a}^{in}_{c,s}$ ($\hat{b}^{in}_{c,s}$) describe small, zero-mean quantum fluctuations and variations due to the signal, and they are related to the creation and annihilation operators through
	\begin{equation}\label{eq:2ph_quad_def}
	  \hat{a}_c = \frac{\hat{a}+\hat{a}^\dag}{\sqrt{2}}\,,\quad \mbox{and}\quad\hat{a}_s = \frac{\hat{a}-\hat{a}^\dag}{i\sqrt{2}}\,,
	\end{equation}
and similarly for outgoing fields. Note that we do not specify time as an argument in Eq.\,\eqref{eq:2ph_quad_def}, as the same definition holds in the frequency domain, which is assumed in the rest of this article. The time and frequency domain are related through
the following Fourier transform:
	\begin{equation}\label{eq:FourierQuads}
	  \hat{a}_{c,s}(t) = \int_{-\infty}^{\infty} \frac{d\Omega}{2\pi} \hat{a}_{c,s}(\Omega)e^{-i\Omega t}\,.
	\end{equation} 

To fully describe signal and noise in a (lossless) GW interferometer, we shall quantify how the quadrature operators  of the input field transform when propagating through the interferometer to the output. Mathematically, the transformation can be represented as a matrix operating on the two-dimensional vectors $\vq{a} = \{\hat{a}_c,\,\hat{a}_s\}^{\rm T}$ and $\vq{b} = \{\hat{b}_c,\,\hat{b}_s\}^{\rm T}$ and GW signal $h(\Omega)$. Note that one needs to calculate both, the propagation of the carrier field mean amplitudes (denoted by capital letters) and of the zero-mean fluctuational sideband fields defined above. The former ones are needed to calculate the response of the interferometer to the mirrors' displacement as well as the effects of quantum back-action, as both depend on the value of the classical laser field amplitude at the mirror (cf. \textit{e.g.} Eqs. (245) and (257) of \cite{Liv.Rv.Rel.15.2012}).
 We  assume that the interferometer is working in a small perturbations regime where all the transformations of the signal and noise can be considered as linear ones, and all the noise sources under study are Gaussian and stationary, which can be quantified by using the frequency domain spectral 
density. 

For a GW detector, the transformation, which is also 
called the input-output relation, can be written in the general form as:
	\begin{equation}\label{eq:IOlossless}
	  \vq{b} = \tq{T}\cdot\vq{a} + \vs{t}\frac{\mathcal{X}}{\mathcal{X}_{\rm SQL}}\,,
	\end{equation}
where
	\begin{equation}\label{eq:T_def}
	  \tq{T} \equiv
	  \begin{bmatrix}
	    T_{cc}(\Omega) & T_{cs}(\Omega)\\
	    T_{sc}(\Omega) & T_{ss}(\Omega)
	  \end{bmatrix}
	\end{equation}
is the optical transfer matrix of the interferometer (including the optomechanical back-action effects),
	\begin{equation}\label{eq:t_h_def}
	  \vs{t} \equiv
	  \begin{bmatrix}
	    t_c(\Omega)\\
	    t_s(\Omega)
	  \end{bmatrix}
	\end{equation}
is the optomechanical (OM), SQL-normalised response of the interferometer to a general signal. 
The signal is denoted as $\mathcal{X}$ and describes only the change in the physical state of the interferometer caused by the signal in question, \textit{e.g.}, GW, and $\mathcal{X}_{\rm SQL}$ is the corresponding free-mass \textit{standard quantum limit (SQL)} for the mechanical 
degree of freedom expressed in the unit of $\mathcal{X}$, which is a normalisation 
factor and will be explained later in more details (see Sec.~\ref{sec:2}). In precision interferometry, $\mathcal{X}$ is either the signal displacement of the test mass, 
$x$, or an external signal force, $F$, that causes this displacement, or, more
specific for GW interferometry, the GW strain, $h$. In each case, the corresponding SQL applies. The relation between these three quantities is discussed in Sec.~4.3 of Ref.\,\cite{Liv.Rv.Rel.15.2012}.

The interferometer's readout quantity depends on the implemented readout scheme, but in all cases it invariably involves measuring the photocurrent $\hat{i}^{out}(t)$ derived from the photodetectors that sense the light leaving the readout port of the interferometer. Assuming that all the future GW interferometers will use the \textit{balanced homodyne detection (BHD)} (see Sec. 2.3.1 of \cite{Liv.Rv.Rel.15.2012} for basics description of BHD, or \cite{Fritschel:14} for more in-depth analysis thereof) one can project to an arbitrary quadrature $\hat{o}_{\phi_{\rm LO}}$ of the outgoing light, varying the homodyne phase $\phi_{\rm LO}$:
	\begin{equation}\label{eq:o_zeta_lossless}
	  \hat{o}_{\phi_{\rm LO}} \equiv \hat{b}_c\cos\phi_{\rm LO} + \hat{b}_s\sin\phi_{\rm LO} \equiv \vs{H}_{\phi_{\rm LO}}^{\rm T}\cdot\vq{b}\,,\
	  \vs{H}_{\phi_{\rm LO}}\equiv
	  \begin{bmatrix}
	    \cos\phi_{\rm LO}\\
	    \sin\phi_{\rm LO}
	  \end{bmatrix}\,.
	\end{equation}
The corresponding quantum noise spectral density in the unit of the observable of interest, $\mathcal{X}$ reads:
	\begin{equation}\label{eq:SpDens_h}
	  S^\mathcal{X}(\Omega) = \mathcal{X}^2_{\rm SQL}\frac{\vs{H}^{\rm T}_{\phi_{\rm LO}}\cdot\tq{T}\cdot\tq{S}_{a}^{in}\cdot\tq{T}^\dag\cdot\vs{H}_{\phi_{\rm LO}}}{|\vs{H}^{\rm T}_{\phi_{\rm LO}}\cdot\vs{t}|^2}
	\end{equation}
where $\tq{S}_a^{in}$ stands for spectral density matrix of input field and components thereof is defined as:
	\begin{multline}\label{eq:SpDens_a}
	  \pi\delta(\Omega-\Omega') \, \tq{S}_{a, ij}^{in}(\Omega) \equiv 
	    \frac12\bra{in}\hat{a}_i(\Omega)(\hat{a}_j(\Omega'))^\dag+(\hat{a}_j(\Omega'))^\dag\hat{a}_i(\Omega)\ket{in}\,,
	\end{multline}
where $\ket{in}$ is the quantum state of the field injected in the dark port of the interferometer and $(i,j) = \{c,s\}$  (see Sec. 3.3 in \cite{Liv.Rv.Rel.15.2012} for more details).
In this article, we deal with \emph{single-sided} spectral densities $S$ and hence in 
the case of input vacuum state:
	\begin{equation*}
	  \ket{in} = \ket{vac} \qquad\Rightarrow\qquad \tq{S}_a^{in} = \mathbb{I}\,,
	\end{equation*}
where $\mathbb{I}$ is the \textit{identity matrix}.

\subsubsection{Case of multiple input/output channels.}
This formalism can be easily extended to a more general case of an interferometer with more than one input and output channel. Two examples of such a schemes will be discussed in more detail in Sections, where ~\ref{ssec:3-3} and \ref{ssec:6-2}two-mode squeezed states are used as the input fields of the interferometer. Another situation, when one needs to take into account more optical degrees of freedom arises in the case of loss and imperfection analysis as we discuss below, in Sec.~\ref{ssec:2-6}. In any of these situations, one simply needs to extend the number of dimensions of the model from 2, for two quadratures of a single optical degree of freedom, to $2N$ with $N$ being the number of the input and output channels of interferometer. Then the vectors $\vq{a}$ and $\vq{b}$ are defined as:
\begin{align}
\vq{a} \equiv \{a_{c}^{(1)},\,a_{s}^{(1)},\,\ldots a_{c}^{(i)},\,a_{s}^{(i)},\,\ldots a_{c}^{(N)},\,a_{s}^{(N)}\}^{\rm T}\ \mathrm{with}\ i=\{1,N\}\\ 
\vq{b} \equiv \{b_{c}^{(1)},\,b_{s}^{(1)},\,\ldots b_{c}^{(i)},\,b_{s}^{(i)},\,\ldots b_{c}^{(N)},\,b_{s}^{(N)}\}^{\rm T}\,, \mathrm{with}\ i=\{1,N\}
\end{align}
and the corresponding transfer matrix and response vector read:
\begin{align}\label{eq:Nch-TrMat}
\tq{T}_{2N\times2N} \equiv 
\begin{bmatrix}
\tq{T}^{(11)} & \cdots & \tq{T}^{(1j)} & \cdots & \tq{T}^{(1N)}\\
\vdots & \ddots & \vdots &  & \vdots\\
\tq{T}^{(i1)} & \cdots & \tq{T}^{(ij)} & \cdots & \tq{T}^{(iN)}\\
\vdots &  & \vdots & \ddots & \vdots\\
\tq{T}^{(N1)} & \cdots & \tq{T}^{(Nj)} & \cdots & \tq{T}^{(NN)}
\end{bmatrix} \quad \mathrm{and} \quad 
\vs{t}_{2N} \equiv \begin{bmatrix}
\vs{t}^{(1)}\\
\vdots\\
\vs{t}^{(i)}\\
\vdots\\
\vs{t}^{(N)}
\end{bmatrix}\,,
\end{align}
where each term $\tq{T}^{(ij)}$ and $\vs{t}^{(i)}$ in the above expressions stands for a $2\times2$-matrix block or a 2-dimensional response vector described by Eqs.~\eqref{eq:T_def} and \eqref{eq:t_h_def}, respectively. Naturally, $\tq{T}^{(ij)}$ describes the contribution of the $j$-th input field $\vq{a}^{(j)}$ to the $i$-th output field $\vq{b}^{(i)}$, while $\vs{t}^{(i)}$ stands for the SQL-normalised response of the $i$-th output channel to the signal influence $\mathcal{X}$. \SD{Transformation $\tq{T}_{2N\times2N}$ on the light quadrature operators $\vq{a}$ is unitary and represents the Bogolyubov transformation that conserves commutation relations for the outgoing quadrature operators. This, along with the fact that we consider in this review only Gaussian quantum states of light, means that $\tq{T}_{2N\times2N}$ is a \textit{symplectic} matrix, \textit{i.e.} the one keeping the fundamental commutator invariant \cite{2007_JoPA.40.7821_Adesso_illuminati}.}

Another consequence of Gaussianity of the states of light and operations under study is that any entangled and/or squeezed multimode state injected in the GW detectors to boost its QN-limited sensitivity can be effectively represented as an additional symplectic transformation, $\tq{T}^{\rm sqz}_{2N\times2N}$, on a set of vacuum fields $\vq{a}^{\rm vac}$, \textit{i.e.}:
\begin{equation}
\vq{a}^{\rm sqz} = \tq{T}^{\rm sqz}_{2N\times2N}\vq{a}^{\rm vac}\ \Rightarrow\ \tq{S}^{in,\,sqz}_{a} =  \tq{T}^{\rm sqz}_{2N\times2N}\cdot\tq{I}_{2N\times2N}\cdot\bigl(\tq{T}^{\rm sqz}_{2N\times2N}\bigr)^{\dag}\,,
\end{equation}
where $\tq{I}_{2N\times2N}$ is an identity matrix standing for the power spectral density of the $2N$-mode vacuum state. By definition, $\tq{T}^{\rm sqz}_{2N\times2N}$ stands for all the manipulations that are performed on the input vacuum fields before they enter the main interferometer, which includes, for instance, squeezing and passage through the filter cavities for optimal frequency-dependent rotation of squeezing noise ellipse (see Sec.~\ref{ssec:4-2}). 

To conclude, we need to generalise the treatment of multiple readout channels. In the N-dimensional case, readout observable $  \hat{o}_{\phi_{\rm LO}}$ of Eq.~\eqref{eq:o_zeta_lossless} transforms into a vector of $N$ outputs, $\vq{o}_{N}$, where each output can have its own homodyne readout phase $\phi_{\rm LO}^{(i)}$ and a corresponding homodyne vector $\vs{H}^{(i)}_{\phi_{\rm LO}}$ as defined in \eqref{eq:o_zeta_lossless}. Finally, all the readout channels comprising the readout vector $\vq{o}_{N}$ which contain information about the GW signal and has added Gaussian noise needs to be processed so that the signal is extracted with the highest signal-to-noise ratio (SNR) possible. This is usually achieved by combining the readouts with some optimal weight functions, chosen so as to maximise the SNR, or any other chosen figure of merit. In general, this will require to define a vector of coefficient functions $\vec{\alpha}_{N}$ (generally, frequency dependent) that has to be found as a result of optimisation procedure of a chosen figure of merit, \textit{e.g.} the SNR, in which case $\alpha_{i}(\Omega)$ are known as Wiener filters. The resulting combined readout then reads:
\begin{equation*}
\hat{o}_{\rm opt} = \sum_{i=1}^{N} \alpha_i\bigl\{\hat b_c^{(i)}\cos\phi_{\rm LO}^{(i)}+\hat b_s^{(i)}\sin\phi_{\rm LO}^{(i)}\bigr\} \equiv \sum_{i=1}^{N} \alpha_i\, \vs{H}^{\rm T}_{\phi^{(i)}_{\rm LO}}\cdot\vq{b}^{(i)}\,,
\end{equation*}
which gives the following estimate for the signal observable $\mathcal{X}$:
\begin{equation}
\tilde{\mathcal{X}}_{\rm opt} = \mathcal{X}_{\rm SQL}\hat{o}_{\rm opt}/\bigl(\sum_{i=1}^{N} \alpha_i\, \vs{H}^{\rm T}_{\phi^{(i)}_{\rm LO}}\cdot\vs{t}^{(i)}\bigr)\,,
\end{equation}
where the sum in the denominator stands for the effective response function for a multi-channel interferometer.
Gathering all the definitions of this section together, the noise power spectral density for the noise power spectral density in the units of signal $\mathcal{X}$ reads:
\begin{multline}\label{eq:o_multichan_PSD}
S^{\mathcal{X}_{\rm opt}}(\Omega) = \frac{\mathcal{X}^2_{\rm SQL}}{\Bigl|\sum_{i=1}^{N} \alpha_i\, \vs{H}^{\rm T}_{\phi^{(i)}_{\rm LO}}\cdot\vs{t}^{(i)}\Bigr|^2}\times\\
\sum_{i=1}^{N}\sum_{j=1}^{N}\alpha_i\alpha_j\, \vs{H}^{\rm T}_{\phi^{(i)}_{\rm LO}}\cdot\bigl[\tq{T}_{2N\times2N}\cdot\tq{T}^{\rm sqz}_{2N\times2N}\cdot(\tq{T}^{\rm sqz}_{2N\times2N})^\dag\cdot(\tq{T}_{2N\times2N})^\dag\bigr]_{ij}\cdot\vs{H}_{\phi^{(j)}_{\rm LO}}
\end{multline}
with $[\ldots]_{ij}$ denoting the $2\times2$ subblock with the indices $ij$ within a large $2N\times2N$ matrix product written inside the brackets.

%
%

\subsection{Transfer functions of the quantum-noise-limited interferometer}\label{ssec:2-3}


The internal structure of the above expressions might be rather complex for given advanced interferometer schemes, but the underlying physics is rather simple and comes from the 
following two facts:
 \begin{itemize}
 \item
 mirrors can move when subject to the action of an external force, thus making the interferometer sensitive to the GW, and \footnote{Strictly speaking, there are two possible ways of looking at the action of GW on the light in the interferometer. In this review, we will follow the point of view that the test masses move in a Local Lorentz (LL) frame of a central beam splitter, and GWs act akin to tidal forces on the test masses of the interferometer making them move w.r.t. the defined LL-frame of the detector\cite{04BookBlTh}. Another way to describe GW action is to consider the interferometer in a so-called \textit{transverse-traceless (TT) gauge}, where test masses are assumed to remain at rest and GW action leads to the modulation of the effective index of refraction of the space interval between the test masses. Interested readers are invited to read an excellent course book by Blandford and Thorne.} 
 \item 
light interacts with the mirrors, which manifests in two ways, \textit{i.e.} the mirror motion modulating the phase of light and the light exerting a radiation pressure force on the mirror.  \end{itemize}
 
Quantitatively, these two facts are described by means of corresponding transfer functions (TF)\footnote{The rigorous mathematical treatment of the linear quantum measurement and of all transfer functions is given in Sec.~4.2 of  \cite{Liv.Rv.Rel.15.2012}}:
\begin{enumerate}
\item
Force-to-displacement TF is described by the mechanical susceptibility, $\chi_{m}$ of the 
centre of mass motion of the test mass mirror;
\item
Displacement-to-field TF, $\vb{R}_{x}\equiv \{\partial a_{c}/\partial x,\,\partial a_{s}/\partial x\}$, reflects how much the two quadratures of the outgoing field are changed by the displacement of the mirror $x$, and
\item
Field-to-force TF, $\vb{F} \equiv \{\partial \hat F_{\rm r.p.}/\partial a_{c},\,\partial \hat F_{\rm r.p.}/\partial a_{s}\}$, describes how much the radiation pressure force depends on 
the sine and cosine quadrartures of the ingoing field;
\item
Displacement-to-force TF,  $\vb{K} \equiv -\partial F/\partial x$, describes the \textit{dynamic back-action} or \textit{optical spring} that manifests as restoring force created by the part of the optical field dependent on the mirror displacement $x$.
\end{enumerate}

\begin{figure}[htbp]
\begin{center}
\includegraphics[width=.7\textwidth]{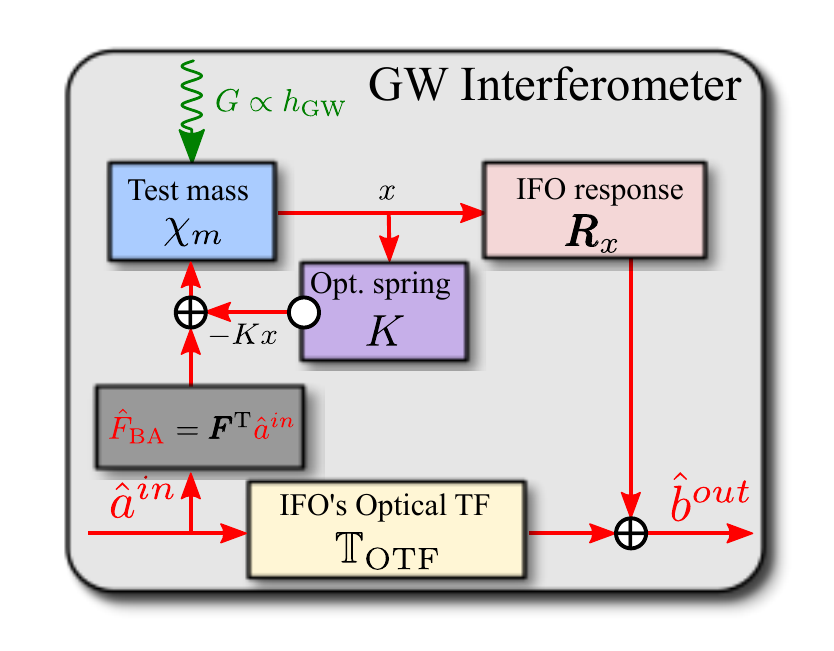}
\caption{Schematics of the input-output relations of the GW interferometer in a form of a flowchart.}
\label{fig:IFO_flowchart}
\end{center}
\end{figure}

The basic operation of any interferometer can be described by means of a simple flowchart diagram including the above TFs, as shown in Fig.~\ref{fig:IFO_flowchart}. Here the external signal force (GW) interacts with the mechanical degree of freedom (DoF), displacing its mirrors by $x$. The magnitude of this displacement is defined by the \textit{mechanical susceptibility} $\chi_m(\Omega)$,
which can be read off from the Fourier domain solution to the Newtonian equation of motion 
(the same as the Heisenberg equation of motion due to linearity of the system):
	\begin{equation*}
	m \ddot x(t) = \mathcal{F}(x(t),\,\dot x(t)) + \sum_{k}F_k^{\rm ext}(t)\,\ \overset{\rm Fourier}{\underset{\rm domain}{\Longrightarrow}}\ x(\Omega) = \chi_m(\Omega)\sum_kF_k^{\rm ext}(\Omega)\,.
	\end{equation*}
Here $m$ is the reduced mass of the mechanical DoF,  $\mathcal{F}(x,\,\dot x)$ is the sum of the internal forces of the system (\textit{e.g.} restoring force of the suspensions, dissipative forces), and $F_k^{\rm ext}$ stand for all the external forces acting on the mirror, including the GW signal force $G$. For GW with the strain amplitude $h(t)\leftrightarrow h(\Omega)$, this effective differential force reads:
\begin{equation*}
G(t) = mL\ddot h(t)\ \overset{\rm Fourier}{\underset{\rm domain}{\Longrightarrow}}\ G(\Omega) = -mL\Omega^2h(\Omega)\,,
\end{equation*}  
 
Displacement of the mirrors modulates the light reflected off from the mirrors. 
This results in additional variation of the outgoing light quadratures, which is proportional to $x$. The \textit{displacement-to-field} TF $\vb{R}_{x}$ essentially defines the strength of the interaction of light with the mechanics, \textit{i.e.} the \textit{optomechanical coupling}.

The other end of the \textit{optomechanical coupling} is given by the \textit{field-to-force} TF. It stems from the radiation pressure (RP) that light exerts on the mirrors. Thus the TF in question is a vector of coefficients at the corresponding quadratures of the input fields in the expression for a back-action force, $F_{\rm BA}$. This force contributes to the actual displacement of the mirrors and thus mimics the signal displacement. Noteworthy is that the radiation pressure may depend on the displacement of the mirror, if the interferometer is detuned. This creates a feedback loop and results in a restoring force. This light-induced restoring force is known as \textit{dynamical back-action} or \textit{optical rigidity}, represented by a violet box in Fig.~\ref{fig:IFO_flowchart}.

Finally, there is also the \textit{field-to-field} TF that describes how the input light fields would be transformed by the interferometer, were its mirrors fixed. This is an optical TF shown as a yellow block in the flowchart.

Note that all these considerations apply equally to a system with an arbitrary number of inputs and outputs. 

\subsection{I/O-relations for tuned interferometers}

We can use the developed formalism to derive the input-output (I/O) relation of a given 
interferometer configuration and the quantum noise. And quite astonishingly, a very broad class of so called \textit{tuned} interferometers turns out to have the I/O-relations of the same general shape that depends on the two frequency dependent parameters, the optomechanical coupling strength $\mathcal{K}(\Omega)$ and the phase $\beta(\Omega)$:
\begin{equation}\label{eq:I/O_KLMTV}
  \vq{b} = e^{2i\beta(\Omega)} 
  \begin{bmatrix}
  1 & 0\\
  -\mathcal{K}(\Omega) & 1
  \end{bmatrix} \vq{a} +
    e^{i\beta(\Omega)}
    \begin{bmatrix}
    0\\
    \sqrt{2\mathcal{K}(\Omega)}
    \end{bmatrix}\frac{h}{h_{\rm SQL}} \,.
\end{equation}
Interferometers that are described by the above relations are \textit{tuned} in the sense that the cosine quadrature of an incident light would be transformed into the cosine quadrature of an outgoing light, and likewise would the sine quadrature do, if the mirrors were fixed. Optomechanical coupling factor $\mathcal{K}$ was introduced by Kimble \textit{et al.} \cite{02a1KiLeMaThVy} to describe the strength of interaction between light and the mechanical degrees of freedom of the test masses. By construction, $\mathcal{K}$ is an absolute value of the product of \textit{force-to-displacement} TF $\times$ \textit{displacement-to-field} TF $\times$  \textit{field-to-force} TF. It shows the fraction of light intensity modulation transformed into phase modulation at sideband frequency $\Omega$ mediated by the radiation pressure force. As for $\beta$, it is an extra phase shift.

Hence the optical transfer matrix, $\tq{T}$, of the tuned interferometer and its optomechanical response, $\vb{t}$, read:
\begin{equation}\label{eq:OTM_MI}
	\tq{T} = 
	e^{2i\beta(\Omega)} 
	  \begin{bmatrix}
	  1 & 0\\
	  -\mathcal{K}(\Omega) & 1
	  \end{bmatrix}\,,\ \quad\ 
	  \vb{t} = e^{i\beta(\Omega)}
    \begin{bmatrix}
    0\\
    \sqrt{2\mathcal{K}(\Omega)}
    \end{bmatrix}\,.
\end{equation} 
We ought to mention that for the long-arm interferometric detectors where travel time of light in the arms become comparable with the GW half-period (as it is planned for all the designs of the next generation GW interferometers) the assumption of stationarity of the GW strain within the detection frequency band breaks. To account for the resulting reduction of 
response of the interferometer to GW signal, the following correction factor has to be applied to the above expression for the response \cite{1997_CQG.14.6.1513_Schilling_LISA_response,PhysRevD.96.084004}:
\begin{equation}
\vb{t} \to \vb{t}D(\Omega)\ \mbox{where}\ D(\Omega)= \mathrm{sinc}(\Omega L/c)\,,
\end{equation} 
with $\mathrm{sinc}(x)\equiv\sin x/x$. In general, factor $D(\Omega)$ depends on the mutual orientation of the detector and the source of GWs \cite{PhysRevD.96.084004}, but in the simple case of normal incidence with optimal polarisation it can be approximated as shown above.

Using Eq.~\eqref{eq:SpDens_h}, we can obtain the general expressions for the power spectral density of the quantum noise for tuned interferometers in unit of GW strain $h$. Given 
an arbitrary readout quadrature defined by the homodyne angle $\phi_{\rm LO}$, it reads:
\begin{equation}\label{eq:QNLS_tuned_IFO}
	S^h = \frac{h_{\rm SQL}^2}{2 D^2}\left[\dfrac{(\mathcal{K}-\cot\phi_{\rm LO})^{2}+1}{\mathcal{K}}\right]\,.
\end{equation}
In the special case of phase quadrature readout, $\phi_{\rm LO}=\pi/2$, this expression simplifies as
\begin{equation}\label{eq:QNLS_phase_quad}
	S^h = \frac{h_{\rm SQL}^{2}}{2D^2}\left[\dfrac{1}{\mathcal{K}}+\mathcal{K}\right]\,,
\end{equation}
which clearly shows two components of the quantum noise, namely the quantum shot noise represented by the first term inside the brackets, and the quantum radiation pressure noise given by the last term. In Sec.~\ref{sec:2}, we use this expression to derive the SQL.

\subsection{Quantum noise of a tuned Michelson interferometer}

\begin{figure}[htbp]
\begin{center}
\includegraphics[width=.7\textwidth]{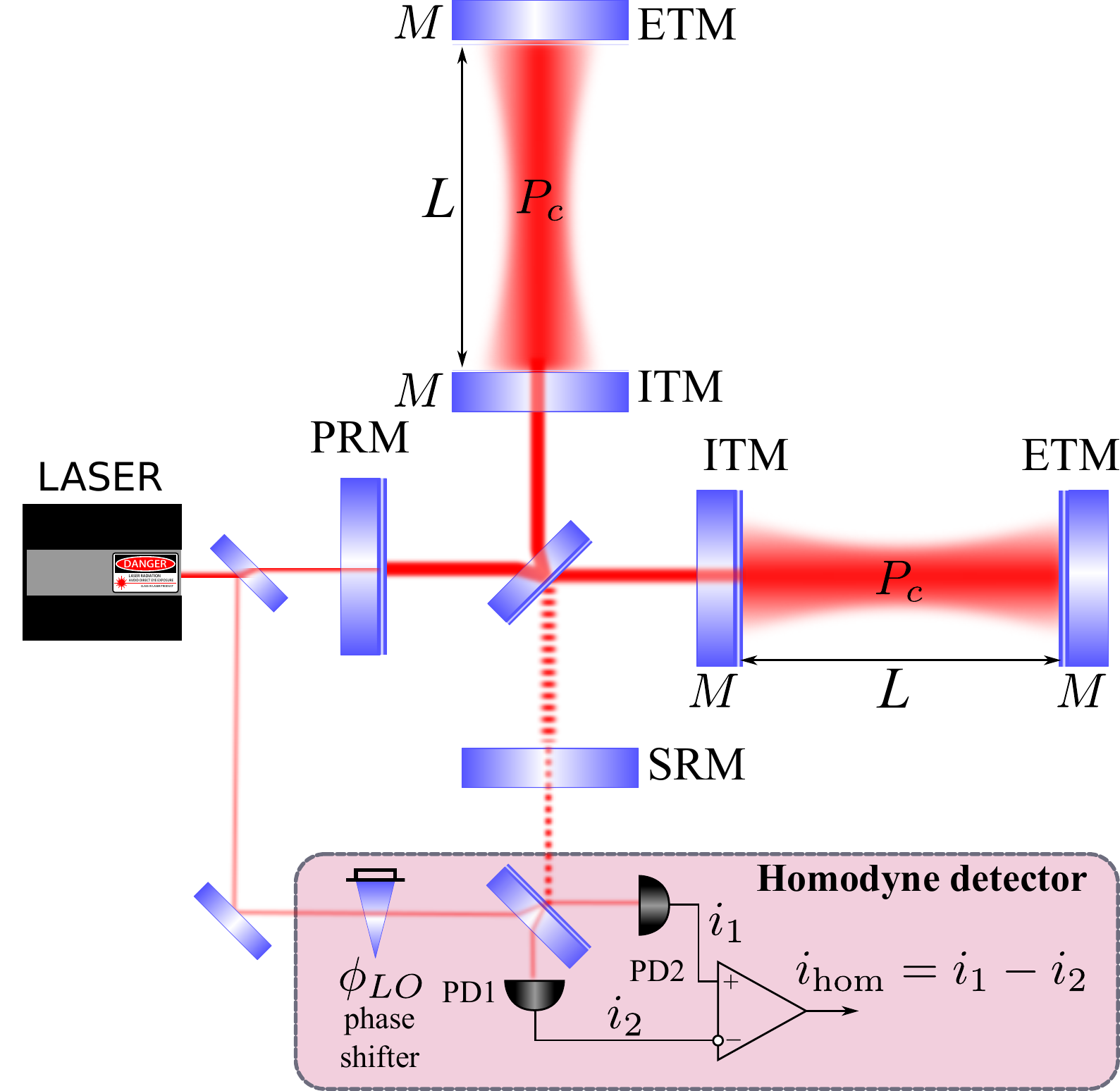}
\caption{Schematics of a dual-recycled Fabry-Perot--Michelson interferometer with balanced homodyne readout.} 
\label{fig:DRFPMI}
\end{center}
\end{figure}

\begin{figure}[htbp]
\begin{center}
\includegraphics[width=\textwidth]{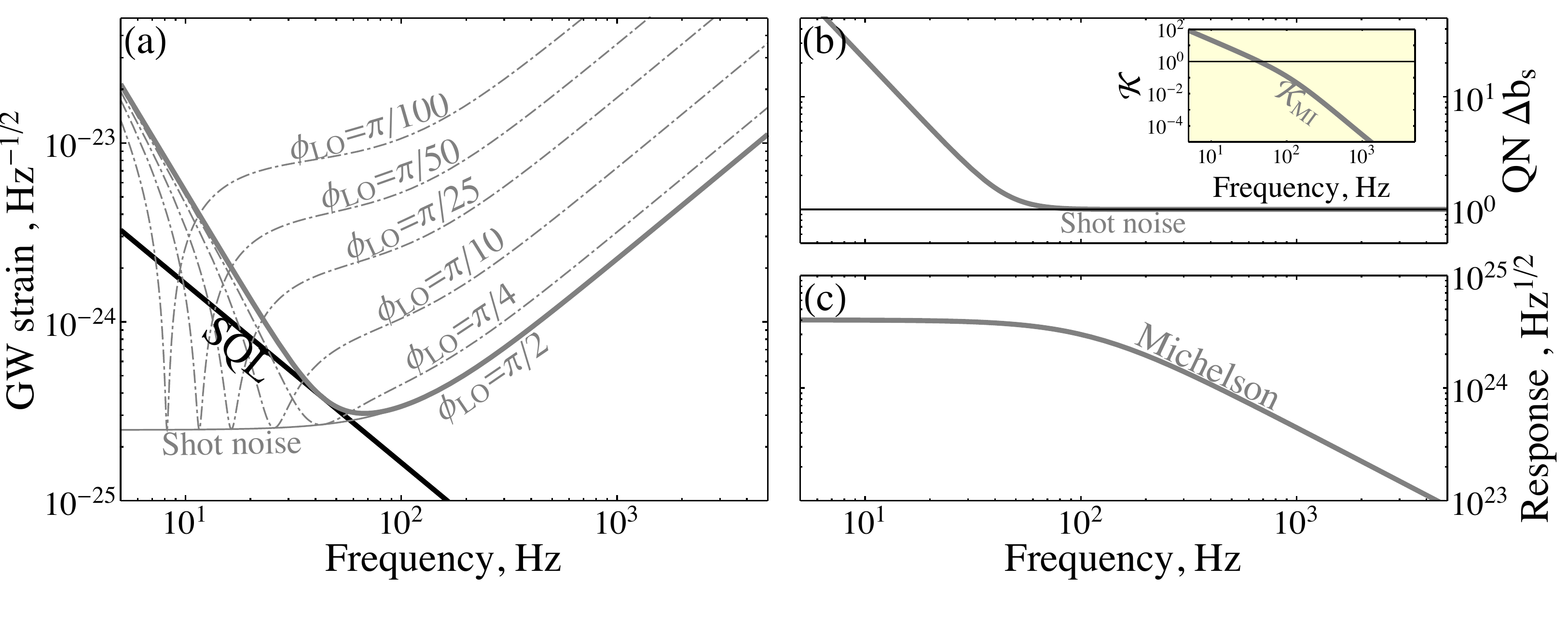}
\caption{Quantum noise of a \textit{Michelson interferometer}: \textit{(a):} QNLS of the Michelson interferometer with phase quadrature readout (solid grey trace) for parameters given in the table \ref{tab:benchmark_3G}. Thin dash-dotted grey lines show the effect of change of readout quadrature (homodyne angle); \textit{(b)} quantum fluctuations of the phase quadrature of the readout light of the Michelson interferometer (grey trace); \textit{(c)} response functions of the Michelson (grey trace) interferometer to the GW strain.} 
\label{fig:MI_QNLS}
\end{center}
\end{figure}

It will be instructive for our review to present here the relevant expressions for a conventional Michelson interferometer with Fabry-Perot cavities in the arms, a signal-recycling 
mirror and a power recycling mirror, as shown in Fig.~\ref{fig:DRFPMI}. In the Appendix~\ref{app:FPMI_lossy}, we derive rigorous expressions for the I/O-relations of 
such a Fabry-Perot--Michelson interferometer (FPMI) with optical loss and shall refer the interested reader to Sec.~5.3 of \cite{Liv.Rv.Rel.15.2012} where even more detailed step-by-step derivation is performed. Here we merely write down the final expressions for the OM coupling factor $\mathcal{K}_{\rm MI}$ and sideband phase shift $\beta_{\rm MI}$ in the 
ideal case without optical losses:
\begin{align}
 \mathcal{K}_{\rm MI} &= \dfrac{\Theta_{\rm MI}\tau}{\Omega^2} \dfrac{1-R^2_{\rm ITM}}{1-2\sqrt{R_{\rm ITM}} \cos 2\Omega\tau + R_{\rm ITM}}\simeq \frac{2 \Theta_{\rm MI} \gamma_{\rm arm}}{\Omega^2(\gamma_{\rm arm}^2+\Omega^2)}\,,\label{eq:KMI}\\
\beta_{\rm MI} &= \atan{ \dfrac{1+\sqrt{R_{\rm ITM}}}{1-\sqrt{R_{\rm ITM}}} \tan\Omega\tau }\simeq  \atan{\Omega/\gamma_{\rm arm}}\label{eq:BMI}\,,
\end{align}
with $\Theta_{\rm MI} = 4\omega_0 P_c/(McL)$ , where $P_c$ is the optical power circulating in the interferometer
and $\gamma_{\rm arm} = T_{\rm ITM}/(4\tau)$ is the half bandwidth of the arm cavity. 
Given the parameters listed in Table~\ref{tab:benchmark_3G}, the signal-referred 
noise spectral density Eq.\,\eqref{eq:QNLS_phase_quad} with $\cal K$ replaced by 
${\cal K}_{\rm MI}$ 
is shown in Fig.~\ref{fig:MI_QNLS}a. We also show the
noise spectrum of the quantum fluctuation $\delta b_{s}^{out}$ in the phase quadrature (see Fig.~\ref{fig:MI_QNLS}b), and the detector response to the GW signal  (see Fig.~\ref{fig:MI_QNLS}c).

The above equations can be generalised to the case of signal-recycled interferometer, using the \textit{``scaling law''} approach of Chen and Buonanno \cite{Buonanno2003}. As shown in detail in Sec.~5.3.4 of \cite{Liv.Rv.Rel.15.2012}, if the distance between the SRM and the ITMs $l_{\rm SRC}\ll L$ (see Fig.~\ref{fig:DRFPMI}), the frequency-dependent phase shift, $\Omega l_{\rm SRC}/c$, acquired by light sidebands in the signal-recycling cavity can be neglected, and one can introduce an effective compound input mirror made of the SRM and the ITMs with effective complex reflectivity and transmissivity, leading to the following modification of the initial bandwidth and detuning of the arms:
\begin{subequations}
\begin{align}
\gamma &= \gamma_{\rm arm}\mathrm{Re}\left[\dfrac{1-\sqrt{R_{\rm SRM}}e^{{2i\phi_{\rm SR}}}}{1+\sqrt{R_{\rm SRM}}e^{{2i\phi_{\rm SR}}}}\right] = \dfrac{\gamma_{\rm arm}T_{\rm SRM}}{1+2\sqrt{R_{\rm SRM}\cos 2\phi_{\rm SR}+R_{\rm SRM}}}\,\\
\delta &= \delta_{\rm arm}-\gamma_{\rm arm}\mathrm{Im}\left[\dfrac{1-\sqrt{R_{\rm SRM}}e^{{2i\phi_{\rm SR}}}}{1+\sqrt{R_{\rm SRM}}e^{{2i\phi_{\rm SR}}}}\right] =\nonumber \\
&=  \delta_{\rm arm}+\dfrac{2\gamma_{\rm arm}\sqrt{R_{\rm SRM}}\sin2\phi_{\rm SR}}{1+2\sqrt{R_{\rm SRM}\cos 2\phi_{\rm SR}+R_{\rm SRM}}}\,
\end{align}
\end{subequations}
with $\delta_{\rm arm}$ the differential detuning of the arms (zero for the tuned case considered here), $\phi_{\rm SR} = \omega_pl_{\rm SR}/c$ the signal-recycling cavity single-pass phase shift, $T_{\rm SRM}$ and $R_{\rm SRM}$ the signal-recycling mirror transmissivity and reflectivity. The general formulas for signal-recycled interferometer are derived in Appendix~\ref{app_ssec:FPMIgeneral}.

In the special case of $\phi_{\rm SR} = 0\,(\pi/2)$ these formulas take particularly simple form, namely $\delta = \delta_{\rm arm}$ and 
\begin{equation}
\gamma_{\rm SR\,(RSE)} = \gamma_{\rm arm}\dfrac{1\mp\sqrt{R_{\rm SRM}}}{1\pm\sqrt{R_{\rm SRM}}}
\end{equation}
where the upper signs in the numerator and denominator correspond to the so called \textit{``resonant signal recycling''} configuration, where resonant tuning of the SR cavity makes an effective bandwidth of the interferometer narrower, proportionally increasing the signal sideband amplitude in this narrow band, whereas the lower signs in the numerator and denominator give the case of \textit{``resonant sideband extraction''}, where effective bandwidth of the interferometer is increased with respect to $\gamma_{\rm arm}$ at the expense of proportional loss of signal. In section~\ref{ssec:6-3}, we discuss the ways to increase the effective bandwidth without loss of peak sensitivity. 

The approximate expressions above are obtained assuming that cavity linewidth and signal frequency are much smaller than the cavity free spectral range $FSR=c/2L$, which is known as a \textit{single-mode approximation}.  For the next generation GW detectors with longer arms where $FSR$ may be close to the detection band, one normally needs to use the exact expressions, although the effect of factor $D(\Omega)$ is usually stronger and covers up any effects of departure of the interferometer response from the ones written in the single-mode approximation.

%
%

From Eq.~\eqref{eq:QNLS_tuned_IFO} one can immediately notice that setting homodyne angle $\phi_{\rm LO}$ such that $\mathcal{K} = \cot\phi_{\rm LO}$, the  second term in the brackets vanishes, which means one evades the back-action noise this term is standing for. This is the manifestation of the principle of variational readout, first proposed in \cite{96a2eVyMa} and later generalised in \cite{02a1KiLeMaThVy} that prescribes to read out not the phase quadrature of the outgoing light where GW signal strength is maximal, rather the one that does not contain back-action noise. This technique, in an absence of loss, allows to completely get rid of the back action noise where the above match of homodyne phase to OM coupling strength could be satisfied. However, since $\mathcal{K}_{\rm MI}$ is strongly frequency dependent, the total back action cancellation is only possible at a single frequency, as demonstrated by a series of thin dash-dotted traces in Fig.~\ref{fig:MI_QNLS}a with an envelope of these curves being the quantum shot noise-limited sensitivity. We show in Sec.~\ref{sec:3} a fundamental relation of this shot noise-limited sensitivity and variational readout concept to the \textit{fundamental quantum limit} for precision interferometry.

\subsection{Quantum back-action and ponderomotive squeezing}
\label{ssec:1-3}

The optical transfer matrix \eqref{eq:OTM_MI} allows an interesting interpretation from the point of view of the quantum state of the outgoing light. As shown in \cite{02a1KiLeMaThVy}, the optomechanical  transfer matrix \eqref{eq:OTM_MI} can be interpreted as a transformation of the phase space amounting to a sequence of rotations and squeezing. They showed that the initial quantum state $\ket{in}$ of the vacuum fields entering the readout port of the interferometer light gets ponderomotively squeezed and rotated by the radiation pressure effects embodied by the off-diagonal term in the transfer matrix in \eqref{eq:I/O_KLMTV}:
\begin{equation}\label{eq:gen_pond_sqz_state}
  \ket{out} = e^{2i\beta}\hat{R}(u_{\rm pond})\hat{S}(r_{\rm pond})\hat{R}(v_{\rm pond})\ket{in},
\end{equation}
where $\hat{R}(\alpha)$ is a rotation operator and $\hat{S}(r)$ is a squeezing operator, defined, \textit{e.g} in Section~3.2 of \cite{Liv.Rv.Rel.15.2012}.
Mathematically this means that transfer matrix $\tq{T}$ can be represented, using singular value decomposition, as the following product\footnote{In fact, the symplectic nature of $\tq{T}$ requires a more restrictive Bloch-Messiah Decomposition \cite{2016_Phys.Rev.A.94.062109_Bloch-Messiah_Decomposition} that ensures singular values which include their own reciprocals.}:
\begin{equation}\label{app.eq: pond sqz matr transformation}
  \vb{b} = \tq{T}\,\vq{a} = e^{2i\beta}\,\mathbb{R}[u_{\rm pond}]\,\mathbb{S}[r_{\rm pond}]\,\mathbb{R}[v_{\rm pond}]\,\vq{a}\,,
\end{equation}
with $\tq{R}$ the rotation matrix and $\tq{S}$ the squeezing matrix that are defined as:
\begin{align}\label{eq:Rot+Sqz}
 \tq{R}[\phi] &= \begin{bmatrix}
  \cos\phi & -\sin\phi\\
  \sin\phi & \cos\phi
  \end{bmatrix}\,, &
  \tq{S}[r] &= \begin{bmatrix}
  e^r & 0\\
  0 & e^{-r}
  \end{bmatrix}\,.
\end{align}
In a tuned case, transformed quantum state at the output port of the interferometer is described by the two numbers - ponderomotive squeezing factor $r_{\rm pond}$ and squeezing angle, $u_{\rm pond}$, that are expressed in terms of $\mathcal{K}$ as follows:
\begin{equation}\label{eq:r_and_phi_pond}
  e^{r_{\rm pond}} = \sqrt{1+\Bigl(\frac{\mathcal{K}}{2}\Bigr)^2}+\frac{\mathcal{K}}{2}\,,\quad
  u_{\rm pond} = \frac\pi2+v_{\rm pond} =  -\frac12\arctan\frac{\mathcal{K}}{2}-\frac\pi4\,.
\end{equation}

\begin{figure}[htbp]
\begin{center}
\includegraphics[width=\textwidth]{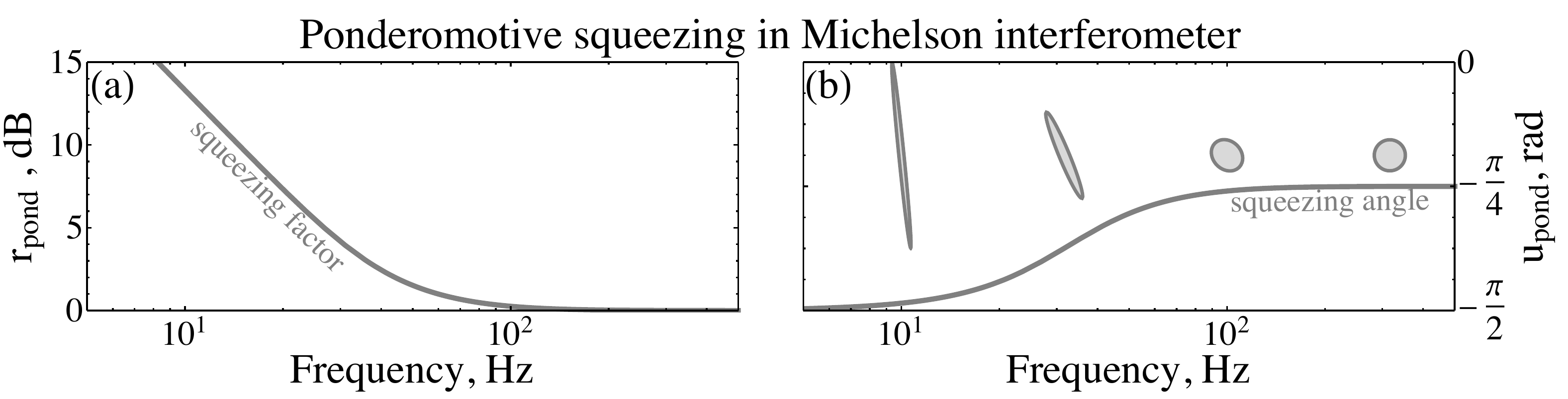}
\caption{Ponderomotive squeezing in the Michelson interferometer. \textit{Left panel} shows dependence of squeezing factor, $r_{\rm pond}(\Omega)$, on signal frequency, and \textit{right panel} shows how the squeezing angle, $u_{\rm pond}(\Omega)$ depends on signal frequency. The noise ellipses at different signal frequencies are shown (not to scale!) to illustrate the effect that interferometer with movable mirrors imposes on the quantum state of the outgoing light.} 
\label{fig:MI_PondSqz}
\end{center}
\end{figure}

Ponderomotive squeezing is the direct consequence of quantum back action, since it is through this non-linear mechanism amplitude fluctuations of light are transformed into the additional fluctuations of phase with the frequency dependent gain given by the OM coupling factor $\mathcal{K}$. Understanding quantum back-action in terms of squeezing of the state of light leaving the interferometer comes very useful when one tries to figure out why one needs frequency dependent squeezing injection to achieve broadband quantum noise suppression, and why injection of phase-squeezed light in the readout port does not suffice. We discuss these topics in Section~\ref{sec:3}. One can also gain additional understanding of noise transformations in more complicated schemes, like, \textit{e.g.}, the scheme of the EPR-speed meter that we consider in Section~\ref{sec:4}.

In Appendix~\ref{app:pond_sqz}, we consider a more general case of a detuned interferometer and derive general formulas for ponderomotive squeezing.

\subsection{Losses and imperfections}\label{ssec:2-6} 
In a real experiment, the idealised situation where the interferometer can be described solely by the I/O-relations \eqref{eq:IOlossless} with one input and one output channel can never work. According to the Fluctuation-Dissipation Theorem of Callen and Welton \cite{PhysRev.83.34}, in a lossy system, there are always additional channels through which a part of the signal-carrying light field leaves the interferometer unobserved, while the incoherent vacuum fields from the environment enter and admix with the non-classical light travelling through the interferometer, thereby curtailing quantum correlations contained therein and increasing noise. Generally, there are many places in the interferometer where loss can occur and therefore, there are many loss channels and vacuum fields associated with them.

These vacuum fields propagate through the interfrometer and couple to the readout channel very similar to the input field $\vq{a}$ with the only difference in the frequency dependence of the optical transfer matrix $\tq{N}_k$ that reflects the fact that the optical path of loss vacuum fields differs from that of $\vq{a}$ (see, \textit{e.g.}, treatment of a lossy Fabry-Perot-Michelson interferometer in Appendix~\ref{app:FPMI_lossy}). 

Thus one can describe the lossy interferometer as the multiple input/output device outlined above save to the fact that the loss channels are not measured, and the corresponding information is thereby lost. Input fields of a lossy interferometer thus can be written as:
\begin{equation*}
\vq{a} \equiv \{a_{c}^{(1)},\,a_{s}^{(1)},\,n_{c}^{(2)},\,n_{s}^{(2)},\,\ldots n_{c}^{(i)},\,n_{s}^{(i)},\,\ldots n_{c}^{(N)},\,n_{s}^{(N)}\}^{\rm T}\ \mathrm{with}\ i=\{1,N\}
\end{equation*}
and the corresponding transfer matrix reads:
\begin{align}
\tq{T}_{loss} \equiv 
\begin{bmatrix}
\tq{T} & \cdots & \tq{N}^{(1j)} & \cdots & \tq{N}^{(1N)}\\
\vdots & \ddots & \vdots &  & \vdots\\
\tq{N}^{(i1)} & \cdots & \tq{N}^{(ij)} & \cdots & \tq{N}^{(iN)}\\
\vdots &  & \vdots & \ddots & \vdots\\
\tq{N}^{(N1)} & \cdots & \tq{N}^{(Nj)} & \cdots & \tq{N}^{(NN)}
\end{bmatrix}\,.
\end{align}
As only one channel of the interferometer is measured, all the rows of the above transfer matrix but the first two (recall that $\tq{T}$ is a $2\times2$-matrix) are irrelevant. Hence the corresponding general expression for total quantum noise PSD of a lossy interferometer reads:
\begin{equation}\label{eq:SpDens_h_loss_gen}
  S^h_{\rm PD\,loss}(\Omega) = h^2_{\rm SQL}\frac{\vs{H}^{\rm T}_{\phi_{\rm LO}} \cdot 
    \left[ \tq{T}\cdot\tq{S}^{in}_{a}\cdot\tq{T}^\dag+ \sum_{k=2}^{N}\tq{N}^{(1k)}\cdot(\tq{N}^{(1k)})^{\dag}\right]\cdot\vs{H}_{\phi_{\rm LO}}}{|\vs{H}^{\rm T}_{\phi_{\rm LO}}\cdot\vs{t}_h|^2} \, ,
\end{equation}

The exact frequency dependence of the loss-related transfer matrices $\tq{N}^{(1i)}$ depends on the location of the element of the interferometer, where loss originates from. This means that the optical path of a specific loss-related vacuum field $\vq{n}^{(i)}$ cannot be generalised. Below we consider several most common sources of loss and describe how they enter the final expression for the quantum noise PSD, which allows to categorise loss into a few types in regard to their place of origin.

As for the imperfections, by which we mean here departure of the parameters of key components of the interferometer from the assumed uniformity (\textit{e.g.}, perfect overlap of the signal and local oscillator beams, perfect mode matching on the beam splitter \textit{etc.}) and symmetry (\textit{e.g.} perfect 50/50 beam splitting ratio, equal mass of all test masses, equal length/tuning of the arms, equal absorption and photon loss in the arms \textit{etc.}), it is hard to give a general recipe how to account for their influence on quantum noise. However, these studies are crucial for the design of the next generation GW interferometers, and there are several studies that attempted rigorous treatment of imperfections for selected configurations. Nonideal FPMI with frequency dependent squeezing injection (see Sec.~\ref{sec:3}) was studied in \cite{Miao14}. An in-depth comparison of FPMI and Sagnac speed meters (see Sec.~\ref{sec:4} with account for imperfections was done in \cite{2015arXiv150301062V}. Influence of imperfections on Sagnac speed meter performance was the topic of \cite{2015_NJP17.043031_asymSag}. Impact of optical path stability and mode matching in balanced homodyne readout was the topic of \cite{2015_PhysRevD.92.072009_Steinlechner,2017_PhysRevD.95.062001}.

\subsubsection{Losses in the readout train.}\label{ssec:loss} 
The sources of loss in the readout train are quite diverse, ranging from non-unity quantum efficiency of the photodiodes to the imperfect mode matching of the local oscillator beam with the signal beam in the balanced homodyne detector \cite{2017_PhysRevD.95.062001}. In most cases loss may be reduced to a single, frequency independent coefficient of an effective quantum efficiency, $\eta_d = 1 - \epsilon_d < 1$, where $\epsilon_d < 1$ can be thought of as a fractional photon loss at the photodetector \cite{02a1KiLeMaThVy,Miao14,Liv.Rv.Rel.15.2012}. Frequency dependence can be safely omitted here, for any resonant optical element in the readout train, including output mode cleaners (OMC), has bandwidth much larger than the detection band of the main interferometer. 

The expression \eqref{eq:o_zeta_lossless} for an output observable of the GW interferometer is modified in the presence of readout losses as follows:
\begin{multline}\label{eq:o_zeta_lossy}
  \hat{o}_{\phi_{\rm LO}}^{\rm loss} \equiv
    \sqrt{1-\epsilon_d} \left(\hat{b}_c\cos\phi_{\rm LO} + \hat{b}{}_s\sin\phi_{\rm LO}\right) \\
    + \sqrt{\epsilon_d} \left(\hat{n}_{d;\;c}\cos\phi_{\rm LO} + \hat{n}_{d;\;s}\sin\phi_{\rm LO}\right) \\
  \equiv
    \sqrt{1-\epsilon_d}\,\vs{H}_{\phi_{\rm LO}}^{\rm T}\cdot\vq{b}+\sqrt{\epsilon_d}\,\vs{H}_{\phi_{\rm LO}}^{\rm T}\cdot\vq{n}{}_d \, ,
\end{multline}
where $\vq{n}{}_d = \{\hat{n}_{d;\;c},\,\hat{n}_{d;\;s}\}^{\rm T}$ stands quadrature vector of loss-associated vacuum fields with unity spectral density matrix.

Spectral density formula \eqref{eq:SpDens_h} in lossy readout case will read:
\begin{equation}\label{eq:SpDens_h_loss_PD}
  S^h_{\rm PD\,loss}(\Omega) = h^2_{\rm SQL}\frac{\vs{H}^{\rm T}_{\phi_{\rm LO}} \cdot 
    \left[ \tq{T}\cdot\tq{S}^{in}_{a}\cdot\tq{T}^\dag+\xi^2_d \right]\cdot\vs{H}_{\phi_{\rm LO}}}{|\vs{H}^{\rm T}_{\phi_{\rm LO}}\cdot\vs{t}_h|^2} \, ,
\end{equation}
where $\xi_d = \sqrt{\epsilon_d/(1-\epsilon_d)}$.


%

\subsubsection{Optical loss in the arms and in filter cavities.}

Optical loss in Fabry-Perot cavities, such as arm cavities and filter cavities, is known to have frequency dependence with the major impact at low sideband frequencies within the cavity optical bandwidth. A  very illuminating discussion on this subject is given in \cite{Miao14} where optical loss in filter cavities is studied in detail. The main source of such loss in large suspended cavities is the scattering of light off the mirror surface imperfections of microscopic (micro-roughness) and relatively macroscopic ("figure error") size \cite{2013_OE.21.30114_Loss_in_FC_Isogai,Miao14}. 

In general, optical loss in the cavity depends on the cavity length in an involved way (see, \textit{e.g.}, Appendix C in \cite{Miao14}). However, if we consider a cavity 
of a fixed length the single value of total photon loss per metre ($\epsilon_f $ in ppm/m) will fully define the total optical loss and the conventional description of Fabry-Perot cavity with one lossy mirror (usually, an ETM one) and another lossless one (an ITM, respectively), works perfectly fine. A detailed derivation of lossy cavity I/O-relation is given in Appendix~\ref{app:lossy FP I/O-rels}. Here we only present its general form which reads:
\begin{equation}\label{eq:lossyFP}
  \vq[arm]{b} = \tq[arm]{T}\vq[arm]{a} + \tq[arm]{N}\vq[arm]{n} + \vs[arm]{t}\frac{h}{h_{\rm SQL}}.
\end{equation}
with $\tq[arm]{T} = \tq[arm]{T}^{\rm s.n.}+\tq[arm]{T}^{\rm b.a.}$ a transfer matrix for input fields, $\vq[arm]{a}$,  $\tq[arm]{N} = \tq[arm]{N}^{\rm s.n.}+\tq[arm]{N}^{\rm b.a.}$  is a transfer matrix for loss-associated vacuum fields, $\vq[arm]{n}$, and $\vs[arm]{t}$ is an optomechanical response function of the cavity defined by Eq.~\eqref{eq:t_h_def}. We wrote transfer matrices $\tq[arm]{T} $ and $\tq[arm]{N} $ as sums of shot-noise component, $\tq[arm]{T}^{\rm s.n.} (\tq[arm]{N}^{\rm s.n.})$, and back-action component, $\tq[arm]{T}^{\rm b.a.} (\tq[arm]{N}^{\rm b.a.})$ (cf. Eqs.~\eqref{eq_app:FP_Tsn} and \eqref{eq_app:FP_Nsn}), to discern Fabry-Perot cavities with strong classical carrier light circulating inside, as in the arms, from the ones with no, or very weak classical light inside, as in the filter cavity. In the latter case, the back-action components can be set to zero, as well as the optomechanical response function $\vs[arm]{t}=0$.
%
%


\section{Quantum limits} 
\label{sec:2}
\subsection{Standard Quantum Limit}

The Standard Quantum Limit (SQL) was firstly pointed out by Braginsky when 
studying the quantum limit of continuous position measurements\,\cite{92BookBrKh}. 
In the context of laser interferometric GW detectors, it constraints the
detector sensitivity in the nominal operation mode 
with tuned optical cavities and phase quadrature measurement.  
It comes from a trade-off between the shot noise and radiation pressure 
noise---the former is inversely proportional to the optical power while the latter is 
proportional to the power. There is an optimal power for achieving the 
maximum sensitivity at each frequency which defines the SQL. 

\begin{figure}[b]
\begin{center}
\includegraphics[width=0.52\textwidth]{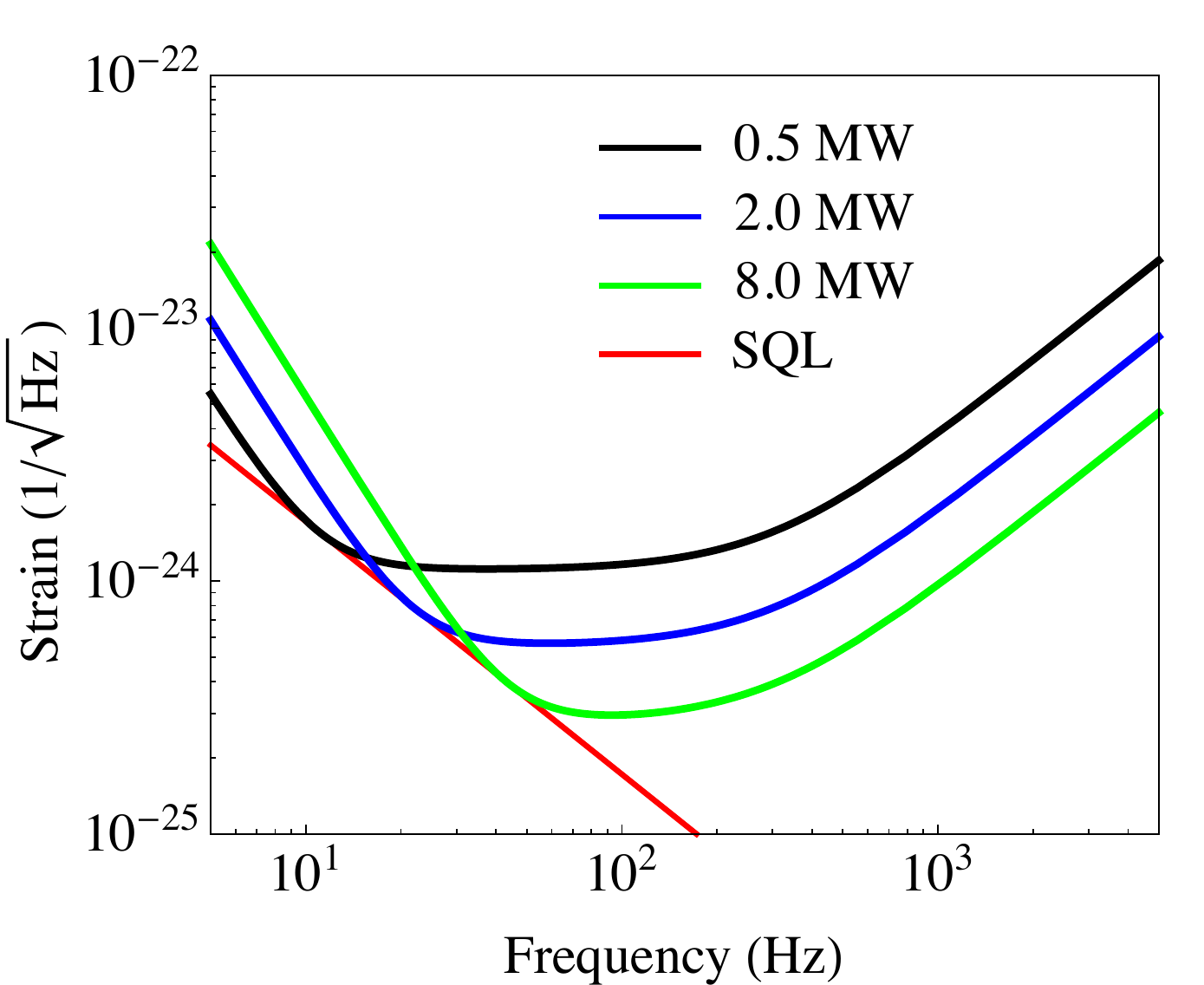}
\caption{Plot showing the
SQL for a tuned dual-recycled Michelson 
interferometer. It is defined as the locus of those points 
where the shot noise is equal to the 
radiation pressure noise at different powers. }
\label{fig:SQL}
\end{center}
\end{figure}

We can derive the SQL explicitly for a tuned dual-recycled 
Michelson interferometer by using Eq.\,\eqref{eq:QNLS_phase_quad}. 
In particular focusing on frequencies lower than 
the free spectral range $c/(2L)$, we have $D\approx 1$ and 
\begin{equation}\label{eq:hSQL}
S^h = \left(\frac{1}{\cal K} + {\cal K}\right)\frac{h_{\rm SQL}^2}{2} \ge h_{\rm SQL}^2
\equiv S^h_{\rm SQL}\,.
\end{equation}
The SQL is defined as 
\begin{equation}\label{eq:SQL}
h_{\rm SQL} =\sqrt{ \frac{8\hbar}{M \Omega^2 L^2}} \approx 2.0\times 10^{-25}\,{\rm Hz}^{-\frac12}\left(\frac{200\,{\rm kg}}{M}\right)^{\frac12}\left(\frac{100\,{\rm Hz}}{\Omega/2\pi}\right)
\left(\frac{20\,{\rm km}}{L}\right) \,.
\end{equation}
In Fig.\,\ref{fig:SQL}, we show the quantum noise curves for different arm cavity 
powers and the SQL. 

The SQL does not just apply to laser interferometric GW detectors, but to general 
linear continuous measurements with a test mass. 
Generally, according to Ref.~\cite{92BookBrKh,Chen2,Chen2013}, 
the output of a linear displacement measurement device 
can be written as
\begin{equation}
\hat Z (\Omega) = \hat Z^{(0)}(\Omega)+\frac{\chi_{ZF}(\Omega) }{1-\chi_m(\Omega)\chi_{FF}(\Omega)} [\chi_m(\Omega)\hat F^{(0)}(\Omega) +x_{\rm sig}(\Omega)] \,.
\end{equation}
Here $\hat Z^{(0)}$ denotes the intrinsic fluctuation of the output port, $\hat F$ 
is the degree of freedom coupled to the probe mass displacement $\hat x$ 
with $\chi_{FF}$ being 
its susceptibility, $\chi_m$ is the mechanical susceptibility of the test mass, and $x_{\rm sig}$
is some displacement signal.  
For an ideal quantum-limited device, the spectral density 
for $\hat Z^{(0)}$ and $\hat F^{(0)}$ satisfies the following Heisenberg 
relation\,\cite{92BookBrKh,Chen2,Miao2017}:  
\begin{equation}\label{eq:uncertainty_relation}
S_{ZZ}(\Omega) S_{FF}(\Omega) - |S_{ZF}(\Omega)|^2 \geqslant \hbar^2 |\chi_{ZF}(\Omega)|^2\,.
\end{equation}
In the special case where $\chi_{FF}=0$ and there is no correlation between 
$\hat Z^{(0)}$ and $\hat F^{(0)}$, i.e. $S_{ZF}=0$, the signal-referred noise 
spectral density is bounded by the general SQL:
\begin{equation}\label{eq:SQL}
S^x(\Omega) = \frac{S_{ZZ}(\Omega)}{|\chi_{ZF}(\Omega)|^2}  
+|\chi_m(\Omega)|^2 S_{FF}(\Omega)\ge 2\hbar |\chi_m(\Omega)|\equiv S^x_{\rm SQL}(\Omega)\,, 
\end{equation}
where we have used $a+b\ge 2\sqrt{ab}$ and $S_{ZZ}S_{FF}=\hbar^2 |\chi_{ZF}|^2$ and $\hat x(\Omega) = \hat Z(\Omega)/\bigl[d\hat Z/dx_{\rm sig}|_{x_{\rm sig}=0}\bigr]$. 

Applying to the Michelson interferometer,  we have 
\begin{equation}
\hat Z^{(0)} = e^{2i\beta}\hat a_2\,,\quad \hat F^{(0)} = \hbar \,\chi_{ZF}\,\hat a_1\,,\quad x_{\rm sig} = L\, h/2\,,
\end{equation}
and the susceptibilities are $\chi_{ZF} = {e^{i\beta}2\sqrt{{2\cal K}}}/({L\,h_{\rm SQL}})$,
$\chi_m = -1/{(m\Omega^2)}$, and $\chi_{FF}=0$. Therefore, 
in the context of laser interferometer, $\hat Z^{(0)}$ 
introduces the shot noise, while $\hat F^{(0)}$ is responsible for the radiation 
pressure noise (quantum backaction). Eq.\,\eqref{eq:hSQL} is simply a special 
case of Eq.\,\eqref{eq:SQL} when normalising to the GW strain. 

\subsection{Fundamental Quantum Limit}

The Fundamental Quantum Limit (FQL) is a sensitivity limit that is more
stringent than the SQL for a given interferometer configuration.
It is also called the Energetic Quantum
Limit\,\cite{00p1BrGoKhTh} or Quantum Cram\'er-Rao
Bound\,\cite{Tsang2011,Miao2017b} in quantum metrology.
In the context of laser interferometric gravitational-wave detectors, 
it can be written as:
\begin{equation}\label{eq:FQL_new}
S^h_{\rm FQL}(\Omega)=\frac{\hbar^2c^2}{S_{PP}(\Omega)L^2} = 
\frac{4\hbar^2}{S_{\cal E E}(\Omega)}\,. 
\end{equation}
Here $S_{PP}$ is the single-sided quantum noise
spectral density for the optical power $P$ inside the arm cavity and 
$S_{\cal EE} = 4 S_{PP} L^2/c^2$ is the energy spectrum. This
means a good sensitivity requires a high fluctuation of the power, or energy,
in the quantum regime---a large energy fluctuation is needed to probe
the spacetime precisely, which is directly related to the energy-time uncertainty 
relation. This is a very beautiful formula involving energy, spacetime, and $\hbar$. 

One point worthy
emphasising is that SQL is the locus of a family of sensitivity curves 
at different power, while the FQL is a sensitivity limit
at different frequencies for a given configuration with fixed parameters including
the power. We can derive the FQL using the same linear-measurement formalism 
for deriving the SQL mentioned above\,\cite{Miao2017b}.
 The key component is the correlation 
between $\hat Z$ and $\hat F$, i.e., $S_{ZF}$. We first consider the special case 
with $\chi_{FF}=0$. When including $S_{ZF}$, Eq.\,\eqref{eq:SQL} becomes 
\begin{align}\nonumber
S^x &= \frac{S_{ZZ}}{|\chi_{ZF}|^2}  + 
2 {\rm Re}\left[\chi_m^* \frac{S_{ZF}}{\chi_{ZF}}\right]
+|\chi_m|^2 S_{FF}\\
& = \frac{\hbar^2}{S_{FF}} + \left| \frac{S_{ZF}}{\chi_{ZF}}+ \chi_m S_{FF}\right|^2
\ge \frac{\hbar^2}{S_{FF}}\,,
\end{align}
where we have used the uncertainty relation Eq.\,\eqref{eq:uncertainty_relation} 
in arriving at the second line. In the most general case with $\chi_{FF}\neq 0$, 
we just need to replace $\hat F$ by $\hat {\cal F}$ which is defined as
$\hat {\cal F}\equiv {\hat F}/{(1-\chi_m \chi_{FF})}$. 
The resulting general FQL for the displacement measurement is given by 
\begin{equation}
S^{x}_{\rm FQL} = \frac{\hbar^2}{S_{\cal FF}}\,.
\end{equation}
In laser interferometric GW detectors, $\hat {\cal F}$ corresponds
to the radiation pressure force on the test mass, which is equal to ${2 P_c}/{c}$, and 
therefore
\begin{equation}
S_{\cal FF} = 4 S_{PP}/c^2\,.
\end{equation}
Converting the FQL for the displacement measurement to that for the strain, 
we obtain Eq.\,\eqref{eq:FQL_new} as the outcome. 
Achieving the FQL 
requires $S_{Z{\cal F}}$ to be equal to $-\chi_m \chi_{Z{\cal F}} S_{\cal FF}$, 
As proven in Ref.\,\cite{Miao2017b}, for $\chi_{FF}=0$,
 this can be realised by 
using the optimal frequency-dependent 
readout, which measures the optimal quadrature at different frequencies using 
the setup proposed by Kimble {\it et al.} with optical filter cavities\,\cite{02a1KiLeMaThVy}. 
For $\chi_{FF}\neq0$ or more specifically ${\rm Im}[\chi_{FF}]\neq0$, this condition  
is not exactly realisable, however, the difference between 
the FQL and the sensitivity achieved by the optimal frequency-dependent readout is 
at most a factor of two. Therefore, the FQL sets a fundamental benchmark for the 
sensitivity limit of a given configuration. 

\begin{figure}[t]
\begin{center}
\includegraphics[width=0.505\textwidth]{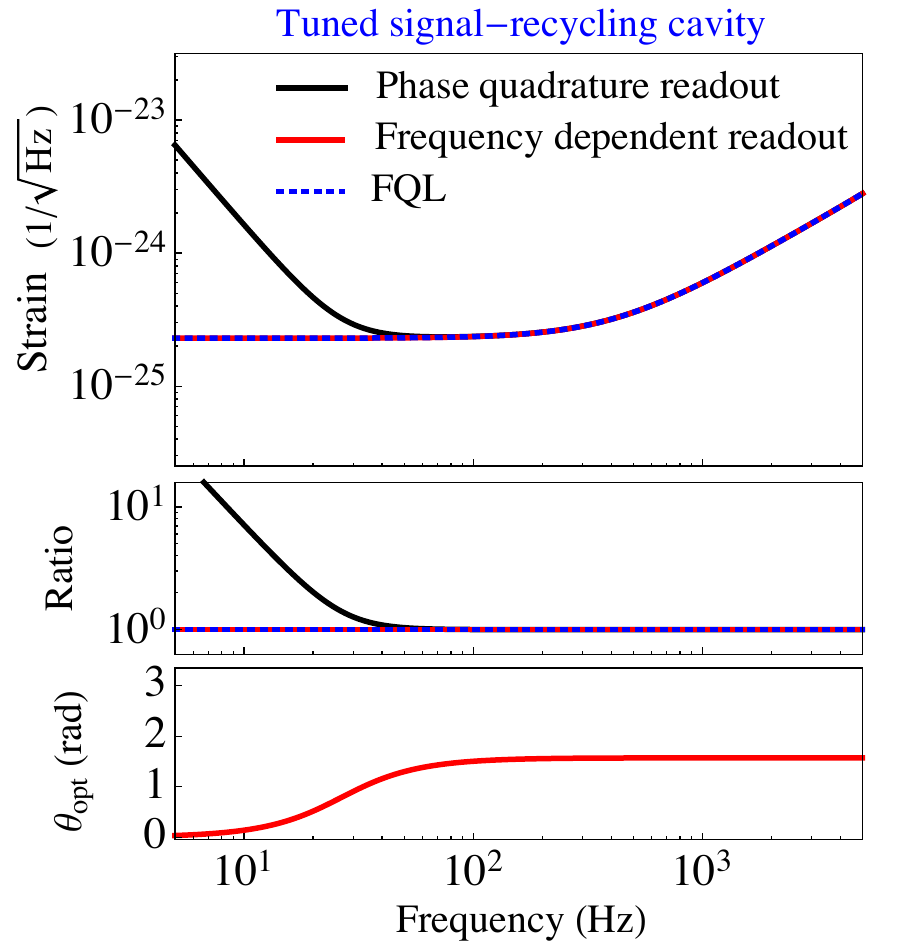}
\includegraphics[width=0.42\textwidth]{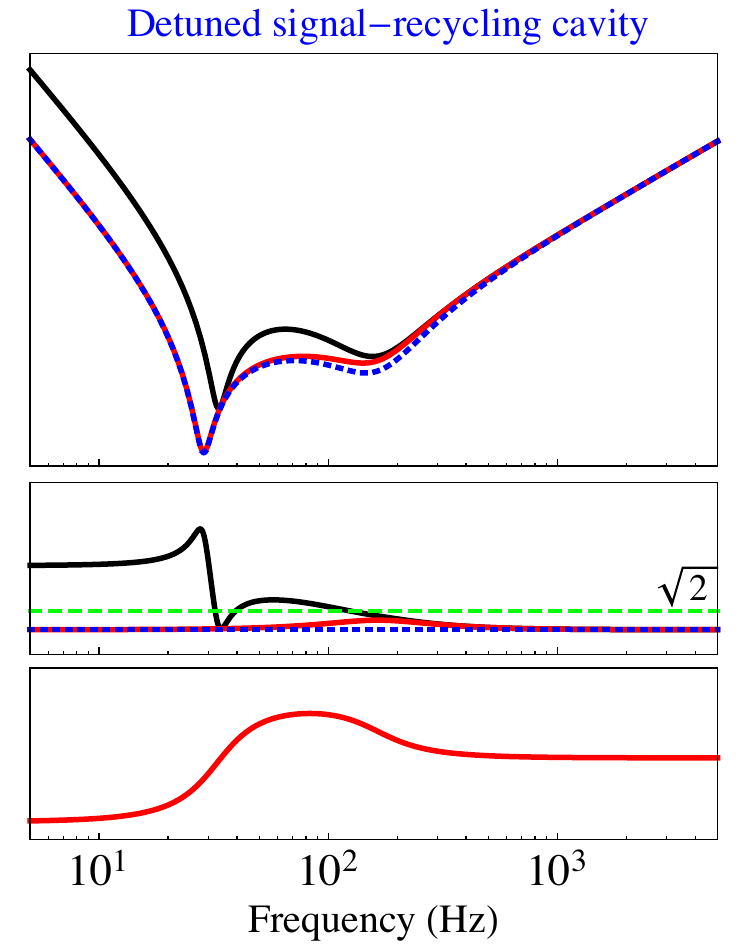}
\caption{The top panel shows the quantum noise curves
for reading out the phase quadrature, the optimal frequency
dependent readout, and the FQL. The middle panel is the ratio 
of these curves to the FQL. The bottom panel is the optimal 
frequency-dependent readout angle. The left column is the 
case of a tuned signal-recycled Michelson interferometer 
while the right one is the detuned case. 
\label{fig:FQL}}
\end{center}
\end{figure}

Again using the tuned dual-recycled Michelson 
interferometer as an example, the power fluctuation 
inside the arm cavity is given by  
\begin{equation}
S_{PP}(\Omega)= \frac{2 c P_c\hbar \gamma\omega_0}{L(\gamma^2+\Omega^2)}\,.
\end{equation}
The resulting FQL is
\begin{equation}
S^h_{\rm FQL}(\Omega) = \frac{\hbar c (\gamma^2 +\Omega^2)}{ 2 L P_c \gamma \omega_0} = \frac{h_{\rm SQL}^2(\Omega)}{2{\cal K}_{\rm MI}}\,. 
\end{equation}
Compared with Eq.\,\eqref{eq:hSQL}, this  simply  
corresponds to the shot-noise only sensitivity 
without contribution from the radiation pressure noise. Indeed, we know that 
such a sensitivity is achievable using the optimal frequency-dependent readout in the lossless case\,\cite{02a1KiLeMaThVy}, as mentioned earlier. A similar result applies to the speed meter configuration 
that we will discuss in Sec.\,\ref{sec:4} by replacing ${\cal K}_{\rm MI}$ with the corresponding 
optomechanical coupling strength ${\cal K}_{\rm SM}$ for a speed meter. The only difference is that ${\cal K}_{\rm SM}$ is approximately constant (frequency independent) at low frequencies, and, therefore, 
a constant quadrature readout is sufficient to reach the FQL at those frequencies.  

In Fig.\,\ref{fig:FQL}, we illustrate the FQL for both the 
tuned and detuned dual-recycled Michelson interferometer, together with the 
quantum noise curves for constant phase quadrature readout and the 
optimal frequency-dependent readout. In the tuned case, the sensitivity 
with the optimal readout is identical to the FQL. In the detuned case, however, 
they overlap for most of the frequencies, but not at the detune frequency. The 
difference is less than $\sqrt{2}$ in amplitude (a factor of two in power), which matches 
the general theorem in Ref.\,\cite{Miao2017b}. 

\begin{figure}[t]
\begin{center}
\includegraphics[width=0.8\textwidth]{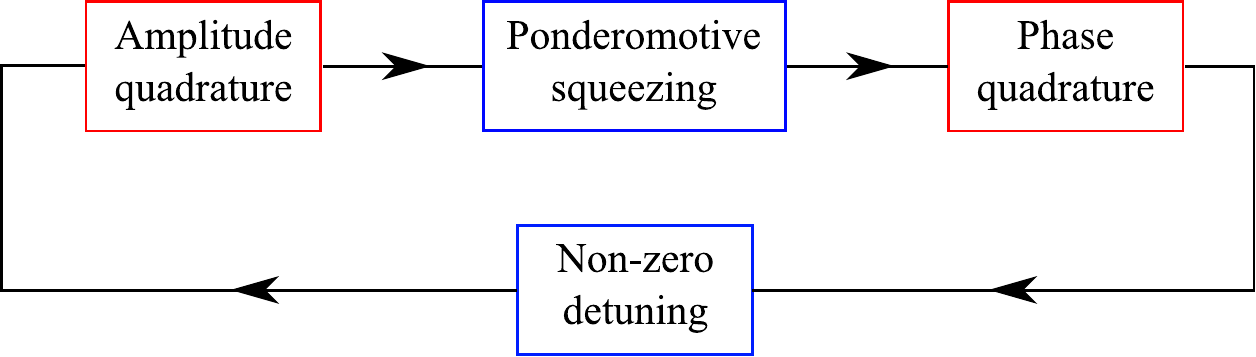}
\caption{A schematics showing the optical feedback in the detuned case of 
a dual-recycled Michelson interferometer. This is to explain
the enhancement of quantum fluctuation in the amplitude quadrature (or
equivalently the power fluctuation).
\label{fig:feedback_FQL}}
\end{center}
\end{figure}

Worthy of highlighting, there are two noticeable
dips in the FQL for the detuned case. The
high-frequency one simply coincides with the detune frequency which 
defines the optical resonance of the interferometer. 
The low-frequency one,  as discussed
by Buonanno and Chen\,\cite{Chen2}, is attributable to the 
so-called optical
spring effect which shifts the test-mass centre-of-mass frequency 
due to the position-dependent
radiation pressure. In sight of the FQL, we can provide an
alternative point of view: the sensitivity is better around such a frequency
implies that the optical power fluctuation is
significantly larger than other frequencies, according to Eq.\,\eqref{eq:FQL_new}.
It can be explained using the positive feedback as illustrated in Fig.\,\ref{fig:feedback_FQL}. The
quantum fluctuation in the amplitude quadrature is converted into that of phase quadrature
due to the ponderomotive squeezing (amplification) effect. In the presence of 
non-zero detuning and the signal-recycling mirror, 
the phase quadrature fluctuation is feeding back to the
amplitude one. With the round-trip feedback gain approaching unity, the amplitude
quadrature fluctuation, or equivalently the power fluctuation, is significantly enhanced
and leads to the dip in the sensitivity curve that we observe. 

The above insight provides a new perspective on how the arm cavity power
fluctuation can be enhanced, i.e.,
achieving a better sensitivity. In addition to increasing power or external squeezing injection,
we can also take advantage of the internal ponderomotive squeezing. If we can
insert proper optical filters such that the simple detuning in Fig.\,\ref{fig:feedback_FQL}
is replaced by more sophisticated feedback, we could achieve a broadband resonant enhancement of the power fluctuation. We can,
therefore, combine different techniques in a coherent way to optimise the 
the signal-to-noise ratio (SNR) for the signal of interest:  
\begin{equation}
{\rm SNR}^2_{\rm FQL}=
\int \frac{{\rm d}\Omega}{2\pi}
\frac{|h_{\rm sig}(\Omega)|^2}{S^h_{\rm FQL}(\Omega)}= \frac{2 L^2}{\hbar^2 c^2} \int  \frac{{\rm d}\Omega}{2\pi}  |h_{\rm sig}(\Omega)|^2 S_{PP}(\Omega) \,. 
\end{equation}
We can shape the power fluctuation with different 
techniques such that it has a good spectral 
overlap with the signal, which implies a high SNR according to the above formula. 
This idea is now under study in the 
GW community. 
The only limitation to this idea comes from the optical loss 
which could set a more stringent bound
if the FQL is made sufficiently low\,\cite{Miao2017}. Coming up with schemes
with low FQL and robust against is one of the challenges.

\section{Interferometers using non-classical light} 
\label{sec:3}
\subsection{Squeezed vacuum injection}

\begin{figure}[b]
\centering  
\includegraphics[width=.45\textwidth]{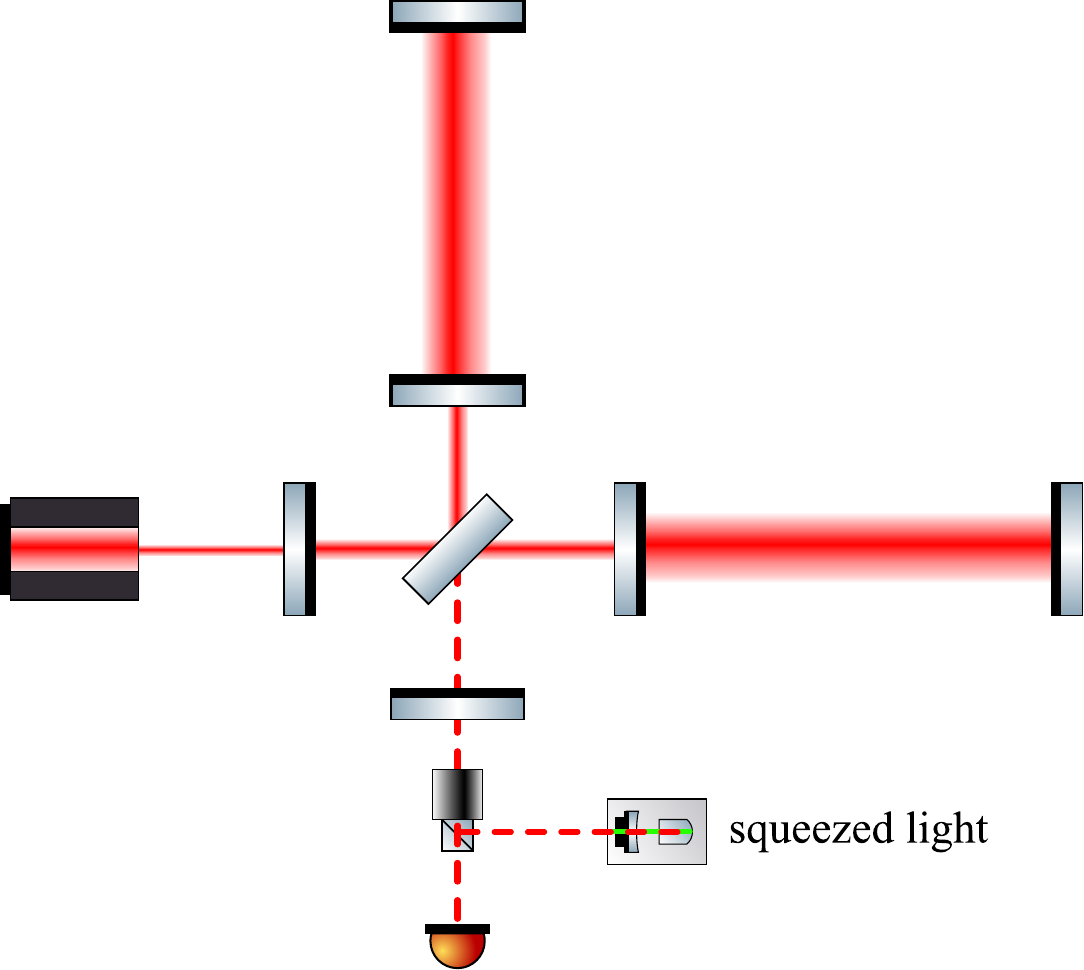}
\includegraphics[width=.51\textwidth]{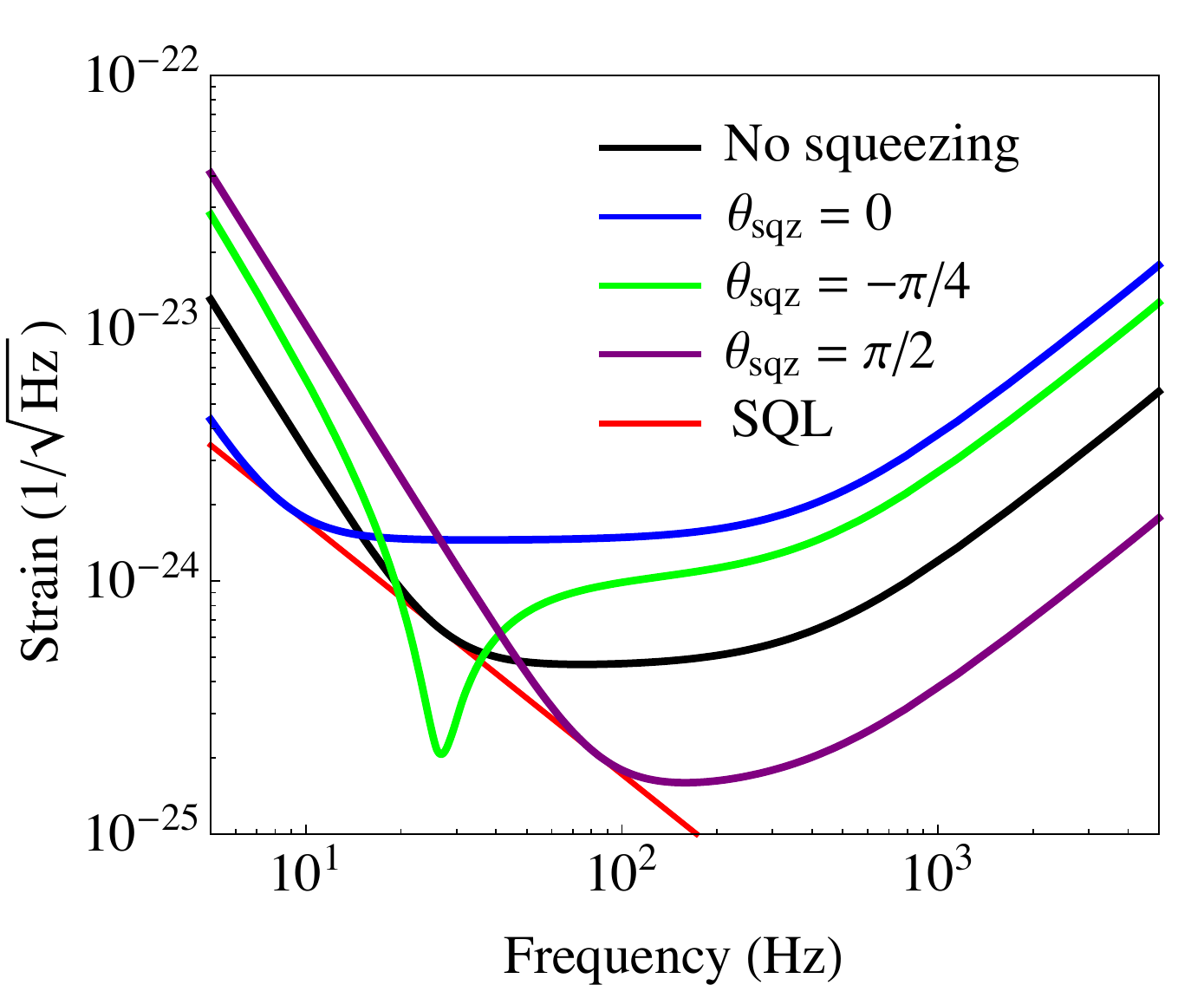}
\caption{Sketch of a laser interferometric GW detector with squeezed 
light injection (left) and the quantum-noise curve $\sqrt{S_{hh}}$
for different squeezing angles (right). The SQL is shown as a reference.}
\label{fig:freq_indep_sqz} 
\end{figure}

One approach to reducing quantum noise is using the non-classical state of light---the
squeezed vacuum state, which is produced by non-linear optical processes mentioned
earlier. This approach is originally proposed by Caves when analysing the quantum 
limit of laser interferometers\,\cite{1981_PRD.23.1693_Caves}. 
The basic setup is shown schematically 
in Fig.\,\ref{fig:freq_indep_sqz}. The squeezed light is injected into the dark port 
of the interferometer using an optical isolator (circulator). After several pioneering 
experimental works on the generation of squeezed state, it
has been successfully demonstrated in GEO 600\,\cite{2011_Nat.Phys.7.962_LSC}, and LIGO\,\cite{Aasi2013NatPhot} for reducing the high-frequency
shot noise (see a recent review article by Schnabel\,\cite{Schnabel2017}). 

The detector sensitivity with squeezed light depends on the 
the squeezing quadrature (angle). The latter is determined by the relative phase
between the carrier of the main interferometer and the pump field which 
produces the squeezed light. We again use
the tuned dual-recycled Michelson interferometer as an example. Assuming 
that we measure the output phase quadrature, the quantum noise
spectral density is 
\begin{equation}\label{eq:Shh_tuned}
S^h(\Omega)=\frac{h_{\rm SQL}^2}{2{\cal K}_{\rm MI}} 
\left[e^{-2r_{\rm s}}(\sin\theta_{\rm s}-{\cal K}_{\rm MI} \cos\theta_{\rm s})^2+
 e^{2r_{\rm s}}({\cal K}_{\rm MI}\sin\theta_{\rm s}+\cos\theta_{\rm s})^2 
\right]\,, 
\end{equation}
where $r_{\rm s}$ is the squeezing factor and $\theta_{\rm s}$ is the 
squeezing angle. We show the resulting noise 
curves for different squeezing angles in the right panel of 
Fig.\,\ref{fig:freq_indep_sqz}. The phase squeezing with $\theta_s=0$ gives 
\begin{equation}
S^{h}(\Omega) = \frac{h_{\rm SQL}^2}{2}\left[\frac{e^{-2r_{\rm s}}}{{\cal K}_{\rm MI}} + 
e^{2r_{\rm s}} {\cal K}_{\rm MI}\right]\,,
\end{equation}
which implies that we reduce the shot-noise term at a price of increasing the 
radiation-pressure-noise term proportional to $\cal K$. To reduce the shot 
noise and the radiation pressure noise
simultaneously, the squeezing angle $\theta_s$ needs to be frequency
dependent, which will be discussed in the next section. 

\subsection{Frequency-dependent squeezing}\label{ssec:4-2}

As we have learnt from the previous section, a fixed squeezing angle only improves the sensitivity for some frequencies but not all. This is because the fluctuation in the amplitude quadrature and the phase quadrature contribute to the quantum noise differently at different frequencies. We can, therefore, optimise the sensitivity by making the squeezing angle frequency dependent.

\begin{figure}
\centering  
\includegraphics[width=.42\textwidth]{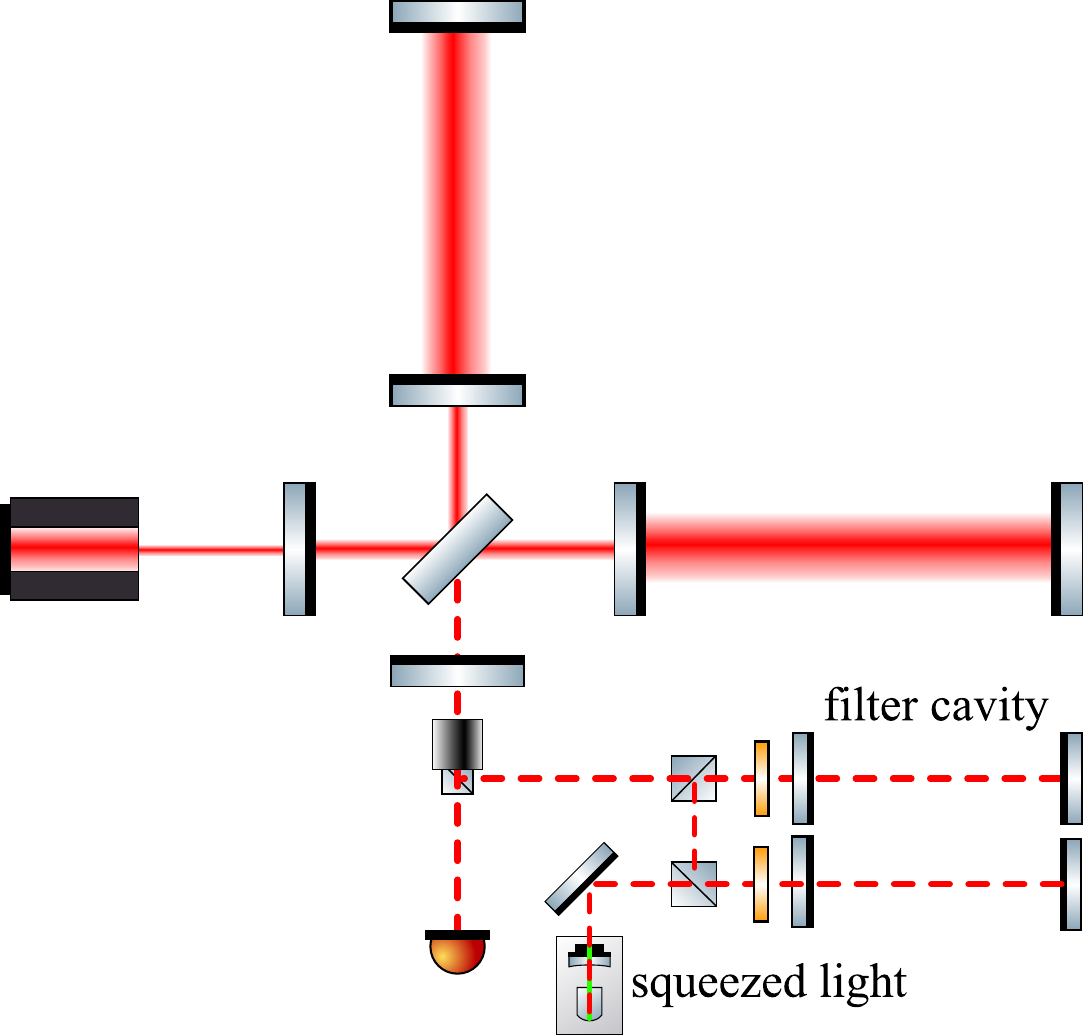}
\includegraphics[width=.53\textwidth]{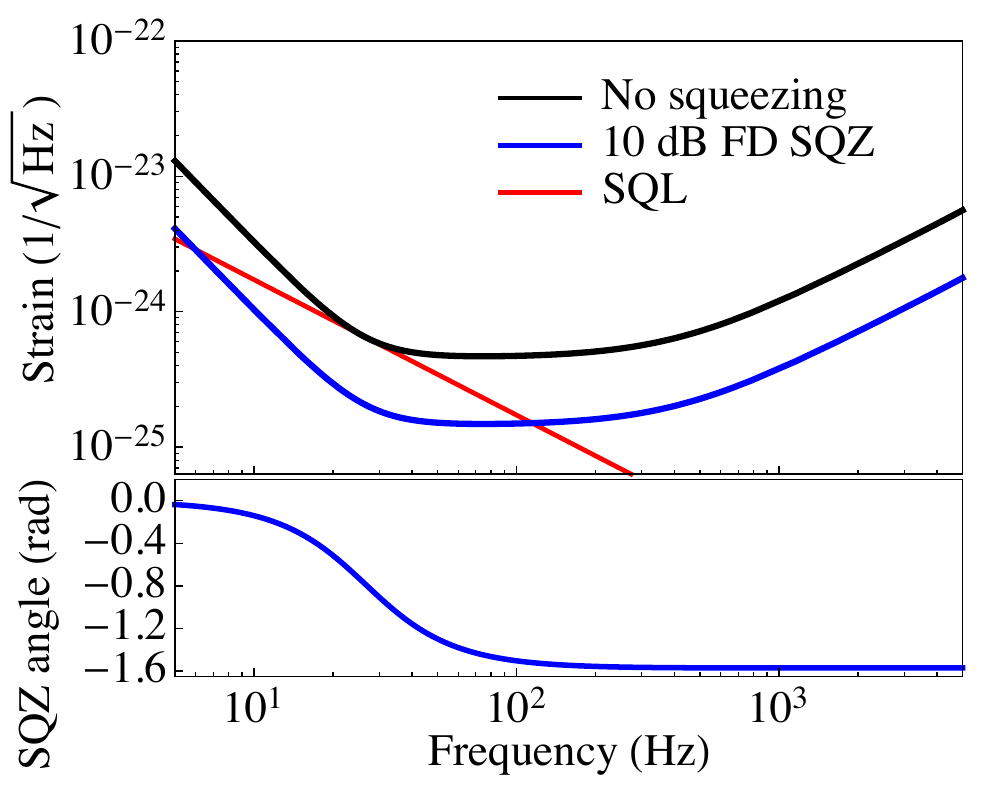}
\caption{The left panel is a schematic of 
 the dual-recycled Michelson interferometer with
10\,dB frequency-dependent (FD) squeezing. The top right shows
the quantum-noise curve with and without squeezing. The bottom 
right shows the optimal squeezing (SQZ) angle as a function of frequency.}
\label{fig:freq_dep_sqz} 
\end{figure}

Again using the tuned dual-recycled interferometer for illustration, 
the optimal frequency-dependent 
squeezing angle is equal to 
\begin{equation}\label{eq:sqz_angle_tuned}
\tan\theta_s = -1/{\cal K}_{\rm MI} \propto \Omega^{2}(\Omega^2+\gamma^2)\,
\end{equation}
such that the anti-squeezing term, proportional to $e^{2r_{s}}$, 
in Eq.\,\eqref{eq:Shh_tuned}  vanishes
and the quantum noise is reduced over the entire frequency band. The 
frequency-dependent squeezing is realised by sending the squeezed 
light through a cascade of so-called filter cavities,
which are Fabry-P\'erot cavities with proper bandwidth and detuning, 
as illustrated in Fig.\,\ref{fig:freq_dep_sqz}.

For a general detector configuration with the input-output relation given 
by Eq.\,\eqref{eq:IOlossless}, the optimal squeezing angle is determined by
\begin{equation}
\tan\theta_s(\Omega) = \frac{T_{cs}(\Omega) \cos\zeta+ 
   T_{ss}(\Omega) \sin\zeta}{T_{cc}(\Omega) \cos\zeta + T_{sc}(\Omega)\sin\zeta}\,, 
\end{equation}
where $\zeta$ is the measured quadrature angle at the output. 
It is equal to $-{\cal K}$ in the special case mentioned above. 
The number of filter cavities is determined by the order of $\Omega$ 
in the frequency dependence of $\tan\theta_s$. As shown in Ref.\,\cite{Purdue2002}, 
 if $\tan\theta_s$ is a rational function of $\Omega$ 
with the highest order equal to $\Omega^{2n}$ for its numerator and denominator, 
the number will be equal to $n$ and we can derive the bandwidth $\gamma_k$ and detuning $\Delta_k$ 
for the individual filter cavity analytically:
\begin{equation}
\frac{1+i \tan\theta_s(\Omega)}{1-i\tan\theta_s(\Omega)} = e^{2i\bar\theta} \prod_{k=1}^n 
\frac{\gamma_k+i(\Omega+\Delta_k)}{\gamma_k-i(\Omega+\Delta_k)}\frac{\gamma_k+i(-\Omega+\Delta_k)}{\gamma_k-i(-\Omega+\Delta_k)}\,, 
\end{equation}
where $\bar \theta$ defines the global constant phase of the 
filter cavity chain at $\Omega\rightarrow \infty$. However, if $\tan\theta_s$ is 
not a rational function of $\Omega$ or one wish to approximately realise $\theta_s$ 
using the number of cavities less than $n$, one can use a numerical algorithm to obtain the filter cavity parameters by fitting to the angle. The authors find that, when proper physical constraints on the parameters are imposed, using 
a minimisation routine to minimise the following cost function leads to a good 
answer: 
\begin{equation}
{\cal J}= \left\{\theta_s - \bar \theta- \sum_{k=1}^n \arctan[(\Omega+\Delta_k)/\gamma_k] + \sum_{k=1}^n\arctan[(\Omega-\Delta_k)/\gamma_k] \right\}^2\,.
\end{equation}

\begin{figure}[t]
\centering  
\includegraphics[width=.48\textwidth]{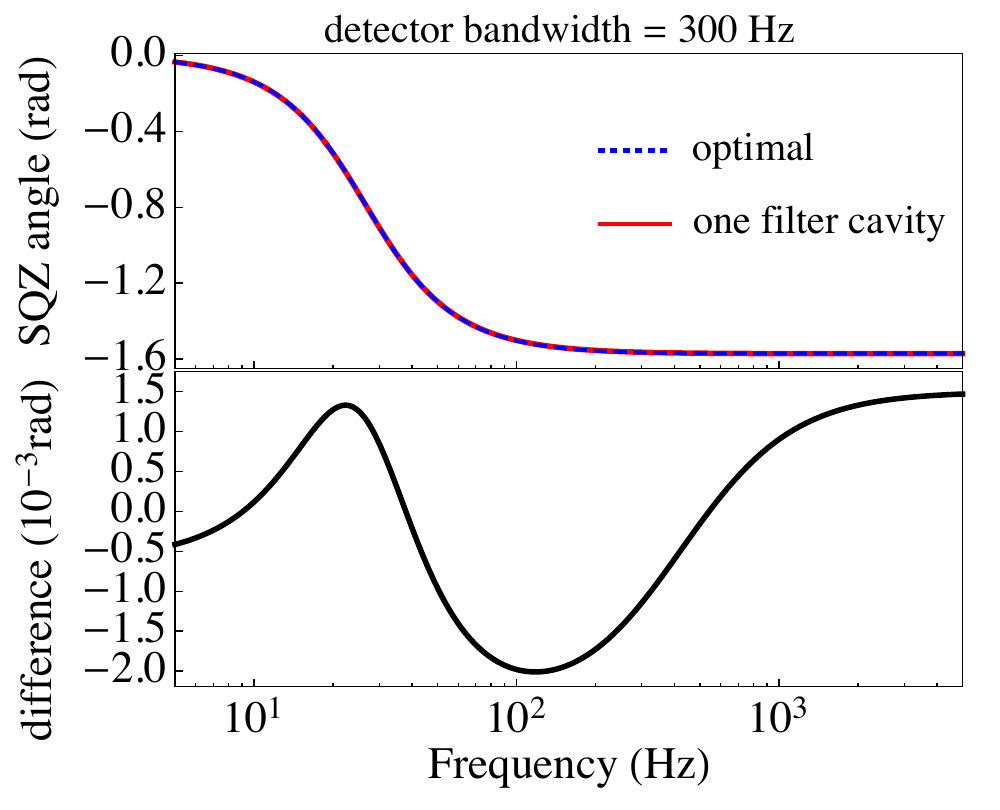}
\includegraphics[width=.459\textwidth]{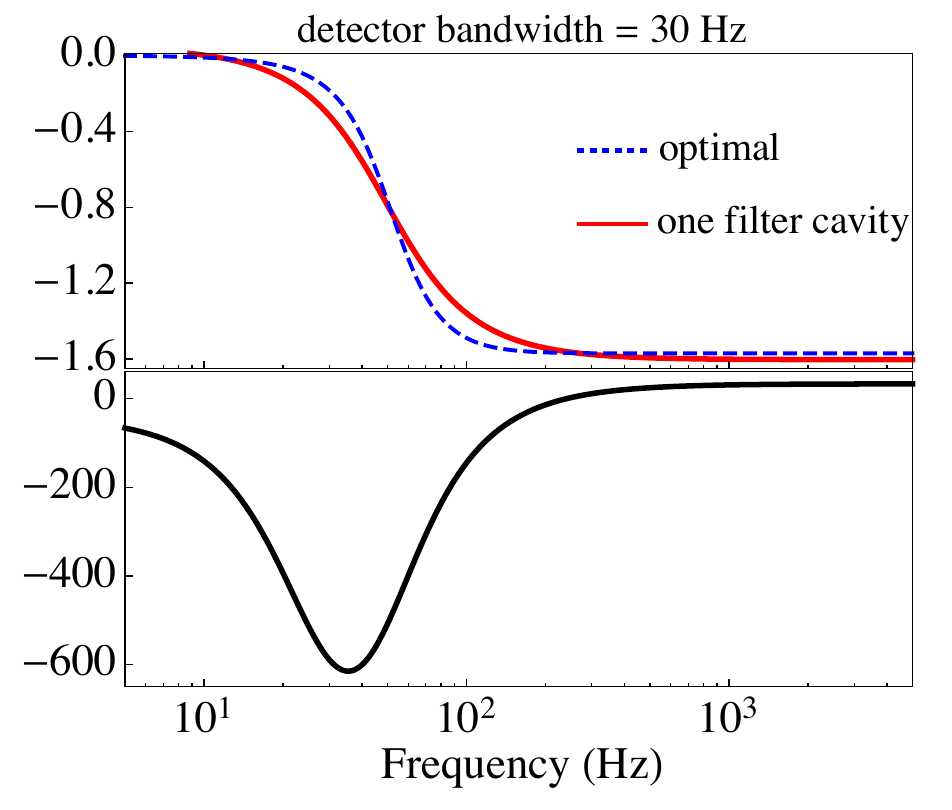}
\caption{The difference in the squeezing angle between the optimal angle 
and the approximation realised by using only one filter cavity for 
different values of detector bandwidth.}
\label{fig:sqz_angle_approx} 
\end{figure}

For example, in the tuned dual-recycled Michelson, two filter cavities are needed 
to achieve the optimal squeezing angle, as the highest order of $\Omega$ in
$\tan\theta_s$ is four, cf. Eq.\,\eqref{eq:sqz_angle_tuned}. When the detector
bandwidth $\gamma$ is much larger than the frequency for the transition 
from the radiation-pressure-noise dominated to the shot-noise dominated, one filter 
cavity can approximately realise the optimal squeezing angle. This is the case
for the resonant-sideband-extraction mode of detectors. Given 
the default parameters that we assumed,  the bandwidth 
is of the order of a few hundred Hz and the transition frequency is around 30\,Hz. 
Indeed, as shown in the left panel of Fig.\,\ref{fig:sqz_angle_approx}, the 
difference between the optimal 
angle and the one realised with one filter cavity is less than one milliradian, and
the projection noise from the anti-squeezing is smaller than 0.4\,dB for 10\,dB 
squeezing. However, when the detector bandwidth is narrow, e.g., around 100Hz, 
as shown in the right panel of Fig.\,\ref{fig:sqz_angle_approx}, one filter cavity 
is not able to produce the optimal angle which has a steeper change than the 
case of having a broad detector bandwidth. 

One may also estimate how many filter cavities is sufficient for an interferometer from the perspective of loss, as discussed in Sec.~\ref{ssec:3-4}. The price to pay in the case of imperfect rotation angle is the extra quantum noise that comes from the projection of anti-squeezed quadrature on the readout one. If this contribution, that can be estimated as $\Delta_{\rm imp.\, \theta_s} = s_+\delta\theta_{s}$ (cf. Eq.~\eqref{eq:sqz_angle_jitter}) is smaller than the contribution to the phase fluctuations due to loss in the squeezing injection optics (see Eq.~\eqref{eq:r_eff_def}), which yields:
\begin{equation*}
\delta\theta_s \lesssim \sqrt{\epsilon_{\rm sqz}e^{-2 r_+}}\,.
\end{equation*}

\subsection{Conditional frequency-dependent
 squeezing via EPR entanglement }\label{ssec:3-3}

As mentioned in the previous section, the canonical setup for realising the 
frequency-dependent squeezing for a broadband detector
 involves at least one additional filter cavity. 
In contrast, the recently proposed idea based upon the
Einstein-Podolsky-Rosen (EPR) entanglement of light shows 
a new approach without a need of the external long filter cavity\,\cite{Ma_NPhys_13_776_2017}. 
This idea takes advantage of the entanglement (correlation) 
between fields around the half of the frequency $\omega_p$ 
of the pump field that drives the nonlinear crystal. 

\begin{figure}[t]
\begin{center}
\includegraphics[width=0.6\columnwidth]{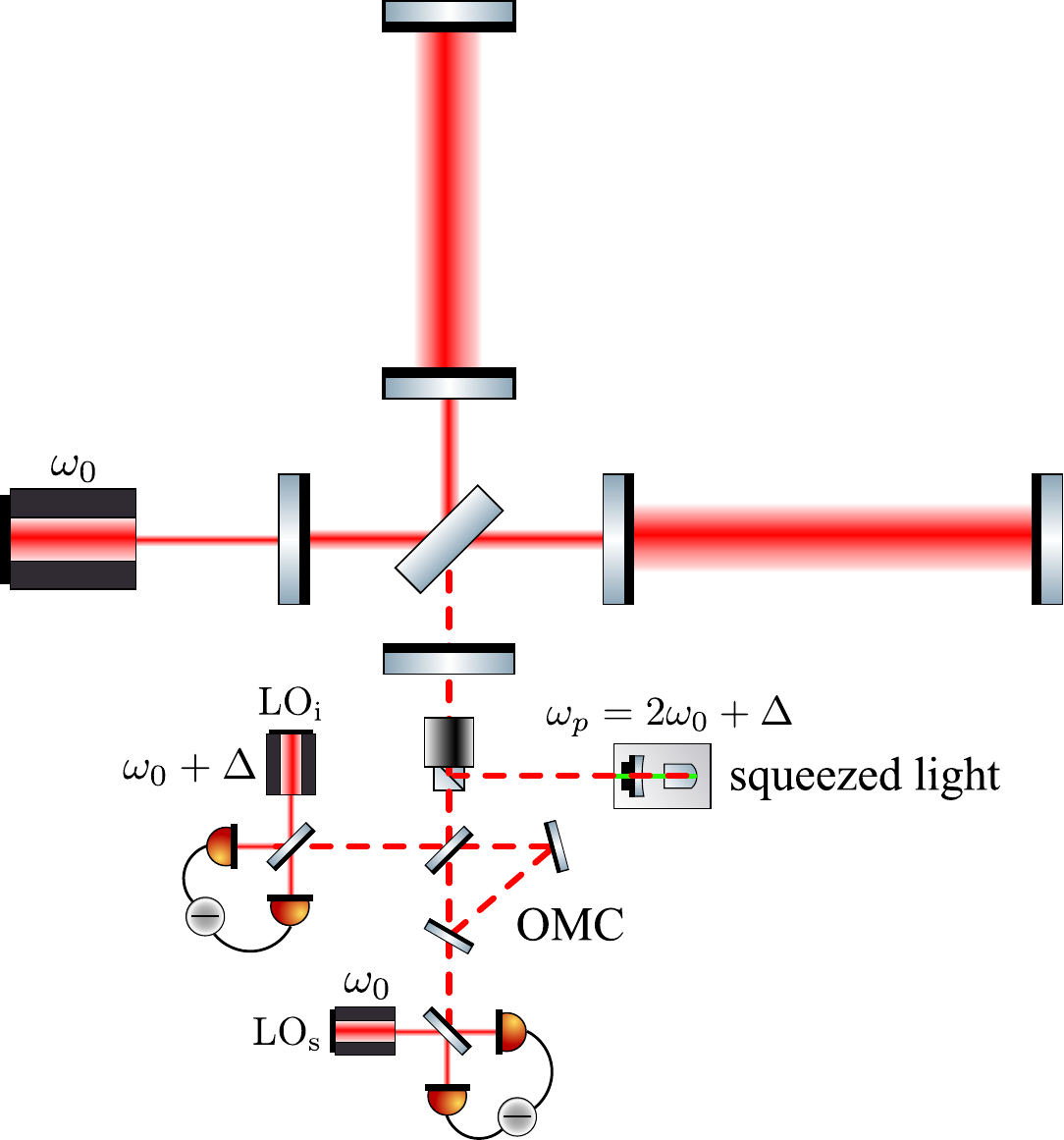}
\caption{A schematic showing the configuration of realising frequency-dependent
squeezing using the idea of EPR entanglement. The basic setup is the 
same as frequency-independent squeezing but with the pump frequency 
slightly shifted away twice the carrier frequency $\omega_0$ of the interferometer by 
$\Delta$. }
\label{fig:configuration_EPR}
\end{center}
\end{figure}

Compared to
the canonical setup, where $\omega_p/2$ coincides with the carrier 
frequency $\omega_0$ of the interferometer, this scheme slightly shifts the pump 
frequency by, e.g., tens of MHz, denoted as $\Delta$,
 which needs to much larger than 
the GW frequency but smaller than the bandwidth of the squeezed light
source. The field around $\omega_0$, which contains the GW signal, 
are called signal field (mode); 
that around $\omega_0+\Delta$ is called the idler 
field. They are correlated due to the nonlinear process in the squeezed light 
source; measuring one will allow us to reduce our uncertainty of the other, which 
is so-called conditional squeezing. Since the idler field is separated with 
the signal fields by tens of MHz, it will not mix with the strong carrier at $\omega_0$
to produce a radiation pressure on the test masses. The interferometer will 
just behave like an optical filter cavity for the idler field; the conditional squeezing
can gain the desired frequency dependence by properly tuning $\Delta$, and 
no external filter cavity is needed. 
The setup is shown in Fig.\,\ref{fig:configuration_EPR}. The input path is 
the same as the frequency-independent squeezing. The additional 
complication comes from the output path. It requires 
a short (tens of centimetre scale) cavity similar to the output 
mode cleaner (OMC) to separate the signal field and the idler field. Two sets of 
balanced homodyne detection are needed to measure these two fields. 

\begin{figure}[b]
\begin{center}
\includegraphics[width=0.85\columnwidth]{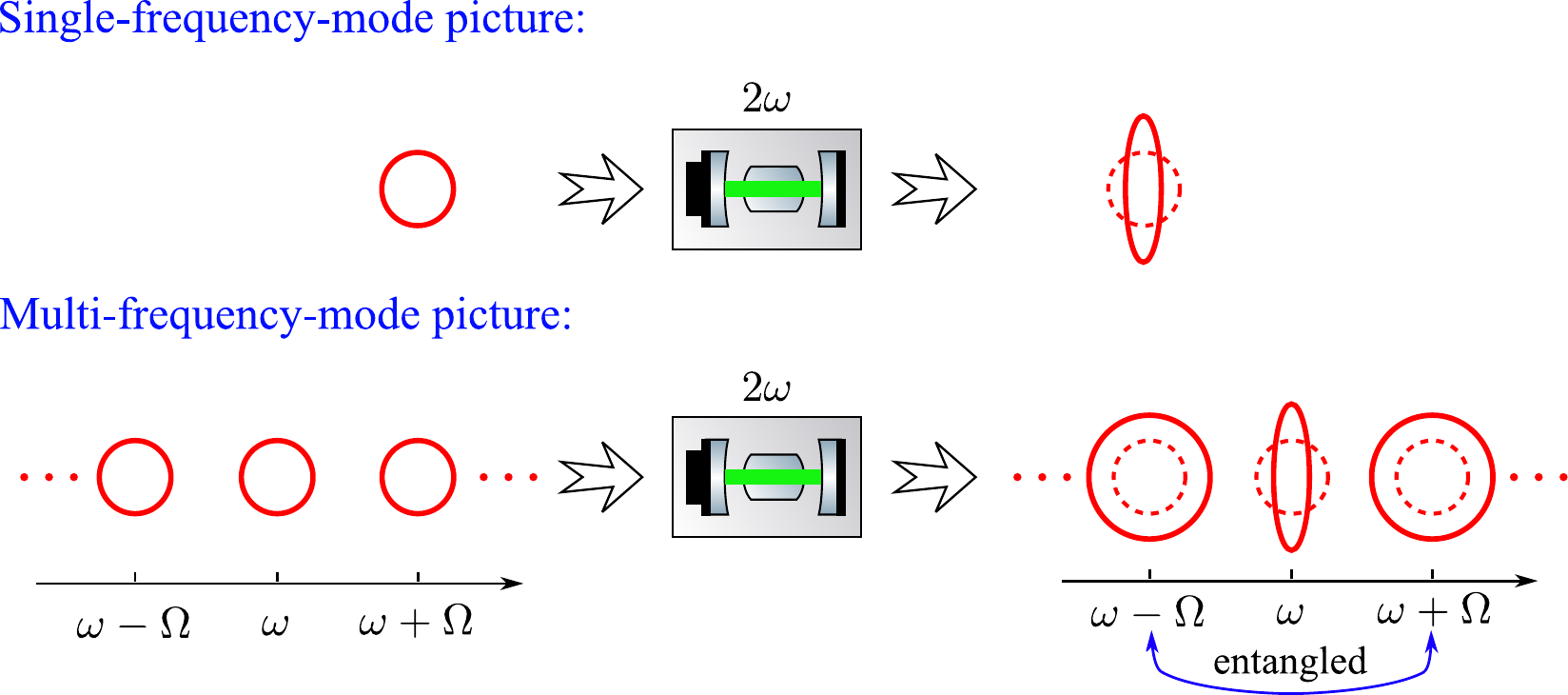}
\caption{The squeezer in the single-frequency-mode picture (top) and 
multi-frequency-mode picture (bottom). In the latter picture, the field 
is only squeezed precisely at $\omega$, the half of the pump frequency.
The upper and lower sidebands around $\omega$ will have fluctuations 
larger than that of a 
vacuum state, but are entangled (correlated) if their sum frequency 
is equal to $2\omega$.}
\label{fig:squeezing}
\end{center}
\end{figure}

\begin{figure}[t]
\begin{center}
\includegraphics[width=0.9\columnwidth]{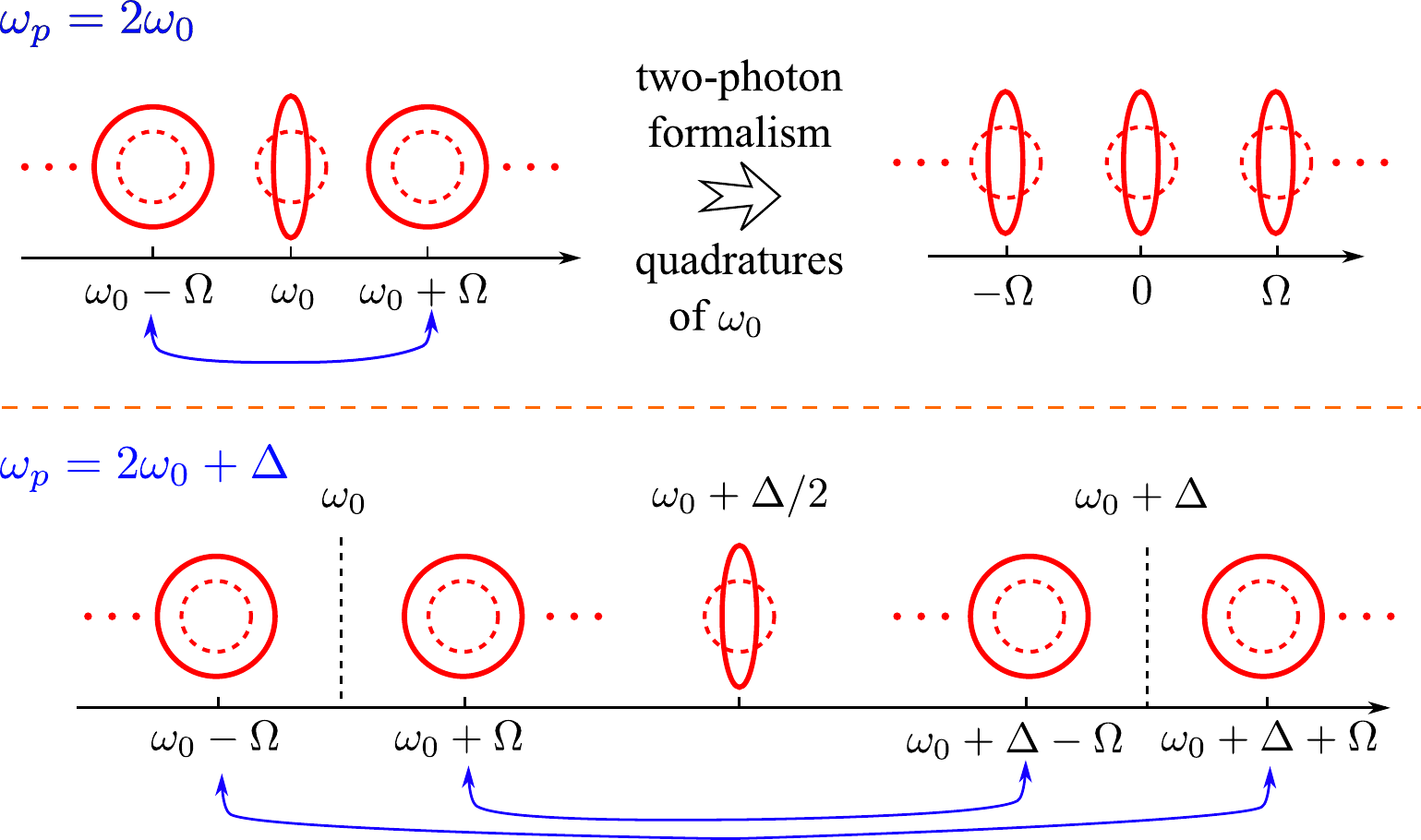}
\caption{The upper panel illustrates the standard case 
with $\omega_p=2\omega_0$. We can transform the entanglement between 
the upper and lower sidebands into quadrature squeezing without entanglement,
 by using the two-photon 
formalism. The lower panel shows the EPR squeezing idea where such an
entanglement is explicitly explored to produce the conditional squeezing.
}
\label{fig:squeezing_tuned_detuned}
\end{center}
\end{figure}

To understand this idea, we need to look at the structure of EPR entanglement 
in the multi-frequency-mode picture, as illustrated in Fig.\,\ref{fig:squeezing}. 
The upper sideband at $\omega+\Omega$ and the lower sideband at 
$\omega-\Omega$ are entangled in the sense that their quantum fluctuations 
are not independent but correlated. In the standard case with $\omega_p$
equal to twice the carrier frequency $\omega_0$ of the interferometer, 
we often do not need to consider such an entanglement in the two-photon 
formalism after introducing the amplitude 
and phase quadratures which are linear combinations of the upper 
and lower sidebands. This is because the test-mass-light interaction inside 
the interferometer and the homodyne readout only involve these quadratures 
rather than individual sidebands. It turns out that the entanglement between 
upper and lower sidebands can be converted into quadrature squeezing in
the frequency reference with respect to $\omega_0$, 
as illustrated by Fig.\,\ref{fig:squeezing_tuned_detuned}.

In the EPR squeezing idea, the entanglement between the upper and lower 
sidebands is the key to create the conditional squeezing. Let us go through 
the math behind the illustration shown in the 
lower panel of Fig.\,\ref{fig:squeezing_tuned_detuned}. 
With the offset $\Delta$ of the pump frequency, 
the sidebands around $\omega_0$ and those
around $\omega_0+\Delta$ are correlated. Specifically, 
the optical field 
$\hat o(\omega_0 -\Omega)$ is correlated with 
$\hat o(\omega_0+\Delta+\Omega)$, and 
$\hat o(\omega_0 +\Omega)$ is correlated with 
$\hat o(\omega_0+\Delta-\Omega)$. To distinguish 
between the sidebands around $\omega_0$ and those 
around $\omega_0+\Delta$, we introduce
\begin{equation}\label{eq:a_b_sidebands}
\hat a_{\pm}\equiv \hat o(\omega_0\pm \Omega)\,,\quad 
\hat b_{\pm}\equiv \hat o(\omega_0+\Delta\pm \Omega)\,. 
\end{equation}
Their correlations can be quantified by the cross spectrum in the 
frequency domain. Specifically, given the squeezing factor $r_s$ and
angle $\theta_s$ of the squeezed light source, we have 
\begin{align}\label{eq:sqz_spec1}
S_{a_+a_+}&=S_{a_-a_+}
=S_{b_+b_+}=S_{b_-b_-}=\cosh 2r_s\,, \\
\label{eq:sqz_spec2}
S_{b_-a_+}&= S^*_{a_+b_-} = S_{b_+a_-} =S^*_{a_-b_+} = - e^{2i\theta_s}\sinh2r_s\,,\\
S_{a_-a_+}&=S_{a_-b_-}=S_{a_+b_+}=S_{b_-b_+} = 0\,.
\label{eq:sqz_spec3}
\end{align}
In terms of the amplitude and phase quadratures for $\hat a$ and $\hat b$, 
we can obtain the covariance matrix for 
$(\hat a_c\; \hat a_s\; \hat b_c\; \hat b_s)$:
\begin{equation}
{\bf S}=\left[\begin{array}{cccc}
  \cosh2r_s & 0 & -\cos2\theta_s\sinh2r_s & \sin2\theta_s\sinh2r_s \\
  0 & \cosh2r_s & \sin2\theta_s\sinh2r_s & \cos2\theta_s\sinh2r_s \\
  -\cos2\theta_s\sinh2r_s & \sin2\theta_s\sinh2r_s & \cosh2r_s & 0 \\
  \sin2\theta_s\sinh2r_s & \cos2\theta_s\sinh2r_s & 0 & \cosh2r_s
\end{array}\right]\,.
\label{eq:cov_mat2}
\end{equation}
In the special case when $\theta_s=\pi/2$ (phase squeezing 
injection), the covariance matrix becomes 
\begin{equation}\label{eq:S_0}
{\bf S}|_{\theta_s=\pi/2} = \left[
\begin{array}{cccc}
  \cosh2r_s & 0 & \sinh2r_s & 0 \\
  0 & \cosh2r_s & 0 & -\sinh2r_s \\
 \sinh2r_s & 0 & \cosh2r_s & 0 \\
  0 & -\sinh2r_s& 0 & \cosh2r_s
\end{array}
\right]\,. 
\end{equation}
We can see that 
$\hat a$ and
 $\hat b$ are mutually correlated, or equivalently forming
quantum entanglement, manifested by the 
 nonzero off-diagonal terms in the covariance matrix. 
 It is such a correlation that allows 
 us to reduce the uncertainty (variance) of $\hat a$ by 
 making a measurement on $\hat b$, or vice versa. 
This is the main principle behind the conditional 
squeezing. 

To show the conditional squeezing explicitly, suppose we use the 
homodyne detection scheme to measure the quadrature $\hat b_{\phi}$: 
\begin{equation}\label{eq:HDb}
\hat b_{\phi}\equiv \hat b_c \cos\phi +\hat b_s\sin\phi\,. 
\end{equation}
The remaining
uncertainty of $\hat a_{1,2}$ conditional 
on the measurement of $\hat b_{\phi}$, i.e., the conditional 
variance can be derived by using the definition of 
conditional probability: 
\begin{equation}
P(\hat {\bf a}|\hat b_{\phi}) =\frac{P(\hat {\bf a}, \hat b_{\phi})}{P(\hat b_{\phi})}\,.
\end{equation}
Here $P(\hat {\bf a},\hat b_{\phi})$ is  the joint probability distribution of
$\hat {\bf a} \equiv  (\hat a_1\; \hat a_2)$ and $\hat b_{\phi}$
a three-dimensional Gaussian distribution with 
mean equal to zero and covariance matrix derived from Eq.\,\eqref{eq:S_0}. 
The resulting covariance matrix for the conditional probability is 
\begin{equation}\label{eq:cond_sqz}
{\bf S^{\rm cond}_{aa}} = {\bf S_{aa}} - \frac{{\bf S}_{{\bf a}b_{\phi}}
{\bf S}_{b_{\phi}{\bf a}}}{S_{b_{\phi}b_{\phi}}}
={\bf R}_{-\phi}\left[\begin{array}{cc}
e^{-2 r_{\rm eff}} & 0 \\ 0 & e^{2r_{\rm eff}}
\end{array}\right]{\bf R}_{\phi}\,, 
\end{equation}
where the effective squeezing factor $r_{\rm eff}$ is defined through
\begin{equation}
e^{2r_{\rm eff}} \equiv \cosh 2r_s\,.
\end{equation}
Therefore, the signal field $\hat a$ is a squeezed state conditional on
the measurement of $\hat b_{\phi}$\footnote{In some sense, Eq.~\eqref{eq:cond_sqz} is another way to derive a Wiener filter for a 2-channel interferometer, as described in Sec.~\ref{ssec:2-1}. In this case the 2 quadratures of the signal field are combined with the idler-channel readout multiplied by a frequency-dependent coefficients $\vs{K} = \{K_{c}(\Omega),\,K_{s}(\Omega)\}$ that minimise the spectral density of the difference: $(\vq{a}-\vs{K}\hat b_{\phi})$, \textit{i.e.} $\min\limits_{\vs{K}}\Bigl[\mean{(\vq{a}-\vs{K}\hat b_{\phi})\circ(\vq{a}-\vs{K}\hat b_{\phi})^\dag}\Bigr]$}.

The squeezing angle is $-\phi$
and the magnitude of the conditional squeezing is
around 3\,dB less than the squeezing level directly 
measured using a local oscillator at $\omega_0+\Delta/2$. 
For example, given 10\,dB squeezed light source, i.e. $e^{2r_s}=10$, 
the observed conditional squeezing
is approximately equal to 7dB:
\begin{equation}\label{eq:dB}
10 \log_{10}(e^{2r_{\rm eff}})=10
 \log_{10}(\cosh2r_s)\approx 7\,. 
\end{equation}
One would need to have 13\,dB squeezing as the input to obtain 10\,dB 
squeezing using this approach. 

\begin{figure}[t]
\begin{center}
\includegraphics[width=0.6\columnwidth]{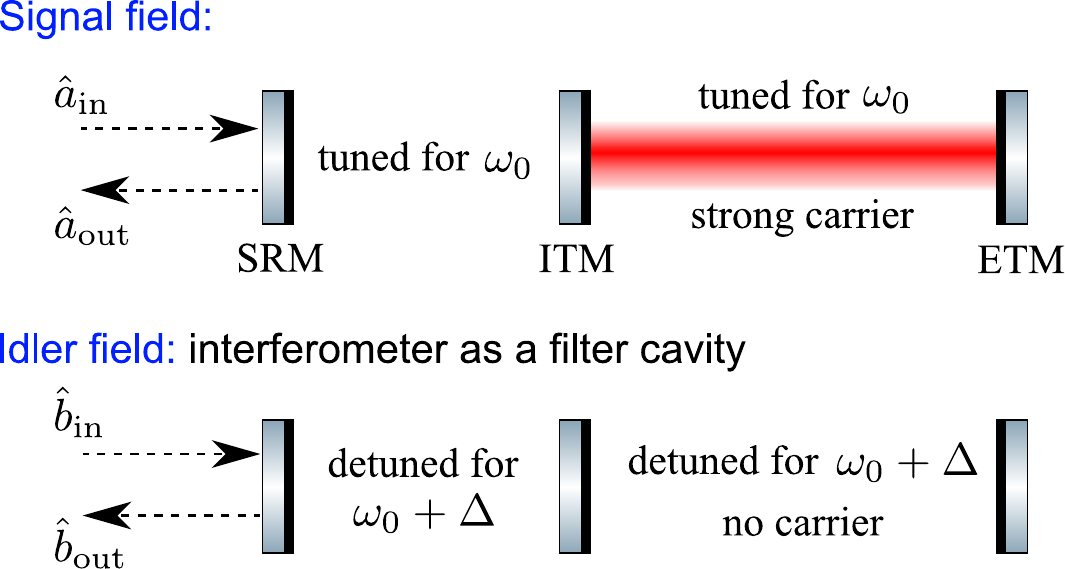}
\caption{The dual role of the interferometer (schematics showing 
the differential mode): it acts as a signal extraction device and a ponderomotive
squeezer for the signal field (top panel), while acting as a filter cavity for the idler 
field (bottom panel). Measuring the output of the idler field will project the signal 
field into a squeezed state. 
}
\label{fig:IFO_dual_role}
\end{center}
\end{figure}

Fig.\,\ref{fig:IFO_dual_role} illustrates how the 
interferometer affects the signal field and the idler field by only looking at 
the differential mode from the dark port (the interferometer is mapped 
into a coupled cavity). For the former, 
the signal-recycling cavity (SRC) formed 
by SRM and ITM is tuned on resonance with respect to $\omega_0$ in the
resonant sideband extraction case. The strong carrier inside the arm cavity 
mixes with the signal field and interacts with the test mass mediated by 
the radiation pressure. This process makes the signal field at the output
squeezed, which is the ponderomotive squeezing effect mentioned earlier. 
It introduces the radiation pressure noise by converting the fluctuation 
of the amplitude quadrature into that of the phase quadrature. For the 
latter, there is no strong carrier at $\omega_0+\Delta$ and there is no 
radiation pressure effect associated with the idler field. The interferometer 
behaves as a passive filter cavity that imprints frequency-dependent 
rotation on the quadratures of the idler field. Since 
measuring $\phi$ quadrature of the idler field will make $-\phi$ quadrature 
of the signal field squeezed, cf. Eq.\,\eqref{eq:cond_sqz}, the 
frequency dependence will be transferred to the squeezing of the signal field. 
As shown in Ref.\,\cite{Ma_NPhys_13_776_2017}, we can achieve the desired frequency-dependent 
squeezing by choosing a proper value of $\Delta$ and fine tuning the length 
of SRC. \SD{One may as well use the I/O-relations formalism of Sec.~\ref{ssec:2-1} to arrive to the above described result. However, since the 2 modes of squeezed light are entangled and thus ought to be considered together, as manifested by Eq.~\eqref{eq:cov_mat2}, the dimensions of the corresponding transfer matrix $\tq{T}$ and the response vector $\vs{t}$ should be expanded to 4x4 and 4x1, respectively.}

There is one last issue worthy of emphasising, which is the optical loss. 
This idea removes the additional filter cavity that is needed in the conventional 
frequency-dependent squeezing. Therefore, the optical loss associated with
the filter cavity is now absent, as the arm cavity is long enough to achieve the 
required filter bandwidth with a low finesse. However, since there are two readout 
channels: one for the signal field and the other for the idler field, the optical loss 
at the output, e.g., from the mode mismatching and finite quantum efficiency of 
the photo detector, is effective doubled compared with the conventional scheme. 
This scheme, if to be implemented, places a more stringent requirement on the output loss. 
In the section below, we will discuss in general how the optical loss influences
the quantum-limited sensitivity of laser interferometers. 

\subsection{Optical losses in interferometers with non-classical light}\label{ssec:3-4}

The performance of the described interferometers with squeezed vacuum injection depends rather strongly on how well the quantum correlations generated by the squeezer are transmitted to the interferometer to counteract the corresponding quantum correlations created by optomechanics (\textit{i.e.} ponderomotive squeezing discussed earlier). As shown by Kimble \textit{et al.} \cite{02a1KiLeMaThVy}, this effect is quite significant and detrimental. There are several mechanisms that cause deterioration of the QNLS of the interferometers using squeezing injection, which we consider below.

\subsubsection{Optical loss in a squeezing injection optics.}
Optical loss in the injection train may be considered the main hindrance for squeezed vacuum to enter the GW detector dark port \cite{2013_OE.21.19047_phase_noise_sqz_LIGO,2015_OE.23.8235_phase_noise_sqz_GEO,2013_OE.21.30114_Loss_in_FC_Isogai}. The mechanism behind is mainly the scattering/mode mismatch and absorption in the auxiliary optical elements used to link the squeezer and the FC input mirror, or the interferometer dark port. As an upper bound estimate, it can be characterised by an integral, frequency-independent injection power loss coefficient, $\epsilon_{\rm sqz}$. Following the same chain of argument as for the readout train loss in Sec.~\ref{ssec:loss}, the I/O-relation for the injection train can be written as:
\begin{equation}
\vq{i}_{\rm dark\ port} = \sqrt{1-\epsilon_{\rm sqz}}\,\vq{i}_{\rm sqz}+\sqrt{\epsilon_{\rm sqz}}\,\vq[sqz]{n}\,,
\end{equation}
where $\vq{i}_{\rm dark\ port}$ stands for the light field, entering the dark port of the detector (or the filter cavity in case of frequency dependent squeezing injection), and $\vq[sqz]{i}$ and $\vq[sqz]{n}$ are the field generated by a squeezer and a vacuum field due to injection loss, respectively. If the squeezer is capable of generating squeezed state with  (anti-)squeezing quadrature variances, $s_- = e^{2r_-},\,(s_+=e^{2r_+})$, the effective (anti-)squeezing factor at the dark port reads:
\begin{eqnarray}\label{eq:r_eff_def}
s_-^{\rm eff}\equiv e^{-2r_-^{\rm eff}} &=& (1-\epsilon_{\rm sqz})e^{-2r_-}+\epsilon_{\rm sqz}\,,\nonumber\\
\Bigl(s_+^{\rm eff}\equiv e^{2 r_+^{\rm eff}} &=& (1-\epsilon_{\rm sqz})e^{2r_+}+\epsilon_{\rm sqz}\Bigr)\,.
\end{eqnarray}
Here we took into account that the real squeezer produces not a pure squeezed vacuum state, for which $s_+ = 1/s_- = e^{2 r}$, rather a mixed state that can be described by a diagonal spectral density matrix:
\begin{equation}\label{eq:sqz_mat}
\tq{S}_{i}^{\rm sqz} = 
\begin{bmatrix}
s_+ & 0\\
0 & s_-
\end{bmatrix}
\end{equation}
with $s_+$ usually larger than $1/s_-$ (see e.g. \cite{2006_App.Phys.Lett.89.6.061116_Furusawa_squeezing}). 

\subsubsection{Squeezing angle fluctuations}

Another source of noise is known as `phase quadrature noise', or `squeezing angle jitter' \cite{2015_OE.23.8235_phase_noise_sqz_GEO}. It comes from the random fluctuations of the optical path length between the squeezer and the dark port of the interferometer. 

Although the fluctuation may happen anywhere along the squeezing injection train, the absence of active nonlinear components between the squeezer and the interferometer justifies viewing it as a random rotation of a squeezed vacuum state at the output of the squeezer. In this case, the effect can be described by a random angle of rotation, $\lambda$, normally distributed around the  zero mean with an r.m.s. uncertainty $\sigma_{\lambda}$: $w(\lambda) = \frac{1}{\sqrt{2\pi\,\sigma_\lambda^2}} \exp\left[ -\frac{\lambda^2}{2\,\sigma_\lambda^2} \right]$. Provided that the r.m.s. uncertainty $\sigma_\lambda$ is quite small ($\sim 10$ mrad), one can assume that the resulted quantum state of light remains Gaussian to a good precision and therefore only the transformation of the field second moments, \textit{i.e.} of the PSD matrix \eqref{eq:sqz_mat}, under these random rotations is of interest. The averaged over $\lambda$ squeezed state PSD matrix read:
	\begin{multline}\label{eq:sqz_angle_jitter}
		\mean{\tq{S}^{\rm sqz}_i}_\lambda= \int_{-\infty}^\infty d\lambda\,w(\lambda) \tq{R}[\lambda]\cdot\tq{S}_{i}^{\rm sqz}\cdot\tq{R}[-\lambda] = \\
		= \frac{s_++s_-}{2}\begin{bmatrix}
		1+\frac{s_+-s_-}{s_++s_-}e^{-2\sigma_\lambda^2} & 0\\
		0 & 1-\frac{s_+-s_-}{s_++s_-}e^{-2\sigma_\lambda^2}
		\end{bmatrix}
		\simeq \begin{bmatrix}
		s_+ & 0\\
		0 & s_-+\sigma_\lambda^2s_+
		\end{bmatrix}\,,
	\end{multline}
	where the last approximate inequality takes into account that $\sigma_\lambda\ll1$ and $s_+\gg s_-$. So we see that the phase quadrature fluctuations lead to a contamination of the squeezed quadrature, $s_-$, by the noise contained in the anti-squeezed quadrature, $s_+$. 

\subsubsection{Losses in filter cavities}

Filter cavities used for frequency-dependent squeezing have a bandwidth that is smaller than the detection band of the interferometer. Hence, the influence of extra vacuum fields associated with loss in the FC's mirrors has a distinct frequency dependence that can be accounted for using the model of a lossy Fabry-P\'erot cavity derived in Appendix~\ref{app:lossy FP I/O-rels}.
As there is no carrier light propagating in the FC, the general I/O-relations can be simplified by omitting back-action and signal parts in \eqref{app. matr.: i-o realtion for arm}:
\begin{equation}\label{eq:IOlossyFC}
  \vq{o}_f(\Omega) =  \tq{T}_{\rm FC}\vq{i}(\Omega) + \tq{N}_{\rm FC}\,\vq{n}\,,
\end{equation}
where $\vq{i}$ and $\vq{o}$ stand for input and output fields of the FC, respectively, and $\vq{n}$ represents vacuum fields due to loss. Transfer matrices for filter cavity are defined as:
\begin{equation}
\tq{T}_{\rm FC} = \tq[arm]{T}^{\rm s.n.}(\Omega)\,,\quad\tq{N}_{\rm FC} = \tq[arm]{N}^{\rm s.n.}(\Omega)\,,
\end{equation}
with expressions for $\tq[arm]{T}^{\rm s.n.}$ and $\tq[arm]{N}^{\rm s.n.}$ given by Eqs.~\eqref{eq_app:FP_Tsn} and \eqref{eq_app:FP_Nsn}, respectively.  

Using this simplified formula, quantum noise spectral density for interferometer with lossy input filter cavities can be obtained by substituting into \eqref{eq:SpDens_h_loss_PD} the following expression for input field spectral density matrix:
\begin{equation}\label{eq:Sin_FDsqz_loss}
 \tq{S}_{o_f,\,loss}^{in} = \tq{T}_{\rm FC}\cdot\tq{R}_{\lambda}\cdot \tq{S}_i^{\rm sqz}\cdot\tq{R}^\dag_{\lambda}\cdot\tq{T}_{\rm FC}^\dag + \tq{N}_{\rm FC}\cdot\tq{N}_{\rm FC}^\dag\,.
\end{equation}
The last term here peaks near the resonant frequency of the cavity which thereby decreases squeezing of the vacuum fields entering the cavity. But the off-resonant squeezed vacuum fields reflect off the FC without deterioration. This explains why optical loss in the cavities have major impact at low frequencies within the FC linewidth. 

\subsection{Summary and outlook}

After years of developments and researches, squeezing now becomes an
indispensable quantum technique for enhancing the detector sensitivity. 
We can now produce a high level of squeezing, more than 10 dB, at the audio 
band for both 1064 nm and 1550 nm with the goal of expanding to other
wavelengths\,\cite{Schnabel2017}. 
To fully take advantage of the squeezing, efforts are putting in 
minimising the optical loss, due to scattering and mode mismatch, in between 
the squeezed light source and the interferometer output. 
The frequency-dependent squeezing with a filter cavity has already been 
demonstrated on a table-top experiment\,\cite{Oelker2016}, 
and the large scale filter cavity, of the order of hundred meter,
will be implemented in the near term upgrades of current advanced detectors\,\cite{ISWP2017}. The EPR squeezing idea is at an early stage 
and requires table-top demonstrations, which have been started by several 
experimental groups. Since reducing the shot noise of a detuned interferometer also 
requires the frequency-dependent squeezing, this idea equally applies there, which 
has been modelled in details in Ref.\,\cite{Brown2017}. Indeed, the on-going 
experimental demonstrations all use this fact. 

Looking further into the future, more complex frequency-dependent squeezing might 
be needed to optimise the sensitivity of detectors operating 
beyond the current broadband operation. This may require a cascade of filter 
cavities with parameters that can be tuned in situ. For passive optics (without external
energy input), one can achieve the tunability by using compound mirrors. The active 
optomechanical filter cavity idea provides an alternative approach and also can 
achieve narrow cavity bandwidth with a short cavity length\,\cite{Ma2014a}. 
However, it has not yet been investigated experimentally as systematically 
as the passive filter cavity, and more researches are needed.

\section{Speed-meter interferometers} 
\label{sec:4}

Measurement of speed was first proposed by Braginsky and Khalili in \cite{1990_PLA.147.251_Braginsky_SM} as an alternative to a position measurement performed by a conventional Michelson interferometer. The goal was to get rid of the back-action fluctuations of light and thereby drastically improve the sensitivity of GW interferometers at low frequencies. This is possible because, they argued, velocity of the free body is proportional to its momentum, which is a conserved quantity and thus a \textit{quantum-non-demolition (QND)} observable. As such, any measurement of momentum is free from back action by design. The more careful analysis has shown that the dynamics of the test object cannot be considered separately from that of the meter, which is the laser light in the case of GW interferometers. For a combined system `mirrors+light', the generalised momentum is rather a sum of two terms, $\hat P = m\hat v - g_{\rm SM}(t)\hat a_c$ than a simple proportionality to velocity (see, \textit{e.g.} Sec.~4.5.2 in \cite{Liv.Rv.Rel.15.2012}), where $g_{\rm SM}(t)$ is the strength of coupling between the light and the mirrors' mechanical motion, and $\hat a_c = (\hat a+\hat a^\dag)/\sqrt{2}$ is the amplitude quadrature of light. Nevertheless, speed measurement offers a substantial reduction of random back-action force.

\begin{figure}[htbp]
\begin{center}
\includegraphics[width=.9\textwidth]{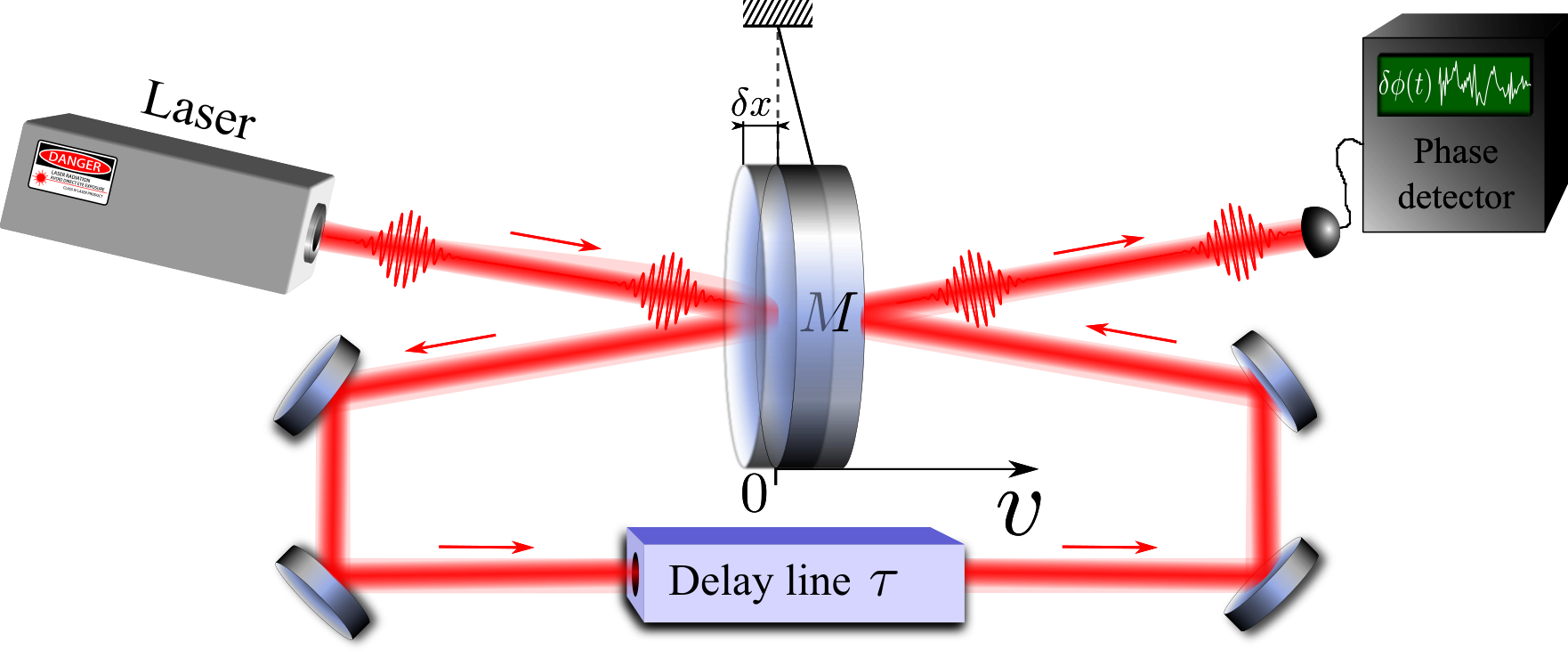}
\caption{Principle scheme of optical measurement of speed.} 
\label{fig:speed_measurement}
\end{center}
\end{figure}

The power of speed-meter interferometer (SI) to reduce back-action noise is nested in its ability to sense the relative rate, or in other words speed of an arm cavities length variation, whereas Michelson interferometer senses arms length variation itself. The simple way to understand how a speed measurement can reduce back-action is to consider a simple thought experiment depicted in Fig.~\ref{fig:speed_measurement}. Here the free mirror is sensed twice by the same laser light that is reflected from both the front and the rear surfaces thereof with a time delay $\tau$ between reflections. The phase of outgoing light is measured by, say homodyne detector, and is proportional to the the difference of the succesive mirror coordinates: $\phi_{\rm out}\propto (x(t+\tau)-x(t))\simeq \bar v\tau$, where $\bar v$ stands for the mean velocity of the mirror over the interval $\tau$. If the signal force one seeks to measure, watching the change of the mirror velocity, has characteristic frequency $\Omega$ much smaller than 
$\tau^{-1}$, the two kicks light gives to the mirror on the consecutive reflection partly compensate each other and the resulting back-action force turns out to be depressed by a factor $\propto\Omega\tau\ll1$:
\begin{equation}\label{eq:F_RP_speedmeter}
  \hat{F}_{\rm b.a.}(\Omega) \simeq -i\Omega \tau\frac{2 \bar P_{\rm pulse}}{c}\,,
\end{equation} 
as compared to the back-action of single light pulse with an average power $\bar P_{\rm pulse}$ which one expects in a single reflection experiment sensitive to the test mass displacement.

\begin{figure}[htbp]
\begin{center}
\includegraphics[width=\textwidth]{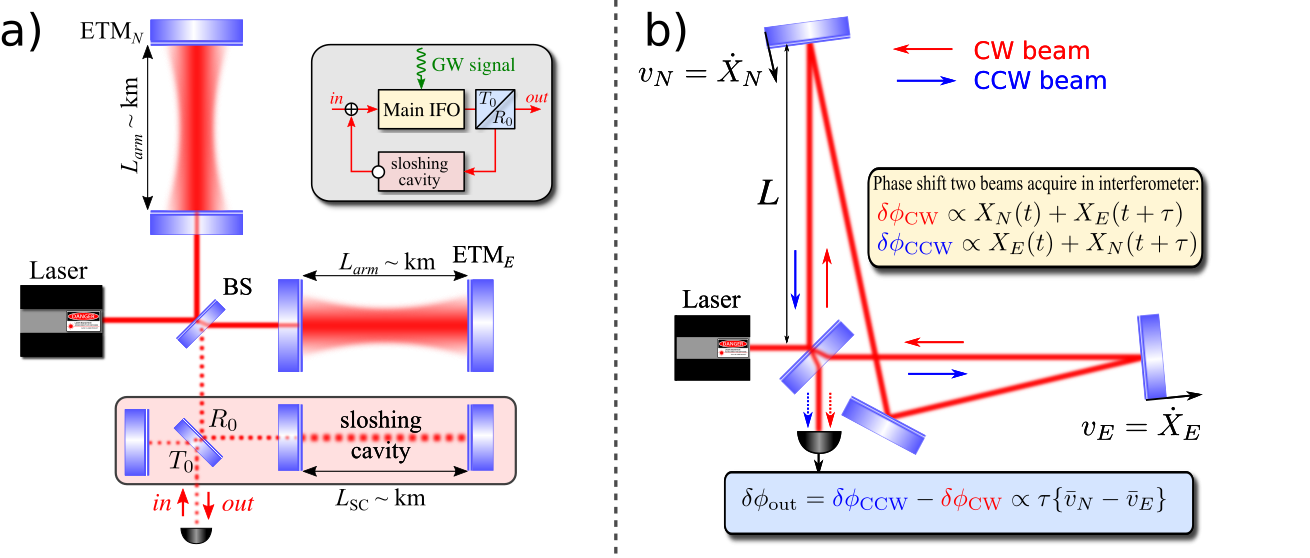}
\caption{Two possible ways of realisation of speed meter in a GW interferometer: a) ``Sloshing'' speed meter scheme based on Braginsky \textit{et al.} \cite{00a1BrGoKhTh}, and b) zero-area Sagnac speed meter based on Chen idea \cite{Chen2003}.} 
\label{fig:sloshing_sagnac}
\end{center}
\end{figure}

\subsection{Speed meters as GW detectors}

Original paper by Braginsky and Khalili \cite{1990_PLA.147.251_Braginsky_SM} considered the microwave speed meter as a readout for bar GW detectors. The first of two proposed schemes was the microwave version of the scheme shown in Fig.~\ref{fig:speed_measurement}. The second one used two coupled microwave cavities with one of them having a movable wall attached to the bar antenna to sense the GW-induced oscillations thereof, and the other cavity served for storing the EM signal with displacement information and sending (``sloshing'') it back to the readout cavity with an opposite sign ($\pi$-phase shift). This allowed sequential measurement of position as described above, thereby yielding speed measurement. In the subsequent years, a lot of new speed-meter interferometer designs were proposed, although it took almost 10 years till the first optical implementation of the original sloshing speed-meter principle has been finally developed by Braginsky \textit{et al.} \cite{00a1BrGoKhTh}. 

\afterpage{%
  \begin{landscape}
    \begin{figure}[h]
\includegraphics[height=.85\textwidth]{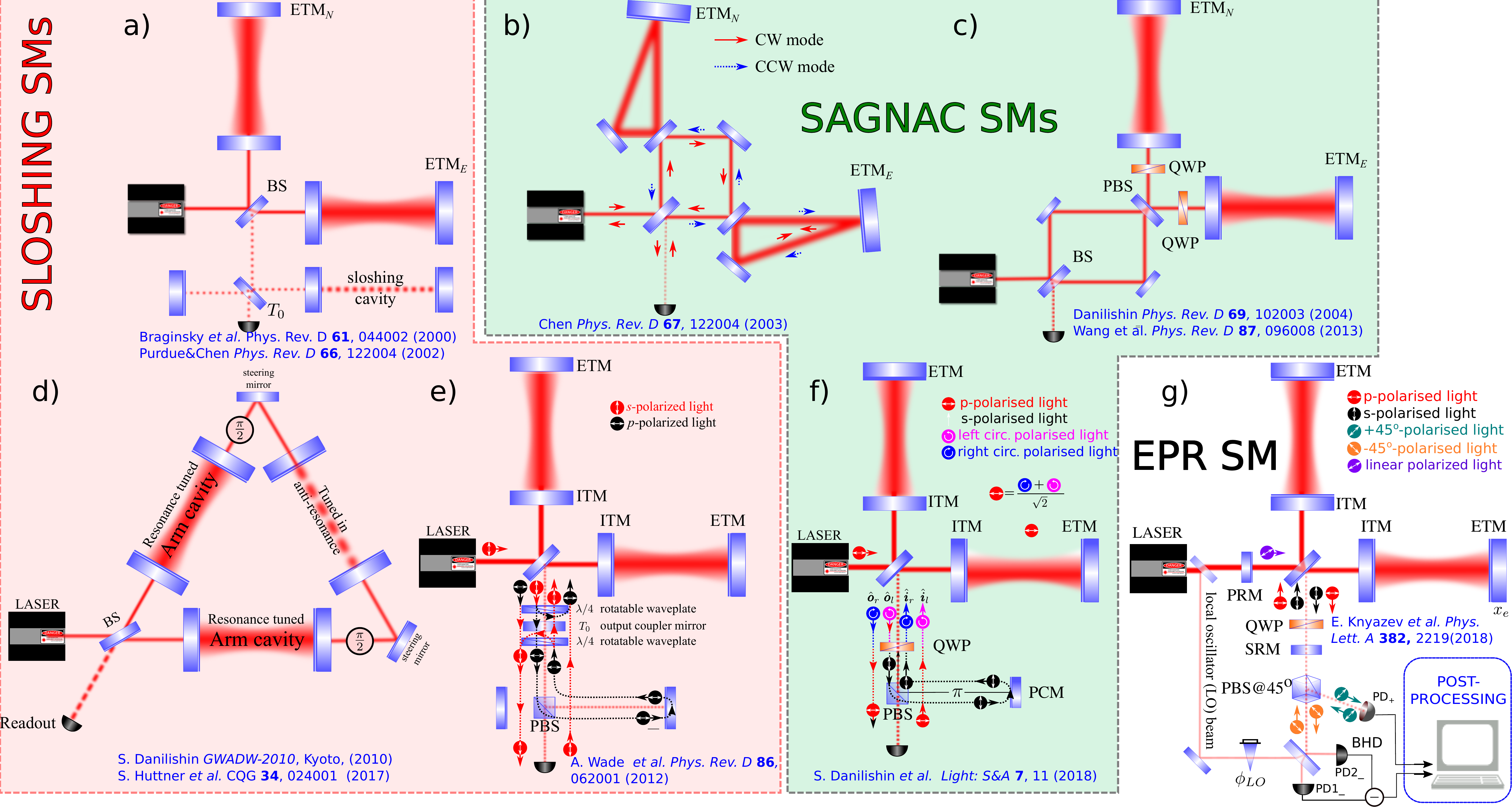}
\caption{Laser interferometers based on speed-meter principle (as of May 2018). All schemes are classified in 3 types --- (i) Sagnac-like speed meters, (ii) sloshing speed meters and (iii) the EPR-speed meter by the mechanism the speed measurement is arranged.} 
\label{fig:SM_Zoo}
\end{figure}

  \end{landscape}%
}

The more or less complete chart of configurations developed so far is shown in Fig.~\ref{fig:SM_Zoo}.  All schemes are classified in 3 types --- (i) Sagnac-like speed meters  \cite{Chen2003,02a2Kh,04a1Da,PhysRevD.87.096008,2018_LSA.7.accepted}, (ii) sloshing speed meters \cite{00a1BrGoKhTh,Purdue2001,Purdue2002,PhysRevD.86.062001,2017_CQG.34.2.024001_Huttner} and (iii) the EPR-type speed meter \cite{2017_Phys.Lett.A_EPR_SM} by the mechanism the speed measurement is arranged. In Sagnac speed meter, signal sidebands interact with the interferometer twice and co-propagate all the time with the carrier light. Sloshing speed meters use an additional not pumped sloshing cavity to store the signal sidebands between the two interactions with the interferometer arms and thus have an extra parameter, the sloshing frequency (defined by the sloshing cavity length and the input coupler mirror reflectivity), that discerns its response function from that of a Sagnac speed meters. And finally, the EPR-type speed meter uses two optically independent position-sensitive interferometers and devise the speed information by combining their outputs into sum and difference combinations with a beam-splitter and then adding the so obtained correlated photocurrents with optimal weights. Let us see how it works in individual schemes.


\subsection{Sloshing speed meter}
In the \textit{sloshing speed meter} proposed by Braginsky \textit{et al.} \cite{00a1BrGoKhTh} (see Fig.~\ref{fig:sloshing_sagnac}a), an auxiliary ``sloshing'' optical cavity was added into the output port of the Fabry-Perot--Michelson interferometer. This makes the GW signal to "slosh" back and forth between the two coupled effective cavities with an alternating sign and the rate $\omega_s = \frac{c}{2}\sqrt{\frac{T_0}{LL_0}}$ defined by the transmissivity $T_0$ of the \textit{input coupler} and the lengths of the arm  $L$ and sloshing cavity  $L_{0}$, respectively. Hence, after the second pass through the interferometer, the outgoing light bears exactly the required combination of position signals, $\propto \hat x(t) - \hat x(t+\tau)\sim \tau\bar{v}$, yielding the speed measurement. 

\begin{figure}[htbp]
\begin{center}
\includegraphics[width=\textwidth]{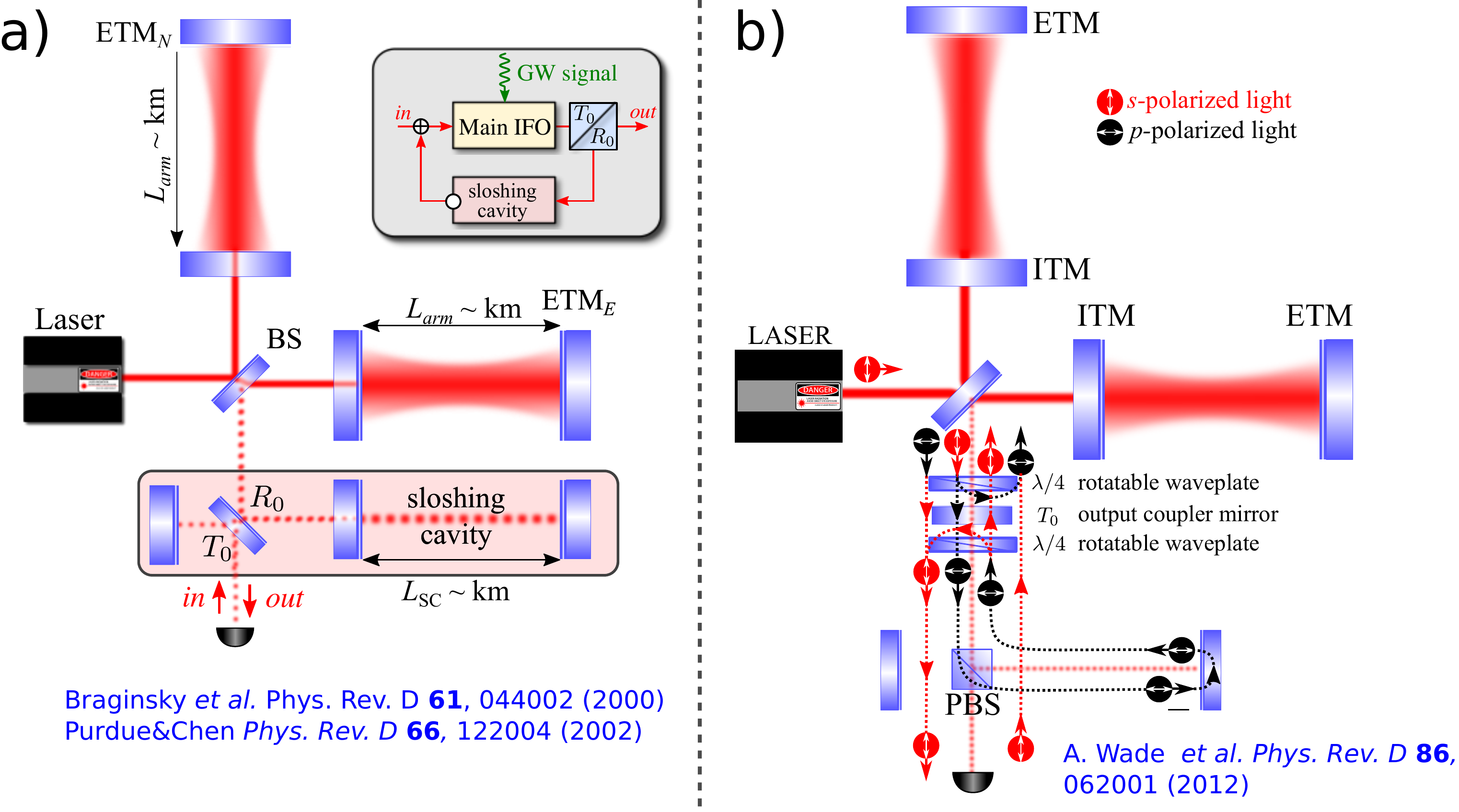}
\caption{Two possible realisations of sloshing speed-meter interferometer - a) using additional sloshing cavity \cite{00a1BrGoKhTh,Purdue2002} and b) using the two orthogonal
 polarisations to make use of the main IFO as a sloshing cavity \cite{PhysRevD.86.062001}} 
\label{fig:SM_sloshing}
\end{center}
\end{figure}

Two possible implementations of such a scheme are shown in Fig.~\ref{fig:SM_sloshing}. The left panel shows the variant with space separation of optical beams used for sequential measurement of arms' differential displacement \cite{00a1BrGoKhTh,Purdue2002}, whereas the right one, proposed by Wade \textit{et al.} \cite{PhysRevD.86.062001} employs two orthogonal polarisations to separate the beams. The latter also gets rid of an extra sloshing cavity by using the orthogonal not pumped polarisation mode of the interferometer. 

In the simple case of no losses and resonantly tuned main IFO and the sloshing cavity, the quantum noise of such interferometer is characterised by the following I/O-relations of the same form as Eqs.~\eqref{eq:I/O_KLMTV}:
\begin{align}\label{eq:I/O_SM_sloshing}
  \hat b_c^{out} &= e^{2i\beta_{\rm SSM}}\hat a_c^{in} \,,\\
  \hat b_s^{out} &= e^{2i\beta_{\rm SSM}}\bigl(\hat a_s^{in}-\mathcal{K}_{\rm SSM}\hat a_c^{in} \bigr) + e^{i\beta_{\rm SSM}}\sqrt{2 \mathcal{K}_{\rm SSM}}\frac{h}{h_{\rm SQL}} \,.
\end{align}
where $\mathcal{K}_{\rm SSM}$ is the sloshing speed meter optomechanical coupling factor. For the general case it can be written as:
\begin{equation}\label{eq:KSSM_gen}
	\mathcal{K}_{\rm SSM}(\Omega) = \dfrac{T_0\mathcal{K}_{\rm MI}\sin^{2}\alpha_{\rm SC}}{\cos^{2}(\beta_{\rm MI}+\alpha_{\rm SC})+T_0R_0\cos^2\beta_{\rm MI}-T_0\cos(\beta_{\rm MI}+2\alpha_{\rm SC})}
\end{equation}
where $\beta_{\rm MI}$ and $\alpha_{\rm SC}$ stand for the frequency-dependent phase shifts gained by the sidebands at frequency $\Omega$ as they pass through the main Michelson interferometer and the sloshing cavity (see Eq.~\eqref{eq:BMI} for definition).
It can be simplified, if one uses a single-mode approximation where all the sideband frequencies of interest are much smaller than the arm cavity $FSR=c/2L$ \cite{Miao14}:
\begin{equation}\label{eq:KSSM_NB}
\mathcal{K}_{\rm SSM}(\Omega) \simeq \dfrac{4 \Theta \gamma}{(\Omega^2-\Omega_s^2)^2+\gamma^2\Omega^2}
\end{equation}
with $\Omega_{s} = \sqrt{c^2T_{0}/(4LL_0)}$ being the \textit{sloshing frequency} that specifies the rate at which the signal sidebands ``slosh'' between the main IFO and the sloshing cavity with length $L_{0}$, and $\beta_{\rm SSM} = \arctan[(\Omega_{s}-\Omega)/(\Omega\gamma)]$ is the frequency dependent phase that a modulation sideband $\Omega$ acquires as it travels through the interferometer. 

\begin{figure}[htbp]
\begin{center}
\includegraphics[width=\textwidth]{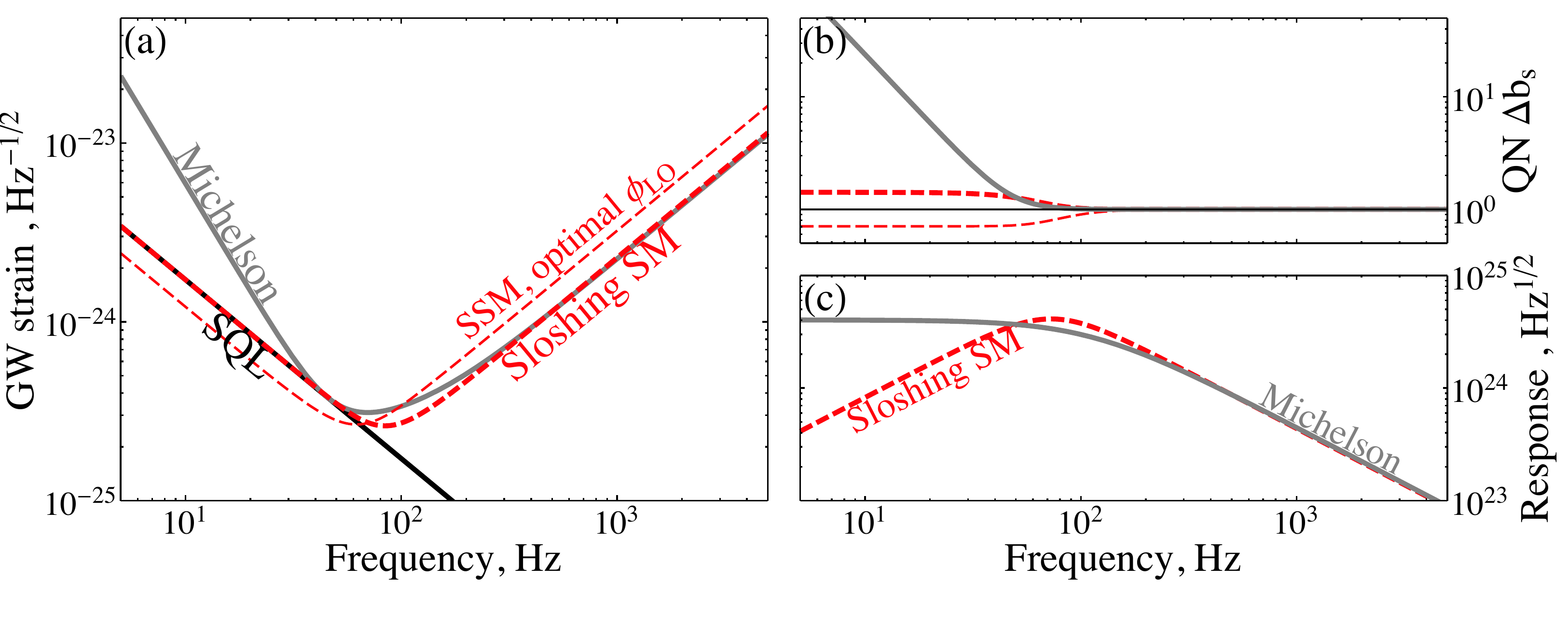}
\caption{Quantum noise of a \textit{sloshing speed meter}: \textit{(a):} QNLS of the sloshing SM for parameters given in the table \ref{tab:benchmark_3G} as compared to an equivalent Michelson (grey trace), with thick red dashed line showing noise in the phase quadrature at readout ($\phi_{\rm LO}=\pi/2$), and the thin red dashed line demonstrating the sub-SQL sensitivity of a SM for optimal readout quadrature $\phi_{\rm LO}=\arccot(\mathcal{K}_{\rm SSM}(0)) = \pi/4$; \textit{(b)} quantum fluctuations of the phase quadrature of the readout light of the SM (red dashed trace) and the equivalent Michelson (grey trace); \textit{(c)} response functions of the sloshing SM (red dashed trace) and Michelson (grey trace) interferometers to the GW strain. For a sloshing cavity, we assumed the same length as the arm cavities $L_{s}=L$ giving $\alpha_{\rm SC} = 2\Omega L_s/c$, no input mirror and chose the transmissivity $T_{0}$ of the coupling mirror from the condition $\mathcal{K}_{\rm SSM}(0)=1$ that yielded $T_{0} = 0.96$.} 
\label{fig:SSM_QNLS}
\end{center}
\end{figure}

It is straightforward to obtain the expression for quantum noise of the sloshing speed meter, using formula \eqref{eq:SpDens_a} that reads:
\begin{equation}\label{eq:QNLS_SSM_gen}
	S^h_{\rm SSM} = \frac{h_{\rm SQL}^{2}}{2}\left[\dfrac{(\mathcal{K}_{\rm SSM}-\cot\phi_{\rm LO})^{2}+1}{\mathcal{K}_{\rm SSM}}\right]\,.
\end{equation}
The corresponding plot of quantum noise limited sensitivity of a lossless sloshing speed meter is shown in the left panel of Fig.~\ref{fig:SSM_QNLS} along with a plot for the QNLS of a Michelson interferometer with similar parameters and a free mass SQL for scaling. 

One can immediately see that the QNLS of speed meter has the same frequency dependence as the SQL at low frequencies, where quantum back-action noise dominates, which is a unique feature of the speed meters in general. It results from the back-action suppression, as expected from the QND speed measurement. However, it does not go parallel to the frequency axis, like, for instance, the frequency-dependent variational readout and the FQL do (see Fig.~\ref{fig:FQL} in Sec.~\ref{sec:2}). One can see why on the two right panels of Fig.~\ref{fig:SSM_QNLS}, where on the top plot, the quantum noise of the outgoing light phase quadrature ($\phi_{\rm LO} = \pi/2$) is plotted (the numerator of Eq.~\eqref{eq:SpDens_a}), whereas on the lower panel we see the response of the interferometer to the signal variation of GW strain (the numerator of Eq.~\eqref{eq:SpDens_a}). Hence the QNLS plot to the left is simply the ratio of the upper and lower plots to the right.

So, one can see in Fig.~\ref{fig:SSM_QNLS}b that quantum back-action noise of speed meter is indeed heavily suppressed as compared to the Michelson interferometer and has the same constant-like frequency dependence as quantum shot noise. The $1/f$-slope in QNLS is coming from the speed response that rolls off as $\propto f$ towards the DC as shown in Fig.~\ref{fig:SSM_QNLS}c.

Mathematically, this suppression comes from the fact the OM coupling factor $\mathcal{K}$ is constant below the cavity pole, \textit{i.e.} at DC: $\mathcal{K}_{\rm SSM}(\Omega\to0)=const$, while for the Michelson interferometer it is $\mathcal{K}_{\rm MI}(\Omega\to0)\propto \Omega^{-2}$. This also means that unlike Michelson the power circulating in the arms of the speed meter must be above a certain threshold value $\Theta_{\rm crit}$, below which the speed meter cannot reach the SQL. Threshold is defined by the condition $\mathcal{K}_{\rm SSM}(0) = 1$. When substituting this condition into the QNLS expression at low frequencies and for phase quadrature readout one gets:
\begin{equation*}
 S^h_{\rm SSM} = \frac{h_{\rm SQL}^{2}}{2}\left[\mathcal{K}_{\rm SSM}(0)+\dfrac{1}{\mathcal{K}_{\rm SSM}(0)}\right]\to h_{\rm SQL}^{2}\,,
\end{equation*} 
Hence the threshold power reads $\Theta_{\rm crit} = \Omega_{s}^{4}/4\gamma$, or $P_{\rm crit} = M c L \Omega_{s}^{4}/(16\omega_0\gamma)$.

Another consequence of the peculiar behaviour of $\mathcal{K}_{\rm SSM}$ for a speed meter is the ability to surpass the SQL at low frequencies if the right quadrature is selected for readout. Indeed, the general expression for the SSM QNLS has a term $\propto[\mathcal{K}_{\rm SSM}-\cot\phi_{\rm LO}]^{2}$ that can be made zero, were $\cot\phi_{\rm LO}]=\mathcal{K}_{\rm SSM}$. This is quite easy to achieve, as $\mathcal{K}_{\rm SSM} = const$ below the cavity pole. The resulting sensitivity at these low frequencies is the FQL for the speed meter, 
as mentioned in Sec.\,\ref{sec:2}. For instance, at the threshold power where $\mathcal{K}_{\rm SSM}(0)=1$ the optimal readout quadrature will equal $\phi_{\rm LO} = \pi/4$. This case is plotted as a thin dashed line in Fig.~\ref{fig:SSM_QNLS}a.

\subsection{Sagnac-type speed meters}

\begin{figure}[htbp]
\begin{center}
\includegraphics[width=\textwidth]{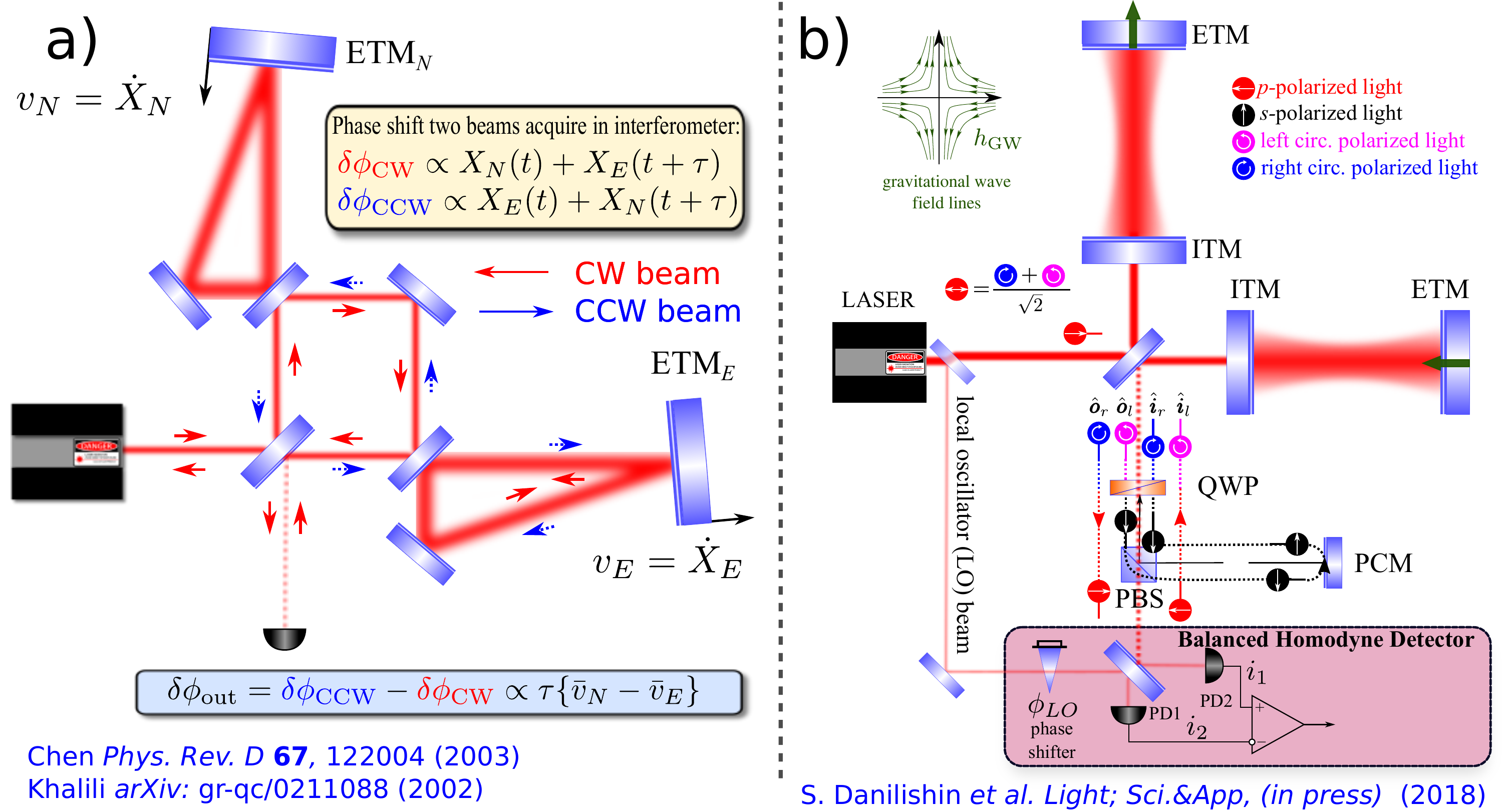}
\caption{Two possible realisations of Sagnac speed-meter interferometer - a) using ring arm cavities to separate the in- and outgoing beams\cite{Chen2003,02a2Kh} and b) using the two orthogonal
polarisations and a $\lambda/4$-plate, PBS and a mirror for the same purpose \cite{2018_LSA.7.accepted}} 
\label{fig:SM_sagnac}
\end{center}
\end{figure}

Another way to make a speed measurement with laser interferometer was suggested independently by Chen and Khalili \cite{Chen2003,02a2Kh}. They showed that the \textbf{zero-area Sagnac} interferometer \cite{1999_JOSAB.16.9.1354_Beyersdorf,1999_Opt.Lett.24.16.1112_Beyersdorf} actually implements the initial double-measurement variant of the quantum speed meter, shown in Fig.\,\ref{fig:speed_measurement}. Indeed, visiting consequently both arms (see Fig.~\ref{fig:SM_sagnac}a), counter propagating light beams acquire phase shifts proportional to a sum of arms length variations $x_{N,E}(t) \equiv \bigl[x_{\rm ETM}^{N,E}(t) - x_{\rm ITM}^{N,E}(t)\bigr]$ (hereinafter I(E)TM stands for Input (End) Test Mass) for  of both cavities taken with time delay equal to average single cavity storage time $\tau_{\rm arm}$:
\begin{eqnarray}
    \delta\phi_R &\propto & x_N(t) + x_E(t+\tau_{\rm arm})\,,\\
    \delta\phi_L &\propto & x_E(t) + x_N(t+\tau_{\rm arm})\,.
\end{eqnarray}
After recombining at the beam splitter and photo detection the output signal will be proportional to the phase difference of clockwise (R) and counter clockwise (L) propagating light beams:
\begin{multline}\label{phi_speedmeter}
 \delta\phi_R - \delta\phi_L \propto [x_N(t) - x_N(t+\tau_{\rm arm})] - [x_E(t) - x_E(t+\tau_{\rm arm})]\propto \\
  \propto \dot{x}_N(t) - \dot{x}_E(t) + O(\tau_{\rm arm})
\end{multline}
that, for frequencies $\Omega \ll \tau_{\rm arm}^{-1}$, is proportional to relative rate of the interferometer arms length variation.

The originally proposed configuration that uses ring cavities for separation of the in- and outgoing beams is not very practical, as the experience of the experimental prototyping of this type interferometer at the University of Glasgow has shown \cite{2014_CQG.31.215009_Graef}. Apart from the infrastructural complexity of placing two large suspended mirrors in the same vacuum tube, the ring arm cavities suffer heavily from the coherent back-scattering of light from one beam to the counter propagating one. This creates an unwanted coupling between the two modes of the ring cavity (associated with clockwise and counterclockwise propagating beams) thereby causing resonance frequency splitting. This means that the arms become detuned with respect to the pump light, which lead to the increase of quantum noise as shown in \cite{Pascucci2018}.

To avoid this problem, a few polarisation-based variants of speed-meter schemes were proposed \cite{04a1Da,PhysRevD.87.096008,2018_LSA.7.accepted}, which relaxed the need for modifications of the main interferometer significantly. The most recent proposal \cite{2018_LSA.7.accepted}, depicted in Fig.~\ref{fig:SM_sagnac}b, no changes to the infrastructure of the main interferometer. It requires, however that all reflective coatings of the core optics have the same properties for both polarisations of light. This is a tough, though not impossible requirement, and some research in this direction is under way already \cite{Hild_personal,Krocker_personal}.

Quantum noise of the Sagnac speed meter can be written exactly in the same way as for the sloshing speed meter before. The only difference will be in the shape of the OM coupling factor that for Sagnac interferometer can be written as:
\begin{equation}
	\mathcal{K}_{\rm Sag} = 4\,\mathcal{K}_{\rm MI} \sin^2\beta_{\rm MI}
  \simeq \frac{8 \Theta \gamma}{(\Omega^2+\gamma^2)^2}\,.
\end{equation}

\begin{figure}[htbp]
\begin{center}
\includegraphics[width=\textwidth]{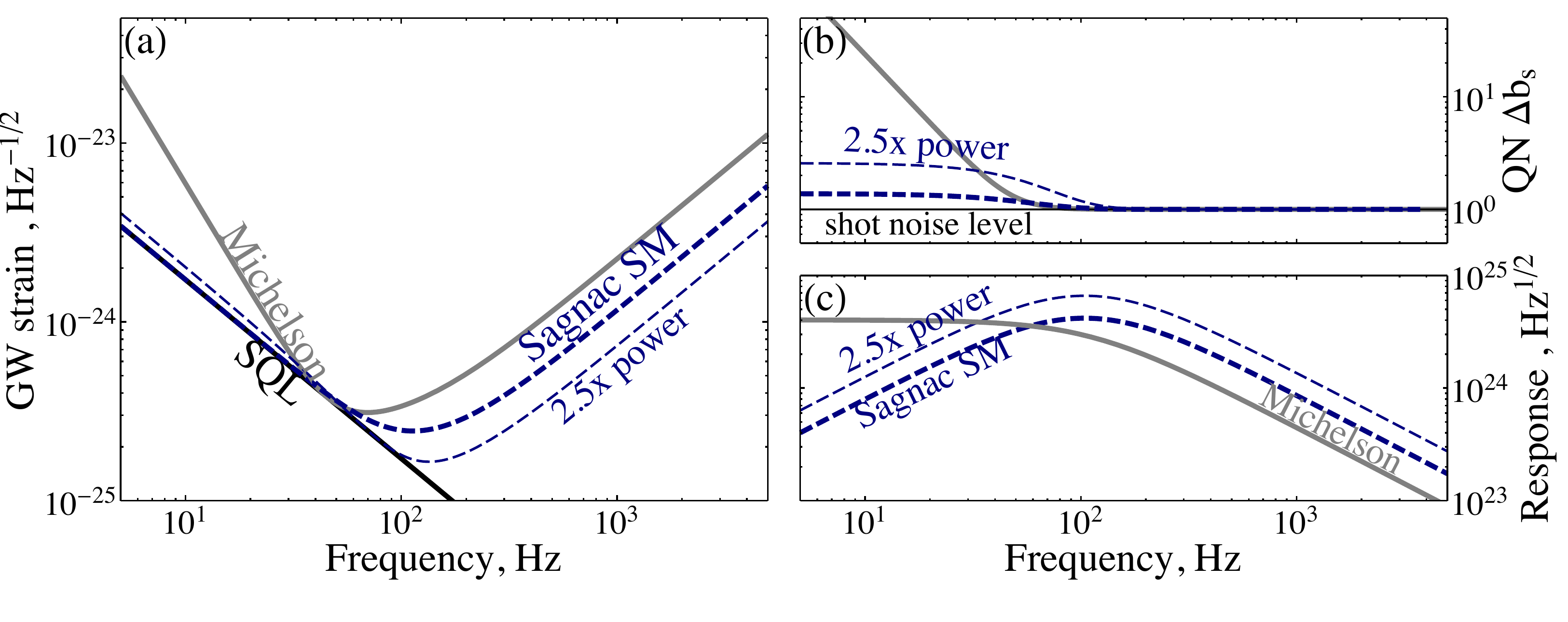}
\caption{Quantum noise of a \textit{Sagnac speed meter}: \textit{(a):} QNLS of the Sagnac SM for parameters given in the table \ref{tab:benchmark_3G} as compared to an equivalent Michelson (grey trace), with thick blue dashed line showing noise in the phase quadrature at readout ($\phi_{\rm LO}=\pi/2$), and the thin blue dashed line demonstrating the effect of increased circulating power (ramped up by 2.5 times to 10 MW); \textit{(b)} quantum fluctuations of the phase quadrature of the readout light of the SM (blue dashed traces) and the equivalent Michelson (grey trace); \textit{(c)} response functions of the Sagnac SM (blue dashed traces) and Michelson (grey trace) interferometers to the GW strain.} 
\label{fig:Sag_QNLS}
\end{center}
\end{figure}

From this expression one can see that Sagnac has an advantage in response as compared to Michelson with the same pump power, as identified by the factor 4 before $\mathcal{K}_{\rm MI}$. The reason is straightforward and comes from the fact that in Sagnac each beam that leaves the main beam splitter visits both cavities in a row. This means that each arm takes twice as much power as that of the equivalent Michelson, thereby producing twice of the optomechanical response. To show this, one just need to substitute $\mathcal{K}_{\rm SSM}\to\mathcal{K}_{\rm Sag}$ in Eq.~\eqref{eq:QNLS_SSM_gen} and calculate the QNLS:
\begin{equation}\label{eq:QNLS_SSM_gen}
	S^h_{\rm Sag} = \frac{h_{\rm SQL}^{2}}{2}\left[\dfrac{(\mathcal{K}_{\rm Sag}-\cot\phi_{\rm LO})^{2}+1}{\mathcal{K}_{\rm Sag}}\right]\,.
\end{equation}
The above QNLS is plotted in Fig.~\ref{fig:Sag_QNLS}a. It is instructive to see how speed meter's sensitivity depends on circulating power. 

\subsection{EPR-type speed meters}

\begin{figure}[htbp]
\begin{center}
\includegraphics[width=\textwidth]{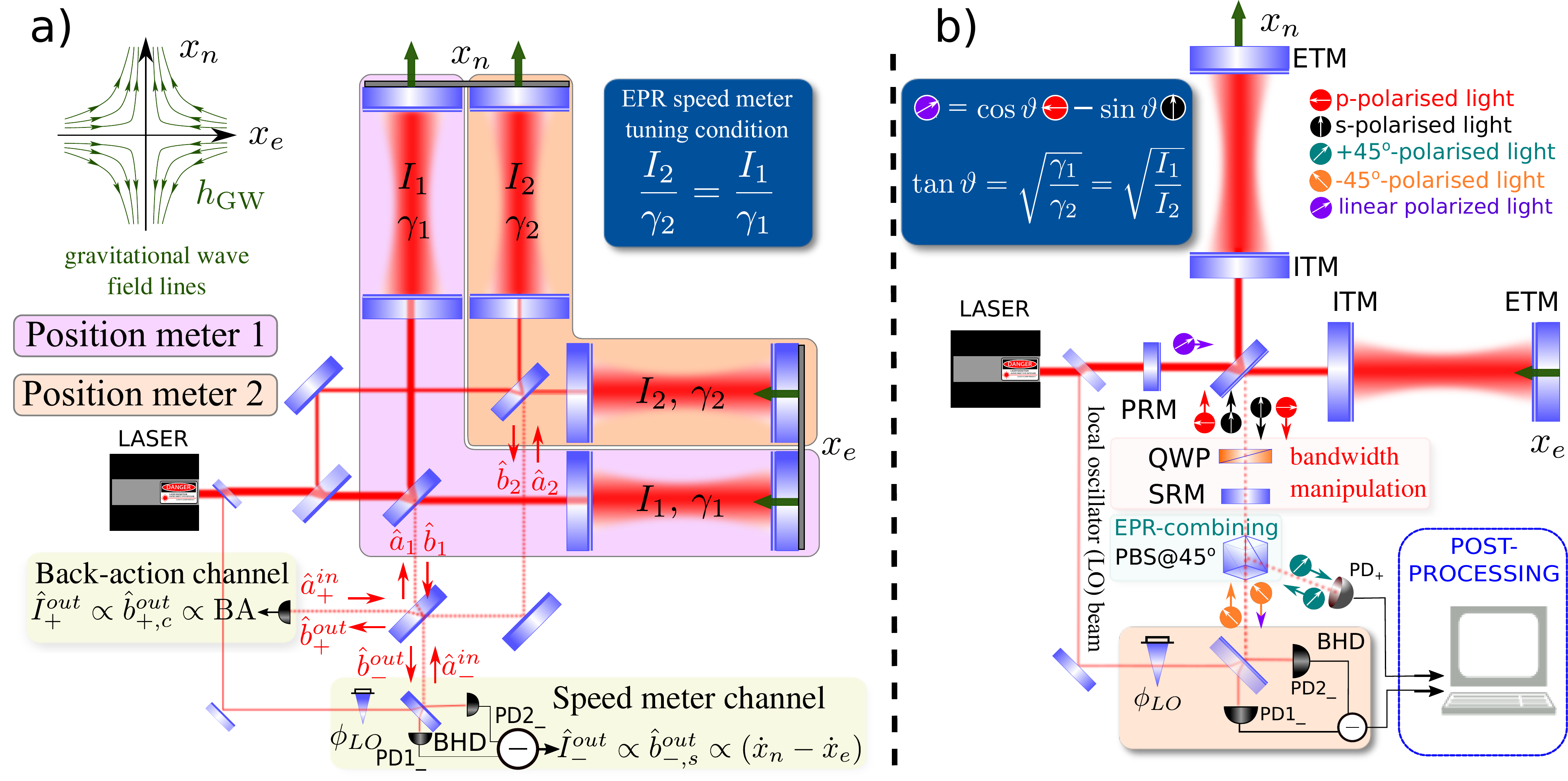}
\caption{EPR-speed meter concept: a) EPR speed meter based on two Fabry-Perot--Michelson interferometers with inputs and outputs combined using a beam-splitter and b) practical EPR-speed meter scheme based on using 2 orthogonal polarisations} 
\label{fig:SM_EPR}
\end{center}
\end{figure}

In 2017, Knyazev \textit{et al.} \cite{2017_Phys.Lett.A_EPR_SM} proposed a third distinct way to realise speed measurement in GW laser interferometer, using 2 position meters (Fabry-P\'{e}rot--Michelson interferometers, see Fig.~\ref{fig:SM_EPR}a that have rigidly connected test masses (or simply share them) but have contrasting light storage times (bandwidths satisfy condition $\gamma_1\gg\gamma_2$). The information about the differential motion of the arms thus comes of the two interferometers at a very different rate given by respective bandwidths. Hence, combining the readout beams of the two interferometers on a beam-splitter and reading out the  ``--''-channel thereof one gets the difference of the two position signals at different times that is, in fact, velocity. There is an additional back-action noise associated with the vacuum fields entering the ``+''-port of the beam-splitter that however can be subtracted from the readout, if one measures the amplitude quadrature at the ``+''-channel and subtracts it, with optimal filter, from the readout of the ``--''-channel. As the two output channels of the readout beam-splitter get entangled, when the two ponderomotively squeezed output fields, $\hat b_{1}$ and $\hat b_{2}$, of the two position meters get overlapped on it, and this entanglement is used to remove the excess back-action noise from the output, this speed meter was dubbed an \textbf{EPR-speed meter}. 

As the design of Fig.~\ref{fig:SM_EPR}a is obviously a nightmare to implement in a real GW detector (it was never intended to be), another one, based on orthogonal polarisation modes of light was proposed in \cite{2017_Phys.Lett.A_EPR_SM} and is shown in Fig.~\ref{fig:SM_EPR}b. The key element here is the quarter-wave plate (QWP) that acts as a $\pi/2$-phase retarder between the two orthogonal polarisation modes of the main interferometer. The QWP placed between the main IFO and the signal-recycling mirror, which position with respect to the arm's ITMs is chosen so that the resulting SR cavity (with the QWP) is tuned resonantly for one of the polarisation modes. The orthogonally polarised light sees the SR cavity as anti-resonant due to the $\pi/2$ phase shift given to it by the QWP. As a consequence of the ``scaling law'' \cite{Buonanno2003}, the polarisation mode that is in resonance with the SRC sees the interferometer with a very narrow effective bandwidth $\gamma_2$ (tuned SR regime, see Sec. 5.3.4 and Eq. (359) with $\phi_{S} = 0$ in \cite{Liv.Rv.Rel.15.2012}), whereas for the orthogonal one the effective bandwidth $\gamma_2\gg \gamma_1$ is greatly increased (resonant sideband extraction (RSE) regime, see Sec. 5.3.4 and Eq. (359) with $\phi_{S} = \pi/2$ in \cite{Liv.Rv.Rel.15.2012}). The polarisation beam splitter (PBS) with a polarisation plane rotated by 45$^{\circ}$ angle with respect to the \textit{s-} and \textit{p}-polarised modes of the main interferometer creates the EPR-type correlations in the ``+'' and ``--'' readout channels. The optimal distribution of circulating powers among the two effective position meters is organised by the proper choice of the angle $\vartheta$ of the carrier light polarisation plane to the vertical direction (see the blue box in Fig.~\ref{fig:SM_EPR}b). 

\begin{figure}[htbp]
\begin{center}
\includegraphics[width=\textwidth]{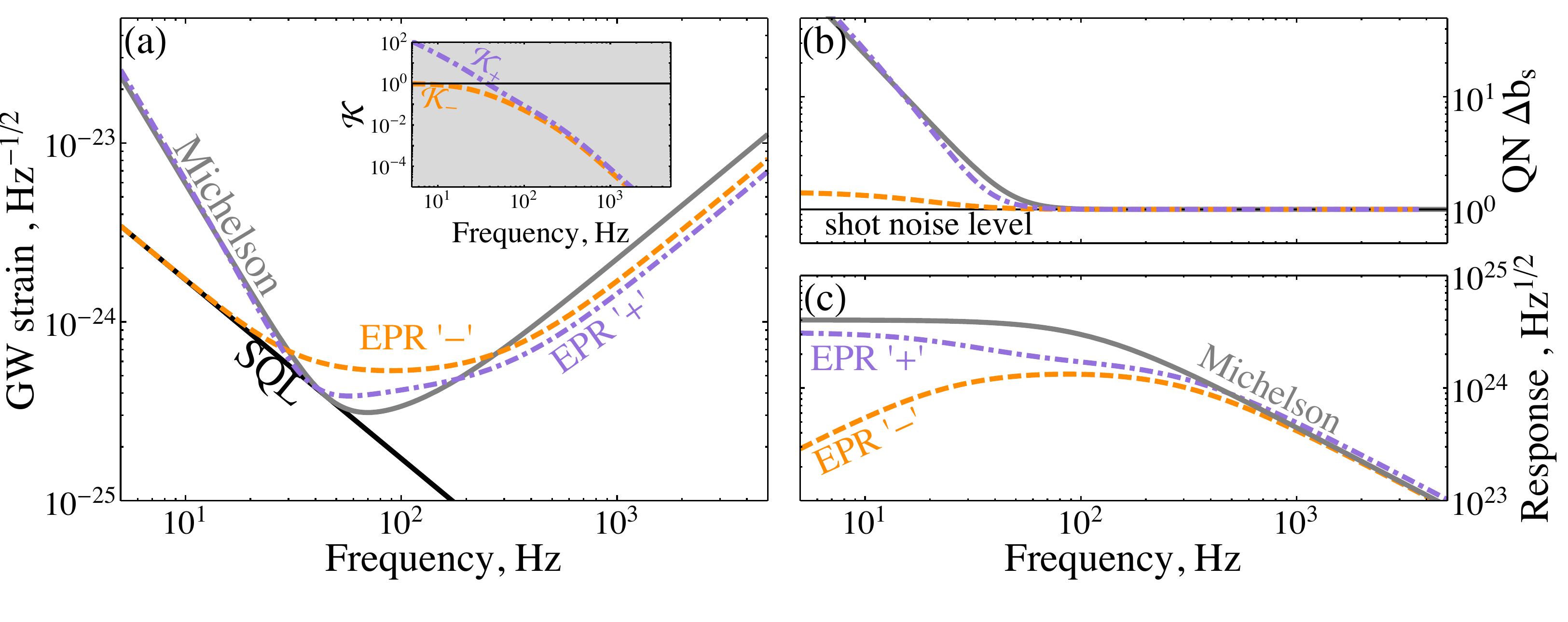}
\caption{Quantum noise of an \textit{EPR speed meter}: \textit{(a):} QNLS of the ``-'' channel (orange dashed trace) realising an EPR speed meter regime and of the ``+''-channel (purple dash-dotted trace) that realises an EPR position meter regime. Parameters for the curves are given in the table \ref{tab:benchmark_3G} and the plot of the QNLS of a  Michelson (grey trace) is given; \textit{(b)} quantum fluctuations of the phase quadrature of the readout light  the equivalent Michelson (grey trace) and of the ``+'' and ``-'' channels of the EPR interferometer; \textit{(c)} response functions of the Michelson (grey trace) interferometer and of the two channels of the EPR interferometer.} 
\label{fig:EPRSM_QNLS}
\end{center}
\end{figure}

Quantum noise of the EPR-speed meter can be calculated using the multi-channel formalism of Sec.~\ref{ssec:2-1} and the I/O-relations \eqref{eq:I/O_KLMTV} for each of the individual Michelson interferometers of the scheme save to the assumption of different bandwidth for each of them and account for the common back action on the test masses imposed by both carriers, \textit{i.e.}:
\begin{subequations}\label{eq:I/O_EPRSM}
\begin{align}
\hat b_c^{(1)} &= e^{2i\beta_1}\hat a_c^{(1)} \,,\\
	        \hat b_s^{(1)} &= \bigl[e^{2i\beta_1}\hat a_s^{(1)}-\bigl(e^{2i\beta_1}\mathcal{K}_1\hat a_c^{(1)}+ e^{i(\beta_1+\beta_2)}\sqrt{\mathcal{K}_1\mathcal{K}_2}\hat a_c^{(2)} \bigr) \bigr] + \nonumber\\
	         &  + e^{i\beta_1}\frac{\sqrt{2 \mathcal{K}_1}}{h_{\rm SQL}}h \,,\\
         \hat b_c^{(2)} &= e^{2i\beta_2}\hat a_c^{2} \,,\\
         \hat b_s^{(2)} &= \bigl[e^{2i\beta_2}\hat a_s^{(2)}-\bigl(e^{2i\beta_2}\mathcal{K}_2\hat a_c^{(2)} + e^{i(\beta_1+\beta_2)}\sqrt{\mathcal{K}_1\mathcal{K}_2}\hat a_c^{(1)} \bigr) \bigr] +\nonumber \\
         & + e^{i\beta_2}\frac{\sqrt{2 \mathcal{K}_2}}{h_{\rm SQL}}h \,.
\end{align}
Here $\mathcal{K}_{1,2}$ and $\beta_{1,2}$ stand for the OM coupling factors and sideband phase shifts of the two Michelsons, as defined by Eqs.~\eqref{eq:KMI} and \eqref{eq:BMI}.
Note the terms in parentheses in the equations for sine quadratures, which describe radiation pressure contributions to the outgoing light. The EPR entanglement of the outgoing light fields happens at the main beam splitter of the scheme and described by junction equations:
	\begin{align}
		\vq{b}^{(+)} &= \frac{\vq{b}^{(1)}+\vq{b}^{(2)}}{\sqrt{2}}\,, & \vq{b}^{(-)} &= \frac{\vq{b}^{(1)}-\vq{b}^{(2)}}{\sqrt{2}}\\
		\vq{a}^{(1)} &= \frac{\vq{a}^{(+)}+\vq{a}^{(-)}}{\sqrt{2}}\,, & \vq{a}^{(2)} &= \frac{\vq{a}^{(+)}-\vq{b}^{(-)}}{\sqrt{2}}\,.
	\end{align}
\end{subequations}
Solution of the above Eqs.~\eqref{eq:I/O_EPRSM} yields two output channels of the EPR-speed meter, namely the ``+''- and ``--''-channels that each carries an information about the GW-induced signal differential displacement the arms and quantum fluctuations of light:
\begin{align*}
    \vq{b}^{(+)} &= \tq{T}_{++}\vq{a}^{(+)} + \tq{T}_{+-}\vq{a}^{(-)} + \vb{t}_{+}\frac{h}{h_{\rm SQL}}\,,\\
    \vq{b}^{(-)} &= \tq{T}_{+-}\vq{a}^{(+)} + \tq{T}_{--}\vq{a}^{(-)} + \vb{t}_{-}\frac{h}{h_{\rm SQL}}\,.
\end{align*}
where transfer matrices $\mathbb{T}_{\pm\pm}$ are the subblocks of the $4\times4$ full transfer matrix of the form \eqref{eq:Nch-TrMat} read:
\begin{align*}
 \tq[++]{T} &= e^{i\beta_+}
 \begin{bmatrix}
 \cos{\beta_-} & 0\\
 -\frac12(\mathcal{K}_1e^{i\beta_-}+\mathcal{K}_2e^{-i\beta_-}+2\sqrt{\mathcal{K}_1\mathcal{K}_2}\cos\beta_-) & \cos\beta_-
 \end{bmatrix}\\
  \tq[--]{T} &= e^{i\beta_+}
 \begin{bmatrix}
 \cos{\beta_-} & 0\\
 -\frac12(\mathcal{K}_1e^{i\beta_-}+\mathcal{K}_2e^{-i\beta_-}-2\sqrt{\mathcal{K}_1\mathcal{K}_2}\cos\beta_-) & \cos\beta_-
 \end{bmatrix}\,,\\
 \tq[+-]{T} &= \tq[-+]{T} =e^{i\beta_+}\begin{bmatrix}
 i\sin{\beta_-} & 0\\
 -\frac12(\mathcal{K}_1e^{i\beta_-}-\mathcal{K}_2e^{-i\beta_-}) & i\sin\beta_-
 \end{bmatrix}
\end{align*}
and the responses of the ``+'' and ``--'' channels are:	
\begin{align}
     \vb{t}_{+} &=  \vb{t}_{1}+ \vb{t}_{2} =  e^{i\beta_+}\sqrt{2 \mathcal{K}_+}\begin{bmatrix}
     0 \\ 1
     \end{bmatrix}\,, &   \vb{t}_{-} &= \vb{t}_{1}- \vb{t}_{2} =  e^{i\beta_-}\sqrt{2 \mathcal{K}_-}\begin{bmatrix}
     0 \\ 1
     \end{bmatrix}\,,
\end{align}
where we defined the $\pm$-channel OM coupling factors, $\mathcal{K}_{\pm}$, and phase shifts, $\beta_\pm$ as:
\begin{align}
    \mathcal{K}_{\pm} \equiv= \mathcal{K}_1+\mathcal{K}_2\pm\sqrt{\mathcal{K}_1\mathcal{K}_2}\cos(\beta_1-\beta_2)\,, \ \mathrm{and}\ \beta_{\pm} \equiv \beta_1\pm\beta_2\,.
\end{align}

Now, if one looks closely at the structure of the OM factors $\mathcal{K}_{j=1,2}$ in Eq.~\eqref{eq:KMI}, one sees that they can be factorised as follows:
\begin{align}
  \mathcal{K}_j &\simeq  \Bigl[\frac{2 \Theta_j}{ \gamma_j\Omega^2}\Bigr]\frac{\gamma_j^2}{(\gamma_j^2+\Omega^2)}=\kappa_j\cos^2\beta_j\,, & (j=1,2)\,,
\end{align}
with $\kappa_j \equiv \frac{2 \Theta_j}{ \gamma_j\Omega^2}$. Those can be made equal to each other at low enough frequencies ($\Omega\ll\mathrm{min}[\gamma_{1},\,\gamma_{2}]$), if powers and bandwidths of the individual MIs satisfy the following relation:
\begin{equation}\label{eq:EPRSM_cond}
  \frac{\Theta_1}{ \gamma_1} = \frac{\Theta_2}{ \gamma_2}\,,
\end{equation}
which provides $\kappa_1=\kappa_2 \equiv \mathcal{K}_0 = 2\Theta/(\Omega^2(\gamma_1+\gamma_2))$ with $\Theta=\Theta_{1}+\Theta_{2}$ the sum power in both MIs.
In this case, one can get for $\mathcal{K}_-$:
\begin{align}
\mathcal{K}_{-} = \mathcal{K}_{0} \sin^2(\beta_1-\beta_2) = \frac{2\Theta(\gamma_1-\gamma_2)^2}
      {(\gamma_1+\gamma_2)(\gamma_1^2+\Omega^2)(\gamma_2^2+\Omega^2)} \,,
\end{align}
which behaves exactly as one expects from the speed meter, namely it tends to a constant value at low enough frequencies (see orange dashed trace on inset plot in FIg.~\ref{fig:SSM_QNLS}a). Hence, ``-''-channel of the EPR-scheme indeed performs the speed measurement (see orange dashed trace in FIg.~\ref{fig:SSM_QNLS}c). However, there is an additional back-action created by the vacuum fields $\vq{a}^{(+)}$, entering the ``+'' port of the beam splitter that compromise the speed meter's low-frequency advantage. This may be explained by the fact that displacement information flows out of the ``+'' channel with the $\vq{b}^{(+)}$ light fields, as one can see from the plot of the response of the ``+''-channel given by a purple dash-dotted trace in FIg.~\ref{fig:SSM_QNLS}c. This ensues from the shape of the OM coupling factor of the ``+''-channel that reads: 
\begin{multline}
\mathcal{K}_{+} = \mathcal{K}_{0} (\cos^2\beta_1+\cos^2\beta_2-2\cos\beta_1\cos\beta_2\cos(\beta_1-\beta_2)) =\\ 
 \frac{2\Theta [4\gamma_1^2\gamma_2^2 + \Omega^2(\gamma_1+\gamma_2)^2]}
      {\Omega^2(\gamma_1+\gamma_2)(\gamma_1^2+\Omega^2)(\gamma_2^2+\Omega^2)}\,.
\end{multline}
that grows as $\mathcal{K}_{+}(\Omega\ll\gamma_j)\propto \Omega^{-2}$ at low frequencies (see the inset in FIg.~\ref{fig:SSM_QNLS}a).

\begin{figure}[htbp]
\begin{center}
\includegraphics[width=\textwidth]{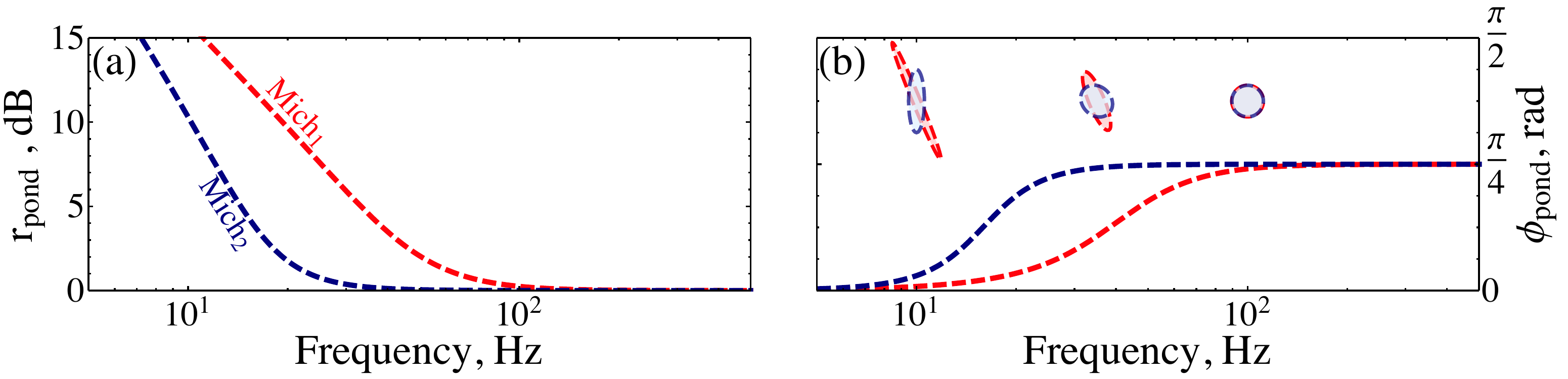}
\caption{Ponderomotive squeezing of the output fields $\vq{b}^{(1)}$ and $\vq{b}^{(2)}$ of the two MIs that compose an EPR-speed meter, before the main beam splitter. \textit{Panel a)} shows the ponderomotive squeezing factor of the outgoing light of the MI 1(red trace) and of the MI 2 (blue trace). \textit{Panel b)} shows the dependence of squeezing angle at different frequencies as well as squeezing ellipses for the corresponding MIs (not to scale!) with phase quadrature uncertainty along the vertical axis. For clarity of representation we chose $\gamma_1/2\pi = 60$~Hz and $\gamma_2/2\pi = 6$~Hz here.} 
\label{fig:EPRSM_pondSQZ}
\end{center}
\end{figure}

It is possible however to remove this additional back-action noise from the readout of the ``-'' channel, by measuring at the ``+''-channel the quadrature (amplitude) that is responsible for this back action, and subtracting it with a proper frequency-dependent weight function from the readout of the ``-''-channel (cf. Sec.~\ref{ssec:2-1} and the Appendix of\cite{2017_Phys.Lett.A_EPR_SM}). Quantum correlations between the two readouts are at its strongest at the lower frequencies, where both MIs output light fields, $\vq{b}^{(1)}$ and $\vq{b}^{(2)}$, are strongly squeezed due to ponderomotive squeezing discussed in Sec.~\ref{sec:1}, as shown in Fig.~\ref{fig:EPRSM_pondSQZ}. After the beam splitter ``+'' and ``-'' channel are highly correlated (entangled), which allows aforementioned subtraction of quantum back action noise. The resulting QN-limited sensitivity for the speed meter channel (see orange trace in Fig.~\ref{fig:EPRSM_QNLS}a) follows the general tuned interferometer pattern of Eq.~\eqref{eq:QNLS_tuned_IFO}:
\begin{equation}\label{eq:S_EPR} 
  S^h_{\rm EPR-} = \frac{h^2_{\rm SQL}}{2}\left[ \frac{(\mathcal{K}_--\cot\phi_{\rm LO}^{(-)})^2+1}{\mathcal{K}_-}
    \right] ,
\end{equation}  
with $\phi_{\rm LO}^{(-)}$ standing for the ``-''-channel readout quadrature angle. Similar expressionfor the ``distilled'' sensitivity of the ``+'' channel that exhibits a vivid position-meter behaviour, is obtained by exchanging ``-'' indices by ``+'' in the above formula, and the resulting plot is shown as a purple trace Fig.~\ref{fig:EPRSM_QNLS}a. Note that both expressions are special cases of the Eq.~\eqref{eq:o_multichan_PSD}.

\subsection{Imperfections and loss in speed-meter interferometers}

Speed-meter interferometers suffer in general from the same sources of noise due to loss as Fabry-Perot--Michelson interferometers, since in most of the speed-meter configurations presented in Fig.~\ref{fig:SM_Zoo} FPMI itself is an integral part. However, loss influence is different from the FPMI and there are also some specific features of speed meters worth mentioning. 

Firstly, cancellation of back-action noise in speed meters comes from the coherent subtraction of back action forces created by the same light beam in two consequent interactions with the test mass. Therefore any admixture of incoherent vacuum due to loss between the two interactions, \textit{e.g.} the loss in the arms, creates an unbalanced back-action force. This loss-associated back action leads to a position-meter-like rise of quantum noise at low frequencies, where it starts to dominate over the suppressed quantum back action of the speed meter. 

Secondly, as any scheme where balancing of the noise contributions between the arms is essential for the noise cancellation, speed meters are very sensitive to the asymmetry of the arms, as discussed in detail in \cite{2015_NJP17.043031_asymSag}. Asymmetry of the beam splitter in Sagnac interferometers, for instance, creates a coupling of laser fluctuations to the readout port through an excess radiation pressure they create on the mirrors which is quite strong as laser noise in the low-frequency range is far from the shot noise limit. This excess noise, however can be cancelled by a wise choice of local oscillator in the balanced homodyne readout as shown in the recent study by Zhang \textit{et al.} \cite{Zhang:2018czu}.

In general, speed-meter interferometers show higher robustness to intracavity loss than Michelson ones due to lower back-action component of quantum noise, which means lower ponderomotive squeezing as discussed in Sec.~\ref{ssec:1-3}. This reduces the effect of loss vacuum fields on the internal squeezing of light since quantum correlation between phase and amplitude fluctuations is already suppressed by the speed meter. For the same reason the requirements on tolerable filter cavity loss and bandwidth in case of frequency-dependent squeezing injection are significantly relaxed for speed meters, The detailed study of loss influence on speed-meter quantum noise is given in \cite{2015arXiv150301062V}.

\subsection{Summary and outlook}

Speed-meter interferometers are arguably the most elaborate and well studied concept alternative to the Michelson interferometer based on position measurement. Their main advantage is a greatly reduced back-action noise that potentially allows to increase the rate of detection of massive binary black-hole systems by up to 2 orders of magnitude compared to the equivalent position meter \cite{2018_LSA.7.accepted} if only quantum noise is considered. Although there is an obvious penalty of vanishing response at low frequencies, the reduction of back-action is still greater to make the overall increase of the SNR worth it. The progress in development of new, more practical topologies of speed meters shown in Fig.~\ref{fig:SM_Zoo} has led to designs that allow to keep the main interferometer intact, yet this comes at a price of using polarisation optics that is prone to imperfections and even more importantly, it requires development of the new all-polarisation type mirror coatings. 

We have considered here all three main genera of speed meters and gave the comparison of their performance. All studies done so far indicate the superiority of speed meters' performance over that of the conventional Michelson interferometers even in the presence of losses and imperfections \cite{PhysRevD.87.096008,Miao14,2015arXiv150301062V,2018_LSA.7.accepted}. However, a thorough and systematic study of losses and imperfections in all the speed-meter schemes is needed as well as experimental prototyping, before any final conclusion can be made.

In the context of FQL, the speed-meter configuration is an approach to shaping the power fluctuation inside
the arm cavity. The FQL can be reached at low frequencies, where the optomechanical 
coupling strength is approximately constant, by using the frequency-independent readout rather than 
the frequency-dependent readout as in the case of a position meter (Michelson interferometer). 
In the tuned case, the price 
we paid is that the power fluctuation gets reduced at low frequencies and the resulting FQL is parallel to 
the SQL rather than flat for the position meter. 

An interesting future direction is to investigate detuned speed-meter configurations with 
additional intra and external filters. Since the optomechanical coupling strength is approximately 
constant at low frequencies, this means the resulting ponderomotive squeezing is frequency independent 
at these frequencies. With detuning, the optical feedback, illustrated in Fig.\,\ref{fig:feedback_FQL}, 
could result in a broadband enhancement of the power fluctuation. Or equivalently, this can be viewed 
as a broadband enhancement of the mechanical response of the test mass, similar to the idea of 
negative inertia to be discussed in the section that follows. 

%
%
%
%
%
%
%
%

\section{Interferometers with optomechanically modified dynamics} 
\label{sec:5}

%
%


\subsection{Introduction}

All schemes of suppression of quantum noise considered so far in this paper are based on the same principle, namely the {\it quantum noise cancellation}, that is the  based on mutual compensation of the measurement noise and the back action noise which is possible by means of introducing the cross-correlation between these noise sources, see  Sec.\,4.4. of \cite{Liv.Rv.Rel.15.2012}. The main problem of this approach is that that the quantum correlations are very fragile and can be easily destroyed by additional noises caused by optical losses in the interferometer and by the non-ideal quantum efficiency of photodetectors. A rule of thumb for the limit of  achievable SQL-beating in this case can be presented as follows \cite{09a1ChDaKhMu}:
\begin{equation}
  S \gtrsim S_{\rm SQL}e^{-r}\sqrt{\frac{1-\eta}{\eta}} \,.
\end{equation}
Here $S$ is the sum quantum noise spectral density of the detector, $S_{\rm SQL}$ is the corresponding SQL, $e^{-r}$ is the squeeze factor and $\eta$ is the unified quantum efficiency of the detector. Even for rather optimistic values of the optical parameters with $\eta=0.95$ and $e^{-2q}=0.1$ (10~dB squeezing), we have $S/S_{\rm SQL}\gtrsim0.07$, which means that sensitivity (in units of the signal amplitude) can surpass the SQL by only a factor of $\sqrt{S_{\rm SQL}/S}\lesssim4$ with the noise-cancellation schemes.

At the same time, the SQL, normalized to the signal force, decreases as the test object susceptibility increases. Because this approach does not require any precise mechanisms for mutual compensation of measurement noise and back-action noise (and, in particular, the SQL is not evaded), it is much more robust with respect to optical losses, than quantum noise cancellation.

A trivial example is just the use of smaller {\it inertial} mass $m_{\rm inert}$. This method can be used, for example, in atomic force microscopes. However, when detecting forces of a gravitational nature, particularly in gravitational-wave experiments, the signal force is proportional to the test-object {\it gravitational} mass $m_{\rm grav}$. Taking into account that, due to the equivalence principle, $m_{\rm inert}=m_{\rm grav}$, the overall sensitivity decreases with the mass, which can be seen, for example, from the expressions for SQL in the $h$-normalization \eqref{eq:SQL} (see, however, Sec.\,\ref{sec:neg_m} below).

Another possibility is to use a harmonic oscillator instead of a free test mass. The susceptibility of a harmonic oscillator rapidly increases near its resonance frequency $\Omega_0$, which improves the $S_{\rm SQL}$ by a factor of $\Omega_0/\Delta\Omega$ in the frequency band $\Delta\Omega$ centered at $\Omega_0$, see Sec.4.3.2 of \cite{Liv.Rv.Rel.15.2012}. This method was demonstrated in several ``table-top'' experiments with mechanical nano-oscillators \cite{Teufel2009,Anetsberger_NPhys_5_909_2009,11a1WeFrKaYaGoMuDaKhDaSc}. In laser gravitational-wave detectors, the characteristic eigenfrequencies of the test mirror pendulum modes are close to 1\,Hz, and in the operating frequency range these mirrors can be considered as almost free masses. Evidently, it is technically impossible to turn the differential mechanical mode of laser detector test mirrors into an oscillator with a frequency in the operating frequency range by using ``ordinary'' springs. However, the {\it optical spring} which arises in detuned interferometer configurations and possesses excellent noise properties can be used for this purpose instead.

The optical spiring is a particular case of the more general electromagnetic rigidity (e.m.) effect, which takes place in any detuned e.m. resonator. This effect, together with the associated e.m. damping, were most probably first discovered and explained in the very early work \cite{64a1eBrMi}, where the low-frequency (sub-Herz) torsional pendulum was used as the mechanical object and the radio-frequency capacitor transducer --- as the position sensor. Few years later, existence of these effects in the optical Fabry-Perot cavities (that is the {\it optical} spring proper) was predicted theoretically \cite{67a1eBrMa}. After that, the e.m. damping was observed in the microwave Fabry-Perot type cavity \cite{70a1eBrMaTi}. In the beginning of 1980s, the first truly optical experiment was done \cite{Dorsel_PRL_51_1550_1983}.

Much later, quantum noise properties of the optical spring and the optical damping were analyzed in the papers \cite{97a1BrGoKh,01a1BrKhVo,99a1BrKh,02a1BrVy} and it was shown that the noise temperature of the optical damping can be very close to zero. This stimulated a series of experimental works where the optical rigidity was observed both in table-top optical setups \cite{03a1eBiSa,Sheard_PRA_69_051801_2004,Corbitt2006,Corbitt2007,Corbitt_PRL_99_160801_2007} and in larger-scale Caltech 40\,m interferometer devoted to prototyping of future GW detectors \cite{Miyakawa_PRD_74_022001_2006}.

It have to be mentioned also that the very low noise temperature of the e.m. damping stimulated also a bunch of optomechanical and electromechanical experiments aimed at preparation of mechanical resonators in the ground state using this cold damping, see \eg the works  \cite{Teufel_Nature_475_359_2011,Chan_Nature_478_89_2011} and the reviews \cite{Aspelmeyer_RMP_86_1391_2014,16a1DaKh}.

Specifically in the context of the large-scale gravitational-wave detectors the optical rigidity was analyzed in the papers \cite{97a1BrGoKh,Buonanno2001,01a2Kh,Buonanno2002,Buonanno2003}. Most notably, it was shown in these works that in very long cavities with the bandwidth $\gamma$ comparable with the or smaller than the characteristic mechanical frequencies $\Omega$, the optical spring has sophisticated frequency dependence which enables some interesting applications, see below.

\subsection{Optical rigidity}\label{sec:emspring}

The e.m. rigidity and the e.m. damping effects were correctly explained in \cite{64a1eBrMi} by respectively, dependence of the e.m. eigen frequency and therefore of the energy $\mathcal{E}$ stored in the e.m. resonator on the mechanical position $x$ and by the time lag between the variation of $x$ and the variation of $\mathcal{E}$. We reproduce below the semi-qualitative, but simple and transparent reasoning of that paper.

Really, if
\begin{equation}
  \omega_0(x) = \omega_0(1-x/L) \,,
\end{equation}
then the effective detuning is equal to
\begin{equation}
  \delta(x) = \omega_p - \omega_0(x) = \delta + \frac{\omega_0 x}{L} \,,
\end{equation}
and the optical energy is equal to
\begin{equation}
  \mathcal{E}(x) = \frac{\gamma^2+\delta^2}{\gamma^2+\delta^2(x)}\,\mathcal{E} \,,
\end{equation}
where $\mathcal{E}$ is the initial (at $x=0$) value of the energy. This, in turn, leads to the $x$-dependence of the ponderomotive force that acts on the
mechanical object:
\begin{equation}\label{F_pond_simple}
  F(x) = \frac{\mathcal{E}(x)}{L} \approx F(0) - Kx + O(x^2)\,.
\end{equation}
where
\begin{equation}\label{em_K_simple}
  K = -\partd{F(x)}{x}\biggr|_{x=0} = \frac{m\Theta\delta}{\gamma^2+\delta^2}
\end{equation}
is the e.m. rigidity.

Note also that the optical energy follows the mechanical motion not instantly, but with some delay $\tau_{\rm delay} \sim 1/\gamma$. Therefore, the force \eqref{F_pond_simple} actually is equal to [we omit the constant term $F(0)$]
\begin{equation}
  F \approx -Kx(t-\tau_{\rm delay}) \approx -Kx(t) + K\tau_{\rm delay}\frac{dx(t)}{dt}
  \approx -Kx(t) - H\frac{dx(t)}{dt} \,,
\end{equation}
where
\begin{equation}\label{em_H_simple}
  H = -K\tau_{\rm delay} \,.
\end{equation}
is the e.m. damping.

The rigorous quantum treatment of the e.m. rigidity was done in the mentioned above articles \cite{97a1BrGoKh,Buonanno2001,01a2Kh,Buonanno2003,Buonanno2003}. It was shown there that it is equal to
\begin{equation}\label{em_KH}
  K(\Omega) = \frac{m\Theta\delta}{\mathcal{D}(\Omega)}
  = \Re K(\Omega) - i\Omega H(\Omega) \,.
\end{equation}
It is easy to see, that Eqs.\,(\ref{em_K_simple}, \ref{em_H_simple}) describe the quasistatic (slow mechanical motion, $\Omega\to0$) particular case of \eqref{em_KH} with the effective delay time
\begin{equation}
  \tau_{\rm delay} = \frac{2\gamma}{\gamma^2+\delta^2} \,.
\end{equation}

According to the fluctuation-dissipation theorem, any damping $H(\Omega)$ is accompanied by the noise force having the spectral density
\begin{equation}
  S_T(\Omega) = 2\varkappa_B|H(\Omega)|\mathcal{T}(\Omega) \,,
\end{equation}
where
\begin{equation}\label{S_FDT}
  \mathcal{T}(\Omega)
  = \frac{\hbar\Omega}{2\varkappa_B}\coth\frac{\hbar\Omega}{2\varkappa_BT}
\end{equation}
is the mean energy of the heatbath modes at the frequency $\Omega$, expressed in units of kelvins, and $T$ is the effective noise temperature.

In the optical spring case, the fluctuational pondermotive force $F_{\rm fl}$ imposed by the quantum fluctuations of the optical energy in the interferometer play the role of the thermal noise. Spectral density of this noise is calculated, inparticular, in the Sec.\,6 of \cite{Liv.Rv.Rel.15.2012}, see Eq.\,(473). Combining this equation with Eqs.\,(\ref{em_KH}, \ref{S_FDT}), we obtain that
\begin{equation}
  \mathcal{T}(\Omega) = \frac{S_{FF}(\Omega)}{2\varkappa_B|H(\Omega)|}
  = \frac{\hbar}{2\varkappa_B}\,\frac{\gamma^2+\delta^2+\Omega^2}{2|\delta|} \,.
\end{equation}
Minimum of this expression at any given frequency $\Omega$ is provided by $\delta = -\sqrt{\gamma^2 +\Omega^2}$ and is equal to
\begin{equation}
  \mathcal{T}(\Omega) = \frac{\hbar\sqrt{\gamma^2+\Omega^2}}{2\varkappa_B} \,.
\end{equation}
In the case of the narrow-band cavity, $\gamma\ll|\Omega|$,
\begin{equation}
  \mathcal{T}(\Omega) \to \frac{\hbar|\Omega|}{2\varkappa_B} \,,
\end{equation}
which corresponds to the noise temperature $T\to0$. The opposite case of the broad-band cavity, $\gamma\gg|\Omega|$ translates to much higher ``temperature''
\begin{equation}
  T \approx \mathcal{T} \approx \frac{\hbar\gamma}{2\kappa_B}
    \gg \frac{\hbar|\Omega|}{2\varkappa_B} \,.
\end{equation}
However, large $\gamma$ means strong flow of information on the mechanical position $x$ from the cavity. This means that the fluctuational pondermotive force $F_{\rm fl}$ in this case has be treated not as the thermal noise of the optical damping, but as the quantum back action due to the measurement.

\subsection{Characteristic regimes of the optical spring}

Non-trivial frequency dependences of the optical rigidity \eqref{em_KH} and of the quantum noise components of the detuned interferometers [see Eqs.\,(376-378) of \cite{Liv.Rv.Rel.15.2012}] lead to very sophisticated shape of the corresponding sum quantum noise spectral density, see Eq.\,(385) of \cite{Liv.Rv.Rel.15.2012}. This shape can be tuned flexibly by varying the interferometer bandwidth $\gamma$ and detuning $\delta$, homodyne and squeezing angles, and the squeezing amplitude, with the optimal tuning depending on many factors, such us the available optical power, intensity of non-quantum (``technical'') noise sources, optical losses \etc. The corresponding exhaustive optimization exceeds the scope of this paper (as well as probably any single paper). Broad set of examples covering the most typical scenarios can be found \eg in the articles \cite{Buonanno2001,Buonanno2003,08a1KoSiKhDa,Liv.Rv.Rel.15.2012}. Therefore, here we concentrate specifically on the modification of the mechanical probe dynamics by the optical spring.

In Fourier domain, mechanical dynamics is described by the response function
\begin{equation}\label{spring_chareq}
  \chi^{-1}(\Omega) = m\left(-\Omega^2 + \frac{\Theta\delta}{\mathcal{D}(\Omega)}\right)
    \,.
\end{equation}
Analysis of the roots of the characteristic equation $\chi^{-1}(\Omega)=0$ shows that this response function can have either two resonance minima or one broader
minimum. If the interferometer bandwidth is sufficiently small, $\gamma\ll\delta$, then
frequencies of these minima can be approximated as
\begin{equation}
  \Omega_{1,2}^2 \approx
      \frac{\delta^2}{2}\pm\sqrt{\frac{\delta^4}{4} - \Theta\delta} \,.
\end{equation}
In the weak pumping case with $\Theta\ll\Theta_{\rm crit}$, where
\begin{equation}
  \Theta_{\rm crit} = \frac{\delta^3}{4}
\end{equation}
is the critical value of the normalize optical power $\Theta$, roots of $\chi^{-1}(\Omega)$ are approximately equal to
\begin{equation}
  \Omega_1 \approx \sqrt{\frac{\Theta\delta}{\gamma^2+\delta^2}}\quad\text{and}\quad
    \Omega_2 \approx \delta \,.
\end{equation}
The first of this root corresponds to the resonance frequency of the ordinary harmonic oscillator created by the static optical rigidity \eqref{em_K_simple}. This is so called {\it mechanical resonance}. The second root, so called {\it optical resonance}, is created by the sharp increase of the optical rigidity at $\Omega\approx\delta$, which allows the second term in \eqref{spring_chareq} cancel the first one even if $\Theta$ is small.

With the increase of the ratio $\Theta/\Theta_{\rm crit}$, these two roots drift toward each other (see Fig.\,44 of \cite{Liv.Rv.Rel.15.2012}), and the area with the reduced $\chi^{-1}(\Omega)$ (that is, with better sensitivity) forms between them \cite{Buonanno2001}. At $\Theta\to\Theta_{\rm crit}$ the roots merge into one broader second-order one. The detailed analysis of this {\it second order pole} regime \cite{01a2Kh} can be found in Sec.\,6.3.4 of \cite{Liv.Rv.Rel.15.2012}. In particular, it is shown there this in essence narrow-band regime can in principle provide an arbitrarily-high signal-to-noise ratio for broadband signals, limited only by the level of the additional noise of non-quantum (technical) origin.

\begin{figure}
  \includegraphics{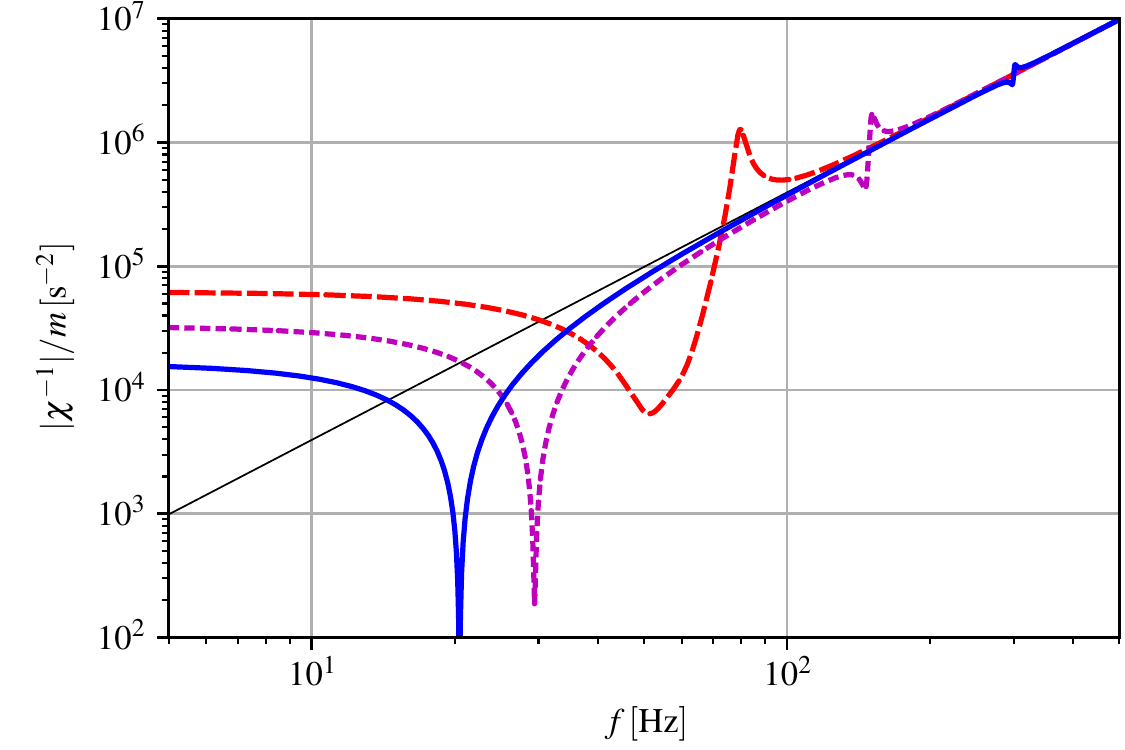}
  \caption{Mechanical response function modified by the optical rigidity. This solid line: free mass ($|\chi^{-1}|/m=-\Omega^2$); thick solid line: $\delta=2\pi\times300\,{\rm s}^{-1}$; short dashes: $\delta=2\pi\times150\,{\rm s}^{-1}$; long dashes: $\delta=(4\Theta)^{1/3}$. In all cases, $\gamma=2\pi\times2\,{\rm s}^{-1}$ and $\Theta=(2\pi\times50)^3\,{\rm s}^{-1}$.
}\label{fig:chi1}
\end{figure}

In Fig.\,\ref{fig:chi1} the absolute value of $\chi^{-1}$ normalized by the mechanical mass $m$ is plotted as a function of the frequency for these characteristic cases. For pedagogical reason (to emphasize the frequency dependencies), a very small value of $\gamma=2\pi\times2\,{\rm s}^{-1}$ is used in these plots. It worth to be mentioned however such a narrow bandwidth actually can be used in configurations with two optical carriers belonging to two free spectral ranges of the interferometer, with one of them having ``standard'' $\gamma\sim10^3\,{\rm s}^{-1}$ and being used for the measurement, and another one, with the small $\gamma$, creating the optical rigidity. Two different values of the bandwidth can be implemented in the signal-recycled configurations of GW detectrors by using the resonant sideband extraction and the ultimate signal recycling regimes for respectively the broadband and the narrow-band carriers.

\begin{figure}
  \includegraphics{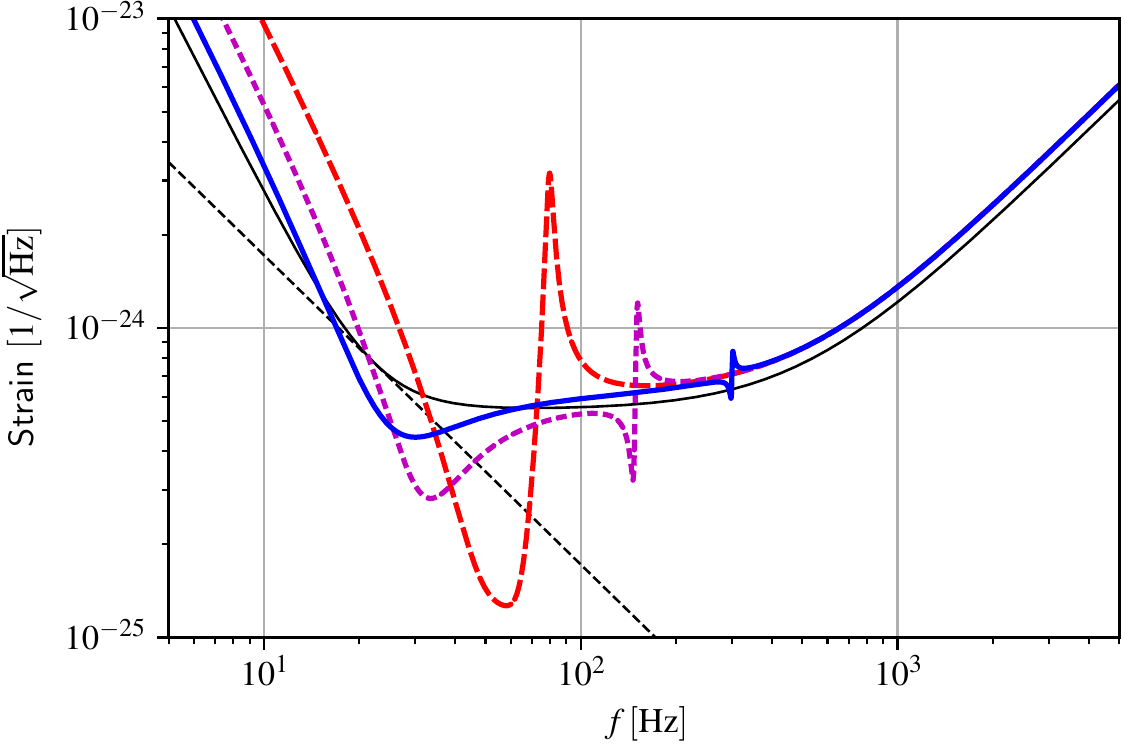}
  \caption{Quantum noise spectral densities for the two-carrier interferometer, with the dedicated second carrier creating the optical spring. Thin dashed line: SQL; thin solid line: lossless SQL-limited interferometer; thick solid line: detuning of the second carrier $\delta=2\pi\times300\,{\rm s}^{-1}$; short dashes: $\delta=2\pi\times150\,{\rm s}^{-1}$; long dashes: $\delta=(4\Theta)^{1/3}$. Bandwidths: $\gamma_1=2\pi\times500\,{\rm s}^{-1}$ (the ``signal'' optical mode) and $\gamma_2=2\pi\times2\,{\rm s}^{-1}$ (the ``spring'' optical mode). For both carriers, $\Theta=(2\pi\times50)^3\,{\rm s}^{-1}$. Quantum efficiency $\eta=0.8$. Other parameters are listed in Table\,1.
}\label{fig:S_spring}
\end{figure}

In Fig.\,\ref{fig:S_spring}, quantum noise spectral densities for this setup are plotted for the same three characteristic values of the detuning of the narrow-band carrier as in Fig.\,\ref{fig:chi1}. It is instructive to compare these plots with the corresponding ones for the case of the single carrier detuned carrier, see {\it e.g.} Fig.\,45 of \cite{Liv.Rv.Rel.15.2012}. It is easy to see that while in the latter case the use of the detuned regime leads to sharp degradation of sensitivity at higher frequencies, in the former one the high-frequency sensitivity remains intact. 

In these plots, we assumed good but not very high value of the overall quantum efficiency of the interferometer $\eta=0.8$ (note that in ``ordinary'' interferometers without squeezed light injection, all optical losses can be absorbed into this unified factor, see Sec.\,6.3.2 of \cite{Liv.Rv.Rel.15.2012}). This resulted only in the barely-visible sensitivity degradation in the shot noise dominated high-frequency area, confirming the above statement about tolerance of the optical spring based schemes to optical losses.

\subsection{Cancellation of mechanical inertia}\label{sec:neg_m}

In the interferometer configurations with two or more optical carriers, more deep modification of the mechanical dynamics is possible, allowing, in some sense, to make the mechnaical inertial mass $m_{\rm inert}$ smaller than the gravitational one $m_{\rm grav}$ by attaching a {\it negative optical inertia} to the former one \cite{11a1KhDaMuMiChZh,Liv.Rv.Rel.15.2012}. Existence of this effect immediately follows from the frequency dependence of the optical spring \eqref{em_KH}.

Assume for simplicity that $\gamma\to0$ and $\Omega\ll\delta$. In this case, Eq.\,\eqref{em_KH} can be approximated as follows:
\begin{equation}
  K(\Omega) \approx K(0) - m_{\rm opt}\Omega^2 \,,
\end{equation}
where $K(0)=\Theta/\delta$ is the static rigidity and
\begin{equation}
  m_{\rm opt} = -\frac{m\Theta}{\delta^3}
\end{equation}
is the optical inertia. which, similar to $K(0)$, can be either positive or negative depending on the sign of the detuning $\delta$.

We now assume that the interferometer is pumped with two detuned carriers having frequencies belonging to different free spectral ranges of the interferometer. In this case, each carrier creates its own optical rigidity. These two rigidities $K_{1,2}$ can be combined in such a way that their static parts would compensate each other and the total optical inertia would compensate the usual mechanical inertia of the test mass:
\begin{subequations}
  \begin{gather}
    K_1(0) + K_2(0) = 0 \,, \\
    m_{\rm opt1} + m_{\rm opt2} = -m \,.
  \end{gather}
\end{subequations}
Obviously, the exact compensation would happen only at zero frequency, but at other sufficiently small frequencies, the responce of such a test object would be significantly stronger than that of the initial test mass.

Let us derive the conditions for this inertia cmpensations. The two optical springs modify the mechanical susceptibility as follows:
\begin{multline}\label{rchi}
  \chi^{-1}(\Omega) = -m\Omega^2 + K_1(\Omega) + K_2(\Omega) \\
  = m\frac{
        -\Omega^2\mathcal{D}_1(\Omega)\mathcal{D}_2(\Omega)
        + \Theta_1\delta_1\mathcal{D}_2(\Omega)
        + \Theta_2\delta_2\mathcal{D}_1(\Omega)
      }{\mathcal{D}_1(\Omega)\mathcal{D}_2(\Omega)}  \,,
\end{multline}
where
\begin{equation}
  \mathcal{D}_{1,2}(\Omega) = (\gamma_{1,2}-i\Omega)^2 + \delta_{1,2}^2
\end{equation}
and the parameters $\gamma_{1,2}$, $\delta_{1,2}$, and $\Theta_{1,2}$ correspond to the respective carriers. The conditions for cancelation of the total inertia and rigidity are equivalent to the cancelation of the terms proportional to $\Omega^2$ and $\Omega^0$ in the numerator of Eq.\,\eqref{rchi}. Calculation gives that this cancelation is provided by
\begin{equation}\label{neg_inertia_cond}
  \Theta_1\delta_1 = \frac{\varGamma_1^4\varGamma_2^2}{\varGamma_2^2-\varGamma_1^2} \,,
    \qquad
  \Theta_2\delta_2 = \frac{\varGamma_1^2\varGamma_2^4}{\varGamma_1^2-\varGamma_2^2} \,,
\end{equation}
where $\varGamma_{1,2}^2=\gamma_{1,2}^2+\delta_{1,2}^2$. It follows from these equations, that since $\Theta_{1,2}$ are, by definition, positive quantities, the signs of the detunings has to be opposite, with negative detuning corresponding to the larger $\varGamma$. Below we assume that $\varGamma_2>\varGamma_1$, $\delta_1>0$, and $\delta_2<0$.

\begin{figure}
  \includegraphics{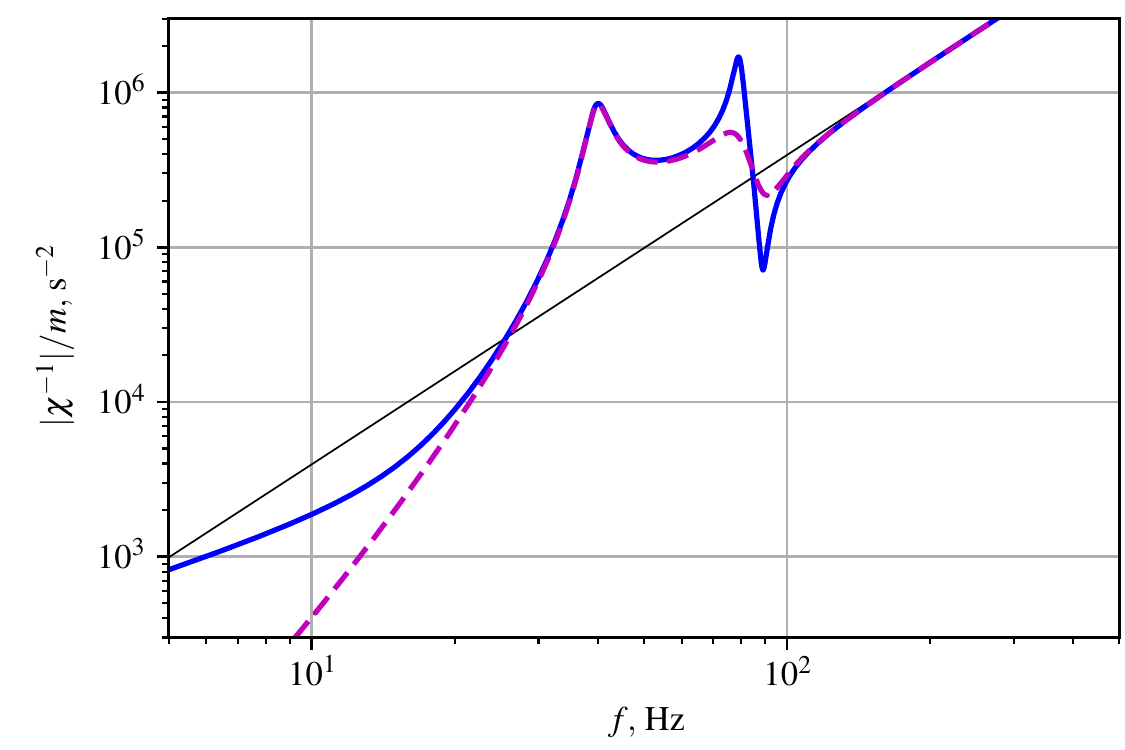}
  \caption{Mechanical response function modified by the negative optical inertia. This solid line: free mass ($|\chi^{-1}|/m=-\Omega^2$); thick solid line: $\gamma_1=\gamma_2=2\pi\times2\,{\rm s}^{-1}$ (the inertia and the static rigidity are canceled); dashed line: $\gamma_1=2\pi\times2\,{\rm s}^{-1}$, $\gamma_2=4\gamma_1$, see Eq.\,\eqref{gammas_ratio} (the inertia, the static rigidity, and the optical damping are canceled).
  In both cases,  $\Theta_1=2/3\times(2\pi\times100)^3\,{\rm s}^{-1}$, $\Theta_2 = 2\Theta_1$, $\delta_1\approx2\pi\times40\,{\rm s}^{-1}$, $\delta_2=-2\delta_1$, see Eqs.\,\eqref{neg_inertia_cond}.
}\label{fig:chi2}
\end{figure}

The resulting mechanical response function is plotted in Fig.\,\ref{fig:chi2} (solid line). It can be seen from this plot that indeed below some threshold frequency (it can be shown that it is equal to the smaller detuning $\delta_1$) the value of $\chi^{-1}$ is noticeably suppressed (that is, the mechanical probe is more responsive) in comparison with the free mass.  The ``residual'' low-freqeuncy value of $|\chi^{-1}|$ is created by optical damping, and for the parameters values used in this example, the gain is limited.

This scheme has also another disadvantage, namely it is dynamically unstable, and this instability could be significant. In paper \cite{11a1KhDaMuMiChZh}, two methods of damping this instability were proposed. First, {\it partial} compensation of the mechanical inertia is possible, with the remaining non-zero inertia stabilizing the system and making the instability time long enough to be damped by an out-of-band feedback system.

The second way is to cancel, in addition to the rigidity and inertia, also the damping. This approach allows also to significantly improve the gain in $|\chi^{-1}|$. This cancellation can be achieved by adjusting the bandwidths $\gamma_{1,2}$ as follows:
\begin{equation}\label{gammas_ratio}
  \frac{\gamma_2}{\gamma_1} = \left|\frac{\Theta_1\delta_1}{\Theta_2\delta_2}\right| .
\end{equation}
The corresponding response function is also plotted in Fig.\,\ref{fig:chi2} (dashed line), demonstrating much more significant low-frequency gain.

Similar to the previous (single optical spring) case, two strategies of
implementation of the negative inertia are possible. In the first one, two carriers have to be used in order to both create the negative inertia and also measure the test mirrors motion. This strategy was analyzed in detail in the work \cite{Liv.Rv.Rel.15.2012} in the context of the Advanced LIGO parameters set. The second one require three dedicated carriers: one for the measurement and additional two for creation of the negative inertia. Unfortunately, in both cases the results can not be considered as satisfactory ones. Within the optical power constrains of existing and planned GW detectors, they can provide only very  moderate low frequency sensitivity gain which accompanied by strong sensitivity degradation at higher frequencies. The reason for this is simple: indeed the negative inertia strongly increase the mechanical response, but only in the frequency band where the radiation pressure dominates and the therefore the sensitivity does not depend on the the mechanical susceptibility. 

\subsection{Summary and outlook}

The method of increasing the GW detectors sensitivity by means of optical modification of the test masses dynamics was proposed two decades ago and looks very simple and elegant. It does not require any sophisticated quantum states of light or radical alterations in the GW detectors core optics and also tolerant to the optical losses. However, as long as we aware, no specific plans of implementing this method in future GW detectors exist. This probably can be attributed to the following two reasons: first, technical problems associated with the detuned regime of GW interferometers, and second, the optical power constraints. The rule-of-thumb estimates show that in broad-band configurations, in order to shift the mechanical resonance up to some frequency $f_m$ by means of the optical spring, about the same optical power is required as make the back action noise equal to the shot noise at this frequency $f_m$. This means that using a single carrier, it is impossible to shift $f_m$ into the shot-noise dominated area where the increase of the mechanical response could provide a significant effect, and in the two-carriers configuration, the carrier which create the optical spring has to be more powerful than the one which do the measurement. Taking into account the tight optical power budget of the contemporary GW detectors and even more tight of the future ones (with much more heavy test masses and longer arms), implementation of this regime could be problematic. 

A possible solution to this problem was proposed recently in the papers \cite{Somiya2014,17a1KoKhSc}. In was shown in these works, that using the parametric amplification of the optical field inside the interferometer, it is possible to amplify the optical spring without increase of the optical power. This approach, in principle, can be combined with other applications of the intracavity parametric amplification (white-light cavity, back action evasion), see in particular Sec.\,7.2.



\section{Hybrid schemes}\label{sec:6}

In this section, we review a relatively novel approach that seeks to enhance the sensitivity of the GW interferometer by coupling it to another, generally nonlinear, quantum system. Depending on the nature of the nonlinearity and on the way it is coupled to the interferometer, one can suppress back-action noise  or reshape the optomechanical response of the interferometer so as to increase its bandwidth without sacrificing peak sensitivity. 

The first effect, known as \textit{coherent quantum noise cancellation (CQNC)} was pioneered by Tsang and Caves in \cite{2010_PhysRevLett.105.123601_CQNC_Tsang}. They suggested to use a combination of a nonlinear Kerr crystal and an unbalanced beam-splitter to couple the optomechanichal system under study (a GW interferometer, in our case) and an \textit{ancilla} optical mode, where the frequency offset of the ancilla to the main interferometer, the splitting ratio of the beam-splitter and the nonlinear gain of the crystal are tailored so as to perfectly counteract the effect of ponderomotive squeezing due to optomechanical back-action. In this work, it was also shown that an all optical ancilla system interacts with the signal light as if it was an optomechanical system with negative mass mechanical oscillator. Wimmer \textit{et al.} \cite{2014_PhysRevA.89.053836_CQNC_Wimmer} have developed this idea further to the level of a practical experiment that is currently being built at the University of Hannover. They also performed a thorough analysis of imperfections and their influence on this system ability for coherent cancellation of quantum back-action noise. This analysis has shown that it is problematic to realise this scheme in a GW detector due to stringent constraints on the ancilla's optical bandwidth and frequency offset that must both be much smaller than the mechanical resonance frequency, which is $\sim 1$~Hz for Advanced LIGO mirrors. However, another physical implementation of the \textit{negative mass oscillator} principle based on the interaction of the collective spin of Cesium vapours in magnetic field with light was proposed by Polzik and Hammerer in \cite{Polzik_AnnPhys_527_A15_2014}, and the back-action cancellation effect in such systems was demonstrated experimentally by M{\o}ller \textit{et al.} \cite{Moeller_Nature_547_191_2017}. As we discuss in the following Sec.~\ref{ssec:6-2}, such spin-based systems might be used in GW detectors.

Another way to use nonlinear system coupled to the optical degree of freedom is for creation a so-called \textit{white-light-cavity (WLC)} effect \cite{Wicht1997}, that is to introduce in the interferometer an active element that compensates the positive dispersion of the arm cavities by its own negative dispersion and thereby increase the effective band of a high response to the GW signal. Original idea by Wicht \textit{et al.} \cite{Wicht1997} proposed to use atomic medium with electromagnetically induced transparency effect providing the desired negative dispersion, which suffered from the internal loss in the gas cell. In the following Sec.~\ref{ssec:6-3}, we discuss more promising variants based on active nonlinear optical and optomechanical negative dispersion elements. These solutions are less lossy and thus stand a good chance to be a part of the next generation GW detectors, which might benefit from the additional astrophysical output the improved high-frequency sensitivity of such schemes may offer \cite{Miao2017c}.

\subsection{Negative-mass spin oscillator} 
\label{ssec:6-2}
%
%
%
 
\subsubsection{The negative-frequency system}
 
Multi-atomic spin ensembles proposed in the article \cite{Duan_PRL_85_5643_2000} and demonstrated experimentally in the work \cite{Julsgaard_Nature_413_400_2001} (see also the review papers \cite{Hammerer_RMP_82_1041_2010,Polzik_AnnPhys_527_A15_2014}) possess a set of unique features which make them attractive for use in quantum optomechanical experiments. Under certain conditions (see below), the dynamics of collective spin of such a system with high precision models the one of the ordinary harmonic oscillator, which eigen frequency can be made both positive and negative. Moreover, interaction of this spin system with light can be made similar to the ordinary pondermotive interaction of a movable mirror with the probing light.

The collective spin of the atomic ensemble can be described by the angular momentum vector $\hbar\times\{\hat{J}_x, \hat{J}_y, \hat{J}_z\}$. Suppose that this system is placed in a strong  external magnetic field $B$, which we assume to be pointed along the x-axis. The minimum energy state in this case corresponds to the large negative value $-\hbar J_x$ of the $x$ component of the angular momentum, with $J_x\gg1$ and the energy equal to $-\hbar\Omega_SJ_x$, where $\Omega_S$ is the Larmor frequency. Relatively weak (with the number of inverted spins much less that $J_x$) excitations can be described by the effective Hamiltonian \cite{Holstein_PR_58_1098_1940,Moeller_Nature_547_191_2017}
\begin{equation}\label{hamilt_S_pos} 
  \hamilt_S = -\hbar\Omega_SJ_x + \frac{\hbar\Omega_S}{2}(\hat{X}_S^2 + \hat{P}_S^2) \,,
\end{equation} 
where
\begin{equation}
  \hat{X}_S = \frac{\hat{J}_z}{\sqrt{J_x}} \,, \qquad 
  \hat{P}_S = \frac{\hat{J}_y}{\sqrt{J_x}}
\end{equation} 
are the effective (dimensionless) position and momentum of the spin ensemble obeying the standard commutation relation
\begin{equation}
  [\hat{X}_S, \hat{P}_S] = i \,.
\end{equation} 
Up to the the irrelevant $c$-number term, the Hamiltonian \eqref{hamilt_S_pos} describes a harmonic oscillator with the eigen frequency $\Omega_S$.

In a similar way, if all atoms are optically pumped to the energetically inverted spin state, then the collective spin is given by the positive value $\hbar J_x$. Weak de-excitations around this maximal value can be described by the effective Hamiltonian 
\begin{equation}\label{hamilt_S_neg} 
  \hamilt_S = \hbar\Omega_SJ_x - \frac{\hbar\Omega_S}{2}(\hat{X}_S^2 + \hat{P}_S^2) \,,
\end{equation}
where in this case 
\begin{equation}
  \hat{X}_S = \frac{\hat{J}_z}{\sqrt{J_x}} \,, \qquad 
  \hat{P}_S = -\frac{\hat{J}_y}{\sqrt{J_x}}
\end{equation} 
which corresponds to a Harmonic oscillator with the {\it negative eigen frequency} $\Omega_S$. Note that the term {\it negative mass} is used in the works is used in the works \cite{Julsgaard_Nature_413_400_2001,Hammerer_PRL_102_020501_2009,Polzik_AnnPhys_527_A15_2014,Moeller_Nature_547_191_2017} instead. However, in order to implement the dynamics \eqref{hamilt_S_neg}, the effective rigidity also have to be negative, which corresponds to the negative frequency $-\Omega_S$. 

The Hamiltonian \eqref{hamilt_S_neg} gives the following equation of motion of the position $X_S$:
\begin{equation}
  \hat{X}_S(t) = \hat{X}_S(0)\cos\Omega_St - \hat{P}_S(0)\sin\Omega_St \,,
\end{equation} 
while evolution of position $X_m$ of an ordinary positive-frequency harmonic oscillator is described by the following equation:
\begin{equation}
  \hat{X}_m(t) = \hat{X}_m(0)\cos\Omega_St + \hat{P}_m(0)\sin\Omega_St \,,
\end{equation} 
where $P_m$ is the corresponding momentum. Note that the sum of these positions autocommutes:
\begin{multline}
  [\hat{X}_S(t) + \hat{X}_m(t), \hat{X}_S(t') + \hat{X}_m(t')] \\
  = [\hat{X}_S(t), \hat{X}_S(t')] + [\hat{X}_m(t), \hat{X}_m(t')] = 0 \,,
\end{multline} 
that is, $\hat{X}_S+\hat{X}_m$ is a QND variable which can be continuously monitored with precision not limited by the uncertainty relation. During such a measurement, both $\hat{X}_S$ and $\hat{X}_m$ are perturbed, but these perturbations, being equal by absolute values and having opposite signs, cancel each other.

Consider now interaction of the atomic spin system with the probing light, which allows to implement such a measurement. Following \cite{Moeller_Nature_547_191_2017}, we assume that the light propagates in $z$-direction. If the light is far detuned from the atomic resonance, then the interaction Hamiltonian can be presented as follows (see details in \cite{Hammerer_PRL_102_020501_2009,Moeller_Nature_547_191_2017}):
\begin{equation}
  \hamilt_{\rm int} = -\hbar\varkappa\hat{S}_3\hat{J}_z \,,
\end{equation} 
where $\varkappa$ is the coupling constant, 
\begin{equation}
  \hat{S}_3 
  = i(\hat{{\rm a}}_x^\dagger\hat{{\rm a}}_y - \hat{{\rm a}}_y^\dagger\hat{{\rm a}}_x)
\end{equation} 
is the Stokes operator, and $\hat{{\rm a}}_{x,y}$ are the annihilation operators of the two linear polarizations of the optical beam. This is so-called Faraday interaction \cite{Hammerer_RMP_82_1041_2010}, which describes mutual rotation of the collective atomic spin and the optical polarization. Suppose then that the light is linearly polarized in $x$-direction, and the corresponding classical amplitude is equal to ${\rm a}_x = i\alpha$, where $\alpha$ is real. In this case, 
\begin{equation}
  \hat{S}_3 \approx \sqrt{2}\alpha\hat{{\rm a}}_S^c \,,
\end{equation} 
where 
\begin{equation}
  \hat{{\rm a}}_S^c = \frac{\hat{{\rm a}}_y + \hat{{\rm a}}_y^\dagger}{\sqrt{2}}  
\end{equation} 
is the cosine quadrature of the $y$-polarized light, and
\begin{equation}\label{hamilt_S_int} 
  \hamilt_{\rm int} \approx -\hbar\sqrt{2J_x}\varkappa\alpha\hat{X}_S\hat{{\rm a}}_S^c \,.
\end{equation} 
This Hamiltonian is identical to the standard Hamiltonian describing the dispersive coupling of an optical mode and a mechanical object, with the factor $\sqrt{2J_x}\varkappa$ playing the role of the vacuum optomechanical coupling strength $g_0$ \cite{Aspelmeyer_RMP_86_1391_2014}.

\subsubsection{Sequential scheme}\label{sec:scheme_seq} 

The QND measurement discussed above was first demonstrated in the work \cite{Julsgaard_Nature_413_400_2001} using two atomic spin systems having, respectively, positive and negative effective eigenfrequencies $\Omega_S$ and $-\Omega_S$ and consisting of $\sim 10^{12}$ Cesium atoms. Later in the article \cite{Hammerer_PRL_102_020501_2009} the idea of combining the negative-frequency atomic spin system with the positive-frequency optomechanical system was put forward. Recently, this idea was implemented experimentally using silicon-nitride nanomembrane as the mechanical resonator \cite{Moeller_Nature_547_191_2017}. 

\begin{figure}
  \centerline{\includegraphics[scale=0.8]{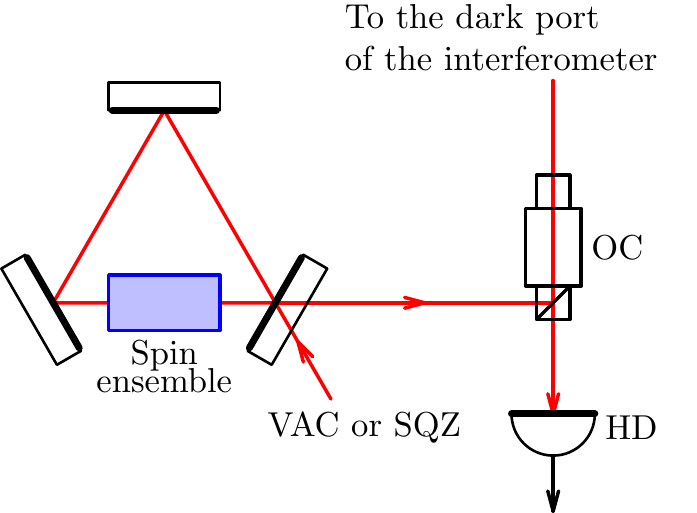}}
  \caption{Scheme of back-action evading measurement using light which sequentially probes the atomic spin system and the interferometer test mass(es). OC --- optical circulator; HD --- homodyne detector; ``VAC or SQZ'' --- incident optical field in vacuum or squeezed state.}\label{fig:scheme_seq}     
\end{figure} 

The sketch of this class of measurement schemes is shown in Fig.\,\ref{fig:scheme_seq}. Here the probing light interacts first with the atomic spin system and then is injected into the main interferometer, which measures the probe object position. The light leaving the interferometer is measured by the homodyne detector. It is easy to note similarity of this scheme with the one which uses the frequency-dependent squeezed light prepared by means of an additional filter cavity, see \cite{02a1KiLeMaThVy}, Sec.\,6.1 of \cite{Liv.Rv.Rel.15.2012}, and Sec.\,4.2 of this paper. Another option is to put the atomic system system after the main interferometer, similar to the variational-output scheme of \cite{02a1KiLeMaThVy}. In the case of the atomic spin system (and opposite to the filter cavity based schemes), both layouts provide identical results \SD{in the ideal (loss-free) case}. Therefore, we consider here only the one shown in Fig.\,\ref{fig:scheme_seq}. 

In order to demonstrate the basic features of this scheme while keeping the equations length within reasonable limit, we ignore the optical losses both in the atomic spin system and in the main interferometer. The full analysis with account of the optical losses can be found in \cite{17a1KhPo}. We would like to mention however, that while the problem of optical losses is a very serious one, it is generic for all interferometric schemes which use non-classical light, see Sec.\,4.4. At the same time, the atomic spin system introuces a new source of imperfection, namely the noise associated with the imaginary part of its effective succeptibility. We take into account this noise source here.

Using the analogy with the ordinary optomechanical systems, equations (\ref{hamilt_S_neg}, \ref{hamilt_S_int}) can be recast into the Heisenberg equations of motion for the atomic spin system interacting with the continuous traveling optical wave (see \eg \cite{Aspelmeyer_RMP_86_1391_2014}):
\begin{subequations}\label{spin_io} 
  \begin{gather}
    \fulldd{\hat{X}_S(t)}{t} + 2\gamma_S\fulld{\hat{X}_S(t)}{t} + \Omega_S^2\hat{X}_S(t)
      = -\Omega_S\sqrt{2\Gamma_S}\,\hat{{\rm a}}_S^c(t) - \sqrt{\Omega_S}\hat{f}_S(t) \,,
      \label{spin_io_spin}  \\
    \hat{{\rm b}}_S^c(t) = \hat{{\rm a}}_S^c(t) \,, \\
    \hat{{\rm b}}_S^s(t) = \hat{{\rm a}}_S^s(t) + \sqrt{2\Gamma_S}\,\hat{X}_S(t)
      \label{spin_io_b_s} \,,
  \end{gather}
\end{subequations}
where $\hat{{\rm a}}_S^{c,s}$ are the cosine and sine quadratures of the incident light, $\hat{{\rm b}}_S^{c,s}$ are the corresponding quadrature of the outgoing light, and $\Gamma_S$ is the readout rate (see details in \cite{Moeller_Nature_547_191_2017}). In Eq.\,\eqref{spin_io_spin}, the internal damping in the atomic spin system, with the damping rate $\gamma_S$, is taken into account, together with the corresponding (normalized) thermal force $\hat{f}_S$. It worth to be noted that the spin degree of freedom is very well isolated from the mechanical motion of the atoms and therefore can be prepared in an almost pure (\eg ground) quantum stated, even if the motional degree of freedom has the room temperature. This corresponds to the spectral density of $\hat{f}_S$ equal to
\begin{equation}
  S_S = 4|\Omega|\gamma_S \,.
\end{equation} 
In order to increase interaction with the probing light, the atomic spin system can be placed into the optical cavity, as shown in Fig.\,\ref{fig:scheme_seq}. In this case, the factor $\Gamma_S$ scales up by the effective number of the light passes $2\mathcal{F}/\pi$, where $\mathcal{F}$ is the cavity finesse \cite{17a1KhPo}.

Rewriting Eqs.\,\eqref{spin_io} in Fourier picture and combining Eqs.\,(\ref{spin_io_spin}, \ref{spin_io_b_s}), we obtain that
\begin{subequations}\label{spin_io_Omega} 
  \begin{gather}
    \hat{{\rm b}}_S^c(\Omega) = \hat{{\rm a}}_S^c(\Omega)  \,, \\ 
    \hat{{\rm b}}_S^s(\Omega) = \hat{{\rm a}}_S^s(\Omega) 
      + 2\theta\chi_S(\Omega)\hat{{\rm a}}_S^c(\Omega) 
      + \sqrt{2\theta}\,\chi_S(\Omega)\hat{f}_S(\Omega) \,,
  \end{gather}
\end{subequations}
where
\begin{equation}\label{chiS} 
  \chi_S(\Omega) = \frac{1}{\Omega^2 - \Omega_S^2 +2i\Omega\gamma_S}
\end{equation} 
is the effective susceptibility of the atomic spin system and $\theta=\Omega_S\Gamma_S$.

In the simplest case of the resonance tuned interferometer without optical losses, its  input/output relations look as follows (see Sec.\,2.3):
\begin{subequations}\label{ifo_io_Omega}
  \begin{gather}
    \hat{{\rm b}}_I^c(\Omega) = \hat{{\rm a}}_I^c(\Omega)  \,, \\ 
    \hat{{\rm b}}_I^s(\Omega) 
      = \frac{\ell^*(\Omega)}{\ell(\Omega)}\,\hat{{\rm a}}_I^s(\Omega) 
        - \frac{2\gamma\Theta}{\Omega^2\ell^2(\Omega)}\,\hat{{\rm a}}_I^c(\Omega) 
        + \frac{1}{\ell(\Omega)}\sqrt{\frac{2m\gamma\Theta}{\hbar}}\,
            x_{\rm sign}(\Omega) \,, 
  \end{gather}
\end{subequations}
where $x_{\rm sign}(\Omega)$ is the signal displacement of the free test mass(es) of the interferometer and $\hat{{\rm a}}_I^{c,s}$, $\hat{{\rm b}}_I^{c,s}$ are, respectively, the cosine and sine quadratures of the light at the input and the output light of the interferometer.

In the scheme of Fig.\,\ref{fig:scheme_seq}, 
\begin{equation}
  \hat{{\rm a}}_I^{c,s} = \hat{{\rm b}}_S^{c,s} \,.
\end{equation} 
We assume also that the sine quadrature $\hat{{\rm b}}_I^s$ of the interferometer output is measured by the homodyne detector. With account of this, combination of Eqs.\,(\ref{spin_io_Omega}, \ref{ifo_io_Omega}) gives that 
\begin{equation}
  \hat{{\rm b}}_I^s(\Omega) = \frac{1}{\ell(\Omega)}\sqrt{\frac{2m\gamma\Theta}{\hbar}}
    \left[x_{\rm sign}(\Omega) + \hat{x}_{\rm sum}(\Omega)\right] ,
\end{equation} 
where
\begin{multline}
  \hat{x}_{\rm sum}(\Omega) = \ell^*(\Omega)\sqrt{\frac{\hbar}{2m\gamma\Theta}}\Bigl\{
      \hat{{\rm a}}_S^s(\Omega) + \sqrt{2\theta}\,\chi_S\hat{f}_S(\Omega) \\
    + \bigl[2\theta\chi_S(\Omega) - \mathcal{K}_{\rm MI}(\Omega)\bigr]
        \hat{{\rm a}}_S^c(\Omega)  
  \Bigr\} 
\end{multline} 
is the position-normalized sum quantum noise.

Suppose that the squeezed light in injected into the atomic spin system. In this case, the double-sided spectral densities of the quadratures $\hat{{\rm a}}_S^c(\Omega)$ and $\hat{{\rm a}}_S^s(\Omega)$ are equal to, respectively,
\begin{equation}
  S[\hat{{\rm a}}_S^c] = e^{2r} \,,\qquad 
  S[\hat{{\rm a}}_S^s] = e^{-2r} \,,
\end{equation} 
and spectral density of the sum noise $\hat{x}_{\rm sum}$ is equal to 
\begin{equation}\label{S_sum_seq} 
  S^x(\Omega) = \frac{\hbar}{m\Omega^2\mathcal{K}_{\rm MI}(\Omega)}\bigl\{
      e^{-2r} + 4\theta|\Im\chi_S(\Omega)|
      + |2\theta\chi_S(\Omega) - \mathcal{K}_{\rm MI}(\Omega)|^2e^{2r}
  \bigr\} .
\end{equation} 
In order to cancel the back action, which corresponds to the last term in the curly brackets, the following condition has to be satisfied:
\begin{equation}\label{BAE_cond_raw} 
  2\theta\chi_S(\Omega) = \mathcal{K}_{\rm MI}(\Omega) \,.
\end{equation} 
It can be seen from Eqs.\,(\ref{chiS}, \ref{eq:KMI}) that this requirement can not be fulfilled at all frequencies. However, in all planned GW detectors, the quantum back action will be significant only well within the interferometer bandwidth, $\Omega\ll\gamma$. In this frequency band, $\mathcal{K}_{\rm MI}(\Omega)\propto1/\Omega^2$. On the other hand, if $\Omega\gg\Omega_S,\gamma_S$ then dynamics of the atomic spin system is close to the one of a free mass, $\chi_S(\Omega)\propto1/\Omega^2$. Therefore, in the frequency band $\Omega_S,\gamma_S\ll\Omega\ll\gamma$ frequency dependencies of $\mathcal{K}_{\rm MI}$ and $\xi_S$ match to each other, allowing to satisfy \eqref{BAE_cond_raw} by setting
\begin{equation}\label{BAE_cond} 
  \theta = \frac{\Theta}{\gamma} \,.
\end{equation} 

\begin{figure}
  \includegraphics{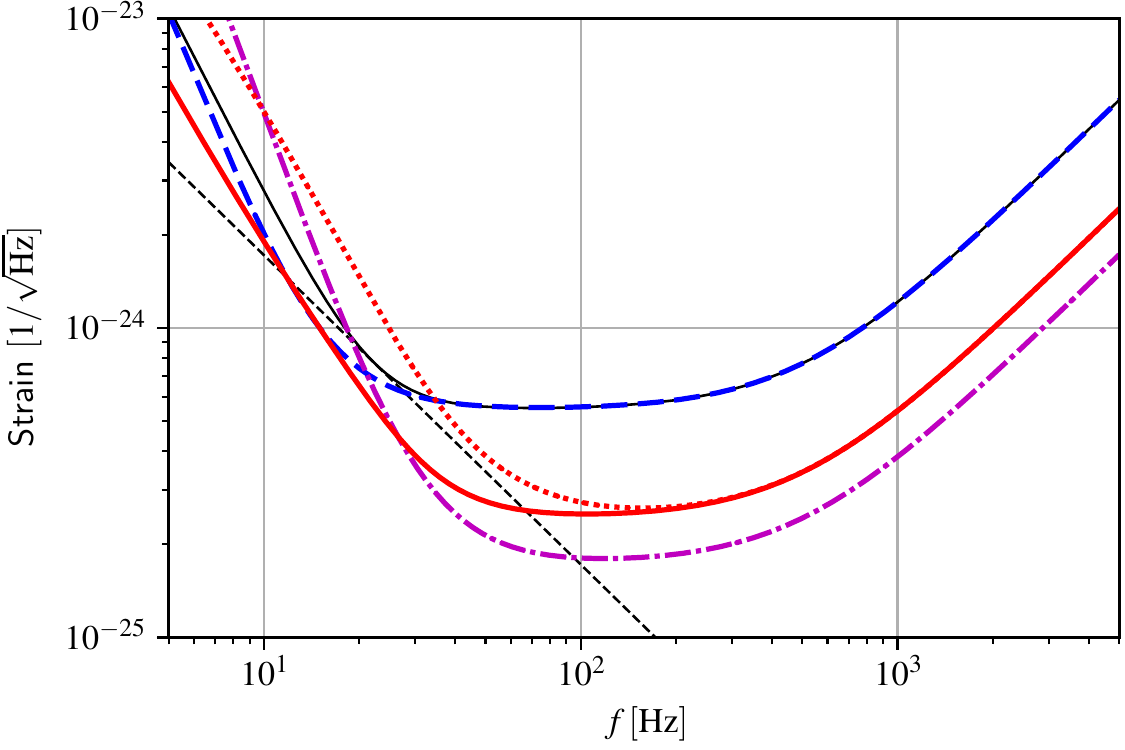}
  \caption{Quantum noise spectral densities for the measurement schemes using spin systems with negative effective frequency. Thin dashed line: SQL; thin solid line: SQL-limited interferometer; thick dashed line: sequential scheme of Fig.\,\ref{fig:scheme_seq}, no squeezing, $\gamma_S=2\pi\times3\,{\rm s}^{-1}$; thick dash-dotted line: sequential scheme of Fig.\,\ref{fig:scheme_seq}, 10\,db squeezing, $\gamma_S=2\pi\times3\,{\rm s}^{-1}$; thick solid line: parallel scheme of Fig.\,\ref{fig:scheme_par}, 10\,db of squeezing, $\gamma_S=2\pi\times3\,{\rm s}^{-1}$; dotted line: parallel scheme of Fig.\,\ref{fig:scheme_par}, 10\,db of squeezing, $\gamma_S=2\pi\times30\,{\rm s}^{-1}$. In all cases, $\gamma=2\pi\times500\,{\rm s}^{-1}$, $\Omega_S=\times3\,{\rm s}^{-1}$, $\theta$ is given by Eq.\,\eqref{BAE_cond}, and all other parameters are listed in Table\,1.}\label{fig:plots_negmass} 
\end{figure} 

In Fig.\,\ref{fig:plots_negmass}, quantum noise spectral densities of the considered scheme is plotted for two particular cases: no input squeezing and 10\,db of squeezing. In these plots, parameters of the main interferometer (the normalized optical power and the bandwidth) correspond to the ones listed in Table\,1. 
For the atomic spin spin system, the same quite demanding but realistic values
\begin{equation}\label{Omega_g_S} 
  \Omega_S = \gamma_S = 2\pi\times3\,{\rm s}^{-1}
\end{equation} 
as in \cite{17a1KhPo} are used. 

\subsubsection{Parallel (or EPR) scheme}

A serious problem of the scheme discussed in the previous section is the disparity of the typical optical wavelengths used in the GW detectors and the atomic spin systems. Optical transition of the cesium atoms used in the works \cite{Julsgaard_Nature_413_400_2001,Moeller_Nature_547_191_2017} corresponds to the wavelength $\approx850\,{\rm nm}$. Light with this wavelength can be used in the table-top interferometers, as it was done in \cite{Moeller_Nature_547_191_2017}. However, the contemporary GW detectors use light with the wavelength 1064\,nm, and longer wavelengths are planned for future interferometers. 

\begin{figure}
  \centering\includegraphics[scale=0.8]{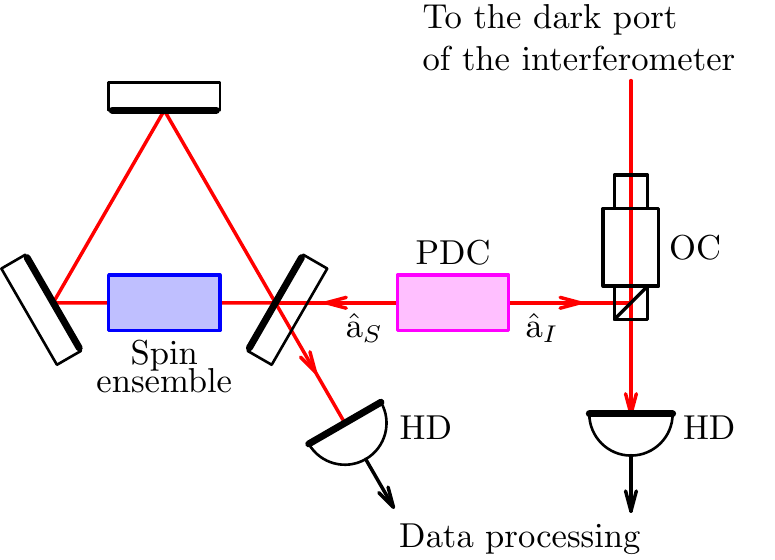}
  \caption{Scheme of back-action evading measurement using two entangled beams probing the atomic spin system and the interferometer test mass(es). OC --- optical circulator; HD --- homodyne detectors; PDC --- parametric down conversion; $\hat{{\rm a}}_{I,S}$ --- two entangled light beams.}\label{fig:scheme_par}     
\end{figure} 

This problem can be avoided by using another ``parallel'' optical layout, see Fig.\,\ref{fig:scheme_par}. It relies on high degree of cross-correlation between quantum fluctuations in the two entangled ``signal'' and ``idler'' light beams generated in the parametric down-conversion conversion (PDC) process. These two beams could have different wavelengths (the non-degenerate case), which should match the working frequency of the GW detector and the atomic transition frequency. Each of the beam has to interact with the respective subsystem, as shown in see Fig.\,\ref{fig:scheme_par}. Then both output signals have to be combined using optimal weight factors. Due to the above-mentioned cross-correlation, both the shot noise and the radiation pressure noise contributions will be suppressed in the combined output signal.

Note that a similar scheme was proposed initially in the paper \cite{Ma_NPhys_13_776_2017} for another purposes, namely, as a method of generation of effective frequency dependent squeezing without the use of an additional filter cavity (as in \cite{02a1KiLeMaThVy}); see details in Sec.\,4.3.

Consider quantum noise of this scheme, using the same assumptions as in Sec.\,\ref{sec:scheme_seq}. Quadratures of the two optical beams generated by the PDC are equal to 
\begin{subequations}\label{quad_PDC} 
  \begin{gather}
    \hat{{\rm a}}_{I,S}^c = \hat{{\rm z}}_{I,S}^c\cosh r + \hat{{\rm z}}_{S,I}^c\sinh r
      \,,\\ 
    \hat{{\rm a}}_{I,S}^s = \hat{{\rm z}}_{I,S}^s\cosh r - \hat{{\rm z}}_{S,I}^s\sinh r\,,
  \end{gather}
\end{subequations}
where $\hat{{\rm z}}_I^{c,s}$ and $\hat{{\rm z}}_S^{c,s}$ correspond to two independent vacuum fields and their (two-sided) spectral densities are equal to $1/2$. Correspondingly, spectral densities of the PDC beams and their only non-zero cross-correlation spectral densities are equal to
\begin{subequations}\label{S_a} 
  \begin{gather}
    S[\hat{{\rm a}}_I^c] = S[\hat{{\rm a}}_I^s] 
      = S[\hat{{\rm a}}_S^c] = S[\hat{{\rm a}}_S^s] = \cosh2r \,, \\
    S[\hat{{\rm a}}_I^c\hat{{\rm a}}_S^c] = \sinh2r\,, \qquad 
    S[\hat{{\rm a}}_I^s\hat{{\rm a}}_S^s] = -\sinh2r \,.
  \end{gather} 
\end{subequations}
Input/output relations for the interferometer and the atomic spin system are given by the same equations Eqs.\,(\ref{spin_io_Omega}, \ref{ifo_io_Omega}) and in the scheme of Sec.\,\ref{sec:scheme_seq}, but with the input optical fields defined by Eqs.\,\eqref{quad_PDC}. The outgoing fields are be measured by two independent homodyne detectors, which output photocurrents are data-processed together. We assume that both detectors measure the sine quadratures of the respective output beams $\hat{{\rm b}}_I^s$ and $\hat{{\rm b}}_S^s$, which gives the following equation for the combined output signal:
\begin{equation}
  \hat{{\rm b}}_I^s(\Omega) + \alpha(\Omega)\hat{{\rm b}}_S^s(\Omega)
  = \frac{1}{\ell(\Omega)}\sqrt{\frac{2m\gamma\Theta}{\hbar}}
      \left[x_{\rm sign}(\Omega) + \hat{x}_{\rm sum}(\Omega)\right] ,
\end{equation} 
where $\alpha(\Omega)$ is the weight factor which has to be optimized, 
\begin{multline}
  \hat{x}_{\rm sum}(\Omega) = \ell^*(\Omega)\sqrt{\frac{\hbar}{2m\gamma\Theta}}\Bigl\{
      \hat{{\rm a}}_I^s(\Omega) - \mathcal{K}_{\rm MI}(\Omega)\hat{{\rm a}}_I^c(\Omega) \\
      + \beta(\Omega)\bigl[
            \hat{{\rm a}}_S^s(\Omega) + 2\theta\chi_S(\Omega)\hat{{\rm a}}_S^c(\Omega)
            + \sqrt{2\theta\chi_S(\Omega)}\hat{f}_S(\Omega)  
          \bigr]  
    \Bigr\}
\end{multline} 
is the position-normalized sum quantum noise, and
\begin{equation}
  \beta(\Omega) = \frac{\ell(\Omega)}{\ell^*(\Omega)}\,\alpha(\Omega) \,.
\end{equation} . 

With account of Eqs.\,\eqref{quad_PDC}, spectral density of this noise is equal to
\begin{equation}
  S^x(\Omega) = \frac{\hbar}{m\Omega^2\mathcal{K}_{\rm MI}(\Omega)}\bigl[
      \sigma_I(\Omega) - 2\Re(\beta(\Omega)\sigma_{IS}(\Omega)) 
      + |\beta(\Omega)|^2\sigma_S(\Omega)
    \bigr] ,
\end{equation} 
where
\begin{subequations}
  \begin{gather}
    \sigma_I(\Omega) = \bigl[1 + \mathcal{K}_{\rm MI}^2(\Omega)\bigr]\cosh2r \,, \\
    \sigma_S(\Omega) = \bigl[1 + 4\theta^2|\chi_S(\Omega)|^2\bigr]\cosh2r
      + 4\theta|\Im\chi_S(\Omega)| \,, \\
    \sigma_{IS}(\Omega) 
      = \bigl[1 + 2\mathcal{K}_{\rm MI}(\Omega)\theta\chi_S(\Omega)\bigr]\sinh2r\,.  
  \end{gather}
\end{subequations}
It is easy to see that the optimal value of $\beta$ is equal to
\begin{equation}\label{beta_opt} 
  \beta(\Omega) = \frac{\sigma_{IS}^*(\Omega)}{\sigma_S(\Omega)} \,,
\end{equation} 
which gives that
\begin{multline}\label{S_sum_par} 
  S^x(\Omega) = \frac{\hbar}{m\Omega^2\mathcal{K}_{\rm MI}(\Omega)}\biggl[
      \sigma_I(\Omega) - \frac{|\sigma_{IS}(\Omega)|^2}{\sigma_S(\Omega)}
    \biggr] \\
  = \frac{\hbar}{m\Omega^2\mathcal{K}_{\rm MI}(\Omega)\sigma_S(\Omega)}\Bigl\{
        \bigl[1+\mathcal{K}_{\rm MI}^2(\Omega)\bigr] 
          \bigl[1 + 4\theta^2|\chi_S(\Omega)|^2 + 4\theta|\Im\chi_S(\Omega)|\cosh2r\bigr] 
          \\
        + |2\theta\chi_S(\Omega) - \mathcal{K}_{\rm MI}(\Omega)|^2\sinh^22r  
      \Bigr\} .
\end{multline} 
Note that leading in $e^{2r}$ term in this equation (the last one in the curly brackets) is similar to the corresponding term for the sequential scheme, see Eq.\,\eqref{S_sum_seq}. Therefore, the same reasoning as in that case can be used here as well, giving the same optimization condition \eqref{BAE_cond}.

In order to provide better insight into the general structure of the obtained quite lengthy equations, it is instructive to consider a simple asymptotic case. First, we neglect the damping in the atomic spin system, assuming that $\Im\chi_S\to0$. Second, we consider the frequency band where the condition \eqref{BAE_cond} is equivalent to the condition \eqref{BAE_cond_raw}. In this case, 
\begin{equation}
  \beta(\Omega) = \tanh2r
\end{equation} 
and
\begin{equation}
  S^x(\Omega) = \frac{\hbar}{m\Omega^2\cosh2r}
    \biggl[\frac{1}{\mathcal{K}_{\rm MI}(\Omega)} + \mathcal{K}_{\rm MI}(\Omega)\biggr] .
\end{equation} 
Taking into account that if $r$ is large, then $\tanh2r\to1$, it follows form these equations, that in the strong squeezing case the optimal strategy is just summing up the outputs of two homodyne detectors, which gives the sensitivity gain, in comparison with the ordinary SQL-limited interferometer, equal to $\cosh2r\approx e^{2r}/2$.

Spectral density \eqref{S_sum_par} optimized by the condition \eqref{BAE_cond} is plotted in Fig.\,\ref{fig:plots_negmass} for the same parameters \eqref{Omega_g_S} as in the previous case. It is easy to see that (in accord with the above reasoning) in the major part of the frequency band the sensitivity is worse by 3\,db than in the sequential scheme (for the same squeeze factor $r$). 

In order to reveal the influence of the internal damping in the atomic spin system, the case with $\gamma_S=2\pi\times30\,{\rm s}^{-1}$ also is presented in Fig.\,\ref{fig:plots_negmass}. It can be seen that this ten-fold increase of $\gamma_S$ noticeably degrade the low-frequency sensitivity, preventing from overcoming the SQL. 

It is interesting, that in the low-frequency band, where the condition \eqref{BAE_cond_raw} starts to deviate form the simplified one \eqref{BAE_cond}. the parallel scheme provide noticeably better sensitivity, than the sequential one (for the same value of $\gamma_S$). This result can be attributed to the fact, that the ``software'' summing of the photodetectors outputs using the optimized frequency-dependent factor \eqref{beta_opt} is more flexible procedure that simple ``hardware'' subtraction of the back actions. 

\subsubsection{Summary}

Using the additional spin systems with negative effective mass, it is possible to suppress the quantum noise in GW detectors across the almost entire frequency bandwidth relevant for gravitational wave observation. In comparison to the most of the other proposals for reducing the quantum noise, the spin system based approach has a significant advantage of being completely compatible with existing and planning GW interferometers thus not requiring complex alterations in the interferometers' core optics. In both ``sequential'' and ``parallel'' variants of this scheme, the only additional elements are the spin system itself, the source of the single-mode or two-mode squeezed light, and the optical scheme of injection the non-classical light into the interferometer. This setup strongly resembles the scheme of injection of ``ordinary'' squeezed light into the interferometer and evidently should has about the same level of complexity and cost. 

It worth to be noted also that this scheme paves the road towards generation of an entangled state of the multi-kilogram GWD mirrors and atomic spins which would be of fundamental interest due to the sheer size of the objects involved.

%

\subsection{Negative dispersion and white-light-cavity schemes} 
\label{ssec:6-3}

\begin{figure}[b]
\begin{center}
\includegraphics[width=0.535\textwidth]{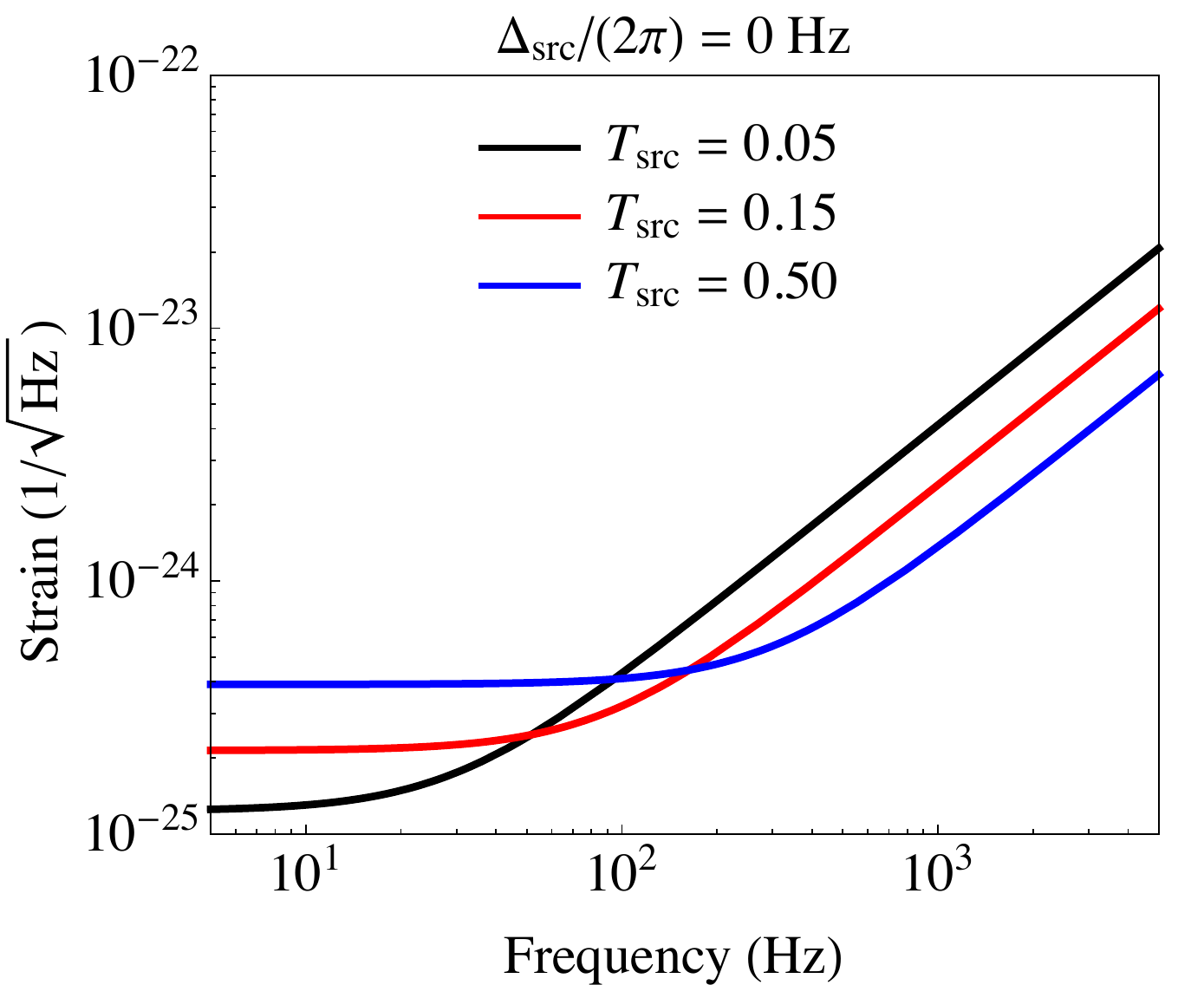}
\includegraphics[width=0.455\textwidth]{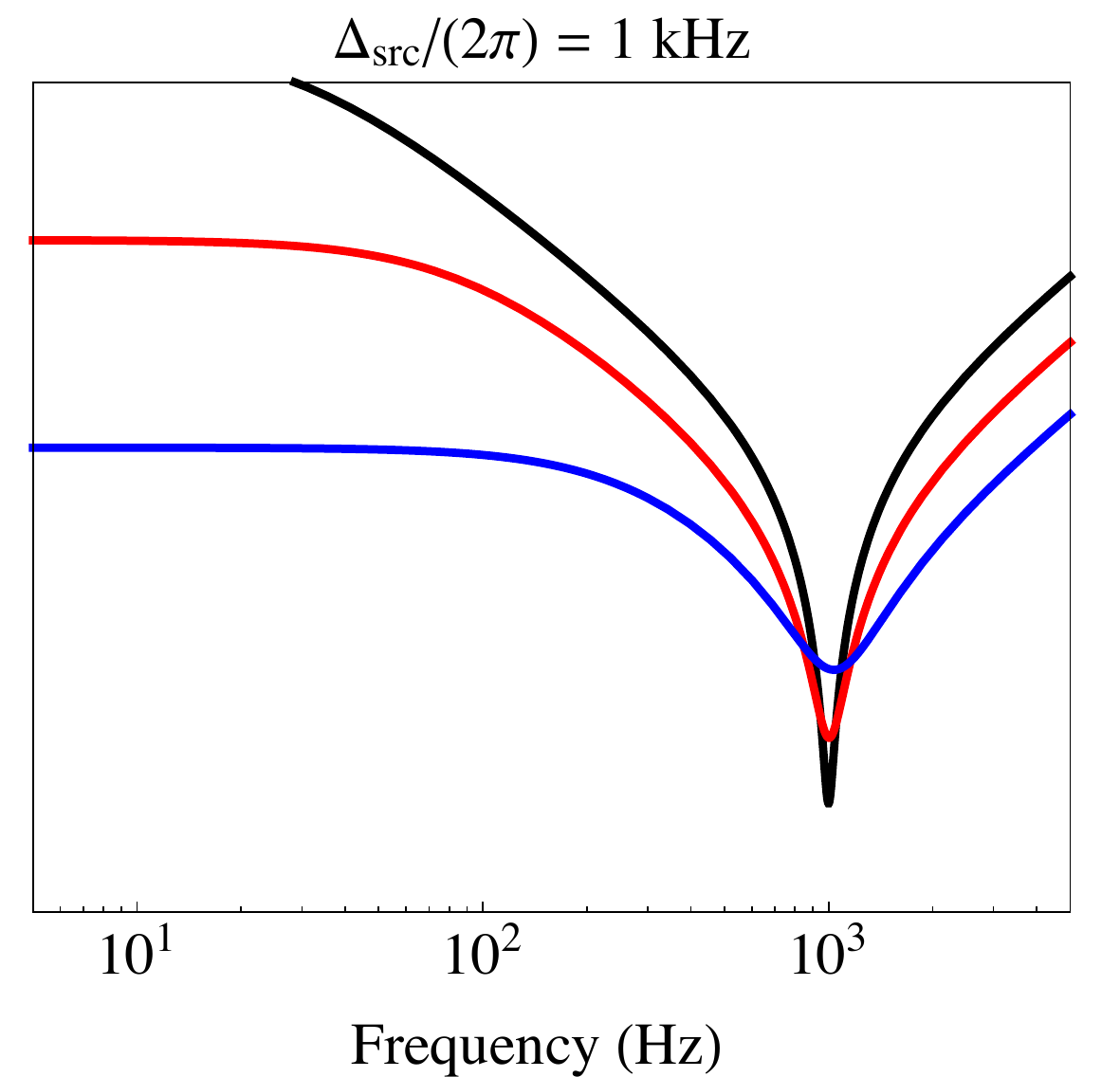}
\caption{The left panel shows the shot noise limited sensitivity 
for tuned signal-recycled Michelson interferometer with different 
effective signal recycling cavity (SRC) transmission; The right panel 
show the counterpart in the detuned case with a detune frequency 
equal to 1\,kHz. 
\label{fig:shotnoise}}
\end{center}
\end{figure}

In the nominal operation mode of Advanced LIGO, the signal-recycling cavity is
tuned to be resonant with respect to the carrier frequency. This is the so-called 
resonant sideband extraction idea, which increases the detector bandwidth \footnote{\SD{It might seem counter intuitive, how a resonantly tuned signal-recycling cavity could result in a broader bandwidth of the combined effective cavity of the arms and the SRC. The reason for that is the sign flip ($\pi$ phase shift) experienced by the light reflected off the resonance-tuned arm cavities. If combined with the SR mirror placed at a distance of an integer number of half-wavelengths of carrier light, it will result in an effectively anti-resonance tuned SRC and therefore will lead to a virtually lower finesse of the combined cavity.}}. However, 
the peak sensitivity limited by the 
shot noise is decreased as a price, as illustrated in Fig.\,\ref{fig:shotnoise}. 
Such a tradeoff between the bandwidth and peak sensitivity 
was firstly discovered by Mizuno when he compares different signal 
recycling schemes. Using the tuned signal-recycled Michelson as an example, 
such a tradeoff is manifested by the following integral of the shot-noise 
spectrum: 
\begin{equation}
\int_0^{\infty}\frac{{\rm d}\Omega}{2\pi} \frac{1}{S_{hh}^{\rm shot}(\Omega)} 
= \int_0^{\infty} \frac{{\rm d}\Omega}{2\pi}
\frac{ 4 L P_c \gamma \omega_0}{\hbar c (\gamma^2 +\Omega^2)} 
= \frac{P_c L\omega_0}{\hbar\, c}\,,
\end{equation}
where we have used a single-mode approximation for the shot-noise spectrum,
otherwise, the upper limit for the integration would be the half of the free spectral
range $\Omega_{\rm fsr}/2=\pi c / L$. Since $1/S_{hh}$ has a Lorentzian profile, the enclosed area is a constant, independent of the detector bandwidth.

To overcome the bandwidth-peak-sensitivity tradeoff, there are two approaches. 
One is keeping the bandwidth and increasing the peak sensitivity with the 
squeezed light, as discussed in Sec.\,\ref{sec:3}. The other is broadening 
the bandwidth while keeping the peak sensitivity, which is the idea of so-called 
white light cavity---a cavity that resonates "all" frequencies. It is motivated by 
the physical origin of the tradeoff, and has to do with the extra phase 
$\phi=\Omega L/c$ 
picked up by the GW sidebands at $\omega_0\pm \Omega$ when propagating 
inside the arm cavity that is tuned on resonance with respect to the 
carrier frequency $\omega_0$. Such a positive dispersion with 
${\rm d}\phi/{\rm d}{\Omega} > 0$ implies that higher the sideband frequency is, 
the more phase it is accumulated and thus is far away from the resonance, 
which leads to a degradation of the signal response. 

\begin{figure}[t]
\begin{center}
\includegraphics[width=0.95\textwidth]{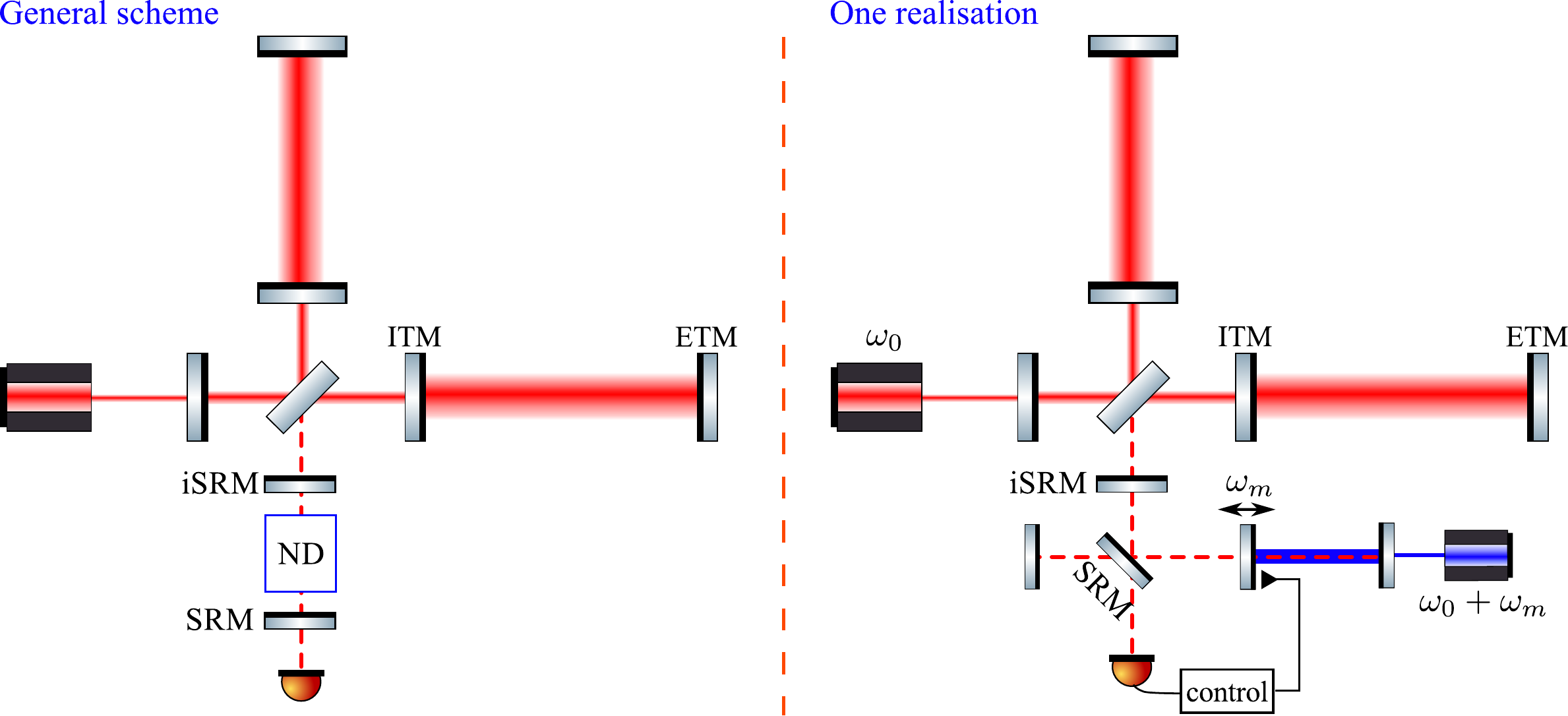}
\caption{The left panel shows the general scheme for the white-light-cavity cavity,
where a negative-dispersion (ND) element is placed inside the signal-recycling cavity. 
The internal signal-recycling mirror (iSRM), which has the 
same transmission as ITM, is introduced to form an impedance match cavity with ITM 
so that the signal sidebands are not affected by the narrow bandwidth of the arm 
cavity. The right panel shows a realisation using the unstable optomechanical filter. 
A global feedback control is needed to stabilise the system. 
\label{fig:unstable}}
\end{center}
\end{figure}

The white-light-cavity idea is introducing an active element, which 
has a negative dispersion ${\rm d}\phi/{\rm d}{\Omega} < 0$ around the 
frequencies of interest, inside the signal recycling cavity. Such a negative 
dispersion compensates the sideband phase and leads to a broadband 
resonance without changing the peak sensitivity. Earlier attempts of realising
the white-light-cavity effect with passive optical elements, which have no external 
energy input, have problems with the absorption associated with negative 
dispersion---a consequence of the Kramers-Kronig relation. Recent studies
instead propose the use of active elements with external pump energy, including 
atomic systems\,\cite{Wicht1997,Zhou2015,Ma2015}, nonlinear crystal (squeezer)\,\cite{Peano2015,Korobko2017} and optomechanical devices\,\cite{Miao2015a,Page2017,Miao2017c}. Here we will focus on the idea of active optomechanical filter 
operating in the unstable regime as the negative dispersion element, of which 
the setup is illustrated in Fig.\,\ref{fig:unstable}. 

The optomechanical filter involves an optomechanical device\,\cite{Chen2013,Aspelmeyer_RMP_86_1391_2014} with a movable oscillator 
as one mirror of the filter cavity. The oscillator has a resonant frequency 
equal to $\omega_m$, and is coupled to the signal field around $\omega_0$ via 
the radiation pressure, which is created by the beating between the signal field 
and an external pump field at $\omega_p=\omega_0+\omega_m$. 
The mechanical resonance frequency 
$\omega_m$ is chosen to be much smaller than the bandwidth of the 
filter cavity---the so-called resolved sideband regime. In such a regime, the 
interaction between the mechanical oscillator and the filter cavity mode can be 
described by a non-degenerate parametric process. In the 
rotating frame at the pump laser frequency, the interaction Hamiltonian is 
\begin{equation}\label{eq:WLC_Hamiltonian}
\hat H_{\rm int} = -\hbar g (\hat a\,\hat b+\hat a^{\dag}\hat b^{\dag})\,,
\end{equation}
where $\hat a$ and $\hat b$ are the annihilation operators for the cavity 
mode and mechanical mode, respectively. The coupling rate $g$ is related 
to the intra-cavity pump power $P_f$ by
\begin{equation}
g=\sqrt{\frac{P_f \omega_p}{m c L_f \omega_m}}\,, 
\end{equation}
in which $m$ is the oscillator mass and $L_f$ is the filter cavity length. 
Solving the following Heisenberg equations of motion leads to the following 
frequency-domain input-output relation for the filter cavity mode:
\begin{equation}\label{eq:io_unstable}
\hat a_{\rm out}(\Omega) = \frac{\Omega+i(\gamma_m +\gamma_{\rm opt})}
{\Omega+i(\gamma_m-\gamma_{\rm opt})}\hat a_{\rm in}(\Omega) + 
\frac{2\sqrt{\gamma_m \gamma_{\rm opt}}}
{\Omega+i(\gamma_m-\gamma_{\rm opt})}\hat b_{\rm th}^{\dag}(-\Omega)\,, 
\end{equation}
where $\gamma_{\rm opt}\equiv g^2/\gamma$ with $\gamma$ being
the filter cavity bandwidth, and $\gamma_m$ is the mechanical damping rate 
with $\hat b_{\rm th}$ being the associated thermal fluctuation in accord with the
fluctuation-dissipation theorem. The last term here comes from the additional noise that any linear phase-insensitive amplifier adds to the amplified signal as quantum uncertainty principle prescribes \cite{PhysRevD.26.1817}. The fact that optomechanical ND cavity acts as an amplifier is clearly seen from its Hamiltonian \eqref{eq:WLC_Hamiltonian} that has the same form as the Hamiltoniam of the non-degenerate optical parametric amplifier, save to that one of the modes here is mechanical rather than an optical one. 

\begin{figure}[t]
\begin{center}
\includegraphics[width=0.49\textwidth]{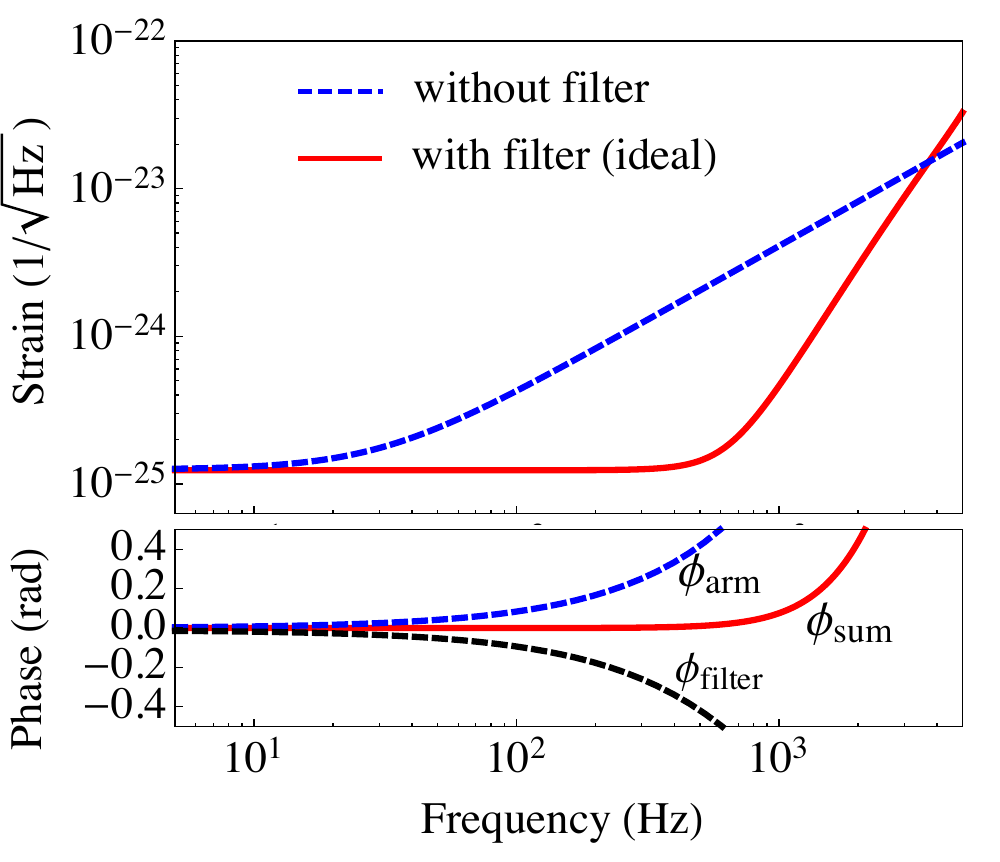}
\includegraphics[width=0.5\textwidth]{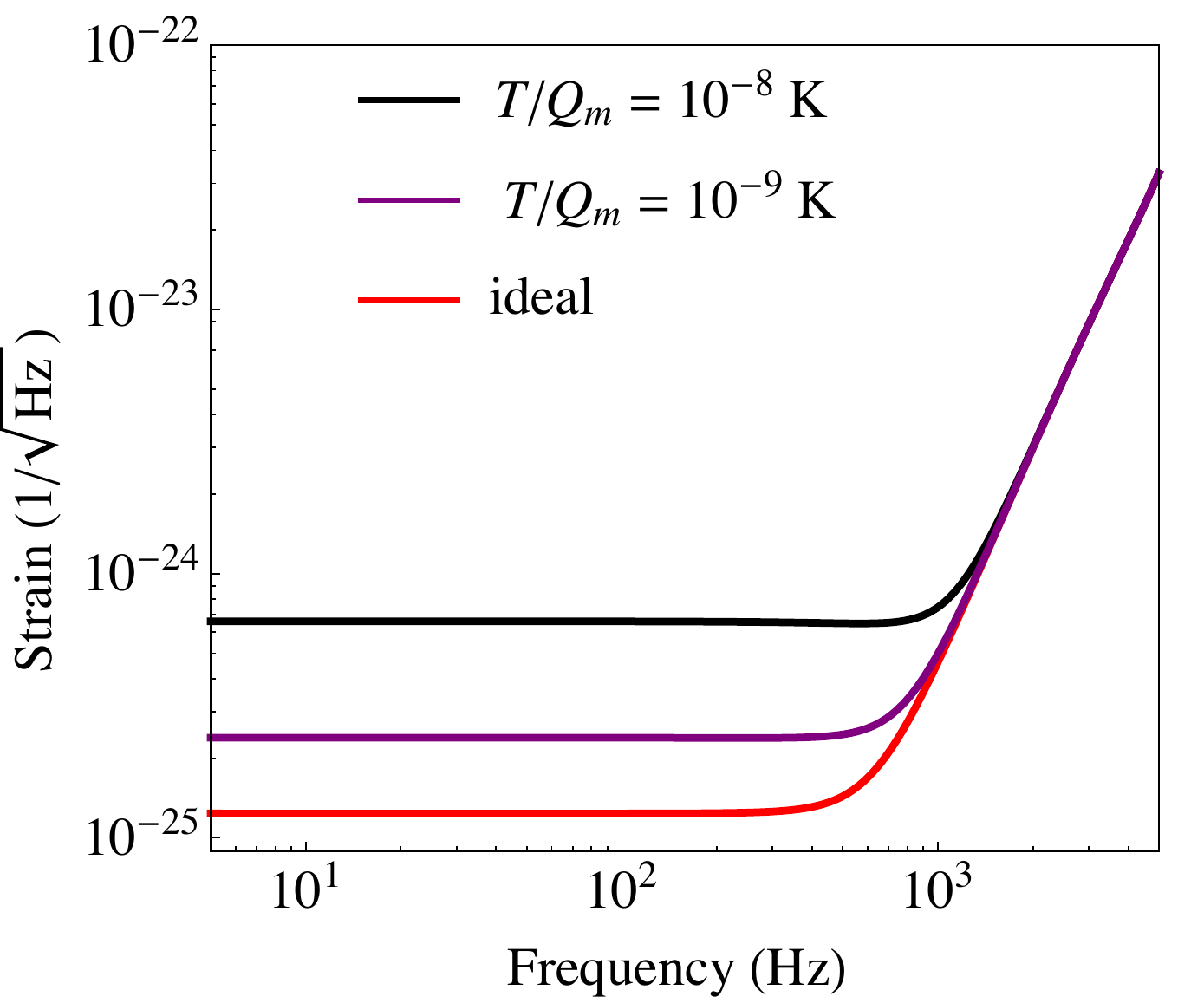}
\caption{The left panel shows the improvement of
 the shot-noise limited sensitivity by adding the unstable optomechanical 
 filter in the ideal case with the cancellation of the propagation phase illustrated at the 
 bottom panel. 
 The right panel illustrates the effect of the thermal noise in the mechanical oscillator 
 of the optomechanical filter. 
\label{fig:shot_noise_unstable_filter}}
\end{center}
\end{figure}

When $\gamma_{opt}$, which is the anti-damping rate due to the optomechanical
coupling, becomes much larger than the intrinsic mechanical damping $\gamma_m$, 
the system will be unstable and deviate from the working 
point if no feedback control is applied. With a proper feedback control engaged, 
the above input-output relation can be interpreted as the open-loop transfer 
function between the input field and the output field of the filter cavity. In the regime
of $\gamma_{\rm opt}\gg \gamma_m$, if ignoring the thermal-fluctuation term at 
the moment, we have \footnote{\SD{Here, the negative dispersion may appear a positive one if one assumes a different sign convention as compared to Eq.~\eqref{eq:FourierQuads}. In that case a special care has to be taken to do ALL the calculations consistent with the chosen sign convention in the definition of the Fourier transform. }}
\begin{equation}\label{eq:io_unstable_filter}
\hat a_{\rm out}(\Omega) \approx \frac{\Omega+i \gamma_{\rm opt}}
{\Omega-i\gamma_{\rm opt}}\hat a_{\rm in}(\Omega) \approx - e^{-2i\Omega/\gamma_{\rm opt}}\hat a_{\rm in}(\Omega)\,.
\end{equation}
Apart from the unimportant $\pi$-phase offset, the filter will 
therefore approximately imprint a negative phase
$\phi_{\rm filter} = - 2\Omega/\gamma_{\rm opt}$ onto those sidebands 
at $\Omega\lesssim \gamma_{\rm opt}$, which can cancel the positive 
round-trip phase $\phi_{\rm arm} = 2\Omega L/c$ when the following condition 
is satisfied 
\begin{equation}\label{eq:cond_opt}
 \gamma_{\rm opt} = \frac{c}{L} = 1.5\times 10^4 \, {\rm s}^{-1}\left( \frac{20\,\rm km}{L}\right)\,.
\end{equation}
Fig.\,\ref{fig:shot_noise_unstable_filter} shows the effect of the unstable filter on 
the shot-noise limited sensitivity of a tuned Michelson interferometer. The propagation 
phase of sidebands is cancelled at low frequencies. At frequencies above 1\,kHz, the 
cancellation starts to become imperfect; this is because the filter approximates $e^{-2i\Omega\tau}$, as shown in Eq.\,\eqref{eq:io_unstable_filter}, only to the second order of $\Omega$. Also in the same 
figure, we have illustrated the effect of the thermal noise in the mechanical 
oscillator on the sensitivity. The thermal noise affects the sensitivity 
similar to the optical loss in the arm cavity, and 
we can define an effective optical loss as follows: 
\begin{equation}
\epsilon_{\rm eff} = \frac{4 k_B}{\hbar \gamma_{\rm opt}}\left( \frac{T_{\rm env}}{Q_m}\right)\,, 
\end{equation} 
where $k_B$ is the Boltzmann constant, $T_{\rm env}$ is the environmental 
temperature and $Q_m$ is the mechanical quality factor. 
The corresponding sensitivity limit from such an effective arm-cavity loss, according to
Ref.\,\cite{Miao2017}, is
\begin{equation}
S^h_{\epsilon} \approx \frac{\hbar\,  c^2 \epsilon_{\rm eff}}{4 L^2 \omega_0
 P_c} = \left(2.0\times 10^{-25}/\sqrt{\rm Hz}\right)^2 \left( \frac{T_{\rm env}/Q_m}{10^{-9 }{\rm K}}\right)\,,
\end{equation}
given the default parameters $L=20\,\rm km$, $P_c = 4\,\rm MW$ and wavelength of 1550\,nm. 

\subsection{Summary and outlook}

Two main issues of active unstable optomechanical filter are the thermal noise and 
the optical loss. Intuitively, the thermal noise is significant because the light is exchanging information with the mechanical oscillator (effectively as a quantum 
memory) at a rate of kHz, one over the light propagation time inside the arm. The thermally-induced decoherence needs to be low enough such that the signal-to-noise
ratio does not degrade at the quantum limit. One feasible approach to mitigating
the thermal noise, instead of brutal-force cryogenic, is using the 
idea of optical dilution\,\cite{Corbitt_PRL_99_160801_2007,Korth2013a}. It takes advantage of the low mechanical dissipation of a suspended optics and increases the stiffness 
by using the optical spring effect. The issue of optical loss comes from both inside the filter and at the interface with the main interferometer due to mode mismatch. In particular, the loss introduced 
inside the signal recycling cavity limits the sensitivity improvement at high frequencies. Because 
of the narrow bandwidth of the arm cavity, the signal strength is suppressed significantly 
at high frequencies, and even a small amount of loss inside the signal recycling cavity is 
important. Specifically, according to Ref.\,\cite{Miao2017,Miao2017c}, 
\begin{align}\nonumber
S_{h}^{\rm SRC}(\Omega) &= \frac{\hbar c
(\gamma_{\rm arm}^2+\Omega ^2)\epsilon_{\rm SRC}}{4  L \omega_0 P_c \gamma_{\rm arm} }\,,\\
&\approx \left(2.4\times 10^{-25} \,{\rm Hz}^{-\frac12}\right)^2 
\left(\frac{\Omega/2\pi}{1\, {\rm kHz}}\right)^2\left(\frac{0.015}{T_{\rm ITM}}\right)
\left(\frac{\epsilon_{\rm SRC}}{10^{-3}}\right)\,, 
\end{align}
where we have assumed the default parameters for the interferometers. 

The atomic based active filter for broadening the detector bandwidth is not suffering the 
from same thermal noise issue as the optomechanical filter; the atomic transition 
involved happens at the optical frequency and the thermal environment can be viewed 
effectively as in the vacuum state. The main issue for the atomic system has to with 
the wavelength being tied to the transition of some specific species of atoms, which 
is different from those used in the current and proposed GW detectors. Exploring 
atomic systems with compatible wavelength or studying coherent frequency conversion 
scheme will be needed. The same issue of optical loss also applies.

\section{Discussion and conclusion} 
\label{sec:7}
We made an attempt to overview in this article the vast body of quantum techniques for suppression of quantum noise that are developed specifically for the field of gravitational-wave astronomy. We are standing now at the moment of inception of the concepts for the next generation of gravitational wave detectors that must have at least 10 times better sensitivity than the existing Advanced LIGO and Advanced Virgo instruments, which are about to be limited by quantum fluctuations of light in the almost entire detection band. The task of building the detector with the best astrophysical output justifies the need to bring some order into the massive collection of quantum noise-mitigation techniques that has beed developed so far. This was the goal of this work along with the aim to put all of those techniques in the same context and measure their merits and downsides against the common ruler. 
This pushed us towards the unified set of parameters for all considered schemes, taking the approach suggested by the GWIC 3G R\&D Committee in LIGO-T1800221 and summarised by Table~\ref{tab:benchmark_3G}.

As an outlook, with the recent understanding of the fundamental quantum limit (FQL), which only 
depends on the power fluctuation inside the arm cavity, it seems to lead to a unified picture of different 
techniques: (1) the external squeezing injection is a direct approach to increasing the power fluctuation; (2) 
Modifying dynamics with the optical spring effect can be viewed as using 
the internal ponderomotive squeezing for enhancing the power fluctuation; (3) 
The white-light-cavity idea is to extend the enhancement over a broad frequency range; (4) 
The speed meter is an approach to shaping the power fluctuation at different 
frequencies such that the FQL can be reached using a frequency-independent 
readout quadrature at those frequencies; (5) The optimal frequency-dependent 
readout is in general needed to attain the FQL at different frequencies. 
Instead of comparing techniques against each other, as in the case for near-term upgrades of 
existing detectors, we may now start to think how we can coherently combine different techniques to 
enhance the power fluctuation at frequency of interest, i.e., lowering
the FQL, and reach the limit. We can then study the susceptibility of different realisations 
to optical loss and other realistic imperfections. Eventually, we may obtain new configurations with 
high sensitivity that goes beyond what can be achieved with the current paradigm of design.

\appendix

\section{Squeezing of light in non-linear medium}\label{ssec:1-2}

\begin{figure}
\centering  \includegraphics[width=.5\textwidth]{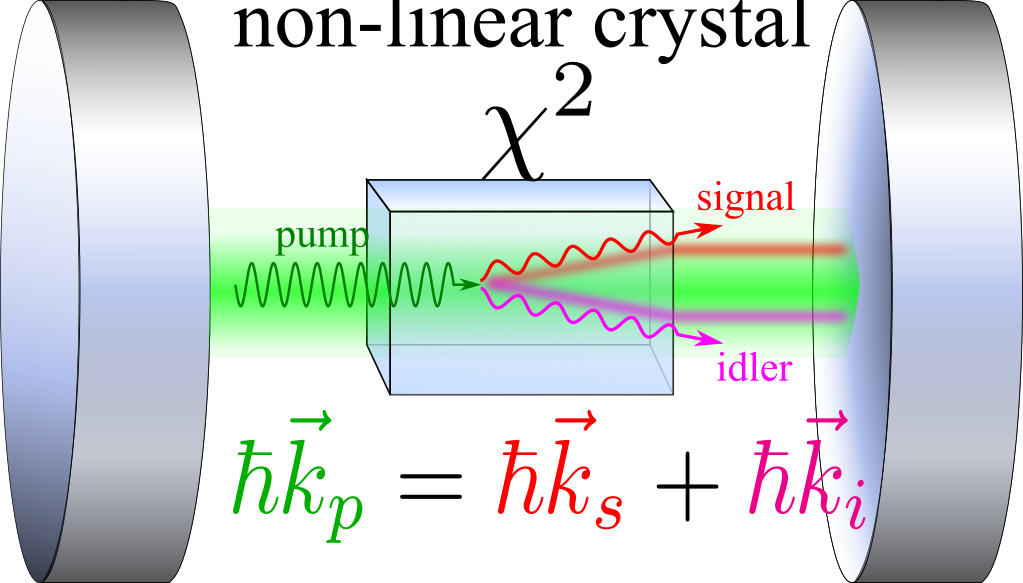}
\caption{Sketch of a parametric down-conversion process in a non-linear crystal that describes the physics of squeezed light generation.}\label{fig:PDC}       
\end{figure}

Squeezing is a well known technique in quantum optics which allows to generate states of light with reduced fluctuations in a chosen quadrature, which is very instrumental for GW detection. Squeezed light can be generated in several different ways \cite{Loudon&Knight} employing quadratic optical non-linearity, or even opto-mechanical non-linearity \cite{PhysRevA.73.023801_2006}. The most successful squeezed light generators\cite{Vahlbruch15dB} are based on the parametric down-conversion (PDC) process that happens in a non-linear medium (\textit{e.g.} PPKTP crystal) with strong enough $\chi^{(2)}$ non-linearity, where photons of the high-frequency pump give birth to a pair of lower frequency entangled photon modes called traditionally \textit{signal} and \textit{idler}, as depicted in Fig.~\ref{fig:PDC}. Pump, \textit{signal} (with frequency $\omega_s$) and \textit{idler} (with frequency $\omega_s$) modes must satisfy energy and momentum conservation laws:
\begin{eqnarray}
2\omega_p &=\omega_s + \omega_i\,, & \vb{k}_p = \vb{k}_s + \vb{k}_i\,.
\end{eqnarray}
where $\vb{k}_p$, $\vb{k}_s$ and $\vb{k}_i$ stand for the wave-vectors of the corresponding beams with lengths $|\vb{k}_p| \equiv 2\omega_p/c$, $|\vb{k}_s| \equiv \omega_s/c$ and $|\vb{k}_i| \equiv \omega_i/c$, respectively. 

The corresponding Hamiltonian of this process, linearised in terms of large classical pump amplitude, can be written in the frame, rotating with the frequencies of the signal and idler modes as follows (see, e.g., Section~5 of~\cite{1995BookWaMi} for details):
\begin{equation}
\label{eq:PDC_hamiltonian}
	\hat{H}_{\mathrm{PDC}} = i\hbar\chi\left[\hat{a}^\dag_s\hat{a}^\dag_ie^{i\phi}-\hat{a}_s\hat{a}_ie^{-i\phi}\right]\,,
\end{equation}
where $\hat{a}_{s,i}$ describe annihilation operators for the \textit{signal} and \textit{idler} photon modes, respectively, and $\chi$ and $\phi$ are the magnitude and phase of the PDC coupling strength that is proportional to the second-order susceptibility $\chi^{(2)}$ of the medium and to the pump power.  

It is straightforward to obtain the evolution of the two modes in the interaction picture (leaving apart the obvious free evolution time dependence $e^{-i\omega_{s,i}t}$) solving the Heisenberg equations:
\begin{align}
\label{eq:pdc_heqs_solutian}
  &\hat{a}_s(t) = \hat{a}_s^{in}\cosh \chi t+(\hat{a}^{in}_i)^\dag e^{i\phi}\sinh \chi t\,,\\
  &\hat{a}_i(t) = \hat{a}_i^{in}\cosh\chi t+(\hat{a}^{in}_s)^\dag
  e^{i\phi}\sinh \chi t\,.
\end{align}
Parameter $t$ here describes the duration of interaction of the pump photons with the nonlinear medium, and $r = \chi t$ is the integral squeezing factor. The above linear relations represent, in fact, the input-output relations for a non-degenerate parametric amplifier (OPA) in time domain. In frequency domain, the expression can be easily obtained using the general 2-photon formalism formulas, which yield:
\begin{eqnarray}
\vq{a}_s(\Omega) &= \mathbb{G}_c\vq{a}_s^{in}(\Omega) + \mathbb{G}_s\vq{a}_i^{in}(\Omega)\,,\\
\vq{a}_i(\Omega) &= \mathbb{G}_s\vq{a}_s^{in}(\Omega) + \mathbb{G}_c \vq{a}_i^{in}(\Omega)\,,\\
\end{eqnarray}
where the corresponding transformation matrices read
\begin{align}
	\mathbb{G}_c = \cosh r
	\begin{bmatrix}
	1 & 0\\
	0 & 1
	\end{bmatrix}\,, & 
	& \mathbb{G}_s = \sinh r
	\begin{bmatrix}
	\cos2\phi &  \sin 2\phi\\
	\sin 2\phi & -\cos 2\phi
	\end{bmatrix}\,.
\end{align}
It has to be noted that the OPA  applied for generating squeezed vacuum in GW interferometers use cavity resonance to enhance $\chi^{(2)}$, as pictured in Fig.~\ref{fig:PDC}, which imposes certain bandwidth limits and makes the above analysis more complicated. Nevertheless, the approximation we used in this section holds pretty well, since the typical bandwidth of the cavities used in squeezers is of the order of the hundreds of MHz, while the GW frequency range spans to maximum a few kHz, one can safely assume $r$ and $\phi$ in the above formulas as frequency independent.

\section{Quantum noise in advanced interferometers}\label{app:QN_adv_IFO}

\subsection{Ponderomotive squeezing in GW interferometers}\label{app:pond_sqz}

Ponderomotive squeezing that takes place in a tuned lossless Michelson interferometer can be written as a sequence of 3 unitary transformations -- rotation, squeezing and second rotation \cite{02a1KiLeMaThVy}:
\begin{equation}\label{app.eq:sqz_state}
  \ket{out} = e^{2i\beta}\hat{R}(u_{\rm pond})\hat{S}(r_{\rm pond})\hat{R}(v_{\rm pond})\ket{in}.
\end{equation}
where $\beta$ is a scheme-specific complex frequency-dependent phase shift which does not change the noise spectral density, the rotation operator $\hat{R}(\alpha)$ and the squeezing operator $\hat{S}(r)$ are defined in Section~3.2 of \cite{Liv.Rv.Rel.15.2012}. Action of these operators on the vector of light quadratures, $\hat{\vb{a}} = \{\hat{a}_1,\,\hat{a}_2\}^{\rm T}$, results in a new vector, $\vq{b} = \{\hat{b}_1,\,\hat{b}_2\}^{\rm T}$, that reads:
\begin{equation}\label{app.eq: pond sqz matr transformation}
  \vb{b} = \tq{T}\,\vq{a} = e^{2i\beta}\,\mathbb{R}[u_{\rm pond}]\,\mathbb{S}[r_{\rm pond}]\,\mathbb{R}[v_{\rm pond}]\,\vq{a}\,,
\end{equation}
with $\tq{R}$ the rotation matrix and $\tq{S}$ the squeezing matrix that are defined by Eq.~\eqref{eq:Rot+Sqz}.

For a general optomechanical system without loss, the transfer matrix (TM) has a specific structure, namely, the optical TM is $\tq{T}^{\rm meas} = e^{2i\beta}\tq{R}[\psi]$ and the radiation pressure one in proportional to $\tq{T}^{\rm b.a.} \propto \vs{t}\left(\sigma_1\vs{t}\right)^{\rm T}$, where $\sigma_1$ is the Pauli`s matrix, also known as $\sigma_x$. This structure of TM preserves covariance matrices of the input and  the output fields, $\tq{V}_a$ and $\tq{V}_b = \tq{T}\,\tq{V}_a\tq{T}^\dag$, from being non-symplectic, \textit{i.e.} it ensures that both are covariance matrices that describe gaussian quantum states. Factoring out common complex phase $e^{2i\beta}$, one ends up with a real matrix $\tq{T}^{\rm Re} = e^{-2i\beta}[\tq{T}^{\rm meas}+\tq{T}^{\rm b.a.}]$, the singular value decomposition of which can be written as:
\begin{equation*}
  \tq{T}^{\rm Re} = \mathbb{R}[u_{\rm pond}]\,\mathbb{S}[r_{\rm pond}]\,\mathbb{R}[v_{\rm pond}] \,,
\end{equation*}
that proves that Eq. \eqref{app.eq: pond sqz matr transformation} is indeed correct.

In order to get the expressions for $r_{\rm pond}$, $u_{\rm pond}$ and $v_{\rm pond}$, one can expand $\tq{T}^{\rm Re}$ in Pauli matrices:
%
\begin{equation*}
  \tq{T}^{\rm Re} = \til{z}_0\tq{I} + \til{z}_1\sigma_1 + \til{z}_2\sigma_2 + \til{z}_3\sigma_3
\end{equation*}
where $\til{z}_{0,1,2,3}$ are complex coefficients.

Symmetries of the TM immediately allow to see that $\til{z}_3 = 0$ and the $\til{z}_0 = \tq{T}^{\rm Re}_{cc} = \tq{T}^{\rm Re}_{ss}$.
Since all elements of $\tq{T}^{\rm Re}$ are real, the following relations hold for the remaining coefficients:
\begin{equation*}
  \til{z}_1 = -\frac{ \tq{T}^{\rm Re}_{cs} + \tq{T}^{\rm Re}_{sc} }{2} = z_1, \quad
  \til{z}_2 = i \frac{ \tq{T}^{\rm Re}_{cs} - \tq{T}^{\rm Re}_{sc} }{2} = i \cdot z_2 \,,
\end{equation*}
which means $z_1,z_2$ are real. 

Then singular values can be calculated:
\begin{equation*}
  s_{1,2} = \abs{ \abs{z_1} \pm \sqrt{z_0^2 + z_2^2} } \,.
\end{equation*}
Assuming $e^{r_{\rm pond}} = \max\{s_1,s_2\}$ and $e^{-r_{\rm pond}} = \min\{s_1,s_2\}$ (i.e. $r_{\rm pond} > 0$) one can get the following expression:
\begin{equation*}
  \sinh r_{\rm pond} = 
  \left\{
  \begin{aligned}
  &\abs{z_1},               & & \text{if } \det\tq{T}^{\rm Re} = 1  \,, \\
  &\sqrt{z_0^2 + z_2^2} \,, & & \text{if } \det\tq{T}^{\rm Re} = -1 \,.
  \end{aligned}
  \right.
\end{equation*} 

The expression for angles $u_{\rm pond}$ and $v_{\rm pond}$ are:
\begin{align*}
  u_{\rm pond} &= - \frac{1}{2} \arctan\frac{z_2}{z_0} - \mathrm{sgn}\left[z_1\right]\frac{\pi}{4} \,, \\
  v_{\rm pond} &= - \frac{1}{2} \arctan\frac{z_2}{z_0} + \mathrm{sgn}\left[z_1\right]\frac{\pi}{4} \,.
\end{align*}

\subsection{I/O-relations of a Fabry-Perot--Michelson interferometer with losses}\label{app:FPMI_lossy}
Here we give the I/O-relations for considered interferometers beyond narrow-band approximation.

\subsubsection{Fabry-Perot interferometer with end moving mirror}\label{app:lossy_FP_I/O-rels}

\begin{figure}[htbp]
\begin{center}
\includegraphics[width=.75\textwidth]{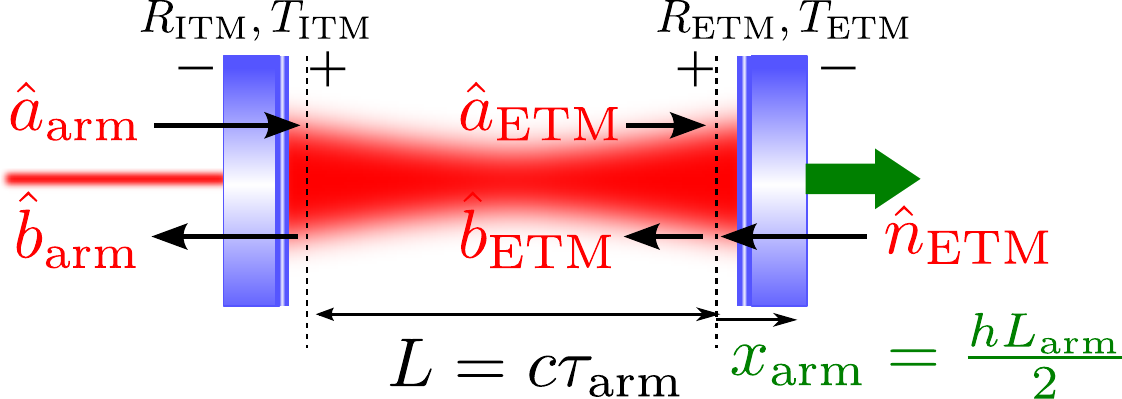}
\caption{Schematics of I/O-relations for a Fabry-Perot  cavity} 
\label{fig:app_FParm}
\end{center}
\end{figure}

I/O relations for a single Fabry-Perot arm cavity (see Fig.~\ref{fig:app_FParm}) without any additional assumptions about its bandwidth can be obtained from the following chain of steps. First, consider the I/O-relations for the ETM:
\begin{equation}\label{app. matr.: i-o realtion for ETM}
  \vq[ETM]{b} = \tq[ETM]{T}\vq[ETM]{a} + \tq[ETM]{N}\vq[ETM]{n} + \vs[ETM]{t}\frac{h}{\sqrt{2}h_{\rm SQL}} \,,
\end{equation}
where the corresponding transfer matrices and the OM response of the mirror read
\begin{equation}\label{app eq: TT, NN and t for ETM}
\begin{aligned}
  \tq[ETM]{T} &= \sqrt{ R_{\rm ETM}} \left( \tq{I} + \tq[ETM]{M} \right) \,, \\
  \tq[ETM]{N} &= \sqrt{ T_{\rm ETM}} \left( \tq{I} + \tq[ETM]{M} \right) \,, \\
  \tq[ETM]{M} &= \matr{0}{0}{- R_{\rm ETM}\mathcal{K}_{\rm TM}}{0} \,,       \\
  \vs[ETM]{t} &= \sqrt{2  R_{\rm ETM} \mathcal{K}_{\rm TM}} \col{0}{1} \,.
\end{aligned}
\end{equation}
and
\begin{equation*}
  \mathcal{K}_{\rm TM} = \frac{8 \omega_p P_c}{M c^2 \Omega^2} = \dfrac{2\Theta_{\rm arm}\tau_{\rm arm}}{\Omega^2}\,,
\end{equation*}
is an optomechanical coupling factor for a single perfectly reflective free mirror and $P_c$ stands for the full light power circulating in the arm. Note also the factor of $\sqrt{2}$ in front of the $h_{\rm SQL}$, as the latter stands for the SQL of a Fabry-Perot cavity with 2 movable mirrors of mass $M$ each.

It has to be noted that the above expressions are derived in the assumption of zero phase of the carrier light at the ETM, namely that only the \textit{cosine} quadrature of carrier light, $A^{c}_{\rm ETM} = \sqrt{2P_c/\hbar\omega_p}$, is not equal to zero, while $A^s_{\rm ETM} = 0$. The general case of arbitrary phase $\Phi = \omega_p\tau$, corresponding to carrier light travel time $\tau = L/c$, can be obtained by means of simple rotation of the corresponding transfer matrices and response vector by $\Phi$:
\begin{equation}\label{app eq: TT, NN and t for ETM}
\begin{aligned}
  \tq[ETM,\,\tau]{T} &= \mathbb{R}[\omega_p\tau] \tq[ETM]{T} \mathbb{R}[-\omega_p\tau]\,, \\
  \tq[ETM,\,\tau]{N} &= \mathbb{R}[\omega_p\tau] \tq[ETM]{N} \mathbb{R}[-\omega_p\tau] \,, \\
  \tq[ETM,\,\tau]{M} &= \mathbb{R}[\omega_p\tau] \tq[ETM]{M} \mathbb{R}[-\omega_p\tau] \,, \\
  \vs[ETM,\,\tau]{t} &= \mathbb{R}[\omega_p\tau] \vs[ETM]{t} \,.
\end{aligned}
\end{equation}

Adding an ITM to the system makes a Fabry-Perot interferometer, described by the system of \eqref{app. matr.: i-o realtion for ETM} and two new equations:
\begin{align*}
  \vq[arm]{b} &= - \sqrt{ R_{\rm ITM}}\vq[arm]{a} + \sqrt{T_{\rm ITM}}\tq[\tau_{\rm arm}]{P}\vq[ETM]{b} \,, \\
  \vq[ETM]{a} &= \tq[\tau_{\rm arm}]{P}\tq[ITM]{N}\vq[arm]{a} + \tq[\tau_{\rm arm}]{P}\tq[ITM]{T}\tq[\tau_{\rm arm}]{P}\vq[ETM]{b} \,.
\end{align*}
Here matrices $\tq[ITM]{T}$ and $\tq[ITM]{N}$ has absolutely the same form as $\tq[ETM]{T}$ and $\tq[ETM]{N}$ correspondingly, provided by \eqref{app eq: TT, NN and t for ETM}. The solution have the following form:
\begin{equation}\label{app. matr.: i-o realtion for arm}
  \vq[arm]{b} = \tq[arm]{T}\vq[arm]{a} + \tq[arm]{N}\vq[arm]{n} + \vs[arm]{t}\frac{h}{\sqrt{2}\,h_{\rm SQL}} \,,
\end{equation}
where transfer matrices and signal response function read:
\begin{equation}\label{app eq: TT, NN and t for arm}
\begin{aligned}
  \tq[arm]{T} &= \sqrt{T_{\rm ITM}}\tq[arm]{M}\tq[\tau_{\rm arm}]{P}\tq[ETM,\,\tau]{T}\tq[\tau_{\rm arm}]{P}\tq[ITM]{N} - \sqrt{ R_{\rm ITM}}\tq{I} \,, \\
  \tq[arm]{N} &= \sqrt{T_{\rm ITM}}\tq[arm]{M}\tq[\tau_{\rm arm}]{P}\tq[ETM,\,\tau]{N} \,, \\
  \vs[arm]{t} &= \sqrt{2\,T_{\rm ITM}}\tq[arm]{M}\tq[\tau_{\rm arm}]{P}\vs[ETM,\,\tau]{t} \,, \\
  &\tq[arm]{M} = \left[ \tq{I} - \tq[\tau_{\rm arm}]{P}\tq[ETM,\,\tau]{T}\tq[\tau_{\rm arm}]{P}\tq[ITM]{T} \right]^{-1} \,,
\end{aligned}
\end{equation}
with
 $\tq[\tau_{\rm arm}]{P} = e^{i\Omega\tau_{\rm arm}} \tq{R}[\omega_p\tau_{\rm arm}]$ standing for the transfer matrix of a free space propagation of light between the mirrors of the arm cavity. 

%
%
\paragraph{Tuned arm cavity:}
In the important special case when the cavity is tuned in resonance, which mathematically means that $\omega_p\tau_{\rm arm} = 2\pi n$ ($n$ integer) the above expressions simplify to:
\begin{align*}
  \tq[arm]{T} &= T e^{2i\beta_{\rm arm}} \matr{1}{0}{-\sqrt{ R_{\rm ETM}}\mathcal{K}_{\rm arm}}{1}, \\
  \tq[arm]{N} &= N e^{i\beta_{\rm arm}}  \matr{1}{0}{-\mathcal{N}}{1}, \\
  \vs[arm]{t} &= t e^{i\beta_{\rm arm}}  \sqrt{\frac{4\, R_{\rm ETM}\mathcal{K}_{\rm arm}}{1+ R_{\rm ITM}}} \col{0}{1},
\end{align*}
where in an assumption of small optical loss ($ T_{\rm ETM} \ll 1$):
\begin{align*}
  \mathcal{K}_{\rm arm} &= \frac{\left(1+ R_{\rm ITM}\right)T_{\rm ITM}\mathcal{K}_{\rm TM}}{1-2\sqrt{ R_{\rm ITM}} \cos 2\Omega\tau +  R_{\rm ITM}} \\
  \beta_{\rm arm}       &= \atan{ \dfrac{1+\sqrt{ R_{\rm ITM}}}{1-\sqrt{ R_{\rm ITM}}} \tan\Omega\tau }, \\
  N           &= \sqrt{\frac{1- R_{\rm ETM}}{1+ R_{\rm ITM}}\frac{\mathcal{K}_{\rm arm}}{\mathcal{K}_{\rm TM}}}, \quad T = t = 1, \\
  \mathcal{N} &= \sqrt{\frac{\mathcal{K}_{\rm arm}\mathcal{K}_{\rm TM} R_{\rm ETM}}{1- R_{\rm ITM}^2}}
                 \frac{1 + e^{2i\Omega\tau_{\rm arm}} R_{\rm ITM}^{3/2}}{e^{-i\beta_{\rm arm}+i\Omega\tau_{\rm arm}}}
\end{align*}

The following expressions for shot-noise and back-action components of the optical transfer matrix of the lossy Fabry-Perot cavity can be finally written:  
\begin{gather}
  \tq{T}^{\rm s.n.} = e^{2i\beta_{\rm arm}} \tq{I}, \quad
  \tq{T}^{\rm b.a.} = e^{2i\beta_{\rm arm}} \matr{0}{0}{-\sqrt{ R_{\rm ETM}}\mathcal{K}_{\rm arm}}{0},\label{eq_app:FP_Tsn}\\
  \tq{N}^{\rm \SD{s.n.}} = e^{i\beta_{\rm arm}} N \tq{I},\quad
  \tq{N}^{\rm b.a.} = e^{i\beta_{\rm arm}} N \matr{0}{0}{-\mathcal{N}}{0}. \label{eq_app:FP_Nsn}
\end{gather}

%
%
\paragraph{Filter cavity I/O-relations}

In case of filter cavities, mirrors can be assumed fixed and no radiation pressure effects are to be considered due to an absence of any significant classical light component therein. One can also make a so called narrow-band approximation, assuming $\Omega L_f / c \ll 1$, where $L_f$ is the filter cavity length and $T_f \ll 1$ is input mirror power transmissivity. Then one can write transfer matrices as:
\begin{align*}
  \tq[FC]{T} &= \frac{1}{\mathcal{D}} \matr{t_1}{t_2}{-t_2}{t_1}, \;
  \begin{aligned}
    t_1 &= \gamma_{f1}^2 - \gamma_{f2}^2 - \delta_f^2 + \Omega^2 + 2i\Omega\gamma_{f2}, \\
    t_2 &= - 2 \gamma_{f1} \delta_f,
  \end{aligned}
  \\
  \tq[FC]{N} &= \frac{2\sqrt{\gamma_{f1}\gamma_{f2}}}{\mathcal{D}} \matr{\gamma_f-i\Omega_1}{-\delta_f}{\delta_f}{\gamma_f-i\Omega},
\end{align*}
where $\mathcal{D} = (\gamma_f-i\Omega)^2 + \delta_f^2$, $\gamma_f = \gamma_{f1} + \gamma_{f2}$ is a full cavity half-bandwidth and $\delta_f$ if its detuning. Here $\gamma_{f1} = c T_f / (4L_f)$ is a half-bandwidth part depending on input mirror transmissivity and $\gamma_{f2} = c A_f / (4L_f)$ is the loss-associated part of bandwidth with $A_f \ll 1$ being the total round-trip fractional photon loss.

\subsection{Fabry-Perot--Michelson interferometer}\label{app_ssec:FPMIgeneral}
\subsubsection{Fabry-Perot--Michelson interferometer w/o signal recycling.}

I/O-relations of a Michelson/Fabry-Perot interferometer can be obtained by completing the above ones for the single arm with junction relations at the beam splitter:
\begin{eqnarray*}
  \vq{a}^{\rm N} = \dfrac{\vq{p}+\vq{i}}{\sqrt{2}}\,, &
  \vq{a}^{\rm E} = \dfrac{\vq{p}-\vq{i}}{\sqrt{2}}\,, &
  \vq{o} = \dfrac{\vq{b}^{\rm N}-\vq{b}^{\rm E}}{\sqrt{2}}\,,
\end{eqnarray*}
where $\vq{a}^{\rm N,E} \equiv \vq[arm]{a}^{\rm N,E}$, $\vq{b}^{\rm N,E} \equiv \vq[arm]{b}^{\rm N,E}$ stand for the input and output fields of the  $N$ and $E$ arms, respectively. Hence, the Michelson interferometer I/O-relations read:
\begin{equation}\label{app. matr.: i-o for tuned Mich}
  \vq{o} = \tq[MI]{T}\vq{i} + \tq[MI]{N}\vq{n} + \vs[MI]{t}\frac{h}{h_{\rm SQL}}\,,
\end{equation}
where
\begin{equation*}
  \tq[MI]{T} = \tq[arm]{T} \,, \;\;
  \tq[MI]{N} = \tq[arm]{N} \,, \;\;
  \vs[MI]{t} = \vs[arm]{t} \,. 
\end{equation*}
Here $\vq{n} = \left(\vq[arm]{n}^{\rm N} - \vq[arm]{n}^{\rm E}\right)/\sqrt{2}$ represents effective vacuum fields associated with optical loss in the arm cavities.

In case of small losses the interferometer is described by opto-mechanical factor $\mathcal{K}_{\rm MI} = \mathcal{K}_{\rm arm}$ and phase $\beta_{\rm MI} = \beta_{\rm arm}$.

\subsubsection{Signal-recycled Fabry-Perot--Michelson (FPM) interferometer.}
\begin{figure}[htbp]
\begin{center}
\includegraphics[width=.75\textwidth]{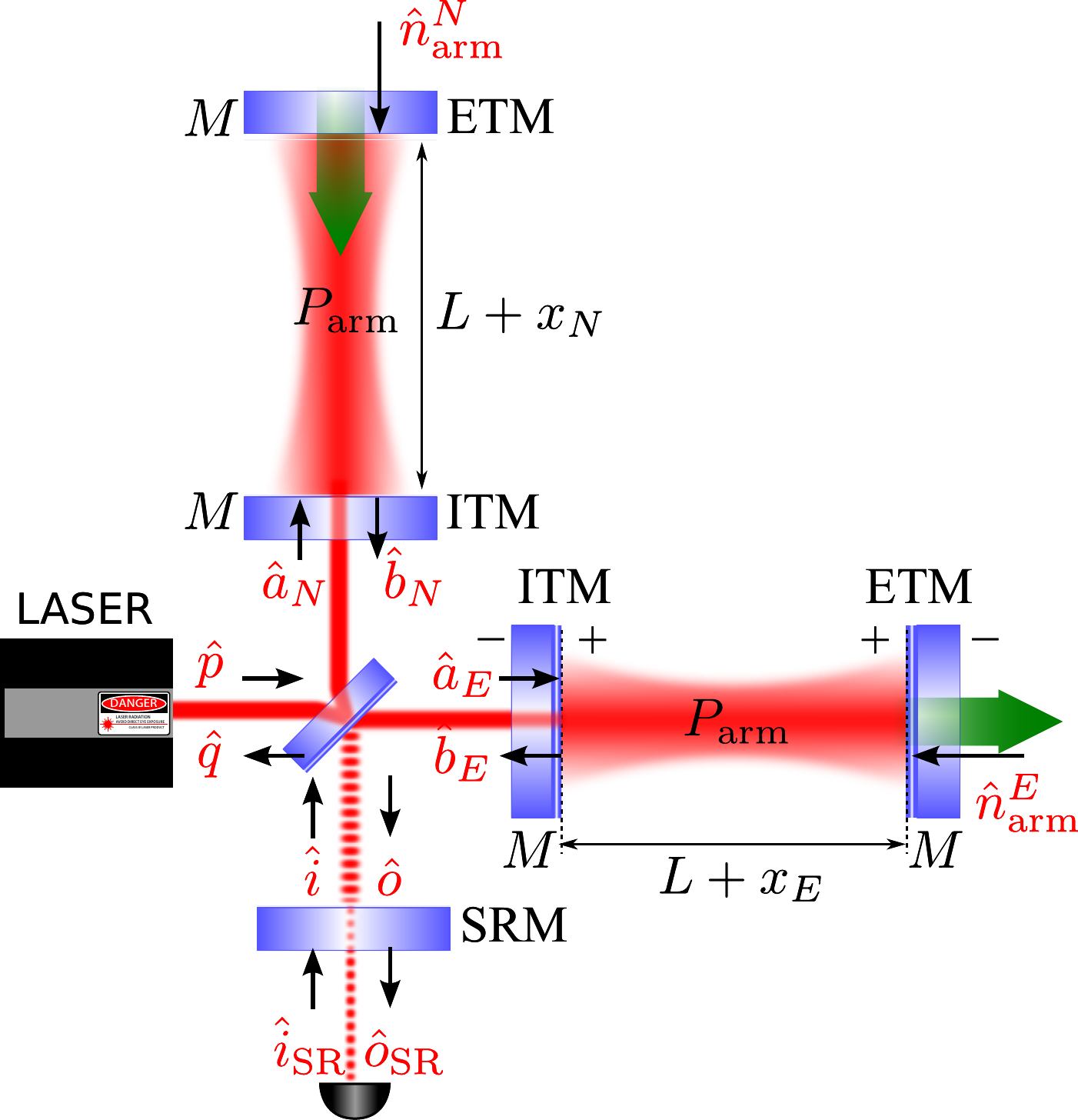}
\caption{Schematics of I/O-relations for a Fabry-Perot--Michelson interferometer} 
\label{fig:app_FPMI}
\end{center}
\end{figure}

I/O-relations of a signal recycled FPMI depicted in Fig.~\ref{fig:app_FPMI} can be obtained from the following equations written for light fields on a signal recycling mirror (SRM):
\begin{equation}\label{app. matr.: SRM i-o in Mich}
  \left\{
  \begin{aligned}
    \vq[SR]{o} &= \tq[\alpha]{P} \left( \sqrt{T_{\rm SR}}\tq[SR]{P}\vq{o} - \sqrt{R_{\rm SR}}\tq[\alpha]{P}\vq[SR]{i} \right) \\
    \vq{i}     &= \tq[SR]{P} \left( \sqrt{R_{\rm SR}}\tq[SR]{P}\vq{o} + \sqrt{T_{\rm SR}}\tq[\alpha]{P}\vq[SR]{i} \right),
  \end{aligned}
  \right.
\end{equation}
where an additional phase shift $\alpha_{\rm SR}$ is introduced to satisfy the Scaling Law of \cite{Buonanno2003} which maps the signal-recycled FPM interferometer and a single detuned Fabry-Perot cavity:
\begin{gather*}
  \tq[SR]{P} = e^{i\frac{\Omega l_{\rm SR}}{c}} \tq{R}[\phi_{\rm SR}] \simeq \tq{R}[\phi_{\rm SR}] \,, \quad
  \phi_{\rm SR} = \frac{\omega_p l_{\rm SR}}{c} \,, \\
  \tq[\alpha]{P} \simeq \tq{R}[\alpha_{\rm SR}] \,, \quad
  \alpha_{\rm SR} = \arctan\left(\frac{\sqrt{R_{\rm SR}}-1}{\sqrt{R_{\rm SR}}+1}\tan\phi_{\rm SR}\right) \,.
\end{gather*}


The solution of \eqref{app. matr.: i-o for tuned Mich} and \eqref{app. matr.: SRM i-o in Mich} gives the following:
\begin{equation*}
  \vq[SR]{o} = \tq[MI\,SR]{T}\vq[SR]{i} + \tq[MI\,SR]{N}\vq{n} + \vs[MI\,SR]{t}\frac{h}{ h{\rm SQL}} \,, \\
\end{equation*}
where
\begin{align*}
  \tq[MI\,SR]{T} &= \tq[\alpha]{P} \bigl[ T_{\rm SR}\tq[SR]{P}\tq[MI\,SR]{M}\tq[MI]{T}\tq[SR]{P}  - \sqrt{R_{\rm SR}}\tq{I} \bigr] \tq[\alpha]{P} \,, \\
  \tq[MI\,SR]{N} &= \sqrt{T_{\rm SR}} \tq[\alpha]{P} \tq[SR]{P} \tq[MI\,SR]{M} \tq[arm]{N} \,, \\
  \vs[MI\,SR]{t} &= \sqrt{T_{\rm SR}} \tq[\alpha]{P} \tq[SR]{P} \tq[MI\,SR]{M} \vs[MI]{t} \,, \\
  &\tq[MI\,SR]{M} = \left[ \tq{I} - \sqrt{R_{\rm SR}} \tq[arm]{T} \tq[SR]{P}^2 \right]^{-1} \,.
\end{align*}

\section{Sagnac interferometer I/O relations.}

First we consider a bare lossless zero-area Sagnac interferometer and derive its input-output (I/O) relations. For definiteness, in this section, we stick to a configuration of Sagnac interferometer that utilises ring arm cavities (as per the left panel of Fig.~\ref{fig:SM_sagnac}), although the results we obtain are applicable to both realisations unless loss is taken into account.

Unlike Michelson interferometer, in Sagnac interferometer light beam visits two arm cavities before recombination with a counter-rotating beam at the beam splitter (see Fig.~\ref{fig:SM_sagnac}). At the same time, two light beams hit the cavity, one coming directly from the beam splitter and the one, that has just left another arm. In notations of Chen's paper \cite{Chen2003} quadrature operators of light enter ing and leaving the arm can be identified with two indices $IJ$, \textit{e.g.} $a_c^{IJ}$, where $I$ stands for the either of two beams, L or R, and $J$ stands for the either of two arms ($J=E,N$). Here $R$ marks the light beam that first enters North arm and then travels the interferometer in the right direction (clockwise), and $L$ marks the beam travelling the interferometer in the opposite (counterclockwise) direction after entering the interferometer through the East arm. Thus, single lossless arm 
I/O relations read, assuming high-finesse arm cavities ($T_{\rm ITM}\ll1$, for general case see Appendix~\ref{app:lossy_FP_I/O-rels}):
\begin{eqnarray}
  b_c^{IJ} &=&   e^{2i\beta_{\rm arm}(\Omega)} a_c^{IJ}\,, \\ 
  b_s^{IJ} &=&   e^{2i\beta_{\rm arm}(\Omega)} [a_s^{IJ} - \mathcal{K}_{\rm arm} (a_c^{IJ}+a_c^{\bar IJ})] \nonumber\\
           & & + e^{i\beta_{\rm arm}(\Omega)} \sqrt{2\mathcal{K}_{\rm arm}} \frac{\sqrt{2}x_J}{h_{\rm SQL}L}
\end{eqnarray}
with $\bar I$ indicating the other beam than $I$, \textit{i.e.} $\bar R = L$ and $\bar L = R$, $h_J = x_J^{\rm ETM}-x_J^{\rm ITM}$ is the arm elongation induced by signal force (\textit{e.g.} gravitational wave tidal force), and
\begin{eqnarray}
  & \mathcal{K}_{\rm arm} &= \dfrac{\Theta_{\rm arm}\tau}{\Omega^2} \dfrac{2\,T_{\rm ITM}}{1-2\sqrt{R_{\rm ITM}} \cos 2\Omega\tau + R_{\rm ITM}} \simeq \nonumber\\
  & &\simeq \frac{2 \Theta_{\rm arm} \gamma_{\rm arm}}{\Omega^2(\gamma_{\rm arm}^2+\Omega^2)}\,,\label{eq:Karm}\\
  & \beta_{\rm arm} &= \atan{ \dfrac{1+\sqrt{R_{\rm ITM}}}{1-\sqrt{R_{\rm ITM}}} \tan\Omega\tau }\simeq \nonumber\\
  & &\simeq  \atan{\Omega/\gamma_{\rm arm}}\label{eq:Betaarm}\,,
\end{eqnarray}
with $\Theta_{\rm arm} = 4\omega_0 P_{\rm arm}/(McL)$ and $P_{\rm arm} = P_c/4$, where $P_c$ is the total optical power circulating in both arms
and $\gamma_{\rm arm} = T_{\rm ITM}/(4\tau)$ is the half-bandwidth of an arm cavity. The final, approximate expressions above are obtained assuming that cavity linewidth and signal frequency are much smaller than cavity free spectral range $\nu_{\rm FSR}=(2\tau)^{-1}$. This approximation nearly breaks down for detectors with arm length $\gtrsim 10$~km, like Einstein Telescope, at frequencies of the order of 10~kHz, therefore we present exact formulae as well.

Then, it is straightforward to derive full Sagnac I/O-relations, using junction equations for the fields at the output beam splitter (ring-cavity topology):
\begin{eqnarray}\label{i-o relations at BS in SI, p.1}
  \vq{a}^{\rm RN} = \dfrac{\vq{p}+\vq{i}}{\sqrt{2}} \,, &
  \vq{a}^{\rm LE} = \dfrac{\vq{p}-\vq{i}}{\sqrt{2}} \,, &
  \vq{o} = \dfrac{\vq{b}^{\rm LN}-\vq{b}^{\rm RE}}{\sqrt{2}} \,,
\end{eqnarray}
as well as continuity relations between the beams that leave one arm and enter the other:
\begin{eqnarray}\label{i-o relations at BS in SI, p.2}
  \vq{a}^{\rm RE} = \vq{b}^{\rm RN} \,, &
  \vq{a}^{\rm LN} = \vq{b}^{\rm LE} \,.
\end{eqnarray}
The resulting I/O-relations for lossless zero-area Sagnac interferometer then read:
\begin{equation}\label{eq:IOSaglossless}
  \begin{bmatrix} \hat{o}_c\\\hat{o}_s\end{bmatrix} =
  e^{2i\beta_{\rm SI}}\begin{bmatrix} 1 & 0\\-\mathcal{K}_{\rm SI} & 1\end{bmatrix}
  \begin{bmatrix} \hat{i}_c\\\hat{i}_s\end{bmatrix}+
  \begin{bmatrix}0\\\sqrt{2\mathcal{K}_{\rm SI}}\end{bmatrix}e^{i\beta_{\rm SI}}
\frac{h}{h_{\rm SQL}}\,,
\end{equation}
with coupling constant $\mathcal{K}_{\rm SI}$ defined as:
\begin{equation}
  \mathcal{K}_{\rm SI} = 4\,\mathcal{K}_{\rm arm} \sin^2\beta_{\rm arm}
  \simeq \frac{4 \Theta_{\rm SI} \gamma_{\rm arm}}{(\Omega^2+\gamma_{\rm arm}^2)^2},
\end{equation}
where $\Theta_{\rm SI} \equiv 4\Theta_{\rm arm}$ and additional phase shift:
\begin{equation}\label{eq:SIphase}
  \beta_{\rm SI} = 2\beta_{\rm arm} + \frac{\pi}{2}
\end{equation}

One can now calculate spectral density of quantum noise of the zero-area Sagnac, using Eq.~\eqref{eq:SpDens_h}, where transfer matrix $\tq{T}$ and response vector $\vs{t}$ read:
\begin{equation}\label{eq:SI_T_and_t}
  \tq{T} = e^{2i\beta_{\rm SI}}\begin{bmatrix} 1 & 0\\-\mathcal{K}_{\rm SI} & 1\end{bmatrix}\,,\quad\vs{t} = e^{i\beta_{\rm SI}} \begin{bmatrix}0\\\sqrt{2\mathcal{K}_{\rm SI}}\end{bmatrix}\,.
\end{equation}
Therefore one gets this simple expression for spectral density (it is the same for all tuned interferometers with balanced homodyne readout of quadrature $b_\zeta$ and vacuum state at the dark port, save to the expression for $\mathcal{K}$):
\begin{equation}\label{eq:Sh_plain}
   S^h = \frac{h^2_{\rm SQL}}{2}\left\{\frac{\left[\mathcal{K}_{\rm SI}-\cot\zeta\right]^2+1}{\mathcal{K}_{\rm SI}}\right\}\,.
\end{equation}

\begin{acknowledgements}
The authors are particularly grateful to Jan Harms for his careful and meticulous reading of the manuscript and for very helpful feedback. We also thank our colleagues from the LIGO-Virgo Scientific Collaboration (LVC) for illuminating discussions and suggestions on how to improve the paper. SLD would like to thank Lower Saxonian Ministry of Science and Culture that supported his research within the frame of the program  “Research Line” (Forschungslinie) QUANOMET – Quantum- and Nano-Metrology . The FYK was supported by the Russian Foundation for Basic Research Grants 14-02-00399 and 16-52-10069.  FYK was also supported by the LIGO NSF Grant PHY-1305863. HM is supported by UK STFC Ernest Rutherford Fellowship (Grant No. ST/M005844/1).
\end{acknowledgements}

\bibliographystyle{spphys}       
\bibliography{LRR_AdvQTechniques_for_GWD}   

\begin{thebibliography}{100}
\providecommand{\url}[1]{{#1}}
\providecommand{\urlprefix}{URL }
\expandafter\ifx\csname urlstyle\endcsname\relax
  \providecommand{\doi}[1]{DOI \discretionary{}{}{}#1}\else
  \providecommand{\doi}{DOI \discretionary{}{}{}\begingroup
  \urlstyle{rm}\Url}\fi

\bibitem{TheLIGOScientific:2014jea}
J.~Aasi, et~al., Class. Quant. Grav. \textbf{32}, 074001 (2015).
\newblock \doi{10.1088/0264-9381/32/7/074001}

\bibitem{TheVirgo:2014hva}
F.~Acernese, et~al., Class. Quant. Grav. \textbf{32}(2), 024001 (2015).
\newblock \doi{10.1088/0264-9381/32/2/024001}

\bibitem{GW_Discovery_Paper_PhysRevLett.116.061102}
{B.~P.~Abbott {\it et al}}, Phys. Rev. Lett. \textbf{116}, 061102 (2016).
\newblock \doi{10.1103/PhysRevLett.116.061102}.
\newblock
  \urlprefix\url{http://link.aps.org/doi/10.1103/PhysRevLett.116.061102}

\bibitem{2016_PhysRevX.6.041015_LVC}
{B.~P.~Abbott {\it et al}}, Phys. Rev. X \textbf{6}, 041015 (2016).
\newblock \doi{10.1103/PhysRevX.6.041015}.
\newblock \urlprefix\url{http://link.aps.org/doi/10.1103/PhysRevX.6.041015}

\bibitem{2016_PhysRevLett.116.241103_LSC_Detection}
{B.~P.~Abbott {\it et al}}, Phys. Rev. Lett. \textbf{116}, 241103 (2016).
\newblock \doi{10.1103/PhysRevLett.116.241103}.
\newblock
  \urlprefix\url{https://link.aps.org/doi/10.1103/PhysRevLett.116.241103}

\bibitem{2017_PhysRevLett.118.221101_LSC_Detection}
{B.~P.~Abbott {\it et al}}, Phys. Rev. Lett. \textbf{118}, 221101 (2017).
\newblock \doi{10.1103/PhysRevLett.118.221101}.
\newblock
  \urlprefix\url{https://link.aps.org/doi/10.1103/PhysRevLett.118.221101}

\bibitem{2017_PRL.119.141101_LVC}
B.P. {Abbott}, R.~{Abbott}, T.D. {Abbott}, F.~{Acernese}, K.~{Ackley},
  C.~{Adams}, T.~{Adams}, P.~{Addesso}, R.X. {Adhikari}, V.B. {Adya}, et~al.,
  Physical Review Letters \textbf{119}(14), 141101 (2017).
\newblock \doi{10.1103/PhysRevLett.119.141101}

\bibitem{2017_ApJ.851.L35_LVC_detection}
B.P. Abbott, et~al., Astrophys. J. \textbf{851}(2), L35 (2017).
\newblock \doi{10.3847/2041-8213/aa9f0c}

\bibitem{Aasi:2013wya}
B.P. Abbott, et~al., Living Rev. Rel. \textbf{21}, 3 (2018).
\newblock \doi{10.1007/s41114-018-0012-9, 10.1007/lrr-2016-1}.
\newblock [Living Rev. Rel.19,1(2016)]

\bibitem{2017_ApJ.848L.12A_LVC}
{B.~P.~Abbott {\it et al}}, \apjl \textbf{848}, L12 (2017).
\newblock \doi{10.3847/2041-8213/aa91c9}

\bibitem{Harry:2018hke}
I.~Harry, T.~Hinderer, Classical and Quantum Gravity \textbf{35}(14), 145010
  (2018).
\newblock \urlprefix\url{http://stacks.iop.org/0264-9381/35/i=14/a=145010}

\bibitem{2017_PRL.119.1101A_LVC_BNS_Discovery}
{B.~P.~Abbott {\it et al}}, Physical Review Letters \textbf{119}(16), 161101
  (2017).
\newblock \doi{10.1103/PhysRevLett.119.161101}

\bibitem{2017ApJ...848L..13A}
{B.~P.~Abbott {\it et al}}, \apjl \textbf{848}, L13 (2017).
\newblock \doi{10.3847/2041-8213/aa920c}

\bibitem{Harms2015}
J.~Harms, Living Reviews in Relativity \textbf{18}(1), 3 (2015).
\newblock \doi{10.1007/lrr-2015-3}.
\newblock \urlprefix\url{https://doi.org/10.1007/lrr-2015-3}

\bibitem{ISWP2017}
{The LIGO Scientific Collaboration}, Collaboration.
\newblock {Instrument Science White Paper} (2018).
\newblock \urlprefix\url{https://dcc.ligo.org/T1800133}

\bibitem{2013_nat.Photon.7.8.644_Crystallyne_coatings}
G.D. Cole, W.~Zhang, M.J. Martin, J.~Ye, M.~Aspelmeyer, Nat. Photon.
  \textbf{7}(8), 644–650 (2013).
\newblock \doi{10.1038/nphoton.2013.174}.
\newblock \urlprefix\url{http://dx.doi.org/10.1038/nphoton.2013.174}

\bibitem{CQG.27.19.194002_2010_Punturo}
M.~Punturo, M.~Abernathy, F.~Acernese, B.~Allen, N.A.K. Andersson, F.~Barone,
  B.~Barr, M.~Barsuglia, M.~Beker, {\it et al}, Classical and Quantum Gravity
  \textbf{27}(19), 194002 (2010).
\newblock \doi{10.1088/0264-9381/27/19/194002}.
\newblock \urlprefix\url{http://stacks.iop.org/0264-9381/27/i=19/a=194002}

\bibitem{2011_CQG.28.9.094013_3rd_gen_design_study}
S.~Hild, M.~Abernathy, F.~Acernese, P.~Amaro-Seoane, N.~Andersson, K.~Arun,
  F.~Barone, B.~Barr, M.~Barsuglia, M.~Beker, {\it et al.}, Class. Quant.
  Gravity \textbf{28}(9), 094013 (2011).
\newblock \doi{10.1088/0264-9381/28/9/094013}.
\newblock \urlprefix\url{http://iopscience.iop.org/0264-9381/28/9/094013}

\bibitem{2017_CQG.34.004001_CEref}
B.P. {Abbott}, R.~{Abbott}, T.D. {Abbott}, M.R. {Abernathy}, K.~{Ackley},
  C.~{Adams}, P.~{Addesso}, R.X. {Adhikari}, V.B. {Adya}, C.~{Affeldt}, {\it et
  al.}, {(LIGO Scientific Collaboration}, Classical and Quantum Gravity
  \textbf{34}, 044001 (2017).
\newblock \doi{10.1088/1361-6382/aa51f4}

\bibitem{1981_PRD.23.1693_Caves}
C.~Caves, Phys. Rev. D \textbf{23}, 1693 (1981).
\newblock \doi{10.1103/PhysRevD.23.1693}

\bibitem{2011_Nat.Phys.7.962_LSC}
{J.Abadie {\it et al}}, Nature Physics \textbf{7}, 962 (2011).
\newblock \doi{10.1038/nphys2083}

\bibitem{Aasi2013NatPhot}
{J.Aasi {\it et al}}, Nature Photonics \textbf{7}, 613 (2013).
\newblock \doi{10.1038/nphoton.2013.177}

\bibitem{Schnabel2017}
R.~Schnabel, Physics Reports \textbf{684}, 1 (2017).
\newblock \doi{10.1016/j.physrep.2017.04.001}.
\newblock
  \urlprefix\url{http://www.sciencedirect.com/science/article/pii/S0370157317300595?via{\%}3Dihub}

\bibitem{1990_PLA.147.251_Braginsky_SM}
V.B. {Braginsky}, F.J. {Khalili}, Physics Letters A \textbf{147}, 251 (1990).
\newblock \doi{10.1016/0375-9601(90)90442-Q}

\bibitem{Chen2003}
Y.~Chen, Phys. Rev. D \textbf{67}, 122004 (2003).
\newblock \doi{10.1103/PhysRevD.67.122004}

\bibitem{Purdue2001}
P.~Purdue, Phys. Rev. D \textbf{66}, 022001 (2002).
\newblock \doi{10.1103/PhysRevD.66.022001}

\bibitem{Purdue2002}
P.~Purdue, Y.~Chen, Phys. Rev. D \textbf{66}, 122004 (2002).
\newblock \doi{10.1103/PhysRevD.66.122004}

\bibitem{04a1Da}
S.L. Danilishin, Phys. Rev. D \textbf{69}, 102003 (2004).
\newblock \doi{10.1103/PhysRevD.69.102003}

\bibitem{PhysRevD.86.062001}
A.R. Wade, K.~McKenzie, Y.~Chen, D.A. Shaddock, J.H. Chow, D.E. McClelland,
  Phys. Rev. D \textbf{86}, 062001 (2012).
\newblock \doi{10.1103/PhysRevD.86.062001}.
\newblock \urlprefix\url{http://link.aps.org/doi/10.1103/PhysRevD.86.062001}

\bibitem{2014_CQG.31.215009_Graef}
C.~{Gr{\"a}f}, B.W. {Barr}, A.S. {Bell}, F.~{Campbell}, A.V. {Cumming}, S.L.
  {Danilishin}, N.A. {Gordon}, G.D. {Hammond}, J.~{Hennig}, E.A. {Houston},
  S.H. {Huttner}, R.A. {Jones}, S.S. {Leavey}, H.~{L{\"u}ck}, J.~{Macarthur},
  M.~{Marwick}, S.~{Rigby}, R.~{Schilling}, B.~{Sorazu}, A.~{Spencer},
  S.~{Steinlechner}, K.A. {Strain}, S.~{Hild}, Classical and Quantum Gravity
  \textbf{31}(21), 215009 (2014).
\newblock \doi{10.1088/0264-9381/31/21/215009}

\bibitem{2015arXiv150301062V}
N.V. {Voronchev}, S.P. {Tarabrin}, S.L. {Danilishin}, ArXiv e-prints  (2015)

\bibitem{Oelker2016}
E.~Oelker, T.~Isogai, J.~Miller, M.~Tse, L.~Barsotti, N.~Mavalvala, M.~Evans,
  Phys. Rev. Lett. \textbf{116}(4), 041102 (2016).
\newblock \doi{10.1103/PhysRevLett.116.041102}.
\newblock
  \urlprefix\url{https://journals.aps.org/prl/abstract/10.1103/PhysRevLett.116.041102}

\bibitem{2013_OE.21.30114_Loss_in_FC_Isogai}
T.~Isogai, J.~Miller, P.~Kwee, L.~Barsotti, M.~Evans, Opt. Express
  \textbf{21}(24), 30114–30125 (2013).
\newblock \doi{10.1364/OE.21.030114}.
\newblock
  \urlprefix\url{http://www.opticsexpress.org/abstract.cfm?URI=oe-21-24-30114}

\bibitem{Ma_NPhys_13_776_2017}
Y.~Ma, H.~Miao, B.H. Pang, M.~Evans, C.~Zhao, J.~Harms, R.~Schnabel, Y.~Chen,
  Nature Physics \textbf{13}(8), 776 (2017).
\newblock \doi{doi:10.1038/nphys4118}

\bibitem{2017_PhysRevD.96.062003_GEO_EPR_squeezing}
D.D. Brown, H.~Miao, C.~Collins, C.~Mow-Lowry, D.~Töyra, A.~Freise, Phys. Rev.
  \textbf{D96}(6), 062003 (2017).
\newblock \doi{10.1103/PhysRevD.96.062003}

\bibitem{Wicht1997}
A.~Wicht, K.~Danzmann, M.~Fleischhauer, M.~Scully, G.~M{\"{u}}ller,
  R.~Rinkleff, Optics Communications \textbf{134}, 431 (1997).
\newblock \doi{10.1016/S0030-4018(96)00579-2}.
\newblock
  \urlprefix\url{http://www.sciencedirect.com/science/article/pii/S0030401896005792}

\bibitem{Zhou2015}
M.~Zhou, Z.~Zhou, S.M. Shahriar, Phys. Rev. D \textbf{92}(8), 082002 (2015).
\newblock \doi{10.1103/PhysRevD.92.082002}.
\newblock \urlprefix\url{https://link.aps.org/doi/10.1103/PhysRevD.92.082002}

\bibitem{Ma2015}
Y.~Ma, H.~Miao, C.~Zhao, Y.~Chen, Phys. Rev. A \textbf{92}(2) (2015).
\newblock
  \urlprefix\url{https://journals.aps.org/pra/abstract/10.1103/PhysRevA.92.023807}

\bibitem{Peano2015}
V.~Peano, H.G.L. Schwefel, C.~Marquardt, F.~Marquardt, Phys. Rev. Lett.
  \textbf{115}(24), 243603 (2015).
\newblock
  \urlprefix\url{http://journals.aps.org/prl/abstract/10.1103/PhysRevLett.115.243603}

\bibitem{Korobko2017}
M.~Korobko, L.~Kleybolte, S.~Ast, H.~Miao, Y.~Chen, R.~Schnabel, Physical
  Review Letters \textbf{118}(14), 143601 (2017).
\newblock \doi{10.1103/PhysRevLett.118.143601}.
\newblock
  \urlprefix\url{http://link.aps.org/doi/10.1103/PhysRevLett.118.143601}

\bibitem{Miao2015a}
H.~Miao, Y.~Ma, C.~Zhao, Y.~Chen, Phys. Rev. Lett. \textbf{115}(21), 211104
  (2015).
\newblock \doi{10.1103/PhysRevLett.115.211104}.
\newblock
  \urlprefix\url{https://link.aps.org/doi/10.1103/PhysRevLett.115.211104}

\bibitem{Page2017}
M.~Page, J.~Qin, J.~{La Fontaine}, C.~Zhao, D.~Blair, arXiv:1711.04469  (2017).
\newblock \urlprefix\url{http://arxiv.org/abs/1711.04469}

\bibitem{Miao2017c}
H.~Miao, H.~Yang, D.~Martynov, Phys. Rev. D \textbf{98}, 044044 (2018).
\newblock \doi{10.1103/PhysRevD.98.044044}.
\newblock \urlprefix\url{https://link.aps.org/doi/10.1103/PhysRevD.98.044044}

\bibitem{Liv.Rv.Rel.15.2012}
S.L. Danilishin, F.Y. Khalili, Living Reviews in Relativity \textbf{15}, 5
  (2012).
\newblock \urlprefix\url{http://www.livingreviews.org/lrr-2012-5}

\bibitem{85a1CaSch}
C.~Caves, B.~Schumaker, Phys. Rev. A \textbf{31}, 3068–3092 (1985).
\newblock \doi{10.1103/PhysRevA.31.3068}

\bibitem{85a2CaSch}
B.~Schumaker, C.~Caves, Phys. Rev. A \textbf{31}, 3093–3111 (1985).
\newblock \doi{10.1103/PhysRevA.31.3093}

\bibitem{Fritschel:14}
P.~Fritschel, M.~Evans, V.~Frolov, Opt. Express \textbf{22}(4), 4224 (2014).
\newblock \doi{10.1364/OE.22.004224}.
\newblock
  \urlprefix\url{http://www.opticsexpress.org/abstract.cfm?URI=oe-22-4-4224}

\bibitem{2007_JoPA.40.7821_Adesso_illuminati}
G.~Adesso, F.~Illuminati, Journal of Physics A: Mathematical and Theoretical
  \textbf{40}(28), 7821 (2007).
\newblock \urlprefix\url{http://stacks.iop.org/1751-8121/40/i=28/a=S01}

\bibitem{04BookBlTh}
R.~Blandford, K.~Thorne.
\newblock Applications of classical physics (an unpublished manuscript of the
  textbook) (2008).
\newblock \urlprefix\url{http://www.pma.caltech.edu/Courses/ph136/yr2008/}

\bibitem{02a1KiLeMaThVy}
H.~Kimble, Y.~Levin, A.~Matsko, K.~Thorne, S.~Vyatchanin, Phys. Rev. D
  \textbf{65}, 022002 (2002).
\newblock \doi{10.1103/PhysRevD.65.022002}

\bibitem{1997_CQG.14.6.1513_Schilling_LISA_response}
R.~Schilling, Classical and Quantum Gravity \textbf{14}(6), 1513 (1997).
\newblock \urlprefix\url{http://stacks.iop.org/0264-9381/14/i=6/a=020}

\bibitem{PhysRevD.96.084004}
R.~Essick, S.~Vitale, M.~Evans, Phys. Rev. D \textbf{96}, 084004 (2017).
\newblock \doi{10.1103/PhysRevD.96.084004}.
\newblock \urlprefix\url{https://link.aps.org/doi/10.1103/PhysRevD.96.084004}

\bibitem{Buonanno2003}
A.~Buonanno, Y.~Chen, Phys. Rev. D \textbf{67}, 062002 (2003).
\newblock \doi{10.1103/PhysRevD.67.062002}.
\newblock \urlprefix\url{http://link.aps.org/doi/10.1103/PhysRevD.67.062002}

\bibitem{96a2eVyMa}
{S.P.Vyatchanin and A.B.Matsko}, Sov. Phys. JETP \textbf{83}, 690 (1996)

\bibitem{2016_Phys.Rev.A.94.062109_Bloch-Messiah_Decomposition}
G.~Cariolaro, G.~Pierobon, Phys. Rev. A \textbf{94}, 062109 (2016).
\newblock \doi{10.1103/PhysRevA.94.062109}.
\newblock \urlprefix\url{https://link.aps.org/doi/10.1103/PhysRevA.94.062109}

\bibitem{PhysRev.83.34}
H.~Callen, T.~Welton, Phys. Rev. \textbf{83}(1), 34–40 (1951).
\newblock \doi{10.1103/PhysRev.83.34}

\bibitem{Miao14}
H.~Miao, H.~Yang, R.X. Adhikari, Y.~Chen, Classical and Quantum Gravity
  \textbf{31}(16), 165010 (2014).
\newblock \doi{10.1088/0264-9381/31/16/165010}.
\newblock \urlprefix\url{http://stacks.iop.org/0264-9381/31/i=16/a=165010}

\bibitem{2015_NJP17.043031_asymSag}
S.L. {Danilishin}, C.~{Gr{\"a}f}, S.S. {Leavey}, J.~{Hennig}, E.A. {Houston},
  D.~{Pascucci}, S.~{Steinlechner}, J.~{Wright}, S.~{Hild}, New Journal of
  Physics \textbf{17}(4), 043031 (2015).
\newblock \doi{10.1088/1367-2630/17/4/043031}

\bibitem{2015_PhysRevD.92.072009_Steinlechner}
S.~Steinlechner, B.W. Barr, A.S. Bell, S.L. Danilishin, A.~Gl\"afke, C.~Gr\"af,
  J.S. Hennig, E.A. Houston, S.H. Huttner, S.S. Leavey, D.~Pascucci, B.~Sorazu,
  A.~Spencer, K.A. Strain, J.~Wright, S.~Hild, Phys. Rev. D \textbf{92}, 072009
  (2015).
\newblock \doi{10.1103/PhysRevD.92.072009}.
\newblock \urlprefix\url{http://link.aps.org/doi/10.1103/PhysRevD.92.072009}

\bibitem{2017_PhysRevD.95.062001}
T.~Zhang, S.L. Danilishin, S.~Steinlechner, B.W. Barr, A.S. Bell, P.~Dupej,
  C.~Gr\"af, J.S. Hennig, E.A. Houston, S.H. Huttner, S.S. Leavey, D.~Pascucci,
  B.~Sorazu, A.~Spencer, J.~Wright, K.A. Strain, S.~Hild, Phys. Rev. D
  \textbf{95}, 062001 (2017).
\newblock \doi{10.1103/PhysRevD.95.062001}.
\newblock \urlprefix\url{http://link.aps.org/doi/10.1103/PhysRevD.95.062001}

\bibitem{92BookBrKh}
V.~Braginsky, F.~Khalili, \emph{{Quantum Measurement}} (Cambridge University
  Press, Cambridge, 1992)

\bibitem{Chen2}
A.~Buonanno, Y.~Chen, Phys. Rev. D \textbf{65}, 42001 (2002)

\bibitem{Chen2013}
Y.~Chen, Journal of Physics B: Atomic, Molecular and Optical Physics
  \textbf{46}, 104001 (2013).
\newblock \doi{10.1088/0953-4075/46/10/104001}.
\newblock \urlprefix\url{http://iopscience.iop.org/0953-4075/46/10/104001}

\bibitem{Miao2017}
H.~Miao, M.~Evans, LIGO DCC-P1700202  (2017).
\newblock \urlprefix\url{https://dcc.ligo.org/LIGO-P1700202}

\bibitem{00p1BrGoKhTh}
{V.B.Braginsky, M.L.Gorodetsky, F.Ya.Khalili and K.S.Thorne}, in
  \emph{Gravitational waves. Third Edoardo Amaldi Conference, Pasadena,
  California 12-16 July}, ed. by S.Meshkov (Melville NY:AIP Conf. Proc. 523,
  2000), pp. 180--189

\bibitem{Tsang2011}
M.~Tsang, H.M. Wiseman, C.M. Caves, Phys. Rev. Lett. \textbf{106}, 090401
  (2011).
\newblock \doi{10.1103/PhysRevLett.106.090401}.
\newblock
  \urlprefix\url{http://journals.aps.org/prl/abstract/10.1103/PhysRevLett.106.090401}

\bibitem{Miao2017b}
H.~Miao, R.X. Adhikari, Y.~Ma, B.~Pang, Y.~Chen, Phys. Rev. Lett.
  \textbf{119}(5), 050801 (2017).
\newblock \doi{10.1103/PhysRevLett.119.050801}.
\newblock
  \urlprefix\url{http://link.aps.org/doi/10.1103/PhysRevLett.119.050801}

\bibitem{2013_OE.21.19047_phase_noise_sqz_LIGO}
S.~Dwyer, L.~Barsotti, S.S.Y. Chua, M.~Evans, M.~Factourovich, D.~Gustafson,
  T.~Isogai, K.~Kawabe, A.~Khalaidovski, P.K. Lam, M.~Landry, N.~Mavalvala,
  D.E. McClelland, G.D. Meadors, C.M. Mow-Lowry, R.~Schnabel, R.M.S. Schofield,
  N.~Smith-Lefebvre, M.~Stefszky, C.~Vorvick, D.~Sigg, Opt. Express
  \textbf{21}(16), 19047 (2013).
\newblock \doi{10.1364/OE.21.019047}.
\newblock
  \urlprefix\url{http://www.opticsexpress.org/abstract.cfm?URI=oe-21-16-19047}

\bibitem{2015_OE.23.8235_phase_noise_sqz_GEO}
K.L. Dooley, E.~Schreiber, H.~Vahlbruch, C.~Affeldt, J.R. Leong, H.~Wittel,
  H.~Grote, Opt. Express \textbf{23}(7), 8235 (2015).
\newblock \doi{10.1364/OE.23.008235}.
\newblock
  \urlprefix\url{http://www.opticsexpress.org/abstract.cfm?URI=oe-23-7-8235}

\bibitem{2006_App.Phys.Lett.89.6.061116_Furusawa_squeezing}
S.~Suzuki, H.~Yonezawa, F.~Kannari, M.~Sasaki, A.~Furusawa, Applied Physics
  Letters \textbf{89}(6), 061116 (2006).
\newblock \doi{http://dx.doi.org/10.1063/1.2335806}.
\newblock
  \urlprefix\url{http://scitation.aip.org/content/aip/journal/apl/89/6/10.1063/1.2335806}

\bibitem{Brown2017}
D.D. Brown, H.~Miao, C.~Collins, C.~Mow-Lowry, D.~T{\"{o}}yr{\"{a}}, A.~Freise,
  Physical Review D \textbf{96}(6), 062003 (2017).
\newblock \doi{10.1103/PhysRevD.96.062003}.
\newblock \urlprefix\url{https://link.aps.org/doi/10.1103/PhysRevD.96.062003}

\bibitem{Ma2014a}
Y.~Ma, S.L. Danilishin, C.~Zhao, H.~Miao, W.Z. Korth, Y.~Chen, R.L. Ward, D.G.
  Blair, Phys. Rev. Lett. \textbf{113}, 151102 (2014).
\newblock \doi{10.1103/PhysRevLett.113.151102}

\bibitem{00a1BrGoKhTh}
V.B. Braginsky, M.L. Gorodetsky, F.Y. Khalili, K.S. Thorne, Phys. Rev. D
  \textbf{61}, 044002 (2000).
\newblock \doi{10.1103/PhysRevD.61.044002}

\bibitem{02a2Kh}
F.~Khalili.
\newblock Quantum speedmeter and laser interferometric gravitational-wave
  antennae (2002)

\bibitem{PhysRevD.87.096008}
M.~Wang, C.~Bond, D.~Brown, F.~Br{\"u}ckner, L.~Carbone, R.~Palmer, A.~Freise,
  Phys. Rev. D \textbf{87}, 096008 (2013).
\newblock \doi{10.1103/PhysRevD.87.096008}.
\newblock \urlprefix\url{http://link.aps.org/doi/10.1103/PhysRevD.87.096008}

\bibitem{2018_LSA.7.accepted}
S.L. {Danilishin}, E.~{Knyazev}, N.V. {Voronchev}, F.Y. {Khalili},
  C.~{Gr{\"a}f}, S.~{Steinlechner}, J.S. {Hennig}, S.~{Hild}, Light: Science \&
  Applications \textbf{7}, 11 (2018).
\newblock \doi{10.1038/s41377-018-0004-2}.
\newblock \urlprefix\url{https://www.nature.com/articles/s41377-018-0004-2}

\bibitem{2017_CQG.34.2.024001_Huttner}
S.H. Huttner, S.L. Danilishin, B.W. Barr, A.S. Bell, C.~Gräf, J.S. Hennig,
  S.~Hild, E.A. Houston, S.S. Leavey, D.~Pascucci, B.~Sorazu, A.P. Spencer,
  S.~Steinlechner, J.L. Wright, T.~Zhang, K.A. Strain, Classical and Quantum
  Gravity \textbf{34}(2), 024001 (2017).
\newblock \urlprefix\url{http://stacks.iop.org/0264-9381/34/i=2/a=024001}

\bibitem{2017_Phys.Lett.A_EPR_SM}
E.~Knyazev, S.~Danilishin, S.~Hild, F.~Khalili, Physics Letters A \textbf{382},
  2219 (2018).
\newblock \doi{https://doi.org/10.1016/j.physleta.2017.10.009}.
\newblock
  \urlprefix\url{http://www.sciencedirect.com/science/article/pii/S0375960117302888}

\bibitem{1999_JOSAB.16.9.1354_Beyersdorf}
P.T. Beyersdorf, M.M. Fejer, R.L. Byer, J. Opt. Soc. Am. B \textbf{16}(9),
  1354–1358 (1999).
\newblock \doi{10.1364/JOSAB.16.001354}.
\newblock \urlprefix\url{http://josab.osa.org/abstract.cfm?URI=josab-16-9-1354}

\bibitem{1999_Opt.Lett.24.16.1112_Beyersdorf}
P.T. Beyersdorf, M.M. Fejer, R.L. Byer, Opt. Lett. \textbf{24}(16), 1112–1114
  (1999).
\newblock \doi{10.1364/OL.24.001112}.
\newblock \urlprefix\url{http://ol.osa.org/abstract.cfm?URI=ol-24-16-1112}

\bibitem{Pascucci2018}
{D. Pascucci {\it et al.}}
\newblock in preparation (2018)

\bibitem{Hild_personal}
S.~Hild.
\newblock personal communication

\bibitem{Krocker_personal}
S.~Krocker.
\newblock personal communication

\bibitem{Zhang:2018czu}
T.~{Zhang}, E.~{Knyazev}, S.~{Steinlechner}, F.Y. {Khalili}, B.W. {Barr}, A.S.
  {Bell}, P.~{Dupej}, J.~{Briggs}, C.~{Gr{\"a}f}, J.~{Callaghan}, J.S.
  {Hennig}, E.A. {Houston}, S.H. {Huttner}, S.S. {Leavey}, D.~{Pascucci},
  B.~{Sorazu}, A.~{Spencer}, J.~{Wright}, K.A. {Strain}, S.~{Hild}, S.L.
  {Danilishin}, New Journal of Physics \textbf{20}, 103040 (2018).
\newblock \doi{10.1088/1367-2630/aae86e}

\bibitem{09a1ChDaKhMu}
{Y.Chen, S.L.Danilishin, F.Y.Khalili, H.M{\"u}ller-Ebhardt}, General Relativity
  and Gravitation \textbf{43}(2), 671 (2011)

\bibitem{Teufel2009}
J.D. Teufel, T.~Donner, M.A. Castellanos-Beltran, J.W. Harlow, K.W. Lehnert,
  Nature Nanotechnology \textbf{4}(12), 820 (2009).
\newblock \urlprefix\url{http://dx.doi.org/10.1038/nnano.2009.343}

\bibitem{Anetsberger_NPhys_5_909_2009}
G.~Anetsberger, O.~Arcizet, Q.P. Unterreithmeier, R.~Riviere, A.~Schliesser,
  E.M. Weig, J.P. Kotthaus, T.J. Kippenberg, Nature Physics \textbf{5}(12), 909
  (2009).
\newblock \doi{10.1038/nphys1425}.
\newblock \urlprefix\url{http://dx.doi.org/10.1038/nphys1425}

\bibitem{11a1WeFrKaYaGoMuDaKhDaSc}
T.~Westphal, D.~Friedrich, H.~Kaufer, K.~Yamamoto, S.~Go{\ss}ler,
  H.~M{\"u}ller-Ebhardt, S.L. Danilishin, F.Y. Khalili, K.~Danzmann,
  R.~Schnabel, Phys. Rev. A \textbf{85}, 063806 (2012).
\newblock \doi{10.1103/PhysRevA.85.063806}.
\newblock \urlprefix\url{http://link.aps.org/doi/10.1103/PhysRevA.85.063806}

\bibitem{64a1eBrMi}
V.~Braginsky, I.~Minakova, Vestnik Moskovskogo Universiteta, Seriya 3
  \textbf{0}(1), 69 (1964)

\bibitem{67a1eBrMa}
V.~Braginsky, A.~Manukin, Sov. Phys. JETP \textbf{25}, 653 (1967)

\bibitem{70a1eBrMaTi}
V.B. {Braginskiǐ}, A.B. {Manukin}, M.Y. {Tikhonov}, Soviet Journal of
  Experimental and Theoretical Physics \textbf{31}, 829 (1970)

\bibitem{Dorsel_PRL_51_1550_1983}
A.~Dorsel, J.D. McCullen, P.~Meystre, E.~Vignes, H.~Walther, Phys. Rev. Lett.
  \textbf{51}, 1550 (1983).
\newblock \doi{10.1103/PhysRevLett.51.1550}.
\newblock \urlprefix\url{http://link.aps.org/doi/10.1103/PhysRevLett.51.1550}

\bibitem{97a1BrGoKh}
V.~Braginsky, M.~Gorodetsky, F.~Khalili, Phys. Lett. A \textbf{232}(5), 340
  (1997).
\newblock \doi{10.1016/S0375-9601(97)00413-1}.
\newblock
  \urlprefix\url{http://www.sciencedirect.com/science/article/pii/S0375960197004131}

\bibitem{01a1BrKhVo}
{V.B.Braginsky, F.Ya.Khalili, S.P.Volikov}, \pla \textbf{287}, 31 (2001)

\bibitem{99a1BrKh}
{V.B.Braginsky, F.Ya.Khalili}, \pla \textbf{257}, 241 (1999)

\bibitem{02a1BrVy}
{V. B. Braginsky and S.P.Vyatchanin}, \pla \textbf{293}, 228 (2002)

\bibitem{03a1eBiSa}
{I.A.Bilenko, A.A.Samoilenko}, Vestnik Moscovskogo Universiteta, series 3
  \textbf{4}, 39 (2003)

\bibitem{Sheard_PRA_69_051801_2004}
B.S. Sheard, M.B. Gray, C.M. Mow-Lowry, D.E. McClelland, S.E. Whitcomb, Phys.
  Rev. A \textbf{69}, 051801 (2004).
\newblock \doi{10.1103/PhysRevA.69.051801}.
\newblock \urlprefix\url{http://link.aps.org/doi/10.1103/PhysRevA.69.051801}

\bibitem{Corbitt2006}
T.~Corbitt, D.~Ottaway, E.~Innerhofer, J.~Pelc, N.~Mavalvala, Phys. Rev. A
  \textbf{74}, 021802 (2006).
\newblock \doi{10.1103/PhysRevA.74.021802}.
\newblock \urlprefix\url{http://link.aps.org/doi/10.1103/PhysRevA.74.021802}

\bibitem{Corbitt2007}
T.~Corbitt, Y.~Chen, E.~Innerhofer, H.~M\"uller-Ebhardt, D.~Ottaway,
  H.~Rehbein, D.~Sigg, S.~Whitcomb, C.~Wipf, N.~Mavalvala, Phys. Rev. Lett.
  \textbf{98}, 150802 (2007).
\newblock \doi{10.1103/PhysRevLett.98.150802}.
\newblock \urlprefix\url{http://link.aps.org/doi/10.1103/PhysRevLett.98.150802}

\bibitem{Corbitt_PRL_99_160801_2007}
T.~Corbitt, C.~Wipf, T.~Bodiya, D.~Ottaway, D.~Sigg, N.~Smith, S.~Whitcomb,
  N.~Mavalvala, Phys. Rev. Lett. \textbf{99}, 160801 (2007).
\newblock \doi{10.1103/PhysRevLett.99.160801}.
\newblock \urlprefix\url{http://link.aps.org/doi/10.1103/PhysRevLett.99.160801}

\bibitem{Miyakawa_PRD_74_022001_2006}
O.~Miyakawa, R.~Ward, R.~Adhikari, M.~Evans, B.~Abbott, R.~Bork, D.~Busby,
  J.~Heefner, A.~Ivanov, M.~Smith, R.~Taylor, S.~Vass, A.~Weinstein,
  M.~Varvella, S.~Kawamura, F.~Kawazoe, S.~Sakata, C.~Mow-Lowry, Phys. Rev. D
  \textbf{74}, 022001 (2006).
\newblock \doi{10.1103/PhysRevD.74.022001}.
\newblock \urlprefix\url{http://link.aps.org/doi/10.1103/PhysRevD.74.022001}

\bibitem{Teufel_Nature_475_359_2011}
J.D. Teufel, T.~Donner, D.~Li, J.W. Harlow, M.S. Allman, K.~Cicak, A.J. Sirois,
  J.D. Whittaker, K.W. Lehnert, R.W. Simmonds, Nature \textbf{475}, 359 (2011).
\newblock \doi{10.1038/nature10261}.
\newblock \urlprefix\url{http://dx.doi.org/10.1038/nature10261}

\bibitem{Chan_Nature_478_89_2011}
J.~Chan, T.P.M. Alegre, A.H. Safavi-Naeini, J.T. Hill, A.~Krause,
  S.~Groblacher, M.~Aspelmeyer, O.~Painter, Nature \textbf{478}(7367), 89
  (2011).
\newblock \doi{10.1038/nature10461}

\bibitem{Aspelmeyer_RMP_86_1391_2014}
M.~Aspelmeyer, T.J. Kippenberg, F.~Marquardt, Rev. Mod. Phys. \textbf{86}, 1391
  (2014).
\newblock \doi{10.1103/RevModPhys.86.1391}.
\newblock \urlprefix\url{http://link.aps.org/doi/10.1103/RevModPhys.86.1391}

\bibitem{16a1DaKh}
F.Y. Khalili, S.L. Danilishin, Progress in Optics \textbf{61}, 113 (2016).
\newblock \doi{http://dx.doi.org/10.1016/bs.po.2015.09.001}.
\newblock
  \urlprefix\url{http://www.sciencedirect.com/science/article/pii/S0079663815000244}

\bibitem{Buonanno2001}
A.~Buonanno, Y.~Chen, Phys. Rev. D \textbf{64}, 042006 (2001).
\newblock \doi{10.1103/PhysRevD.64.042006}

\bibitem{01a2Kh}
F.~Khalili, Phys. Lett. A \textbf{288}, 251 (2001)

\bibitem{Buonanno2002}
A.~Buonanno, Y.~Chen, Phys. Rev. D \textbf{65}, 042001 (2002).
\newblock \doi{10.1103/PhysRevD.65.042001}

\bibitem{08a1KoSiKhDa}
I.S. Kondrashov, D.A. Simakov, F.Y. Khalili, S.L. Danilishin, Phys. Rev. D
  \textbf{78}, 062004 (2008).
\newblock \doi{10.1103/PhysRevD.78.062004}.
\newblock \urlprefix\url{http://link.aps.org/doi/10.1103/PhysRevD.78.062004}

\bibitem{11a1KhDaMuMiChZh}
F.~Khalili, S.~Danilishin, H.~M{\"u}ller-Ebhardt, H.~Miao, Y.~Chen, C.~Zhao,
  Phys. Rev. D \textbf{83}(6), 062003 (2011).
\newblock \doi{10.1103/PhysRevD.83.062003}

\bibitem{Somiya2014}
K.~Somiya, Y.~Kataoka, J.~Kato, N.~Saito, K.~Yano, Physics Letters A
  \textbf{380}(4), 521  (2016).
\newblock \doi{https://doi.org/10.1016/j.physleta.2015.11.010}.
\newblock
  \urlprefix\url{http://www.sciencedirect.com/science/article/pii/S0375960115009883}

\bibitem{17a1KoKhSc}
M.~Korobko, F.Y. Khalili, R.~Schnabel, Physics Letter A  (2017).
\newblock \urlprefix\url{https://doi.org/10.1016/j.physleta.2017.08.008}.
\newblock Available online 12 August 2017

\bibitem{2010_PhysRevLett.105.123601_CQNC_Tsang}
M.~Tsang, C.M. Caves, Phys. Rev. Lett. \textbf{105}, 123601 (2010).
\newblock \doi{10.1103/PhysRevLett.105.123601}.
\newblock
  \urlprefix\url{https://link.aps.org/doi/10.1103/PhysRevLett.105.123601}

\bibitem{2014_PhysRevA.89.053836_CQNC_Wimmer}
M.H. Wimmer, D.~Steinmeyer, K.~Hammerer, M.~Heurs, Phys. Rev. A \textbf{89},
  053836 (2014).
\newblock \doi{10.1103/PhysRevA.89.053836}.
\newblock \urlprefix\url{https://link.aps.org/doi/10.1103/PhysRevA.89.053836}

\bibitem{Polzik_AnnPhys_527_A15_2014}
E.S. Polzik, K.~Hammerer, Annalen der Physik \textbf{527}(1-2), A15 (2015).
\newblock \doi{10.1002/andp.201400099}.
\newblock \urlprefix\url{http://dx.doi.org/10.1002/andp.201400099}

\bibitem{Moeller_Nature_547_191_2017}
C.B. M{\o}ller, R.A. Thomas, G.~Vasilakis, E.~Zeuthen, Y.~Tsaturyan,
  M.~Balabas, K.~Jensen, A.~Schliesser, K.~Hammerer, E.S. Polzik, Nature
  \textbf{547}(7662), 191 (2017).
\newblock \doi{http://dx.doi.org/10.1038/nature22980}

\bibitem{Duan_PRL_85_5643_2000}
L.M. Duan, J.I. Cirac, P.~Zoller, E.S. Polzik, Phys. Rev. Lett. \textbf{85},
  5643 (2000).
\newblock \doi{10.1103/PhysRevLett.85.5643}.
\newblock \urlprefix\url{https://link.aps.org/doi/10.1103/PhysRevLett.85.5643}

\bibitem{Julsgaard_Nature_413_400_2001}
B.~Julsgaard, A.~Kozhekin, E.S. Polzik, Nature \textbf{413}(6854), 400 (2001).
\newblock \doi{http://dx.doi.org/10.1038/35096524}

\bibitem{Hammerer_RMP_82_1041_2010}
K.~Hammerer, A.S. S\o{}rensen, E.S. Polzik, Rev. Mod. Phys. \textbf{82}, 1041
  (2010).
\newblock \doi{10.1103/RevModPhys.82.1041}.
\newblock \urlprefix\url{http://link.aps.org/doi/10.1103/RevModPhys.82.1041}

\bibitem{Holstein_PR_58_1098_1940}
T.~Holstein, H.~Primakoff, Phys. Rev. \textbf{58}, 1098 (1940).
\newblock \doi{10.1103/PhysRev.58.1098}.
\newblock \urlprefix\url{https://link.aps.org/doi/10.1103/PhysRev.58.1098}

\bibitem{Hammerer_PRL_102_020501_2009}
K.~Hammerer, M.~Aspelmeyer, E.S. Polzik, P.~Zoller, Phys. Rev. Lett.
  \textbf{102}, 020501 (2009).
\newblock \doi{10.1103/PhysRevLett.102.020501}.
\newblock
  \urlprefix\url{http://link.aps.org/doi/10.1103/PhysRevLett.102.020501}

\bibitem{17a1KhPo}
F.~Khalili, E.~Polzik, arXiv:1710.10405  (2017)

\bibitem{PhysRevD.26.1817}
C.M. Caves, Phys. Rev. D \textbf{26}, 1817 (1982).
\newblock \doi{10.1103/PhysRevD.26.1817}.
\newblock \urlprefix\url{https://link.aps.org/doi/10.1103/PhysRevD.26.1817}

\bibitem{Korth2013a}
W.Z. Korth, H.~Miao, T.~Corbitt, G.D. Cole, Y.~Chen, R.X. Adhikari, Phys. Rev.
  A \textbf{88}, 033805 (2013).
\newblock \doi{10.1103/PhysRevA.88.033805}.
\newblock
  \urlprefix\url{http://arxiv.org/abs/1210.0309{\%}5Cnhttp://link.aps.org/doi/10.1103/PhysRevA.88.033805}

\bibitem{Loudon&Knight}
R.~Loudon, P.~Knight, Journal of Modern Optics \textbf{34}(6-7), 709 (1987).
\newblock \doi{10.1080/09500348714550721}.
\newblock \urlprefix\url{https://doi.org/10.1080/09500348714550721}

\bibitem{PhysRevA.73.023801_2006}
T.~Corbitt, Y.~Chen, F.~Khalili, D.~Ottaway, S.~Vyatchanin, S.~Whitcomb,
  N.~Mavalvala, Phys. Rev. A \textbf{73}(2), 023801 (2006).
\newblock \doi{10.1103/PhysRevA.73.023801}

\bibitem{Vahlbruch15dB}
H.~Vahlbruch, M.~Mehmet, K.~Danzmann, R.~Schnabel, Phys. Rev. Lett.
  \textbf{117}, 110801 (2016).
\newblock \doi{10.1103/PhysRevLett.117.110801}.
\newblock
  \urlprefix\url{https://link.aps.org/doi/10.1103/PhysRevLett.117.110801}

\bibitem{1995BookWaMi}
D.~Walls, G.~Milburn, \emph{Quantum optics} (Springer, 2008)

\end{thebibliography}

\end{document}